\documentclass[11pt]{article}
\usepackage[utf8]{inputenc}

\usepackage{amsmath}
\DeclareMathOperator*{\argmin}{arg\,min}
\newcommand{\ESG}{\text{ESG}}
\newcommand{\BSM}{\text{BSM}}

\newcommand{\mS}{\mathcal{S}}
\newcommand{\mB}{\mathcal{B}}
\newcommand{\mC}{\mathcal{C}}
\newcommand{\mM}{\mathcal{M}}

\newcommand{\tn}[1][k]{t_{n,#1}}
\newcommand{\dtn}[1][k]{\Delta t_{n,#1}}

\newcommand{\rf}[1][k]{r_{f,n,#1}}

\newcommand{\mukl}[1][e]{\mu_{n,k,\lambda}^{(\text{#1})}}
\newcommand{\sigkl}[1][e]{\sigma_{n,k,\lambda}^{(\text{#1})}}

\newcommand{\pnl}[1][k]{p_{n,#1,\lambda}}
\newcommand{\znl}[1][k]{\varsigma_{n,#1,\lambda}}
\newcommand{\pN}[1][k]{p^{(\aleph)}_{n,#1,\lambda}}
\newcommand{\dN}[1][k]{\delta^{(\aleph)}_{n,#1,\lambda}}
\newcommand{\mnl}[1][k]{\mu_{n,#1,\lambda}}
\newcommand{\snl}[1][k]{\sigma_{n,#1,\lambda}}

\newcommand{\Rkl}[1][M]{r_{n,k,\lambda}^{(\mathcal{#1})}}
\newcommand{\ZMkl}[1][k]{Z_{n,#1,\lambda}^{(\mM)}}
\newcommand{\pMkl}[1][k]{p_{n,#1,\lambda}^{(\mM)}}
\newcommand{\zMkl}[1][k]{\varsigma_{n,#1,\lambda}^{(\mM)}}
\newcommand{\ek}[1][M]{e_{n,k}^{(\mathcal{#1})}}
\newcommand{\rSkl}[1][k]{r_{n,#1,\lambda}^{(\mS,\mM)}}

\usepackage{twoopt}
\newcommandtwoopt{\pMkudl}[2][k][u]{p_{n,#1,\lambda}^{(\mM,\text{#2})}}

\usepackage{amsfonts}
\usepackage{amssymb}
\usepackage{makeidx}
\usepackage{graphicx}
\usepackage{caption}
\usepackage{subcaption}
\usepackage{listings}
\usepackage{indentfirst}
\usepackage{bbm}
\usepackage{color}

\usepackage{apacite}
\usepackage{natbib}

\usepackage{footmisc} 

\usepackage{siunitx}
\usepackage{multirow}
\usepackage{booktabs}

\usepackage{setspace}
\usepackage{authblk}
\usepackage[pdfborder={0 0 0}]{hyperref}
\usepackage[margin=1in]{geometry}

\usepackage[title]{appendix} 

\title{ESG-Valued Discrete Option Pricing in Complete Markets}

\author[1]{Yuan Hu}
\author[2,*]{W. Brent Lindquist}
\author[2]{Svetlozar T. Rachev}

\affil[1]{\small Department of Mathematics, University of California San Diego, La Jolla, CA 92093, U.S.A.\\
yuh099@ucsd.edu}
\affil[2]{\small Department of Mathematics \& Statistics, Texas Tech University, Lubbock, TX 79409-1042, U.S.A.\\
zari.rachev@ttu.edu}
\affil[*]{ Corresponding author, brent.lindquist@ttu.edu}

\begin{document}

\maketitle

\noindent {\textbf {Abstract}}
We consider option pricing using replicating binomial trees, with a two fold purpose.
The first is to introduce ESG valuation into option pricing.
We explore this in a number of scenarios, including enhancement of yield due to trader information
and the impact of the past history of a market driver.
The second is to emphasize the use of discrete dynamic pricing, rather than continuum models,
as the natural model that governs actual market practice.
We further emphasize that discrete option pricing models must use discrete compounding
(such as risk-free rate compounding of $1+r_f \Delta t$)
rather than continuous compounding (such as $e^{r_f \Delta t})$.


\section{Introduction}
\noindent
Socially responsible investing (SRI) is an umbrella term for investment in activities and companies that generate positive social and
environmental impact.
Emerging within the this broad framework (Daugaard, 2020) is a focus on ESG ratings; methodologies under which corporate
and investment structures are evaluated under three performance/risk categories: environmental, social, and governance.
ESG investing is experiencing fast and inexorable growth;
increasing numbers of institutional investors, such as mutual and pension funds, provide socially responsible financial
products (Boffo \& Patalano, 2020); and
increasing numbers of agencies are providing ratings (Amel-Zadeh \& Serafeim, 2018).
Research highlights two main reasons behind the push towards SRI products: ethical beliefs and improved financial performance
(Hartzmark and Sussman, 2019; B\'{e}nabou and Tirole, 2010; Kr{\"u}ger, 2015; Lins et al., 2017;
Liang and Renneboog, 2017; Starks et al., 2017).
How ESG  factors are integrated into investment decisions is varied, including: as (positive or negative) screening tools;
as a thematic investment; or to tilt a portfolio (Amel-Zadeh \& Serafeim, ibid).
There is also evidence (Amel-Zadeh \& Serafeim, ibid; Berry and Junkus, 2013) that investors focus on different aspects
of SRI compared to the developers of socially responsible financial products.
There remains a great deal of subjectivity as to how SRI factors, in particular ESG scores,
are utilized in the investment universe.

In this article we focus on an empirical pricing model that incorporates ESG scores to produce an ESG-valued return and,
consequently, a numeraire that serves as an ESG-valued ``price'' which reflects a value assignment based upon a
combination of financial return and ESG-score.
We explore this ESG-valuation model in the context of pricing European contingency claims, specifically call options.
Our model introduces an ESG-intensity parameter, $\lambda \in [0,1]$, that designates the relative
weight by which a trader values ``ESG return'' against traditional financial return.
In a manner analogous to implied volatility, by comparing model option prices against market data we can extract
ESG-intensity surfaces that can be interpreted as reflecting the implicit weighting given to ESG value in formulating
option prices as functions of moneyness and time to maturity.

The celebrated Black and Scholes (1973) and Merton (1973) option pricing model (BSM), which is based upon the
stochastic differential equation $dS_t = \mu S_t dt + \sigma S_t dB_t$, where $B_t$ is a standard Brownian motion,
naturally leads to the consideration of a continuous logarithmic return (log-return) process.
One consequence of the BSM model is that its option pricing formula only retains the variance of the log-return process,
the drift $\mu$ in the price process is lost.
In addition, the Gaussian nature of the BSM model assumes equal upturn and downturn return probabilities
(i.e. $p^{(\text{u})} = p^{(\text{d})} = 0.5$), disallowing for other possibilities.
The classic Cox-Ross-Rubinstein (1970) and Jarrow-Rudd  (1983) models, as well as more general discrete pricing models
(e.g. Kim et al., 2016, 2019; Hu et al. 2020a, 2020b) have been built to extend the continuum BSM model to capture more of the
stylized facts and microstructure of actual market price processes.

One lasting impact of the BSM model has been the motivation to explore the continuum limit of each discrete model,
in part to explore which stylized facts or microstructural parameters encapsulated within the discrete model survive
into the continuum limit (Kim et al. 2016, 2019; Hu et al. 2020a, 2020b).
Obtaining these continuum limits has required the development of binomial trees based upon non-classical invariance
principles (Cherny et al., 2003; Davydov and Rotar, 2008).
Our motivation to stay within the realm of discrete pricing is two fold.
One is based upon the reality that all trading practices take place in discrete time intervals.\footnote{
	Currently bounded below by $\Delta t \sim 10^{-6}$ seconds, as established by ultra-high frequency traders.
	However this rate is never realized over a full 24-hour period, but generally over a ``working day'' of $\sim 6.5$ hours. }
The second is that taking the continuum limit of these discrete models leads to loss of some of this microstructure,
making continuum limit models less useful.

Log-returns $\ln (S_{t+dt}/S_t)$ are a critical component of continuum pricing theory,
as they make continuum theories analytically tractable.
However, every practical use of returns in the market involves discrete (arithmetic) returns $(S_{t+\Delta t} - S_t)/S_t$
over discrete trading intervals.
In this article we explicitly compare binomial pricing trees that naturally invoke either arithmetic (discretely compounded) or 
logarithmic (continuously compounded) returns.
For daily (and intra-day) returns on stocks, the difference between the arithmetic return process and 
the log-return process is generally small.\footnote{
	However differences compounded over longer time frames may become significant.}
For option pricing, which is a more complex nonlinear process, our results demonstrate that significant differences can arise
between discretely and continuously compounded models (for example, in implied volatility surfaces).
In fact, using continuous compounding for the riskless rate when trading of the underlying stock occurs in
discrete time increments makes a market composed of a stock, a riskless bond, and an option inconsistent with
the fundamental theorem of asset pricing (Delbaen and Schachmayer, 1994; Jarrow et al., 2009).
In section~\ref{sec:ESG_opm} we explicitly note how retention of the no-arbitrage condition requires the use of
discrete compounding when using discrete pricing models.

This paper is organized as follows.
Section~\ref{sec:ESG_opm} introduces the ESG-valued, recombining, binomial tree models
designed for arithmetic and log-returns.
These ESG-valued models are applied to equity data, and implied surfaces for the ESG-intensity parameter are computed.
Section~\ref{sec:inf_opm} extends the ESG-valued models to markets in which there is an informed trader.
Implied surfaces are computed for the trader's information intensity.
These surfaces can be interpreted as indicators of how option traders subjectively weight ESG scores.
In section~\ref{sec:PD_opm} we extend the ESG-valued models of section~\ref{sec:ESG_opm} to the case in which
the direction of price change for the underlying stock is influenced by a market driver.
Using an extended formulation of the Cherny-Shiryaev-Yor (2003) principle, we introduce history dependence into this asset pricing.
Implied surfaces for the ESG-intensity parameter under this formulation are recomputed for the same set of equities
and compared with those from section~\ref{sec:ESG_opm}.
As each model is presented, its continuum limit is presented and the model parameters identified that are contained in the
binomial tree model but lost in the continuum limit.
Final discussion is presented in section~\ref{sec:disc}.

\section{ESG-valued binomial option pricing models}
\label{sec:ESG_opm}
\noindent
Consider a Black-Scholes-Merton market consisting of a risky asset $\mS$, a riskless asset (a bond)  $\mB$,
and a European contingent claim (specifically a call option) $\mC$ having $\mS$ as the underlying.
Over the time interval $[0,T]$, the dynamics of the prices of $\mS,\mB$ and $\mC$ are modeled on a discrete,
recombining binomial tree.
We allow for variable trading times over this interval by a time-partition parameterized by the integer
$n \in \{1,2, \dots \}$.
For a choice of $n$, the partition is $\tn[k];\ k = 1, \dots, k_n$, where
$0 = \tn[0] < \tn[1]  \cdots < \tn[k_n] = T$ denote the trading times.
Let $\dtn[k] = \tn[k] - \tn[k-1]$ denote the duration between successive trades.

We introduce ESG-valuation into option pricing by defining the ESG-valued return,
whether logarithmic or arithmetic,\footnote{
	Using \eqref{eq:ESG_rtn} for both logarithmic and arithmetic returns is justified under the assumption that,
	for  $\dtn[k+1]$ sufficiently small,
	$\ln \left( S_{\tn[k],\lambda} / S_{\tn[k-1],\lambda} \right) \approx S_{\tn[k],\lambda} / S_{\tn[k-1],\lambda} - 1$.
}
 by
\begin{equation}
	r_{\tn[k],\lambda} = \lambda e_{\tn[k]} + (1-\lambda) r_{\tn[k],0}\,.
	\label{eq:ESG_rtn}
\end{equation}
Here $r_{\tn[k],0} = r_{\tn[k],\lambda = 0}$ denotes the usual financial return at trading time $\tn[k]$
and $e_{\tn[k]}$ is a normalized ESG score.
We refer to the weight parameter $\lambda \in [0,1]$ as the \textit{ESG intensity}.
As ESG scores, $\ESG_{\tn[k]}$, are provided in a range $\ESG_{\tn[k]} \in [\ESG_{\text{min}},\ESG_{\text{max}}]$,
where typically $\ESG_{\text{min}} = 0$ and $\ESG_{\text{max}} = 100$,
$e_{\tn[k]}$ is a normalized $\ESG_{\tn[k]}$ value scaled to the range $[-1/c,1/c]$.
The constant $c \in R$ insures that the ranges of $e_{\tn[k]} $ and $r_{\tn[k],0}$ are comparable.\footnote{
	For daily return data, where $\dtn[k]=1$  trading day, we use the value $c = 252$,
	assuming 252 trading days per calendar year.}
We consider two models for the pricing of options pertinent to an ESG-value trader
who holds a short position in the $\mC$-contract and trades at the discrete times $\tn[k]$.

\noindent
\textbf{ESG-valued arithmetic return model.}
Based on arithmetic returns, the price model for stock $\mS$ is given by
\begin{equation}
	S_{n,k+1,\lambda} =
	\begin{cases}
	S^{(\textrm{u})}_{n,k+1,\lambda} = S_{n,k,\lambda} \left( 1 + U_{n,k+1,\lambda} \right), \text{ w.p. } \pnl[k+1]\,, \\
	S^{(\textrm{u})}_{n ,k+1,\lambda} = S_{n,k,\lambda} \left( 1 + D_{n,k+1,\lambda} \right), \text{ w.p. } 1-\pnl[k+1]\,.
	\end{cases}
	\label{eq:ESG_dpm}
\end{equation}
In \eqref{eq:ESG_dpm}, and what follows, we use a condensed time notation.
It is to be understood that $S^{(\text{u})}_{n,k,\lambda} \equiv S^{(\text{u})}_{\tn[k],\lambda}$.
Similarly for $S^{(\text{d})}_{n,k,\lambda}$, $S_{n,k,\lambda}$, $U_{n,k,\lambda}$, $D_{n,k,\lambda}$, $\pnl[k]$,
$r_{n,k,\lambda}$, and all other time dependent variables.
The probability $\pnl[k+1]$ governs the direction of the price change over the interval $[\tn[k],\tn[k+1])$.
The probabilities $\pnl[k]$ are determined by a triangular array of independent Bernoulli random variables
$\znl[1],\znl[2],\ldots,\znl[{k_n}]$, $n =\{1,2,\ldots\}$
satisfying $\mathbb{P}(\znl[k] =1) = 1 - \mathbb{P}(\znl[k] = -1) \equiv \pnl[k] \in (0,1)$, $k =1,\ldots,k_n$.
The pricing tree \eqref{eq:ESG_dpm} is adapted to the discrete filtration
\begin{equation}
	\mathbb{F}^{(n,\lambda)} = \left\{\mathcal{F}^{(n,k,\lambda)}
	= \sigma(\znl[1],\znl[2],\ldots,\znl[k]),\ 
	k = 1, \ldots, k_n,\  \mathcal{F}^{(n,0,\lambda)} = \{\varnothing,\Omega\}\right\}.
	\label{eq:ESG_F}
\end{equation}
The probability $\pnl[k+1]$ is $\mathcal{F}^{(n,k,\lambda)}$-measurable.

The price of the riskless asset $\mB$, generating the risk-free rate of return $\rf[k]$, evolves as
\begin{equation}
	\beta_{n,k+1} = \beta_{n,k} (1+\rf[k] \dtn[k+1])\,.
	\label{eq:ESG_dbond}
\end{equation}
Let $f_{n,k,\lambda}$ denote the price of $\mC$.
The single time step, risk-neutral relationship for the price of $\mC$ is
\begin{equation}
   \begin{aligned}
	f_{n,k,\lambda} &= \frac{1}{1+\rf[k] \dtn[k+1]}
	\left[ q_{n,k+1,\lambda} f_{n,k+1,\lambda}^{\text{(u)}} + (1-q_{n,k+1,\lambda}) f_{n,k+1,\lambda}^{\text{(d)}} \right] , \\
	q_{n,k+1,\lambda} &= \frac{ \rf[k] \dtn[k+1] -  D_{n,k+1,\lambda} }{  U_{n,k+1,\lambda} - D_{n,k+1,\lambda} }\,.
   \end{aligned}
   \label{eq:ESG_dfq}
\end{equation}
The risk neutral probability $q_{n,k+1,\lambda}$ is $\mathcal{F}^{(n,k,\lambda)}$-measurable.
The conditional mean and variance of the arithmetic return process
$r_{n,k+1,\lambda} \equiv (S_{n,k+1,\lambda} - S_{n,k,\lambda})/S_{n,k,\lambda}$ are
\begin{equation}
   \begin{aligned}
	\mathbb{E}\left[ r_{n,k+1,\lambda} | S_{n,k,\lambda} \right]
		&=  \pnl[k+1] U_{n,k+1,\lambda} + (1-\pnl[k+1])  D_{n,k+1,\lambda}\,, \\
	\text{Var}\left[ r_{n,k+1,\lambda} | S_{n,k,\lambda} \right]
		&= \pnl[k+1]  (1-\pnl[k+1]) \left[ U_{n,k+1,\lambda} - D_{n,k+1,\lambda} \right]^2 .
   \end{aligned}
   \label{eq:ESG_EV}
\end{equation}
Defining the instantaneous Sharpe ratio over the time period $\dtn[k+1]$ as
\begin{equation}
	 \Theta_{n,k,\lambda} = \frac{ \mathbb{E} \left( r_{n,k+1,\lambda} |S_{n,k,\lambda} \right) - \rf[k] \dtn[k+1] }
			{ \sqrt{ \text{Var} \left( r_{n,k+1,\lambda} | S_{n,k,\lambda} \right) \dtn[k+1] } }\,,
	\label{eq:ESG_SR}
\end{equation}
from \eqref{eq:ESG_EV} we have
\begin{equation}
	\Theta_{n,k,\lambda} =
		 \frac{\pnl[k+1] - \frac{ \rf[k] \dtn[k+1] -  D_{n,k+1,\lambda} }{  U_{n,k+1,\lambda} - D_{n,k+1,\lambda} } }
				 { \sqrt{ \pnl[k+1] (1-\pnl[k+1]) \dtn[k+1] } }\,,
   \label{eq:ESG_SRUD}
\end{equation}
which leads to the identity
\begin{equation}
	 q_{n,k+1,\lambda} = \pnl[k+1] - \Theta_{n,k,\lambda} \sqrt{ \pnl[k+1] (1-\pnl[k+1]) \dtn[k+1] }\,.
	 \label{eq:ESG_qt}
\end{equation}

Matching the instantaneous mean $\mukl[r]$ and variance $\left( \sigkl[r] \right)^2$ of the
ESG-valued arithmetic return process
\begin{equation}
	\mathbb{E}\left[ r_{n,k+1,\lambda} | S_{n,k,\lambda} \right] = \mukl[r] \dtn[k+1]\,, \qquad
	\text{Var}\left[ r_{n,k+1,\lambda} | S_{n,k,\lambda} \right] = \left( \sigkl[r] \right)^2 \dtn[k+1]\,,
   \label{eq:ESG_dEV}
\end{equation}
leads to the specific forms
\begin{equation}
   \begin{aligned}
	U_{n,k+1,\lambda} &= \mukl[r]\  \dtn[k+1] +\sigkl[r]\  p^{(\text{u})}_{n,k+1,\lambda} \sqrt{\dtn[k+1]}\,, \\
	D_{n,k+1,\lambda} &= \mukl[r]\  \dtn[k+1] - \sigkl[r]\  p^{(\text{d})}_{n,k+1,\lambda} \sqrt{\dtn[k+1]}\,, \\
	p^{(\text{u})}_{n,k+1,\lambda}  &= \sqrt{\cfrac{1-\pnl[k+1]}{\pnl[k+1]}}\,, \qquad \qquad
	p^{(\text{d})}_{n,k+1,\lambda}  = \sqrt{\cfrac{\pnl[k+1]}{1-\pnl[k+1]}}\,.
   \end{aligned}
   \label{eq:ESG_dUD}	
\end{equation}
The conditional expected mean and variance of the stock price are
\begin{equation}
   \begin{aligned}
	\mathbb{E}\left[ S_{n,k+1,\lambda} | S_{n,k,\lambda} \right]
		&=  S_{n,k,\lambda} \left[ \pnl[k+1] \left( 1+ U_{n,k+1,\lambda} \right)
						 + (1-\pnl[k+1])  \left( 1 + D_{n,k+1,\lambda} \right) \right] \\
		&= S_{n,k,\lambda} \left[ 1 + \mukl[r]\  \dtn[k+1]  \right] , \\
	\text{Var}\left[ S_{n,k+1,\lambda} | S_{n,k,\lambda} \right]
		&= S_{n,k,\lambda}^2 \ \pnl[k+1]  (1-\pnl[k+1]) \left[ U_{n,k+1,\lambda} - D_{n,k+1,\lambda} \right]^2 \\
		&= S_{n,k,\lambda}^2 \ \left( \sigkl[r] \right)^2 \dtn[k+1]\,,
   \end{aligned}
   \label{eq:ESG_dSEVUV}
\end{equation}
from which we identify the instantaneous drift coefficient $\mu_{n,k,\lambda}$ and variance $\sigma_{n,k,\lambda}^2$ of $\mS$,
\begin{equation}
	\mu_{n,k,\lambda} = \mukl[r], \qquad \qquad \sigma_{n,k,\lambda}^2 = \left( \sigkl[r] \right)^2\,.
	\label{eq:ESG_dmv}
\end{equation}
Equations \eqref{eq:ESG_dEV} through \eqref{eq:ESG_dSEVUV} can be trivially re-expressed in terms of
$\mu_{n,k,\lambda}$ and $\sigma_{n,k,\lambda}$.
Using \eqref{eq:ESG_dUD}, \eqref{eq:ESG_SRUD} becomes
\begin{equation}
	\Theta_{n,k,\lambda}
	= \frac{\mu_{n,k,\lambda} - \rf[k]}{\sigma_{n,k,\lambda}} \equiv \theta_{n,k,\lambda}\,,
	\label{eq:ESG_dSR}
\end{equation}
where $\theta_{n,k,\lambda}$ denotes the market price of risk.

As in Hu et al. (2020b), a non-standard invariance principle (Davydov and Rotar, 2008) can be used to show that the
pricing tree \eqref{eq:ESG_dpm}, \eqref{eq:ESG_dUD} generates a $D[0,T]$ stochastic process which converges
weakly in Skorokhod $D[0,T]$ topology to the cumulative return process\footnote{
	As defined in Duffie, 2001, section 6D, p. 106.}
$R_{t,\lambda}$, $t \in [0,T]$ determined
by $dR_{t,\lambda} = dS_{t,\lambda}/S_{t,\lambda} = \mu_{t,\lambda} dt + \sigma_{t,\lambda} dB_t$,
where $B_t$ is a standard Brownian motion.
In the risk-neutral world, the limiting process is the same, with $\mu_{t,\lambda}$ replaced by $r_{f,t}$.
We note that the discrete model is much more informative than the continuous time model as it preserves both
$\mu_{\tn[k],\lambda}$ and $p_{\tn[k],\lambda}$.
Secondly, the discrete formulation leads to binomial models which include additional market
microstructure features (Jacod and Ait-Sahaliz, 2014, section 2.2.2; Jarrow et al., 2009).

\noindent
\textbf{ESG-valued log-return model.}
In the second model, which we refer to as the ESG-valued log-return model, the price of $\mS$ is given by
\begin{equation}
	S_{n,k+1,\lambda} =
	\begin{cases}
		S^{(\text{u})}_{n,k+1,\lambda} = S_{n,k,\lambda} e^{U_{n,k+1,\lambda}}, \text{ w.p. } \pnl[k+1]\,, \\
		S^{(\text{d})}_{n,k+1,\lambda} = S_{n,k,\lambda} e^{D_{n,k+1,\lambda}}, \text{ w.p. } 1-\pnl[k+1]\,.
	\end{cases}
	\label{eq:ESG_lpm}
\end{equation}
The terms $U_{n,k+1,\lambda}$ and $D_{n,k+1,\lambda}$ appearing in \eqref{eq:ESG_lpm} will be different from those
in \eqref{eq:ESG_dpm} and hence the price trajectories $S_{n,k},\lambda$ will differ between the two models.
We do not believe the reader will be confused between the similarity in notation when each model is discussed.
The probabilities in \eqref{eq:ESG_lpm} are also given by a triangular array of independent Bernoulli random variables,
consequently the pricing tree \eqref{eq:ESG_lpm} is adapted to the discrete filtration \eqref{eq:ESG_F}.

The price of the riskless asset $\mB$ is now given by
\begin{equation}
	\beta_{n,k+1} = \beta_{n,k} e^{\rf[k] \dtn[k+1]}\,,
	\label{eq:ESG_lbond}
\end{equation}
and the risk-neutral relation for the price of $\mC$ is
\begin{equation}
   \begin{aligned}
	f_{n,k,\lambda} &=  e^{-\rf[k] \dtn[k+1]}
		\left[ q_{n,k+1,\lambda} f_{n,k+1,\lambda}^{\text{(u)}}
			 + (1-q_{n,k+1,\lambda}) f_{n,k+1,\lambda}^{\text{(d)}} \right] , \\
	q_{n,k+1,\lambda} &= \frac{ e^{\rf[k] \dtn[k+1]} -  e^{D_{n,k+1,\lambda}} }
							{  e^{U_{n,k+1,\lambda}} - e^{D_{n,k+1,\lambda}} }\,.
   \end{aligned}
   \label{eq:ESG_lfq}
\end{equation}
The conditional mean and variance of the log-return $r_{n,k+1,\lambda} = \ln (S_{n,k+1,\lambda} / S_{n,k,\lambda} )$
 have exactly the same form as
\eqref{eq:ESG_EV} and the instantaneous Sharpe ratio has the form \eqref{eq:ESG_SRUD}.
However, in this model the identity \eqref{eq:ESG_qt} does not hold.

Matching the instantaneous mean $\mukl[r]$ and variance $\left( \sigkl[r] \right)^2$ of the ESG-valued log-return process
\begin{equation}
	\mathbb{E}\left[ r_{n,k+1,\lambda} | S_{n,k,\lambda} \right] = \mukl[r] \dtn[k+1]\,, \qquad
	\text{Var}\left[ r_{n,k+1,\lambda} | S_{n,k,\lambda} \right] = \left( \sigkl[r] \right)^2 \dtn[k+1]\,,
   \label{eq:ESG_lEV}
\end{equation}
leads to the specific forms,
\begin{equation}
   \begin{aligned}
	U_{n,k+1,\lambda} &= \mukl[r] \dtn[k+1] + p^{(\text{u})}_{n,k+1,\lambda} \sigkl[r] \sqrt{\dtn[k+1]}\,, \\
	D_{n,k+1,\lambda} &= \mukl[r] \dtn[k+1] - p^{(\text{d})}_{n,k+1,\lambda} \sigkl[r] \sqrt{\dtn[k+1]}\,,
   \end{aligned}
   \label{eq:ESG_lUD}	
\end{equation}
where $p^{(\text{u})}_{n,k,\lambda}$ and $p^{(\text{d})}_{n,k,\lambda}$ are as in \eqref{eq:ESG_dUD}.
Using \eqref{eq:ESG_lUD}, the instantaneous Sharpe ratio is
\begin{equation}
	\Theta_{n,k,\lambda} = \frac{\mukl[r] - \rf[k]}{\sigkl[r]}\,.
	\label{eq:ESG_lSR}
\end{equation}
The conditional expected mean and variance of the stock price are
\begin{equation}
   \begin{aligned}
	\mathbb{E}\left[ S_{n,k+1,\lambda} | S_{n,k,\lambda} \right] &=  S_{n,k,\lambda}
		 \left[ \pnl[k+1] e^{U_{n,k+1,\lambda}} + (1-\pnl[k+1])  e^{D_{n,k+1,\lambda}} \right] , \\
	\text{Var}\left[ S_{n,k+1,\lambda} | S_{n,k,\lambda} \right] &= S_{n,k,\lambda}^2 \ \pnl[k+1]  (1-\pnl[k+1])
					 \left[ e^{U_{n,k+1,\lambda}} - e^{D_{n,k+1,\lambda}} \right]^2 .
   \end{aligned}
   \label{eq:ESG_lSEVUV}
\end{equation}

Equations \eqref{eq:ESG_lfq} through \eqref{eq:ESG_lSEVUV} are exact for this log-return model.
From \eqref{eq:ESG_lUD}, expanding \eqref{eq:ESG_lSEVUV} to terms of $O(\dtn[k+1])$ produces
\begin{equation}
   \begin{aligned}
	\mathbb{E}\left[ S_{n,k+1,\lambda} | S_{n,k,\lambda} \right]
		&=  S_{n,k,\lambda} \left[ 1 + \left( \mukl[r]  + \frac{{\sigkl[r]}^2}{2} \right) \dtn[k+1] \right] , \\
	\text{Var}\left[ S_{n,k+1,\lambda} | S_{n,k,\lambda} \right]
		&=   S_{n,k,\lambda}^2 {\sigkl[r]}^2 \dtn[k+1]\,.
   \end{aligned}
   \label{eq:ESG_lSEV}
\end{equation}
Thus, to terms of $O(\dtn[k+1])$, we identify the instantaneous drift coefficient $\mu_{n,k,\lambda}$
and standard deviation $\sigma_{n,k,\lambda}$ of $\mS$ as
\begin{equation}
	\mu_{n,k,\lambda} = \mukl[r] + \frac{ \left( \sigkl[r] \right)^2}{2}\ , \qquad \sigma_{n,k,\lambda} = \sigkl[r]\,.
	\label{eq:ESG_lmv}
\end{equation}
In terms of $\mu_{n,k,\lambda}$ and $\sigma_{n,k,\lambda}$, \eqref{eq:ESG_lSR} becomes
\begin{equation}
	\Theta_{n,k,\lambda} = \frac{\mu_{n,k,\lambda} - \sigma_{n,k,\lambda}^2 / 2 - \rf[k]}{\sigma_{n,k,\lambda}}\,;
	\label{eq:dt_SRd}
\end{equation}
the conditional mean and variance of the log-return are
\begin{equation}
	\mathbb{E}\left[ r_{n,k+1,\lambda} | S_{n,k,\lambda} \right]
		= \left( \mu_{n,k,\lambda} - \frac{\sigma_{n,k,\lambda}^2}{2} \right) \dtn[k+1]\,, \qquad
	\text{Var}\left[ r_{n,k+1,\lambda} | S_{n,k,\lambda} \right]
		= \sigma_{n,k,\lambda}^2 \dtn[k+1]\,;
   \label{eq:dt_EV}
\end{equation}
the appropriate forms for $U_{n,k+1,\lambda}$ and $D_{n,k+1,\lambda}$ in \eqref{eq:ESG_lUD} to generate \eqref{eq:dt_EV} are
\begin{equation}
   \begin{aligned}
	U_{n,k+1,\lambda} &=
		\left[ \mu_{n,k,\lambda} - \frac{\sigma_{n,k,\lambda}^2}{2} \left( p^{(\text{u})}_{n,k+1,\lambda} \right)^2
				\right] \dtn[k+1]
			 + p^{(\text{u})}_{n,k+1,\lambda} \sigma_{n,k,\lambda} \sqrt{\dtn[k+1]}\,, \\
	D_{n,k+1,\lambda} &=
		\left[ \mu_{n,k,\lambda} - \frac{\sigma_{n,k,\lambda}^2}{2} \left( p^{(\text{d})}_{n,k+1,\lambda} \right)^2
				\right] \dtn[k+1]
			 - p^{(\text{d})}_{n,k+1,\lambda} \sigma_{n,k} \sqrt{\dtn[k+1]}\,;
   \end{aligned}
   \label{eq:dt_lUD}	
\end{equation}
and the risk-neutral probability can be written
\begin{equation}
	q_{n,k+1,\lambda} = \pnl[k+1] - \theta_{n,k,\lambda} \sqrt{\pnl[k+1](1-\pnl[k+1])\dtn[k+1]}\,,
	\label{eq:dt_lq}
\end{equation}
where $\theta_{n,k,\lambda}$ is the market price of risk given in \eqref{eq:ESG_dSR}.
Equations \eqref{eq:ESG_lmv} through \eqref{eq:dt_lq} are understood to hold through terms of the appropriate
order of $\dtn[k+1]$, as indicated in each equation.

In the limit $\dtn[k] \downarrow 0$, the binomial tree \eqref{eq:ESG_lpm}, \eqref{eq:dt_lUD}
approximates a cadlag process which converges weakly in the Skorokhod space
$D[0,T]$ to the geometric Brownian motion with time dependent parameters (Hu et al., 2020b)
\begin{equation}
	S_{t,\lambda} = S_0  \exp \left[	\left( \mu_{t,\lambda} - \frac{ {\sigma_{t,\lambda} }^2 }{ 2 } \right) t
							 + \sigma_{t,\lambda} B_t \right] ,
\end{equation}
where $B_t$ is a standard Brownian motion.

For either the ESG-valued arithmetic return or log-return models,
ESG-valuation at intensity $\lambda \in (0,1]$ produces in an ESG-enhanced dividend yield $\mathbb{D}_{n,k,\lambda}$
(relative to $\lambda = 0$) given by
\begin{equation}
	\mathbb{D}_{n,k,\lambda} = \sigma_{n,k,0} \left( \Theta_{n,k,\lambda} - \Theta_{n,k,0} \right) .
   \label{eq:ESG_DY}
\end{equation}
The dividend yield in each case can be computed using the appropriate model values for $\Theta_{n,k,\lambda}$.
When $\lambda = 0$, each ESG-valued model reverts to a binomial tree model based upon standard financial returns.

\noindent
\textbf{ESG-valued option pricing.}
With a final price $S_{n,k_n,\lambda} \equiv S_{T,\lambda}$ at maturity time $T$ on each node of the price tree
obtained by (forward) iteration using either \eqref{eq:ESG_dpm} or \eqref{eq:ESG_lpm},
the initial price $C (S_{T,\lambda},K)$ of the call option can be evaluated from the final payoff
$f_T = g(S_{T,\lambda},K)$ by iterating \eqref{eq:ESG_dfq} or \eqref{eq:ESG_lfq} backwards on the tree
in the usual manner.

\noindent
\textbf{Enforcement of no-arbitrage.}
As noted in the Introduction, use of continuous compounding when trading can only occur at discrete times
violates the fundamental theorem of asset pricing and the condition of no-arbitrage.
This can be seen explicitly by contrasting equations \eqref{eq:ESG_dfq} and \eqref{eq:ESG_lfq} in the two models
considered above.
For a discrete pricing process, ensuring positive asset prices and no arbitrage over any discrete time period requires
the condition (Shreve, 2004, Chapter 1)
\begin{equation}
	0 < 1+D_{n,k+1,\lambda} < 1+\rf[k] \dtn[k+1] < 1+U_{n,k+1,\lambda}\,.
   \label{eq:no_arb}
\end{equation}
This condition guarantees that the value of the risk-neutral probability $q_{n,k+1,\lambda}$ in \eqref{eq:ESG_dfq}
remains bounded in the interval $[0,1]$.
However \eqref{eq:no_arb} does not guarantee that the value of the risk-neutral probability $q_{n,k+1,\lambda}$
in \eqref{eq:ESG_lfq} remains so bounded.
While \eqref{eq:ESG_dfq} is exact, the approximation \eqref{eq:dt_lq} to \eqref{eq:ESG_lfq} only holds if terms
of $\mathcal{O} (\dtn[k+1])$ are ignored.
As will be shown in the results below, the presence of these $\mathcal{O} (\dtn[k+1])$ terms can result in significant
differences between implied volatility surfaces computed for the two models.
Thus, there is a significant risk in using continuum rates
($e^{U_{n,k+1,\lambda}}$, $e^{D_{n,k+1,\lambda}}$, $e^{\rf[k] \dtn[k+1]}$)
in discrete pricing models.
(Such usage of continuum rates in discrete models is commonly presented;
see, e.g., Chapter 12.8 of the text book by Hull (2012).)

\subsection{Application to equity data }
\label{sec:lam_ex}

\noindent
We apply both ESG-valued pricing models to equity data using ESG scores obtained from RobecoSAM
(now S\&P Global).\footnote{
	S\&P Global RobecoSAM ESG ratings data provided through Bloomberg Professional Services.}
We consider daily closing price data for the period 12/31/2020 through 12/31/2021 for three equities, MSFT, AMZN and AAPL.
These three equities were chosen as they have, respectively, high, intermediate and low ESG scores over this time period.
As is currently the case with most data providers, updated RobecoSAM scores are released once per year;
for our time period, new RobecoSAM scores $\ESG^{(\text{R})}$ were released on 11/19/2020 and 11/19/2021.
The raw scores $\ESG^{(\text{R})}$ are on a scale of 0 to 100.
The daily $e_{n,k}$ values required in \eqref{eq:ESG_rtn}, were therefore computed as follows
\begin{equation}
	e_{n,k} = \frac{\ESG^{(\text{R})}  - 50}{50 \cdot 252}\,.
	\label{eq:ESG}
\end{equation}
where $\ESG^{(\text{R})}$ is understood to refer to the most recent value released prior to day $\tn[k]$.
The raw $\ESG^{(\text{R})}$ and scaled $e_{n,k}$ ESG values for 11/19/2020 and 11/19/2021 are given
in Table~\ref{tab:ESG}.
\begin{table}[htb!]
	\caption{ESG values and stock closing prices $S_0$}
	\label{tab:ESG}
   \begin{subtable}{\textwidth} 
	\centering
	\begin{tabular}{l crcr r}
	\toprule
	Equity   & \multicolumn{2}{c}{11/19/2020} & \multicolumn{2}{c}{11/19/2021} & $S_0\ \ \ $\\
	\cline{2-5}
	\rule{0pt}{3ex}
	\strut   &  ESG$^{(\text{R})}$ & $e_{n,k}\ \ \ \ $ &  ESG$^{(\text{R})}$ & $e_{n,k}\ \ \ \ $  & (USD) \\
	\midrule
	MSFT  & 96 & $\ \ 3.65 \cdot 10^{-3}$  & 98 & $\ \ 3.81 \cdot 10^{-3}$ & 336.32\\
	AMZN  & 60 & $\ \ 0.79 \cdot 10^{-3}$  & 71 & $\ \ 1.67 \cdot 10^{-3}$ & 3,334.34\\
	AAPL   & 25 &    $ -1.98 \cdot 10^{-3}$  & 34 &   $ -1.27 \cdot 10^{-3}$ & 177.57\\
	\bottomrule
	\end{tabular}
   \end{subtable}
\end{table}

\begin{figure}[t]
\begin{center}
		\includegraphics[width=1.0\textwidth]{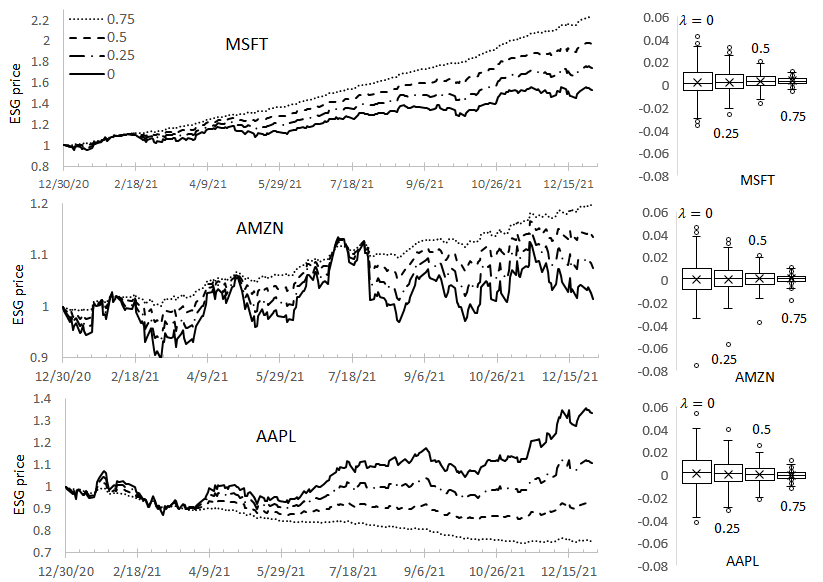}
	\caption{(left) ESG-valued price dynamics for chosen values of $\lambda$.
		For comparison purposes, the initial ESG prices (on 12/30/2020) are scaled to unit value.
		(right) Box-whisker summaries of the values of the arithmetic returns $r_{n,k,\lambda}$.
	}
	\label{fig:ESG_prices}
\end{center}
\end{figure}

We first explore the effects of the addition of ESG-valuation to the financial return in \eqref{eq:ESG_rtn}.
Using the daily closing prices for the three stocks, we compute arithmetic returns $r_{n,k,0}$.
For values of $\lambda \in \{0.25,0.5,0.75\}$ we compute ESG-valued returns $r_{n,k,\lambda}$
(assumed to be arithmetic returns)
from \eqref{eq:ESG_rtn} and subsequently ESG-valued ``prices''.\footnote{
	The ESG-valued returns \eqref{eq:ESG_rtn} effectively generate an ESG-numeraire that we will refer to as an
	ESG-valued price, or more concisely an ESG price.
}
Fig.~\ref{fig:ESG_prices} plots the resultant ESG price time series for the three stocks over the period 
12/31/2020 through 12/31/2021 computed using the values $\lambda \in \{0, 0.25, 0.5, 0.75 \}$.
For comparison purposes, the initial ESG price (on 12/30/2020) for each equity is scaled to unit value.
Thus the $\lambda = 0$ time series correspond to published closing prices rescaled to having a unit value on 12/30/2020.
Price values for the two highest ESG scored stocks, MSFT and AMZN, increase with $\lambda$;
for the lowest ESG rated stock, AAPL, price values decrease as $\lambda$ increases.
This behavior can be explained by examining the box-whisker summaries of the ESG-valued returns presented in
Fig.~\ref{fig:ESG_prices}.
As $\lambda$ increases, the range of return values decreases
due to the higher weighting on $e_{n,k}$ which only changes value once for each stock during this time period.
As $\lambda$ increases, the average value of $r_{n,k,\lambda}$: increases for MSFT; increases very slightly for AMZN;
and decreases for AAPL, becoming negative for $\lambda = 0.5$ and 0.75.

When $r_{n,k,0}$ is computed using log-returns, for a fixed value of $\lambda > 0$, equation \eqref{eq:ESG_rtn}
generates slightly smaller ESG log-return values than when $r_{n,k,0}$ is computed using arithmetic returns.
Thus the price series computed from \eqref{eq:ESG_rtn} using log-returns for $r_{n,k,0}$
(and interpreting the returns in \eqref{eq:ESG_rtn} as log-returns) will be very slightly below\footnote{
	The decrease is greatest for MSFT when  $\lambda = 0.5$, producing a price lower by approximately
	$0.0094$ units of account ($0.47\%$) after one year.
}
those shown in Fig.~\ref{fig:ESG_prices}.

We next consider option pricing based upon the models \eqref{eq:ESG_dpm} and \eqref{eq:ESG_lpm}.
To compute option prices, the parameters $\mu_{n,k,\lambda}$, $\sigma_{n,k,\lambda}$ and $p_{n,k,\lambda}$
must be estimated from the appropriate return data using either \eqref{eq:ESG_dmv} or \eqref{eq:ESG_lmv}.
We consider option prices computed for $\tn[k] =$ 12/31/2021 based upon each of the three stocks as the underlying.
For a given value of $\lambda$ and each stock,
the estimates $\mu_{n,k,\lambda} \equiv \hat{\mu}_\lambda$ and $\sigma_{n,k,\lambda} \equiv \hat{\sigma}_\lambda$
were obtained using the equity's ESG-valued return data over the period 1/1/2021 through 12/31/2021.
The probability estimate $p_{n,k+1,\lambda} \equiv \hat{p}_\lambda$ was computed as the proportion of the number
of days with non-negative values of the equity's ESG-valued return over this period.
The parameter estimates are given in Table~\ref{tab:params} for values of $\lambda \in \{0,0.25,0.5,0.75\}$.
\begin{table}[htb!]
	\caption{Parameter estimates for 12/31/2021 for the ESG-valued models}
	\label{tab:params}
	\centering
	\begin{tabular}{l | cccc | cccc | cccc}
	\toprule
	Equity & $\hat{\mu}_0$ & $\hat{\mu}_{0.25}$ & $\hat{\mu}_{0.5}$ & $\hat{\mu}_{0.75}$
		  & $\hat{\sigma}_0$  & $\hat{\sigma}_{0.25}$ & $\hat{\sigma}_{0.5}$ & $\hat{\sigma}_{0.75}$
		  & $\hat{p}_0$ & $\hat{p}_{0.25}$ & $\hat{p}_{0.5}$ & $\hat{p}_{0.75}$\\
	\omit & \multicolumn{4}{| c}{$\left(\times 10^{-3}\right)$}
		 & \multicolumn{4}{| c |}{$\left(\times 10^{-2}\right)$} &\multicolumn{4}{c}{$\ $} \\
	\midrule
	\omit & \multicolumn{12}{c}{arithmetic return} \\
	MSFT   & 1.76 & 2.24 &  2.72 &  3.19 & 1.33 & 0.99 & 0.66 & 0.33 & 0.52 & 0.57 & 0.66 & 0.85 \\
	AMZN   & 0.21 & 0.38 &  0.55 &  0.72 & 1.52 & 1.14 & 0.76 & 0.38 & 0.51 & 0.52 & 0.54 & 0.60 \\
	AAPL    & 1.31 & 0.50 & -0.30 & -1.11 & 1.58 & 1.19 & 0.79 & 0.40 & 0.52 & 0.51 & 0.46 & 0.36 \\

	\omit & \multicolumn{12}{c}{log-return} \\
	MSFT   & 1.76 & 2.22 &  2.69 &  3.18 & 1.32 & 0.99 & 0.66 & 0.33 & 0.52 & 0.57 & 0.66 & 0.85 \\
	AMZN   & 0.21 & 0.36 &  0.52 &  0.70 & 1.52 & 1.14 & 0.76 & 0.38 & 0.51 & 0.52 & 0.54 & 0.60 \\
	AAPL    & 1.30 & 0.48 & -0.33 & -1.12 & 1.58 & 1.19 & 0.79 & 0.40 & 0.52 & 0.51 & 0.46 & 0.36 \\
	\bottomrule
	\end{tabular}
\end{table}
Over this time period, MSFT had the largest values for $\hat{\mu}_\lambda$ and $\hat{p}_\lambda$,
and the smallest values of $\hat{\sigma}_\lambda$.
As the scaled ESG-values for MSFT and AMZN are positive, $\hat{\mu}_\lambda$ increases with $\lambda$. 
With positive scaled ESG-values, the $\lambda e_{n,k}$ term in \eqref{eq:ESG_rtn} increases the proportion
of positive return values $r_{n,k,\lambda}$ compared to that for $r_{n,k,0}$.
Consequently $\hat{p}_\lambda$ increases with $\lambda$, growing more rapidly for the higher ESG-valued stock.
As the scaled ESG-values for AAPL are negative, both $\hat{\mu}_\lambda$ and $\hat{p}_\lambda$ decrease
as $\lambda$ increases.
For all three equities, as ESG values change only once, on 11/19/2021, $\hat{\sigma}_\lambda$
decreases as $\lambda$ increases.

\medskip\noindent
\textbf{Implied ESG intensity surfaces.}
Using the stock prices  (shown in Table~\ref{tab:ESG}) and the ESG and parameter value estimates for 12/31/2021,
for a given maturity time $T$ and strike price $K$
the implied ESG intensity $\lambda^{(i)} (T,K)$ for stock $i$ was estimated using the relative mean-square minimization
\begin{equation}
	\lambda^{(i)} (T,K) = \argmin_\lambda
		\left(
			\frac{ C^{(i)} (S_T,K;r_f,\hat{\mu}_\lambda, \hat{\sigma}_\lambda, \hat{p}_\lambda) - C^{(i)}(S_T,K) }
				{ C^{(i)}(S_T,K) }
		\right)^2 .
	\label{eq:l_imp_MS}
\end{equation}
Here $C^{(i)} (S_T,K;r,\hat{\mu}_\lambda, \hat{\sigma}_\lambda, \hat{p}_\lambda)$
represents call option prices based upon the appropriate binomial tree
and $C^{(i)}(S_T,K)$ are market call option prices for stock $i$ obtained from Cboe price quotes for 12/31/2021.\footnote{
	On this date, call option prices for these assets had the maturity dates ranging from 01/21/2022 to 07/15/2022.
	The data set consisted of 699 AAPL, 581 MSFT, and 1,213 AMZN valid call option contracts.}
In computing values for
$C^{(i)} (S_T,K;r,\hat{\mu}_\lambda, \hat{\sigma}_\lambda, \hat{p}_\lambda)$,
we used the 10-year U.S. Treasury yield curve rate for 12/31/2021 as the risk-free rate $r_f$.\footnote{
	\url{https://www.treasury.gov.}
	The yearly rate for $r_f$ on 12/31/2021 was $1.52\%$.}
As the call option price data is based upon daily changes, we employed an equally spaced partition over the period $[0,T=252]$
giving $t_{n,k} = k \Delta t$, $k = 1, \dots, 252$ with $\Delta t = 1$.
The minimization \eqref{eq:l_imp_MS} was performed over the set of intensities $\lambda \in \{0, 0.01, 0.02, \ldots, 1\}$
for all $K,T$ values for which the Cboe call option listed a non-zero price.

\begin{figure}[h!]
\begin{center}
    \begin{subfigure}[b]{0.32\textwidth} 
    	\includegraphics[width=\textwidth]{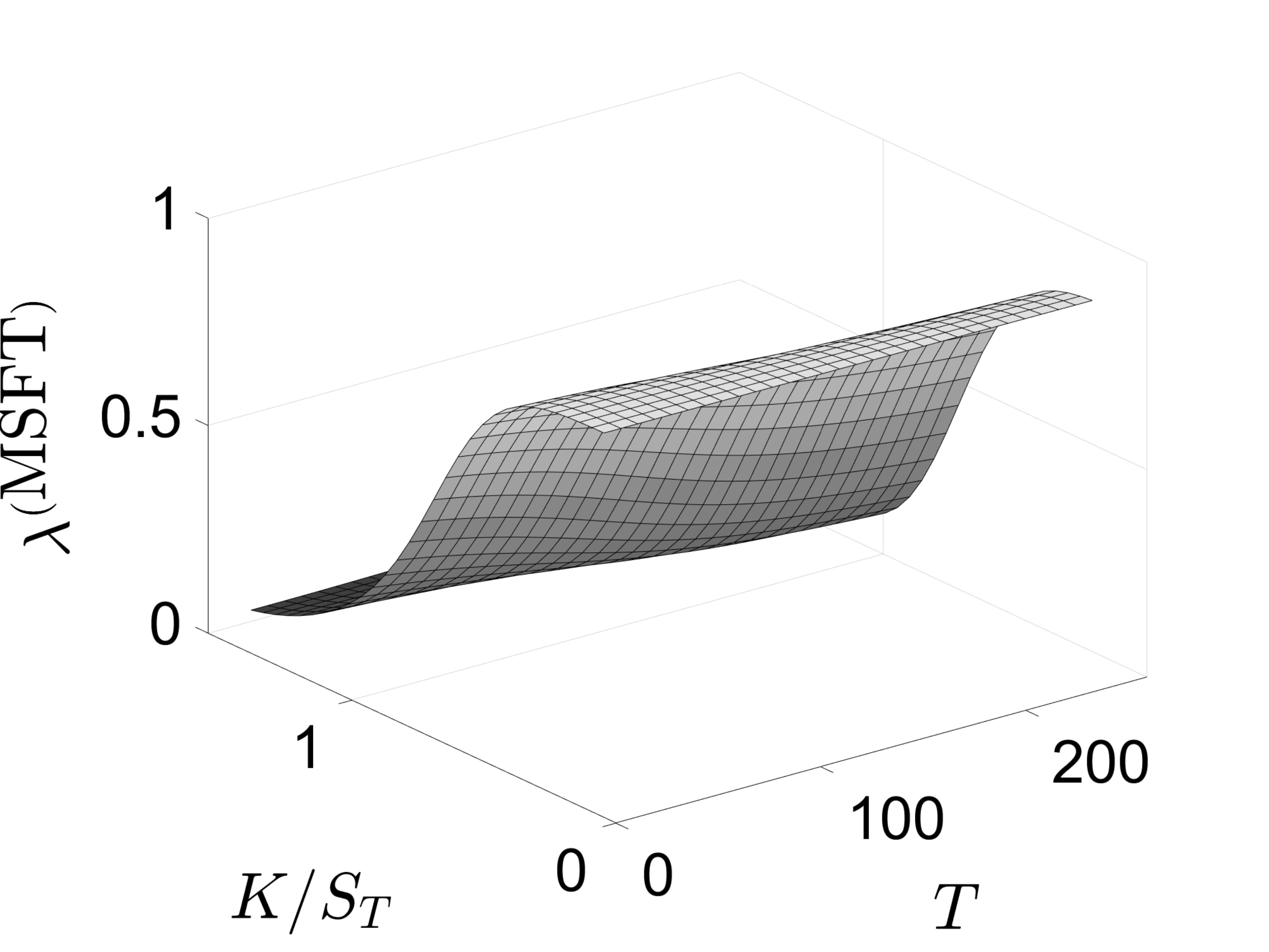}
    \end{subfigure}
    \begin{subfigure}[b]{0.32\textwidth} 
    	\includegraphics[width=\textwidth]{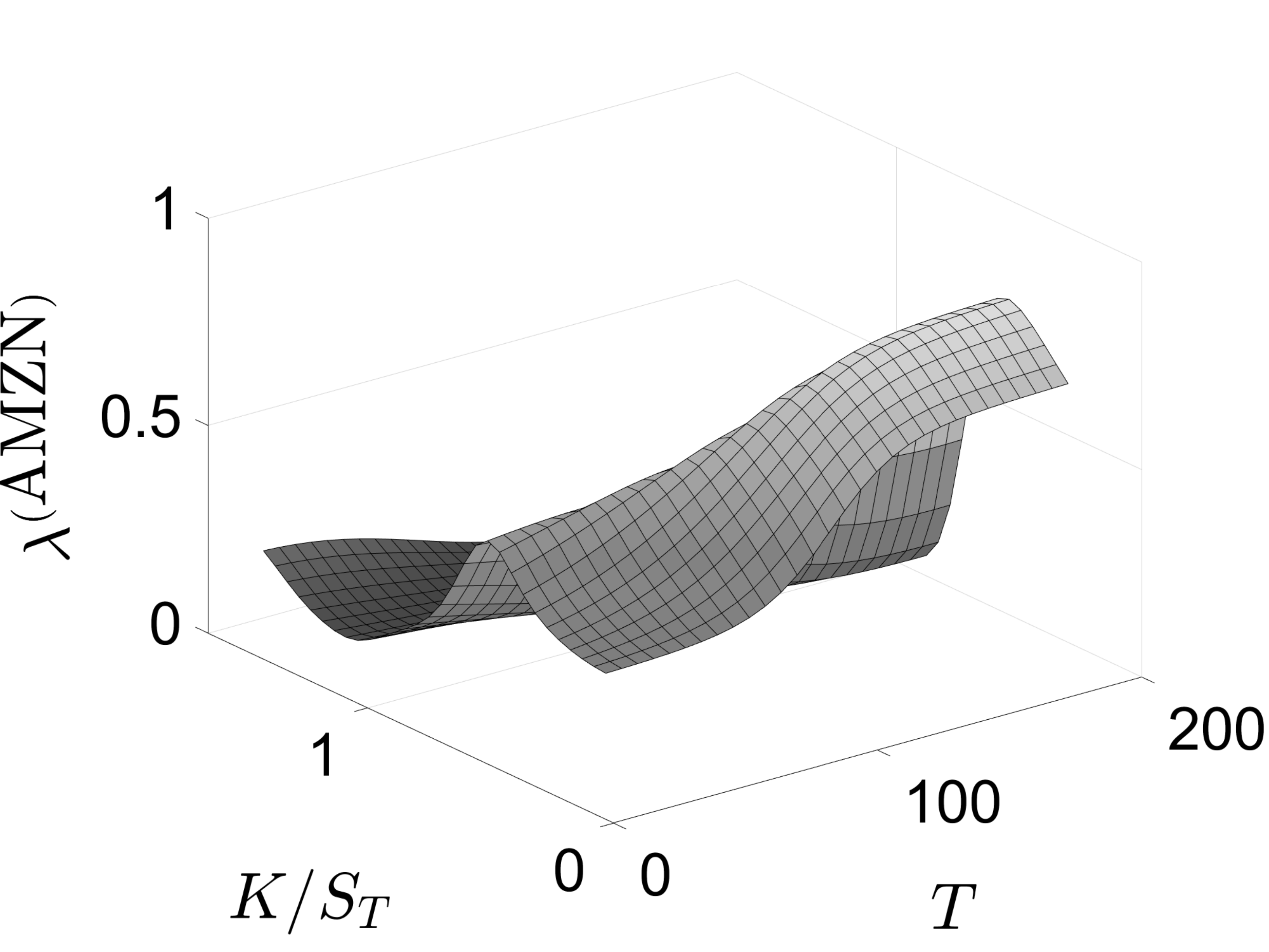}
    \end{subfigure}
    \begin{subfigure}[b]{0.32\textwidth} 
    	\includegraphics[width=\textwidth]{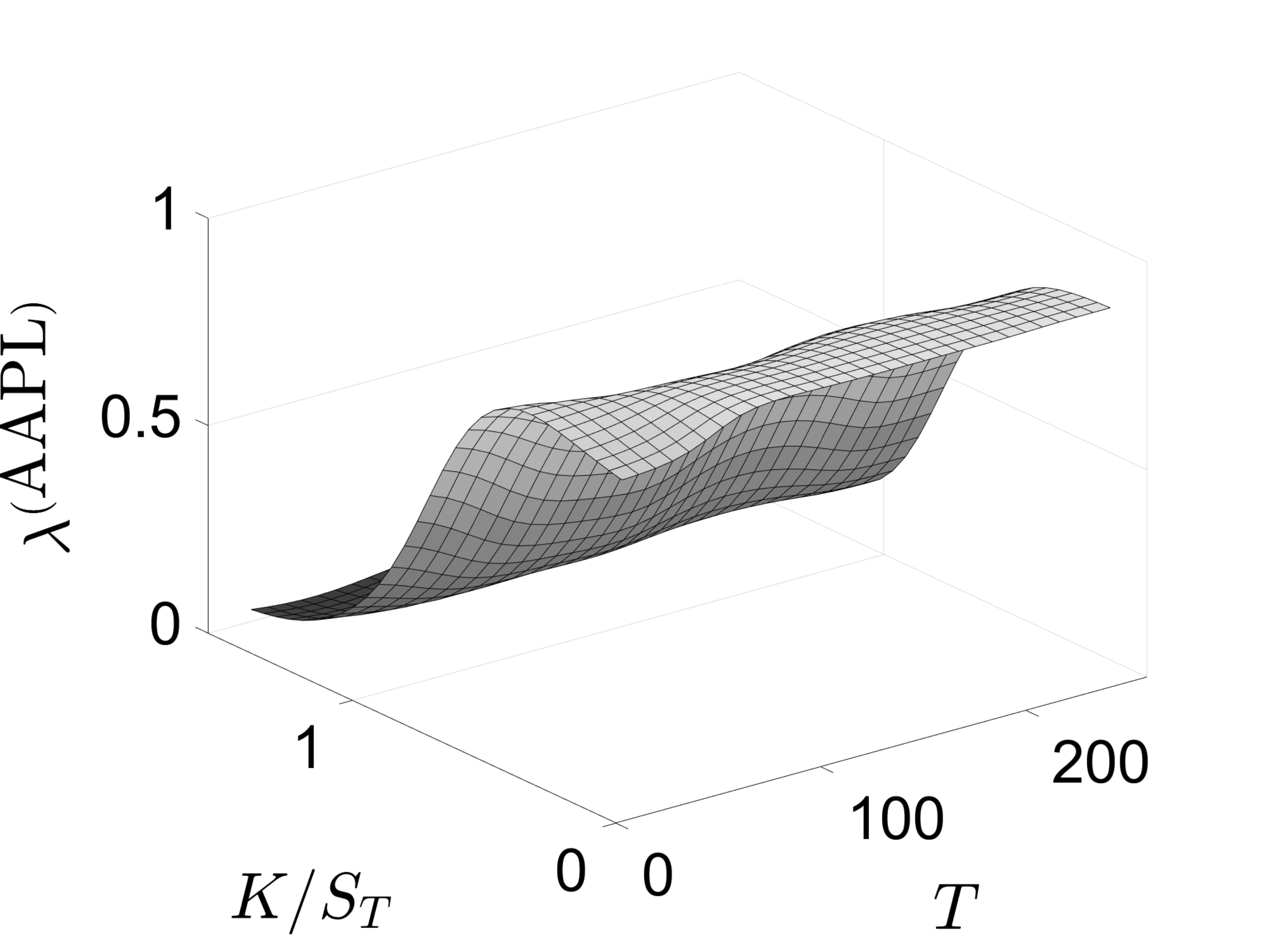}
    \end{subfigure}
    \begin{subfigure}[b]{0.32\textwidth} 
    	\includegraphics[width=\textwidth]{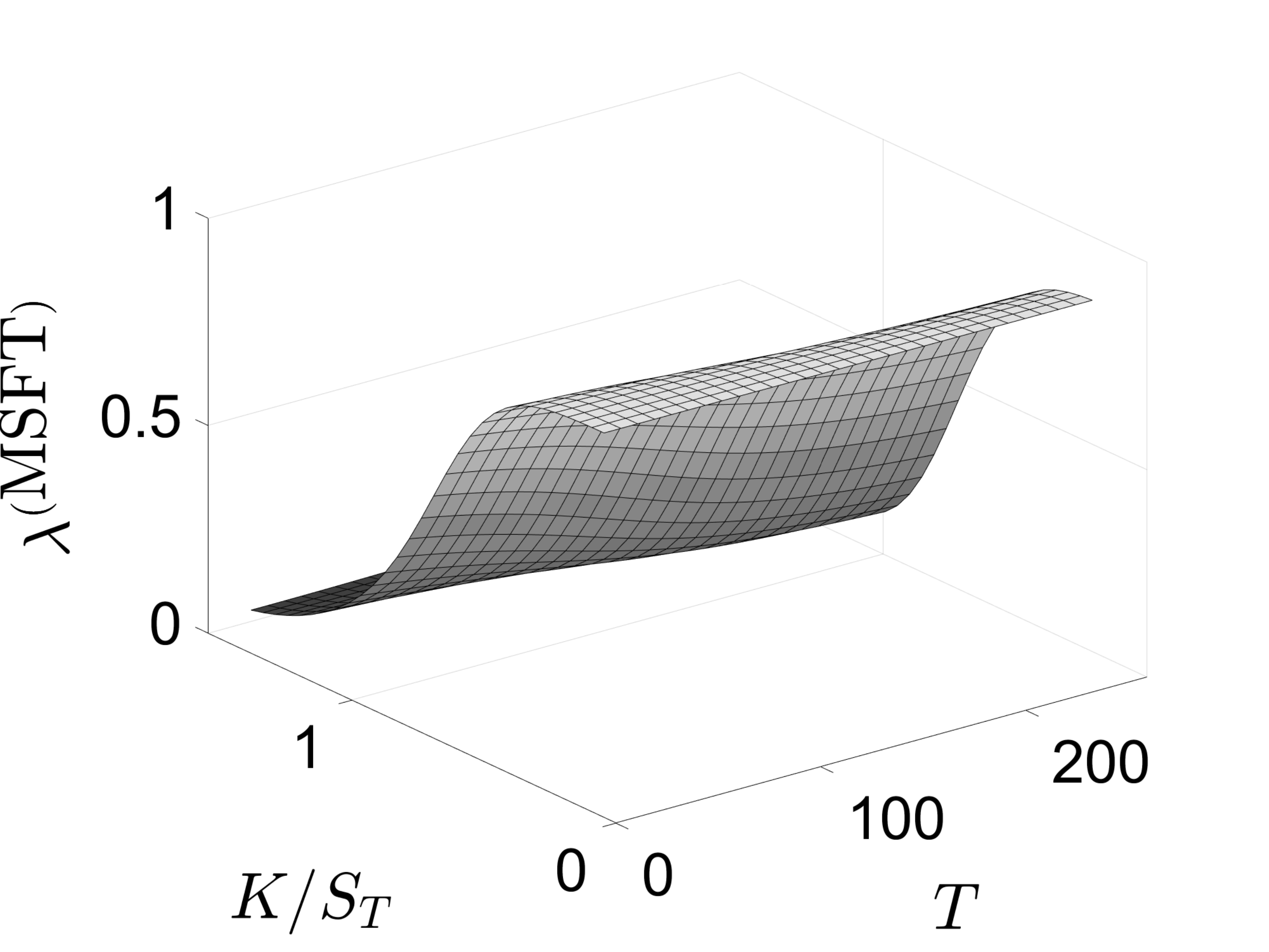}
    \end{subfigure}
    \begin{subfigure}[b]{0.32\textwidth} 
    	\includegraphics[width=\textwidth]{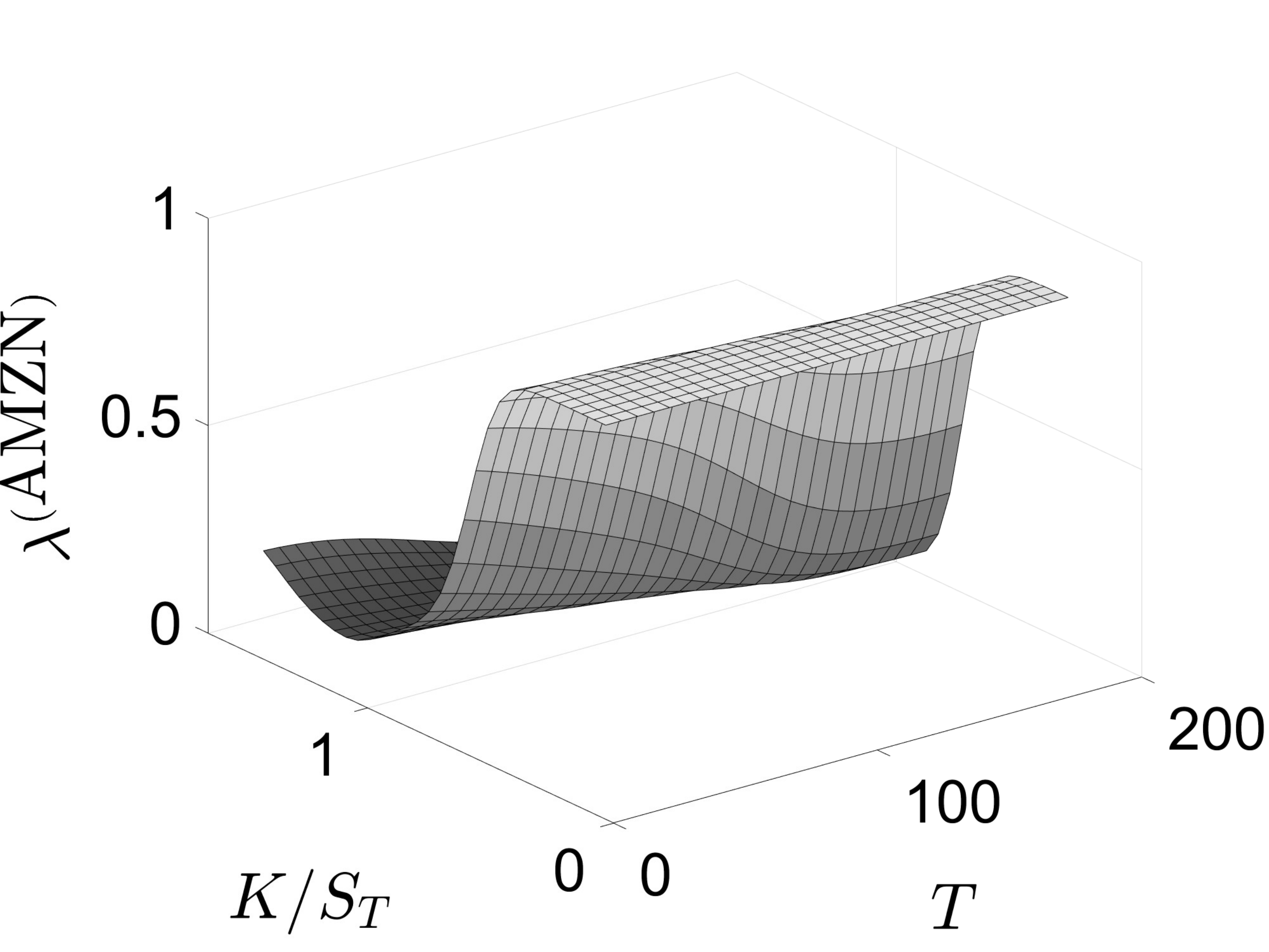}
    \end{subfigure}
    \begin{subfigure}[b]{0.32\textwidth} 
    	\includegraphics[width=\textwidth]{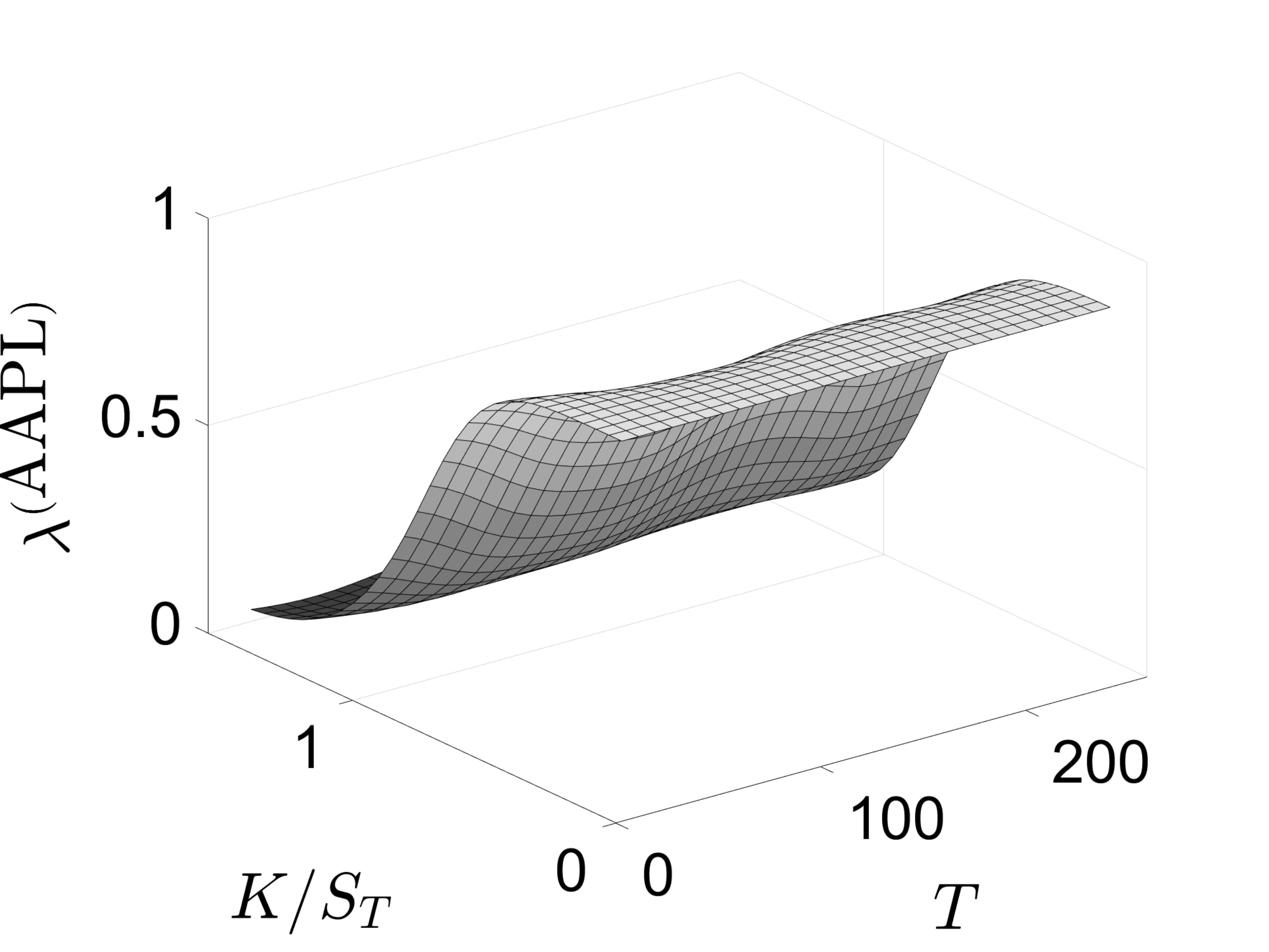}
    \end{subfigure}
    \caption{The implied ESG intensity surfaces $\lambda (T,M)$, plotted as a function of time to maturity $T$ (in days)
    		and moneyness $M=K/S_T$, for the three stocks considered.
    		(top row) Arithmetic return model \eqref{eq:ESG_dpm}; (bottom) log-return model \eqref{eq:ESG_lpm}.
    		}
    \label{fig:impl_lam}
\end{center}
\end{figure}

Fig.~\ref{fig:impl_lam} plots the implied ESG intensity surfaces $\lambda (T,M=K/S_T)$ as functions of
moneyness $M=K/S_T$ and time to maturity $T$ computed from the ESG-valued, arithmetic return 
and log-return models, \eqref{eq:ESG_dpm} and \eqref{eq:ESG_lpm}, respectively.
The surfaces are qualitatively similar for all three stocks under the log-return model;
greater differences are revealed under the arithmetic model.
The value of $\lambda (T,M)$ is small (ESG ratings are a minor consideration) when options are out-of-the-money
and increases (ESG ratings become a greater consideration) as $M$ moves into-the-money.
Generally the dependence of $\lambda$ on time to maturity $T$ is that of a slight increase with $T$.
The strongest difference is displayed for AMZN under the arithmetic return model, which displays a stronger
dependence on $T$ for in-the-money values of $M$.

\medskip\noindent
\textbf{Implied volatility surfaces.}
To evaluate the dependence of implied volatility on ESG intensity $\lambda$,
we computed the implied volatility $\sigma^{(i)} (T,M,\lambda)$ for the values $\lambda \in \{0,0.25,0.5,0.75\}$
and the same values of $T$ and $K$ as for the computations of the implied $\lambda$ surfaces.
The implied volatility for stock $i$ was determined by minimizing the relative mean-square difference
\begin{align}
	\sigma^{(i)} (T,K,\lambda) = \argmin_\sigma
		\left(
			\frac{ C^{(i)} (S_T,K;r_f,\hat{\mu}_\lambda, \hat{p}_\lambda) - C^{(i)}(S_T,K) }
				{ C^{(i)}(S_T,K) }
		\right)^2 .
	\label{eq:impl_vol}
\end{align}
The implied volatilities computed from the arithmetic return model \eqref{eq:ESG_dpm} for the three stocks
are plotted in Fig.~\ref{fig:impl_vol_arith} in Appendix~\ref{app:A}.
The implied volatility surface for AMZN shows a characteristic ``smile'' profile  along the moneyness axis,
progressing to more of a ``smirk'' profile as the time-to-maturity increases.
In contrast, the implied volatility for MSFT and AAPL display strong ``smirk'' profiles that remain relatively constant
as $T$ increases.
For each stock, the volatility surfaces show comparatively small changes as the ESG intensity $\lambda$ increases.
This is quantified in Fig.~\ref{fig:impl_vol_arith_dev}, which plots the percent relative changes
\begin{equation}
	\Delta \sigma^{(i)} (T,K,\lambda)
		= 100 \left( \frac{ \sigma^{(i)} (T,K,\lambda) - \sigma^{(i)} (T,K,0) } { \sigma^{(i)} (T,K,0) } \right),\quad
			\lambda = 0.25, 0.5, 0.75.
\end{equation}
For AAPL and AMZN, and for MSFT when $\lambda \le 0.5$, the percent changes are no more than $\pm 5$ percent.
When $\lambda = 0.75$, relative changes for MSFT approach $\pm 15$ percent.

Fig.~\ref{fig:impl_vol_ln} displays the implied volatility surfaces computed from the log-return model \eqref{eq:ESG_lpm}
for these three stocks.
When $\lambda = 0$, the log-return model surfaces are fairly similar to those for the arithmetic return model
in Fig.~\ref{fig:impl_vol_arith}.
(The viewing angle is slightly different between Figs.~\ref{fig:impl_vol_arith} and \ref{fig:impl_vol_ln}.)
As $\lambda$ increases, the volatility surface for AMZN remains relatively unchanged.
In contrast to the arithmetic return model,
much greater changes in implied volatility under the log-return model are observed for MSFT and AAPL as $\lambda$
increases.
These relative changes for implied volatilities computed for the log-return model are documented in
Fig.~\ref{fig:impl_vol_ln_dev}.
For MSFT and AMZN, any change in implied volatility is toward smaller values as $\lambda$ increases,
while for AAPL, any change is toward large values.
Very strong changes with $\lambda$ are seen far into-the-money at shorter maturity times, with changes
approaching $-40\%$ to $-60\%$ for AMZN and MSFT, while changes approach $200\%$ for AAPL.
 
We also compared the implied volatilities $\sigma^{(i)} (T,K,\lambda=0)$ computed from \eqref{eq:impl_vol}
against the implied volatility computed using the BSM call option formula $C^{(i,\BSM)} (S_T,K;r_f)$,
\begin{align}
	\sigma^{(i,\BSM)} (T,K) = \argmin_\sigma
		\left( \frac{ C^{(i,\BSM)} (S_T,K;r_f) - C^{(i)}(S_T,K) }{ C^{(i)}(S_T,K) } \right)^2 .
	\label{eq:impl_vol_BSM}
\end{align}
Fig. \ref{fig:impl_vol_BSM} shows the BSM implied volatility surfaces $\sigma^{(i,\text{BSM})}$.
In contrast to the discrete models, the implied volatility under BSM shows a ``smile'' profile
for all three stocks (though the range of implied volatility values is small for MSFT under the BSM model).
Also plotted in Fig. \ref{fig:impl_vol_BSM} are the percent relative differences surfaces
\begin{equation}
	\Delta \sigma^{(i,\BSM)}(T,K,0) = 
		100 \left( \sigma^{(i)} (T,K,0) - \sigma^{(i,\BSM)} (T,K) \right) / \sigma^{(i,\BSM)} (T,K)\,,
	\label{eq:impl_vol_BSM_diff}
\end{equation}
computed for both the arithmetic or log-return models.
Strong differences between each discrete model and the BSM model emerge as values for $M=K/S_T$
move further into-the-money.
The discrete model derives larger volatility values for in-the-money options than does the
Black-Scholes-Merton model.
The differences are relatively insensitive to change in maturity date.

\section{ESG-valued option pricing in markets with informed traders}
\label{sec:inf_opm}

\noindent
Following the framework developed in Hu et al. (2020b; section 6),
we extend the ESG-valued option pricing model to markets in which there are informed traders.
Suppose the ESG-value trader $\aleph$ has taken a short position in the option contract.
The risk neutral portfolio used by $\aleph$ in section~\ref{sec:ESG_opm} is adjusted to perfectly hedge
the short position.
We assume a representative investor has complete information on past values of ESG, $\lambda$ and returns $r_{n,k,0}$
governing the underlying asset.
Assume $\aleph$ acquires additional information concerning the direction of movement $\varsigma_{n,k,\lambda}$
of the underlying.
To capitalize on this additional information,
$\aleph$ enters into $N^{(\aleph)}_{n,k,\lambda}$ long (or short) forward contracts if the information leads $\aleph$
to believe the stock price will increase (or decrease).
Let $\pN[k]$ denote the probability that $\aleph$'s information is correct.
We assume $\pN[k] \in (0.5,1)$ and utilize the model
\begin{equation}
	\pN[k+1] = \left( 1 + \dN[k]\sqrt{\dtn[k+1]} \right)/2\,,
	\label{eq:delta}
\end{equation}
where $0 < \dN[k] < (\dtn[k+1])^{-1/2}$ is $\aleph$'s \textit{information intensity}.

$\aleph$'s trading strategy (a combination of trading the stock and forward contracts on the stock) results in the
following modification of either option pricing tree \eqref{eq:ESG_dpm} or \eqref{eq:ESG_lpm}.

\noindent
\textbf{ESG-valued, arithmetic return, informed model.}
Conditionally on $S_{n,k,\lambda}$,
\begin{equation}
	S_{n,k+1,\lambda}^{(\aleph)} = 
	\begin{cases} 
		S_{n,k+1,\lambda}^{(\text{u})}
			+ N^{(\aleph)}_{n,k,\lambda}\left[ S_{n,k+1,\lambda}^{(\text{u})} - S_{n,k,\lambda} (1+\rf[k]\dtn[k+1])\right],\;
			\textrm{w.p.} \; \pnl[k+1]\; \pN[k+1]\,,
	\\
		S_{n,k+1,\lambda}^{(\text{d})}
			- N^{(\aleph)}_{n,k,\lambda}\left[ S_{n,k+1,\lambda}^{(\text{d})} - S_{n,k,\lambda} (1+\rf[k]\dtn[k+1])\right],\;
			\textrm{w.p.} \; (1-\pnl[k+1])\; \pN[k+1]\,,
	\\
		S_{n,k+1,\lambda}^{(\text{u})}
			- N^{(\aleph)}_{n,k,\lambda}\left[ S_{n,k+1,\lambda}^{(\text{u})} - S_{n,k,\lambda} (1+\rf[k]\dtn[k+1])\right],\;
			\textrm{w.p.} \; \pnl[k+1]\; \left(1-\pN[k+1]\right),
	\\
		S_{n,k+1,\lambda}^{(\text{d})}
			+ N^{(\aleph)}_{n,k,\lambda}\left[ S_{n,k+1,\lambda}^{(\text{d})} - S_{n,k,\lambda} (1+\rf[k]\dtn[k+1])\right],\;
			\textrm{w.p.} \; (1-\pnl[k+1])\; \left(1-\pN[k+1]\right),
	\end{cases}
	\label{eq:aleph_dstrat}
\end{equation}
with $S_{n,0,\lambda}^{(\aleph)} = S_0$.
The terms $S_{n,k+1,\lambda}^{(\text{u})}$, and $S_{n,k+1,\lambda}^{(\text{d})}$,
$k = 0,1,...,n-1$, are defined in \eqref{eq:ESG_dpm} and \eqref{eq:ESG_dUD}.
To $O(\dtn[k+1])$, the conditional mean and variance of the return $r^{(\aleph)}_{n,k+1,\lambda}$ on $\aleph$'s
strategy are
\begin{equation}
   \begin{aligned}
	\mathbb{E} \left[ r_{n,k+1,\lambda}^{(\aleph)} |S_{n,k,\lambda}^{(\aleph)} \right]
		&=  \left( \mnl[k] + \snl[k] N^{(\aleph)}_{\delta} N^{(\aleph)}_{n,k,\lambda} \right) \dtn[k+1]\,, \\
	\textrm{Var} \left[ r_{n,k+1,\lambda}^{(\aleph} |S_{n,k,\lambda}^{(\aleph)} \right]
		&= {\snl[k]}^2 \left( 1 + {N^{(\aleph)}_{n,k,\lambda}}^2 \right) \dtn[k+1]\,,
  \end{aligned}
   \label{eq:inf_dEV}
\end{equation}
where
\begin{equation}
	N^{(\aleph)}_\delta = 2\dN[k] \sqrt{ \pnl[k+1](1-\pnl[k+1]) }\,.
	\label{eq:dN_delta}
\end{equation}
A value for $N^{(\aleph)}_{n,k,\lambda}$ can be determined by maximizing $\aleph$'s instantaneous Sharpe ratio,
\begin{equation}
	\Theta_{n,k,\lambda}^{(\aleph)}
	 = \frac{ \mathbb{E} \left[ r_{n,k+1,\lambda}^{(\aleph)} | S_{n,k,\lambda}^{(\aleph)} \right] - \rf[k] \dtn[k+1] }
			{ \sqrt{ \text{Var} \left[ r_{n,k,\lambda}^{(\aleph)} | S_{n,k,\lambda}^{(\aleph)} \right] \dtn[k+1] } }
	= \frac{ \theta_{n,k,\lambda} + N^{(\aleph)}_{\delta} N^{(\aleph)}_{n,k,\lambda} }
			{\sqrt{1 +{N^{(\aleph)}_{n,k,\lambda}}^2 }}\,,
	\label{eq:inf_dSR}
\end{equation}
where $\theta_{n,k,\lambda}$ is defined in \eqref{eq:ESG_dSR}.
The maximizing value is
\begin{equation}
	N^{(\aleph,\text{max})}_{n,k,\lambda} = \frac{ N^{(\aleph)}_{\delta} }{ \theta_{n,k,\lambda} }\,,
	\label{eq:dmax_N}
\end{equation}
giving the expression for the optimal instantaneous Sharpe ratio,
\begin{equation}
	\Theta_{n,k,\lambda}^{(\aleph,\text{opt})}
	 = \sqrt{{\theta_{n,k,\lambda}}^2 + \left( N^{(\aleph)}_{\delta} \right)^2 }\,.
	\label{eq:inf_dSR_opt}
\end{equation}
The optimized conditional mean and variance of the return on $\aleph$'s strategy are
\begin{equation}
   \begin{aligned}
	\mathbb{E} \left[ r_{n,k+1,\lambda}^{(\aleph)} |S_{n,k,\lambda}^{(\aleph)} \right]^{\text{(opt)}}
		&=  \left[
			    \mnl[k] + \frac{  \snl[k] \left( N^{(\aleph)}_{\delta} \right)^2 }{  \theta_{n,k,\lambda} }
			\right] \dtn[k+1]\,,
	\\
	\text{Var} \left[ r_{n,k,\lambda}^{(\aleph} |S_{n,k,\lambda}^{(\aleph)} \right]^{\text{(opt)}}
		&= {\snl[k]}^2 \left[ 1 + \left( \frac{ N^{(\aleph)}_{\delta} }{  \theta_{n,k,\lambda} } \right)^2\right] \dtn[k+1]\,.
   \end{aligned}
   \label{eq:inf_dEV_opt}
\end{equation}
$\aleph$'s information-enhanced dividend yield $\mathbb{D}^{(\aleph)}_{n,k,\lambda}$ is
\begin{equation}
	\mathbb{D}^{(\aleph)}_{n,k,\lambda} = \snl[k] \left( \Theta_{n,k,\lambda}^{(\aleph,\text{opt})} - \theta_{n,k,\lambda} \right)
	= \snl[k]
	\left[ \sqrt{ {\theta_{n,k,\lambda}}^2 +{N^{(\aleph)}_{\delta}}^2 } - \theta_{n,k,\lambda} \right] .
	\label{eq:inf_dDY}
\end{equation}

Based on \eqref{eq:inf_dEV_opt}, $\aleph$'s strategy \eqref{eq:aleph_dstrat} can
be replicated (to $O(\dtn[k+1])$) by the binomial option pricing tree
\begin{equation}
	S^{(\aleph)}_{n,k+1,\lambda} =
	\begin{cases}
	S^{(\aleph)}_{n,k,\lambda} \left[ 1+ U^{(\aleph)}_{n,k+1,\lambda} \right], \quad \text{ w.p. }\pnl[k+1]\,,\\
	S^{(\aleph)}_{n,k,\lambda} \left[ 1+ D^{(\aleph)}_{n,k+1,\lambda}  \right], \quad\text{ w.p. }1-\pnl[k+1]\,,
	\end{cases}
	\label{eq:daleph_binomial}
\end{equation}
where
\begin{equation}
   \begin{aligned}
	U^{(\aleph)}_{n,k+1,\lambda}
		&= \left[ \mnl[k] + \frac{  \snl[k] \left( N^{(\aleph)}_{\delta} \right)^2 }{ \theta_{n,k,\lambda} } \right] \dtn[k+1]
			 + p_{n,k+1,\lambda}^{(\text{u})}\ \snl[k]
			 \sqrt{ \left[ 1 + \left( \frac{ N^{(\aleph)}_{\delta} }{  \theta_{n,k,\lambda} } \right)^2\right] \dtn[k+1]}\,,
	\\
	D^{(\aleph)}_{n,k+1,\lambda}
		&= \left[ \mnl[k] + \frac{  \snl[k] \left( N^{(\aleph)}_{\delta} \right)^2 }{ \theta_{n,k,\lambda} } \right] \dtn[k+1]
			 - p_{n,k+1,\lambda}^{(\text{d})}\ \snl[k]
			 \sqrt{ \left[ 1 + \left( \frac{ N^{(\aleph)}_{\delta} }{  \theta_{n,k,\lambda} } \right)^2\right] \dtn[k+1]}\,,	
   \end{aligned}
   \label{eq:daleph_UD}
\end{equation}
with $p_{n,k+1,\lambda}^{(\text{u})}$ and $p_{n,k+1,\lambda}^{(\text{d})}$ as defined in \eqref{eq:ESG_dUD}.

\noindent
\textbf{ESG-valued, log-return, informed model.}
Conditionally on $S_{n,k,\lambda}$,
\begin{equation}
	S_{n,k+1,\lambda}^{(\aleph)} = 
	\begin{cases} 
		S_{n,k+1,\lambda}^{(\text{u})}
			+ N^{(\aleph)}_{n,k,\lambda}\left[ S_{n,k+1,\lambda}^{(\text{u})} - S_{n,k,\lambda} \exp(\rf[k]\dtn[k+1])\right],\;
			\textrm{w.p.} \; \pnl[k+1]\; \pN[k+1]\,,
	\\
		S_{n,k+1,\lambda}^{(\text{d})}
			- N^{(\aleph)}_{n,k,\lambda}\left[ S_{n,k+1,\lambda}^{(\text{d})} - S_{n,k,\lambda} \exp(\rf[k]\dtn[k+1])\right],\;
			\textrm{w.p.} \; (1-\pnl[k+1])\; \pN[k+1]\,,
	\\
		S_{n,k+1,\lambda}^{(\text{u})}
			- N^{(\aleph)}_{n,k,\lambda}\left[ S_{n,k+1,\lambda}^{(\text{u})} - S_{n,k,\lambda} \exp(\rf[k]\dtn[k+1])\right],\;
			\textrm{w.p.} \; \pnl[k+1]\; \left(1-\pN[k+1]\right),
	\\
		S_{n,k+1,\lambda}^{(\text{d})}
			+ N^{(\aleph)}_{n,k,\lambda}\left[ S_{n,k+1,\lambda}^{(\text{d})} - S_{n,k,\lambda} \exp(\rf[k]\dtn[k+1])\right],\;
			\textrm{w.p.} \; (1-\pnl[k+1])\; \left(1-\pN[k+1]\right),
	\end{cases}
	\label{eq:aleph_lstrat}
\end{equation}
with $S_{n,0,\lambda}^{(\aleph)} = S_0$.
The terms $S_{n,k+1,\lambda}^{(\text{u})}$, and $S_{n,k+1,\lambda}^{(\text{d})}$,
$k = 0,1,...,n-1$, are defined in \eqref{eq:ESG_lpm} and \eqref{eq:dt_lUD}.
To $O(\Delta t)$ the conditional mean and variance of the return $r^{(\aleph)}_{n,k+1,\lambda}$
on $\aleph$'s strategy are\footnote{
	See Hu et al. (2020b).
	Care must be taken in the Taylor series expansions of the terms in \eqref{eq:aleph_lstrat},
	which have the form $\ln \{ (1 \pm N) \exp (U) \mp \exp{r_f \Delta t} \}$,
	to ensure all terms of $O(\Delta t)$ and $O(\sqrt{\Delta t})$ are retained.
	Unfortunately, in Hu et al. (2020b), terms of $O( (N\sqrt{\Delta t})^2 )$ were missed in
	the computations for the conditional mean.}
\begin{align}
	\mathbb{E} \left( r_{n,k+1,\lambda}^{(\aleph)} |S_{n,k,\lambda}^{(\aleph)} \right)
		&=  \left[
			    \mnl[k] - \frac{{\snl[k]}^2}{2}
			    \left( 1 + {N^{(\aleph)}_{n,k,\lambda}}^2 - 2 N^{(\aleph,\delta)}_\mathbb{E} N^{(\aleph)}_{n,k,\lambda} \right)
			\right] \dtn[k+1]\,,
	\nonumber
	\\
	\textrm{Var} \left( r_{n,k+1,\lambda}^{(\aleph)} | S_{n,k,\lambda}^{(\aleph)} \right)
		&= {\snl[k]}^2 \left( 1 + {N^{(\aleph)}_{n,k,\lambda}}^2 \right) \dtn[k+1]\,,
	\label{eq:inf_lEV}
\end{align}
where
\begin{equation}
	N^{(\aleph,\delta)}_\mathbb{E} = \frac{ 2\dN[k] \sqrt{ \pnl[k+1] (1-\pnl[k+1]) } }{ \snl[k] }\,,
	\label{eq:lN_delta}
\end{equation}
and we have used \eqref{eq:delta} to simplify the expressions.
Note that $N^{(\aleph,\delta)}_\mathbb{E}$ is the value of $N^{(\aleph)}_{n,k,\lambda}$ that maximizes
the excess conditional expected log return
$\mathbb{E} \left( r_{n,k+1,\lambda}^{(\aleph)} |S_{n,k,\lambda}^{(\aleph)} \right) - \rf[k] \dtn[k+1]$
with respect to $N^{(\aleph)}_{n,k,\lambda}$.

A value for $N^{(\aleph)}_{n,k,\lambda}$ can be determined by maximizing $\aleph$'s instantaneous Sharpe ratio,
\begin{equation}
	\Theta_{n,k,\lambda}^{(\aleph)}
	 = \frac{\mathbb{E} \left( r_{n,k+1,\lambda}^{(\aleph)} | S_{n,k,\lambda}^{(\aleph)} \right) - \rf[k] \dtn[k+1]}
			{ \sqrt{ \textrm{Var} \left (r_{n,k,\lambda}^{(\aleph)} | S_{n,k,\lambda}^{(\aleph)} \right) \dtn[k+1] } } \quad
	= \frac{ \snl[k ]}{2} \;
		\frac{a_{n,k,\lambda} + 2 N^{(\aleph,\delta)}_\mathbb{E} N^{(\aleph)}_{n,k,\lambda} - {N^{(\aleph)}_{n,k,\lambda}}^2 }
			{\sqrt{1 +{N^{(\aleph)}_{n,k,\lambda}}^2 }}\,,
	\label{eq:inf_lSR}
\end{equation}
where
\begin{equation}
	a_{n,k,\lambda} \equiv \frac{ 2 \theta_{n,k,\lambda} }{ \snl[k] } - 1\,.
	\label{eq:la}
\end{equation}
The maximization reduces to solving the depressed cubic
\begin{equation}
	{N^{(\aleph)}_{n,k,\lambda}}^3 + (2+a_{n,k,\lambda}) N^{(\aleph)}_{n,k,\lambda}
											 - 2 N^{(\aleph,\delta)}_\mathbb{E} = 0\,,
	 \label{eq:dep_cubic}
\end{equation}
for the single real root
\begin{equation}
   \begin{aligned}
	N^{(\aleph,\text{max})}_{n,k,\lambda} &= \left(  N^{(\aleph,\delta)}_\mathbb{E}
								     + b_{n,k,\lambda} \right)^{1/3}
								     + \left(  N^{(\aleph,\delta)}_\mathbb{E} - b_{n,k,\lambda} \right)^{1/3},
	\\
	b_{n,k,\lambda} &= \left[  {N^{(\aleph,\delta)}_\mathbb{E}}^2
								 + \left( \frac{ 2 +a_{n,k,\lambda} }{ 3 } \right)^3 \right]^{1/2} .
   \end{aligned}
   \label{eq:lmax_N}
\end{equation}
From \eqref{eq:dep_cubic},
\begin{equation*}
	2 N^{(\aleph,\delta)}_\mathbb{E} N^{(\aleph,\text{max})}_{n,k,\lambda}
	= {N^{(\aleph,\text{max})}_{n,k,\lambda}}^4 + (2+a_{n,k,\lambda}) {N^{(\aleph,\text{max})}_{n,k,\lambda}}^2,
\end{equation*}
giving the expression for the optimal instantaneous Sharpe ratio,
\begin{equation}
	\Theta_{n,k,\lambda}^{(\aleph,\text{opt})}
	 = \frac{ \snl[k] }{ 2 } \;
		\left( a_{n,k,\lambda} + { N^{(\aleph,\text{max})}_{n,k,\lambda} }^2 \right) 
		\sqrt{ 1 +{N^{(\aleph,\text{max})}_{n,k,\lambda}}^2 }\,.
	\label{eq:opt_mrkt_price}
\end{equation}
The optimized conditional expected return and variance for $\aleph$'s strategy are
\begin{align}
	\mathbb{E} \left( r_{n,k+1,\lambda}^{(\aleph)} |S_{n,k,\lambda}^{(\aleph)} \right)^{\text{(opt)}}
		&=  \left\{
			    \mnl[k] - \frac{{\snl[k]}^2}{2}
			    \left( 1 + {N^{(\aleph,\text{max})}_{n,k,\lambda}}^2
			    		 - 2 N^{(\aleph)}_\mathbb{E} N^{(\aleph,\text{max})}_{n,k,\lambda} \right)
			\right\} \dtn[k+1]\,,
		\nonumber \\
	\textrm{Var} \left( r_{n,k+1,\lambda}^{(\aleph)} | S_{n,k,\lambda}^{(\aleph)} \right)^{\text{(opt)}}
		&= {\snl[k]}^2 \left( 1 + {N^{(\aleph,\text{max})}_{n,k,\lambda}}^2 \right) \dtn[k+1]\,.
	\label{eq:inf_lEV_opt}
\end{align}
$\aleph$'s information-enhanced dividend yield $\mathbb{D}^{(\aleph)}_{n,k+1,\lambda}$ is
\begin{equation}
   \begin{aligned}
	\mathbb{D}^{(\aleph)}_{n,k+1,\lambda}
		&= \snl[k] \left( \Theta_{n,k,\lambda}^{(\aleph,\text{opt})} - \theta_{n,k,\lambda} \right) \\
		&= \snl[k] \left[ \sqrt{ 1 +{N^{(\aleph,\text{max})}_{n,k,\lambda}}^2 }
					\left[ \theta_{n,k,\lambda} + \frac{ \snl[k] }{ 2 }
			 			\left( {N^{(\aleph,\text{max})}_{n,k,\lambda}}^2 - 1 \right)
					\right] - \theta_{n,k,\lambda}
				\right] .
   \end{aligned}
   \label{eq:inf_lDY}
\end{equation}

$\aleph$'s strategy \eqref{eq:aleph_lstrat} can be replicated (to $O(\dtn[k+1])$) by a binomial option pricing tree.
Consider the tree
\begin{equation}
	S^{(\aleph)}_{n,k+1,\lambda} =
	\begin{cases}
	S^{(\aleph)}_{n,k,\lambda}	 \exp \left[ U^{(\aleph)}_{n,k+1,\lambda} \right],
				\text{ w.p. }\;\pnl[k+1]\,,
	\\
	S^{(\aleph)}_{n,k,\lambda} \exp \left[ D^{(\aleph)}_{n,k+1,\lambda}  \right],
				\text{ w.p. }\;1-\pnl[k+1]\,,
	\end{cases}
	\label{eq:laleph_binomial}
\end{equation}
where
\begin{equation}
   \begin{aligned}
	U^{(\aleph)}_{n,k+1,\lambda} &= \left( \mnl[k]
			- \left[p_{n,k+1,\lambda}^{(\text{u})}\right]^2 \cfrac{{\sigma^{(\aleph)}_{1,n,k,\lambda} }^2}{2} \right) \dtn[k+1]
			 + \sigma^{(\aleph)}_{2,n,k,\lambda}\  p_{n,k+1,\lambda}^{(\text{u})} \sqrt{ \dtn[k+1] }\,,
	\\
	D^{(\aleph,\ESG)}_{k+1} &= \left( \mnl[k]
			- \left[p_{n,k+1,\lambda}^{(\text{d})}\right]^2 \cfrac{{\sigma^{(\aleph)}_{1,t_{n,k}} }^2}{2}\right)\dtn[k+1]
			- \sigma^{(\aleph)}_{2,t_{n,k}}\ p_{n,k+1,\lambda}^{(\text{d})} \sqrt{ \dtn[k+1] }\,,			
   \end{aligned}
   \label{eq:laleph_UD}
\end{equation}
with $p_{n,k+1,\lambda}^{(\text{u})}$ and $p_{n,k+1,\lambda}^{(\text{d})}$ as defined in \eqref{eq:ESG_dUD}.
This binomial tree generates the risk-neutral probability
\begin{equation*}
	q_{n,k+1,\lambda}^{(\aleph)} = \frac{ e^{\rf[k] \dtn[k+1]}  - e^{D^{(\aleph)}_{n,k+1,\lambda}} }
							     { e^{U^{(\aleph)}_{n,k+1,\lambda}} - e^{D^{(\aleph)}_{n,k+1,\lambda}} }\,,
\end{equation*}
which is, to $O (\sqrt{ \dtn[k+1] })$,
\begin{equation}
	q_{n,k+1,\lambda}^{(\aleph)}
		= \pnl[k+1]
		- \left[ \frac{ \mnl[k] - \rf[k]
				+ \frac{1}{2} \left( {\sigma^{(\aleph)}_{2,n,k,\lambda}}^2 - {\sigma^{(\aleph)}_{1,n,k,\lambda}}^2 \right) }
			 { \sigma^{(\aleph)}_{2,n,k,\lambda} } \right]
		\sqrt{ \pnl[k+1] (1-\pnl[k+1]) \dtn[k+1] }\,.
	\label{eq:aleph_q}
\end{equation}
The conditional expectation and variance of the log-return for \eqref{eq:laleph_binomial} are
\begin{equation*}
   \begin{aligned}
	\mathbb{E}(r_{n,k+1,\lambda}^{(\aleph)} | S_{n,k,\lambda}^{(\aleph)})
		&= \pnl[k+1]\  U^{(\aleph)}_{n,k+1,\lambda}
		  + (1-\pnl[k+1])\  D^{(\aleph)}_{n,k+1,\lambda}\,, \\
	\text{Var}(r_{n,k+1,\lambda}^{(\aleph)} | S_{n,k,\lambda}^{(\aleph)})
		&= \pnl[k+1]\  (1-\pnl[k+1])
			\left( U^{(\aleph)}_{n,k+1,\lambda} - D^{(\aleph)}_{n,k+1,\lambda} \right)^2 .
   \end{aligned}
\end{equation*}
To  $O (\dtn[k+1])$, these reduce to
\begin{equation}
   \begin{aligned}
	\mathbb{E}(r_{n,k+1,\lambda}^{(\aleph)} | S_{n,k,\lambda}^{(\aleph)})
		&= \left( \mnl[k] - \frac{ {\sigma^{(\aleph)}_{1,n,k,\lambda} }^2 }{ 2 } \right) \dtn[k+1]\,, \\
	\text{Var}(r_{n,k+1,\lambda}^{(\aleph)} | S_{n,k,\lambda}^{(\aleph)})
		&= {\sigma^{(\aleph)}_{2,n,k,\lambda} }^2 \dtn[k+1]\,.
   \end{aligned}
   \label{eq:aleph_EV}
\end{equation}
Comparing \eqref{eq:aleph_EV} with \eqref{eq:inf_lEV_opt}, we see that $\aleph$'s strategy \eqref{eq:aleph_lstrat}
can be replicated (to $O(\dtn[k+1])$) by \eqref{eq:laleph_binomial} under the identification
\begin{equation}
   \begin{aligned}
	\sigma^{(\aleph)}_{1,n,k,\lambda} &= \snl[k] \sqrt{ 1 + {N^{(\aleph,\text{max})}_{n,k,\lambda}}^2
			    		 - 2 N^{(\aleph,\delta)}_\mathbb{E} N^{(\aleph,\text{max})}_{n,k,\lambda} }\,, \\
	\sigma^{(\aleph)}_{2,n,k,\lambda} &= \snl[k] \sqrt{ 1 + {N^{(\aleph,\text{max})}_{n,k,\lambda}}^2 }\,.
   \end{aligned}
   \label{eq:laleph_equiv}
\end{equation}

In the limit $\dtn[k+1] \downarrow 0$, the binomial tree \eqref{eq:laleph_binomial} defined by \eqref{eq:laleph_UD} 
and \eqref{eq:laleph_equiv} approximates a cadlag process that converges weakly
in the Skorokhod space $D[0,T]$ to a continuous diffusion process governed by the stochastic differential equation
(Davydov \& Rotar, 2008) 
\begin{equation}
	dS^{(\aleph)}_{t,\lambda} = \left( 
		\mu_{t,\lambda} - \frac{ {\sigma^{(\aleph)}_{1,t,\lambda} }^2 }{ 2 } 
		\right) S^{(\aleph)}_{t,\lambda} dt
		+ \sigma^{(\aleph)}_{2,t,\lambda} S^{(\aleph)}_{t,\lambda} d B_t\,,
\end{equation}
where $B_t$ is a standard Brownian motion.

\subsection{The implied $\delta^{(\aleph)}$ surface}
\label{sec:Div}

\noindent
Using the values in Table~\ref{tab:params} from fits to return data over the time period
1/1/2021 through 12/31/2021,
for any choices of $\lambda$ and $\aleph$'s information intensity
$\dN[k] \in \left(0, (\dtn[k+1])^{-1/2}\right)$,
option prices for 12/31/2021 based on the underlying stock MSFT, AMZN or AAPL can be computed using the
arithmetic return model \eqref{eq:daleph_binomial}, \eqref{eq:daleph_UD}
or the log-return model \eqref{eq:laleph_binomial}, \eqref{eq:laleph_UD}, \eqref{eq:laleph_equiv}
with the appropriate instantaneous Sharpe ratio maximizing value \eqref{eq:dmax_N} or \eqref{eq:lmax_N}.
For each choice of strike price $K$ and maturity time $T$, an implied value $\delta^{(i,\aleph)} (T,K,\lambda)$
for the information intensity can be computed by minimizing the relative error
\begin{equation}
	\delta^{(i,\aleph)} (T,K,\lambda) = \argmin_{ \dN[k] }
		\left(
			\frac{ C^{(i,\aleph)} (S_{T,\lambda},K;r,\hat{\mu}_\lambda,\hat{\sigma}_\lambda,\hat{p}_\lambda)
				- C^{(i)}(T,K) }
				{ C^{(i)}(T,K) }
		\right)^2\,,
	\label{eq:imp_delta}
\end{equation}
where $C^{(i)}(T,K)$ are the market call option prices for stock $i =$ MSFT, AMZN, AAPL.

Fig.~\ref{fig:dimpl_delta} presents the computed implied information intensity $\delta^{(i,\aleph)} (T,K,\lambda)$
as functions of $T$ and $K$ for the choices $\lambda \in \{0, 0.25, 0.5, 0.75 \}$ for option prices computed using
the arithmetic return model.
A value of $\delta^{(i,\aleph)} (T,K,\lambda) \approx 0$ indicates that the trader has no inside information regarding
the prices of the underlying upon which the options are developed.
Larger values of $\delta^{(i,\aleph)} (T,K,\lambda)$ indicate more effective information.
For $\lambda = 0$, for all three stocks, $\delta^{(i,\aleph)} (T,K,0) \approx 0$ for out-of-the-money values and
rises as $K/S_T$ moves into-the-money.
The increase is greatest for MSFT and smallest for AMZN.
Interestingly $\delta^{(i,\aleph)} (T,K,0)$ increases with time to maturity, indicating the time-importance of information.
For the highest ESG-rated stock, $\delta^{(\text{MSFT},\aleph)} (T,K,\lambda)$ increases with $\lambda$
(even for out-of-the-money values of $K/S_T$), reaching maximum values
of $(\dtn[k+1])^{-1/2} = \sqrt(252) =15.87$ for all in-the-money values when $\lambda = 0.75$.
For the intermediate ESG-rated stock, AMZN, the results are qualitatively similar but involve smaller increases.
As a result, $\delta^{(\text{AMZN},\aleph)} (T,K,\lambda)$ only approaches maximum values when $K/S_T$
is far into-the-money.
For the lowest ESG-rated stock, when  $K/S_T$ is in-the-money,
$\delta^{(\text{AAPL},\aleph)} (T,K,\lambda)$ decreases with $\lambda$.
The dependence of $\delta^{(\text{AAPL},\aleph)} (T,K,\lambda)$ with $\lambda$
is more complicated with the result that for the larger values of $\lambda$, $\delta^{(\text{AAPL},\aleph)} (T,K,\lambda)$
has its highest values when $K/S_T$ is out-of-the-money.
For the two largest values of $\lambda$, $\aleph$ has no inside information when $K/S_T$ is in-the-money.

Fig.~\ref{fig:limpl_delta} presents the analogous computed implied information intensity surfaces for the log-return model.
The optimized value \eqref{eq:lmax_N} results in some significant differences compared to the arithmetic return model.
When $\lambda = 0$, for all three stocks, $\delta^{(i,\aleph)} (T,K,0)$ reaches its maximum value of $\sqrt(252)$
at in-the-money values of $K/S_T$.
For MSFT and AMZN, the surfaces remain qualitatively similar as $\lambda$ increases, indicating that $\aleph$ experiences
roughly little change in inside information.
While the same is true for AAPL when $\lambda$ increases to 0.25,
when $\lambda$ increases to 0.5 and 0.75,  $\delta^{(\text{AAPL},\aleph)} (T,K,\lambda)$ undergoes a significant decrease
for in-the-money values of $K/S_T$.

\section{Path-dependent ESG-valued option pricing model}
\label{sec:PD_opm}

\noindent
We extend the ESG-valued binomial model of section \ref{sec:ESG_opm} to the case where the probabilities
$\pnl[k+1]$ for the stock $\mS$ are determined by a market driver,
which we refer to as an index $\mM$.
We do so under an extended setting formulated from the Cherny-Shiryaev-Yor invariance principle
(CSYIP) (Cherny et al., 2003; Hu et al., 2020b) which introduces history dependence into the asset pricing.
Unlike the models in section \ref{sec:ESG_opm} where the stock prices are adapted to a discrete filtration
$\mathbb{F}^{(n,\lambda)}$ determined by a modeled triangular array of independent Bernoulli random variables,
now the probabilities $\pnl[k+1]$ are based upon the observable returns of $\mM$.

Let $\ek[\mM]$ and $\ek[\mS]$ denote, respectively, the scaled ESG score for the market index and the stock.
The ESG-valued return for $\mM$ is
\begin{equation}
	\Rkl[M] = \lambda \ek[M] + (1-\lambda) r_{n,k,0}^{(\mM)}\,.
	\label{eq:PD_RM}
\end{equation}
Then
\begin{equation}
	 \ZMkl = \frac{ \Rkl[M] - \mathbb{E}[\Rkl[M]]}{ \sqrt{ \text{Var}[\Rkl[M]]} }\,,
	 \label{eq:PD_ZM}
\end{equation}
represents the random component driving the variation in $\Rkl[M]$ and
$\pMkl[k] = \mathbb{P}\left( \ZMkl[k] \ge 0 \right) \in (0,1)$ denotes the probability for an upturn
in the index's return over the time period $[t_{k-1},t_{k})$.
Define
\begin{equation}
	\pMkudl[k][u]  = \sqrt{\cfrac{1-\pMkl}{\pMkl}}\,, \qquad \qquad
	\pMkudl[k][d]  = \sqrt{\cfrac{\pMkl}{1-\pMkl}}\, .
   \label{eq:PD_pud}	
\end{equation}
The sequence
\begin{equation}
	\zMkl = \pMkudl[k][u] \text{I}_{ \left\{ \ZMkl \ge 0 \right\} } -  \pMkudl[k][d] \text{I}_{ \left\{ \ZMkl < 0 \right\} }\,,
		\qquad k = 1, \dots, n\,,
	\label{eq:PD_zM}
\end{equation}
represents an array of independent random variables satisfying
$\mathbb{E}\left[ \zMkl \right] = 0$, $\text{Var}\left[ \zMkl \right] = 1$,
and generates a discrete filtration
\begin{equation*}
	\mathbb{F}^{(\lambda,\mM)} = \left\{ \mathcal{F}^{(n,k,\lambda,\mM)}
	= \sigma \left( \zMkl[1],\ldots,\zMkl[k] \right),\ 
	k = 1, \ldots, k_n,\  \mathcal{F}^{(n,0,\lambda,\mM)} = \{\varnothing,\Omega\}\right\}.
\end{equation*}

We model the return (whether arithmetic or logarithm) of the stock price by
\begin{equation}
   \begin{aligned}
	r_{n,k,\lambda}^{(\mS,\mM)} &= \mukl[r] \dtn[k] + \eta_{k,\lambda}^{(\mM)} \zMkl \sqrt{\dtn[k]}\,,\\
	\eta_{k,\lambda}^{(\mM)} &= c_{1,\lambda} + c_{2,\lambda}\  h\left( X^{(\mM)}_{k/n,\lambda} \right)
						 + c_{3,\lambda}\  g\left( \sum_{j=1}^k X^{(\mM)}_{(j-1)/n,\lambda} \dtn[j] \right) ,\\
	X^{(\mM)}_{(j-1)/n,\lambda} &= \sum_{i=1}^{j-1} \sqrt{ \dtn[i]}\; \zMkl[i]\,.			 
   \end{aligned}
   \label{eq:PD_rtn}
\end{equation}
We assume the coefficients $c_{1,\lambda}$, $c_{2,\lambda}$ and $c_{3,\lambda}$ are independent of time.
The conditional mean and variance of the stock return at $\tn[k+1]$ are
\begin{equation}
	\mathbb{E}\left[ \rSkl[k+1] | \mathcal{F}^{(n,k,\lambda,\mM)} \right] = \mukl[r] \dtn[k+1]\,, \qquad
	\text{Var}\left[ \rSkl[k+1] | \mathcal{F}^{(n,k,\lambda,\mM)} \right] = {\eta_{k,\lambda}^{\mathcal{(M)}}}^2 \dtn[k+1]\,.
   \label{eq:PD_EV}
\end{equation}

\noindent
\textbf{Path dependent, ESG-valued, arithmetic return model.}
We model the dynamics of the stock price by the binomial tree
\begin{equation}
	S_{n,k+1,\lambda}^{(\mS,\mM)} =
	\begin{cases}
	S^{(\mS,\mM,\text{u})}_{n,k+1,\lambda} =
		 S_{n,k,\lambda}^{(\mS,\mM)} \left( 1 + U_{n,k+1,\lambda}^{(\mS,\mM)} \right),
		 	\text{ w.p. } \pMkl[k+1]\,, \\
	S^{(\mS,\mM,\text{u})}_{n,k+1,\lambda} =
		S_{n,k,\lambda}^{(\mS,\mM)} \left( 1 + D_{n,k+1,\lambda}^{(\mS,\mM)} \right),
			\text{ w.p. } 1-\pMkl[k+1]\,,
	\end{cases}
	\label{eq:PD_dpm}
\end{equation}
with
\begin{equation}
   \begin{aligned}
	U_{n,k+1,\lambda}^{(\mS,\mM)} &=
		\mukl[r]\, \dtn[k+1] + \eta_{n,k,\lambda}^{(\mM)}\, \pMkudl[k+1][u]\, \sqrt{\dtn[k+1]}\,, \\
	D_{n,k+1,\lambda}^{(\mS,\mM)} &=
		\mukl[r]\, \dtn[k+1] - \eta_{n,k,\lambda}^{(\mM)}\, \pMkudl[k+1][d]\, \sqrt{\dtn[k+1]}\,.
   \end{aligned}
   \label{eq:PD_dUD}	
\end{equation}
Here $\mukl[r]$ is also the instantaneous drift coefficient of the stock price.
The binomial tree \eqref{eq:PD_dpm} is $\mathbb{F}^{(\lambda,\mM)}$-adapted.
Under this model, the risk-neutral valuation for the price of $\mC$ is
\begin{equation}
   \begin{aligned}
	f_{n,k,\lambda}^{(\mS,\mM)} &= \frac{1}{1+\rf[k] \dtn[k+1]}
		\left[ q_{n,k+1,\lambda}^{(\mS,\mM)} f_{n,k+1,\lambda}^{(\mS,\mM,\text{u})}
			+ \left(1-q_{n,k+1,\lambda}^{(\mS,\mM)}\right) f_{n,k+1,\lambda}^{(\mS,\mM,\text{d})} \right], \\
	q_{n,k+1,\lambda}^{(\mS,\mM)} &= \pMkl[k+1]
				- \theta_{n,k,\lambda}^{(\mS,\mM)}\sqrt{\pMkl[k+1] \left(1-\pMkl[k+1]\right) \dtn[k+1]}\,,
   \end{aligned}
   \label{eq:PD_fq}
\end{equation}
where
\begin{equation}
	\theta_{n,k,\lambda}^{(\mS,\mM)} = \frac{ \mukl[r] - \rf[k] }{\eta_{n,k,\lambda}^{(\mM)}}\,.
	\label{eq:PD_SR}
\end{equation}

\noindent
\textbf{Path dependent, ESG-valued, log-return model.}
The dynamics of the stock price is modeled by the binomial tree
\begin{equation}
	S_{n,k+1,\lambda}^{(\mS,\mM)} =
	\begin{cases}
	S^{(\mS,\mM,\text{u})}_{n,k+1,\lambda} =
		 S_{n,k,\lambda}^{(\mS,\mM)} e^{ U_{n,k+1,\lambda}^{(\mS,\mM)} }\,,
		 	\text{ w.p. } \pMkl[k+1]\,, \\
	S^{(\mS,\mM,\text{u})}_{n,k+1,\lambda} =
		S_{n,k,\lambda}^{(\mS,\mM)} e^{ D_{n,k+1,\lambda}^{(\mS,\mM)} }\,,
			\text{ w.p. } 1-\pMkl[k+1]\,,
	\end{cases}
	\label{eq:PD_lpm}
\end{equation}
with
\begin{equation}
   \begin{aligned}
	U_{n,k+1,\lambda}^{(\mS,\mM)} &=
		\mukl[r]\, \dtn[k+1] + \eta_{n,k,\lambda}^{(\mM)}\, \pMkudl[k+1][u]\, \sqrt{\dtn[k+1]}\,, \\
	D_{n,k+1,\lambda}^{(\mS,\mM)} &=
		\mukl[r]\, \dtn[k+1] - \eta_{n,k,\lambda}^{(\mM)}\, \pMkudl[k+1][d]\, \sqrt{\dtn[k+1]}\,,
   \end{aligned}
   \label{eq:PD_lUD}	
\end{equation}
Note that, to terms of $O(\dtn[k+1])$, the instantaneous drift coefficient of the stock price is
$\mukl[r] + {\eta_{k,\lambda}^{\mathcal{(M)}}}^2/2$.
The risk-neutral valuation for the price of $\mC$ is
\begin{equation}
   \begin{aligned}
	f_{n,k,\lambda}^{(\mS,\mM)} &= e^{\rf[k] \dtn[k+1]}
			\left[ q_{n,k+1,\lambda}^{(\mS,\mM)} f_{n,k+1,\lambda}^{(\mS,\mM,\text{u})}
		 + \left(1-q_{n,k+1,\lambda}^{(\mS,\mM)}\right) f_{n,k+1,\lambda}^{(\mS,\mM,\text{d})} \right],\\
	q_{n,k+1,\lambda}^{(\mS,\mM)} &= \frac{ e^{\rf[k] \dtn[k+1]} -  e^{D_{n,k+1,\lambda}^{(\mS,\mM)}} }
							{  e^{U_{n,k+1,\lambda}^{(\mS,\mM)}} - e^{D_{n,k+1,\lambda}^{(\mS,\mM)}} }\,.
   \end{aligned}
   \label{eq:PD_lfq}
\end{equation}

\subsection{Fitting the ESG-valued path-dependent model to market data.}
\label{sec:PD_fit}

\noindent
Consider the market returns $r^{(\mM)}_{n,k,0}$, market ESG values $\ek[M]$, and a chosen value $\lambda$
over a historical time period denoted by $[0,T]$ comprised of constant time periods
$\dtn[k] = \Delta t = T/n$, $k = 1, ..., n$.
The ESG-valued returns \eqref{eq:PD_RM} generate a given series of values $\ZMkl$,
and consequently of the values $\zMkl$.
We consider two models for the functions $h(x)$ and $g(x)$ where both are described by
the Student's $t$ distribution with (a) $df = 5$ and (b) $df = 50$ (approximating standard normal). 
The coefficient $\mu_{n,T,\lambda}^{\text{(r)}}$ can be fit directly using
\begin{equation}
	\mathbb{E}_{n,k \in [1,n]} \left[ r^{(\mS)}_{n,k,\lambda} \right] = \mu_{n,T,\lambda}^{\text{(r)}} \Delta t\,,
		\label{eq:PD_mu}
\end{equation}
where
\begin{equation}
	r^{(\mS)}_{n,k,\lambda} = \lambda  \ek[S] + (1-\lambda) r^{(\mS)}_{n,k,0}\,,
	\label{eq:PD_RS}
\end{equation}
and $r^{(\mS)}_{n,k,0}$ and $\ek[S]$ represent realized return and ESG values for $\mS$ over the historical period.
Let $S^{(\mS)}_{n,k,\lambda}$ denote the corresponding realized ESG price series.
The model predictions for the return and prices over this historical time period are
\begin{equation}
   \begin{aligned}
	r^{(\mS,\mM)}_{n,k+1,\lambda} 
		& = \mu_{n,T,\lambda}^{\text{(r)}} \Delta t+ \eta_{n,k,\lambda}^{\mathcal{(M)}} \zMkl \sqrt{\Delta t}\,, \\
	S^{(\mS,\mM)}_{n,k,\lambda} & = S_0 
	\begin{cases}
		 \prod_{l=1}^{k}\left(1+ r^{(\mS,\mM)}_{n,l,\lambda}\right),\  (\text{arithmetic return model})\,,\\
		\exp\left(\sum_{l=1}^{k} r^{(\mS,\mM)}_{n,l,\lambda}\right),\  (\text{log-return model})\,,
	\end{cases}
   \end{aligned}
   \label{eq:hist_mdl}
\end{equation}
where $\eta_{n,k,\lambda}^{\mathcal{(M)}}$ is defined in \eqref{eq:PD_rtn} using $\dtn[j] = \Delta t, j = 1, ..., n$.
The unknown coefficients $c_{1,\lambda}, c_{2,\lambda}$ and $c_{3,\lambda}$ can be obtained via the minimization
\begin{equation}
	\min_{c_1, c_2, c_3} || S^{(\mS,\mM)}_{n,k,\lambda}-S^{(\mS)}_{n,k,\lambda} ||_2^2\,.
	\label{eq:c_min}
\end{equation}

We constructed a $\lambda$-dependent market index using the assets comprising the Dow Jones Industrial Average (DJIA).
The RobecoSAM ESG values relevant for the year 2021 for the 30 stocks in the DJIA are given in Table~\ref{tab:ESG_DJIA}.
\begin{table}[ht]
	\caption{ S\&P Global RobecoSAM ESG ratings of Dow Jones stocks}
	\label{tab:ESG_DJIA}
	\begin{center}
	\begin{tabular}{lcc lc clcc}
		\toprule
		Stock  & \ \ 11/19/ & \ \ 11/19/ & Stock  & \ \ 11/19/ & \ \ 11/19/ & Stock & \ \ 11/19/ & \ \ 11/19/ \\
		\omit  & \ \ 2020   & \ \ 2021    & \omit  & \ \ 2020   & \ \ 2021    & \omit & \ \ 2020    & \ \ 2021    \\
		\midrule
		AAPL  & 25   & 34  & 	AMGN & 95  & 97 &	AXP   & 68   & 76 \\
		BA     & 78   & 78   &	CAT   & 94  & 94  & 	CRM   & 91  & 94  \\
	 	CSCO & 100 & 100 &  	CVX   & 54  & 53  &	DIS    & 72  & 76  \\
	  	DOW  & 93  & 93  &	GS     & 71  & 86  & 	HD     & 65  & 79  \\
		HON	  & 48  & 52 &  	IBM   & 69  & 73  &	INTC  & 86  & 90 \\
		JNJ    & 84  & 78  &	JPM   & 55 & 60  & 	KO     & 64 & 68  \\
		MCD  & 76  & 79  &  	MMM & 91 & 91  &	MRK  & 73 & 77 \\
		MSFT & 96  & 98  &	NKE   & 80 & 65 & 	PG     & 64 & 72 \\
		TRV   & 50  & 41 & 	UNH   & 96 & 97 &		V       & 82 & 84 \\
		VZ     & 49  & 48 &		WBA  & 81 & 86 & 	WMT & 68 & 82 \\
		\bottomrule                      
	\end{tabular}
	\end{center}
\end{table}
To avoid any possibility of $\lambda$-arbitrage, we assume the value of the ESG intensity $\lambda$ is determined
by the market and is the same for all stocks in the index as well as for the stock $\mS$ underlying the call option.
The ESG-valued return series for our market index is constructed in a straightforward manner by combining the
ESG-valued returns for each component asset using an equal-weighted strategy.

\begin{figure}[h]
\begin{center}   
    	\includegraphics[width=.75\textwidth]{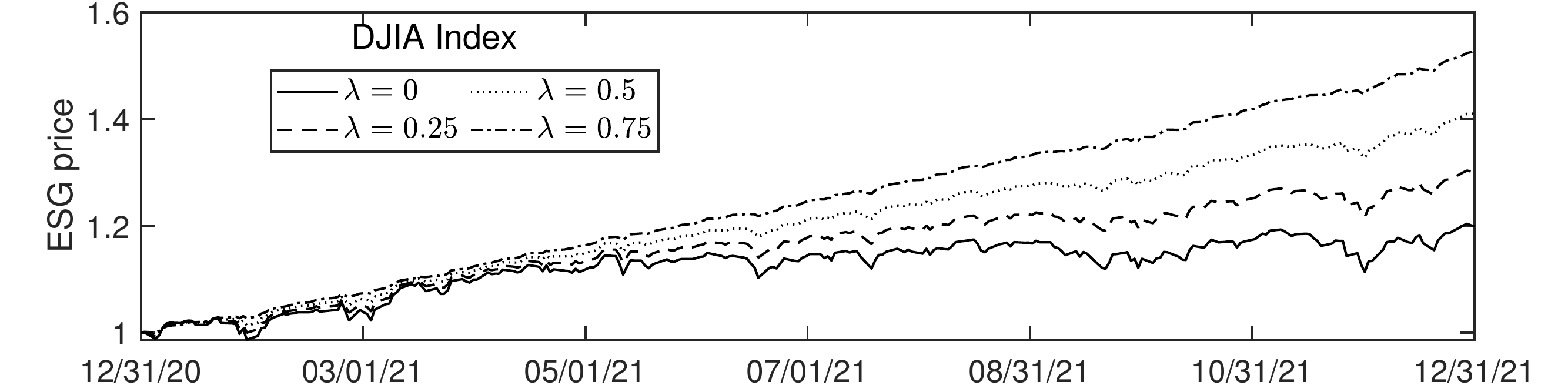}   
    \caption{ESG-valued DJIA index price dynamics for example values of $\lambda$.
    The initial index value is set to one ESG-valued unit of account.}
    \label{fig_DJIAprice}
\end{center}
\end{figure}
Fig. \ref{fig_DJIAprice} exhibits the ESG-valued price dynamics of this $\mM=\text{DJIA}$ index over the period
from 1/1/2021 to 12/31/2021 for $\lambda \in\{0, 0.25, 0.5, 0.75\}$ using log-returns in \eqref{eq:PD_RM}.
As previously, we set the starting value of the ESG-valued numeraire of the index to one unit of account.
The dependence of this index on $\lambda$ is very similar to that for MSFT in Fig.~\ref{fig:ESG_prices}.
Estimates of the expected value and standard deviation of the ESG-valued log-returns $r_{n,k,\lambda}^{(\text{DJIA})}$ 
for this market index over this time period are presented in Table~\ref{tab:DJIinfo}.
\begin{table}[h]
	\caption{Measures based on the return series and price dynamics of the DJIA index over the year 2021.
	SR: Sharpe ratio computed using the 10-year Treasure yield curve rate as the risk-free rate.
	MDD: maximum drawdown.}
	\label{tab:DJIinfo}
	\begin{center}		
	\begin{tabular}{l ccrc ccrc}
		\toprule
		$\lambda$ & \multicolumn{4}{l}{arithmetic return} & \multicolumn{4}{l}{log-return} \\
		\               & $\mathbb{E}\left[ r_{n,k,\lambda}^{(\text{DJIA})} \right]$
				& $\sqrt{ \text{Var}\left[ r_{n,k,\lambda}^{(\text{DJIA})} \right] }$ & SR\ \  & MDD
				& $\mathbb{E}\left[ r_{n,k,\lambda}^{(\text{DJIA})} \right]$
				& $\sqrt{ \text{Var}\left[ r_{n,k,\lambda}^{(\text{DJIA})} \right] }$ & SR\ \  & MDD\\
		\               & \omit & \omit & \omit & (\%) & \omit & \omit & \omit & (\%)\\
		\midrule
		0       & $7.2\cdot 10^{-4}$ & $7.6\cdot 10^{-3}$ & 0.09 & 6.7
			   & $6.2\cdot 10^{-4}$ & $7.6\cdot 10^{-3}$ & 0.08 & 6.8\\
		0.25  & $1.0\cdot 10^{-3}$ & $5.7\cdot 10^{-3}$ & 0.17 & 4.2
			   & $9.5\cdot 10^{-4}$ & $5.7\cdot 10^{-3}$ & 0.16  & 4.4\\
		0.5   & $1.3\cdot 10^{-3}$ & $3.8\cdot 10^{-3}$ & 0.34 & 2.1
			  & $1.3\cdot 10^{-3}$ & $3.8\cdot 10^{-3}$ & 0.33  & 2.1\\
		0.75 & $1.6\cdot 10^{-3}$ & $1.9\cdot 10^{-3}$ & 0.84 & 0.6
			  & $1.6\cdot 10^{-3}$ & $1.9\cdot 10^{-3}$ & 0.83  & 0.6\\
		\bottomrule
	\end{tabular}
	\end{center}
\end{table}
Also shown are computed values of the $\lambda$-dependent Sharpe ratio over this 1-year period,
as well as the $\lambda$-dependent maximum drawdown seen during the year.
As the ESG scores for each component of the index are essentially constant (changing only on 11/19/2021), 
the variance $\mathbb{E}\left[ r_{n,k,\lambda}^{(\text{DJIA})} \right]$ decreases as $\lambda$ increases,
as does the maximum drawdown.
In contrast, the expected value $\mathbb{E}\left[ r_{n,k,\lambda}^{(\text{DJIA})} \right]$ increases with $\lambda$,
consequently so does the Sharpe ratio.
There are minor differences between these measures based upon arithmetic and log-returns.
The differences decrease as $\lambda$ increases as the weighting shifts to the normalized ESG score.

\begin{figure}[h!]
\begin{center}
		\includegraphics[width=0.49\textwidth]{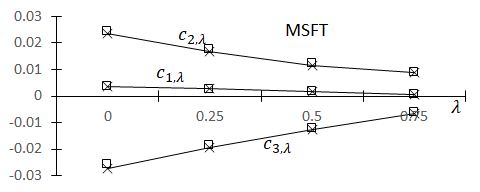}
		\includegraphics[width=0.49\textwidth]{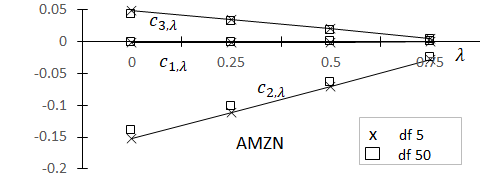}
		\includegraphics[width=0.49\textwidth]{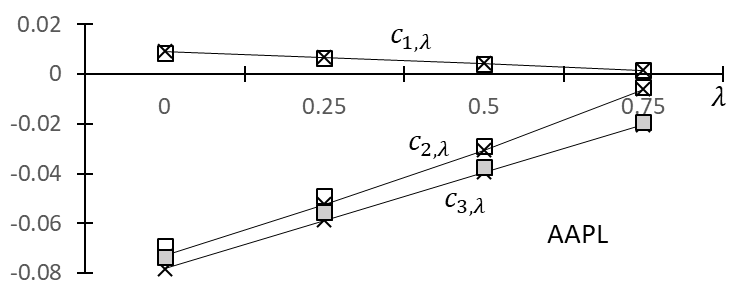}
	\caption{Estimated coefficients $c_{i,\lambda}, i = 1, 2, 3$ computed from the minimization  \eqref{eq:c_min},
		based on the path dependent, arithmetic return model. 
		The functions $h(\cdot)$ and $g(\cdot)$ are student's $t$ with indicated degrees of freedom (df).
		The solid lines serve to ``guide the eye''.
	}
	\label{fig:c123}
\end{center}
\end{figure}

The parameter estimates for $\mu_{n,T,\lambda}^{\text{(r)}}$ can be determined from the data in Table \ref{tab:params}.
The estimated coefficients from the minimization \eqref{eq:c_min} using loarithmetic returns are plotted in Fig.~\ref{fig:c123}.
The coefficient estimations using log-returns are virtually identical to that shown in Fig.~\ref{fig:c123}.
Interestingly, the values of the coefficients are not significantly different when the functions
$h(\cdot)$ and $g(\cdot)$ are modeled as Student's $t$ with 5 or 50 degrees of freedom
and what differences are seen are stock dependent.
The magnitude of every coefficient decreases as $\lambda$ increases;
the decrease in magnitude is approximately linear with $\lambda$.
The $c_{1,\lambda}$ coefficient has the smallest magnitude in all cases.
For MSFT, $c_{2,\lambda}$ is positive while $c_{3,\lambda}$ is negative,
both having approximately the same magnitude (for the same value of $\lambda$).
For AMZN, $c_{2,\lambda}$ is negative while $c_{3,\lambda}$ is positive,
with the magnitude of $c_{2,\lambda}$ approximately three times that of  $c_{3,\lambda}$.
For AAPL both are negative, with  $c_{2,\lambda}$ having somewhat smaller magnitude.
\begin{figure}[h!]
\begin{center}
    \begin{subfigure}[b]{0.6\textwidth} 
	\includegraphics[width =\textwidth]{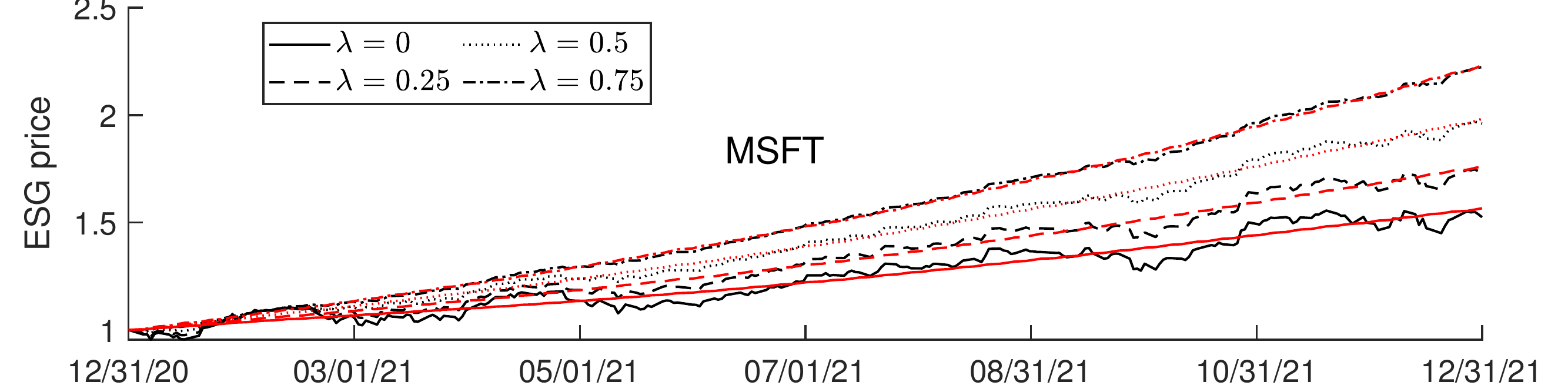}
    \end{subfigure}
    \begin{subfigure}[b]{0.6\textwidth} 
	\includegraphics[width=\textwidth]{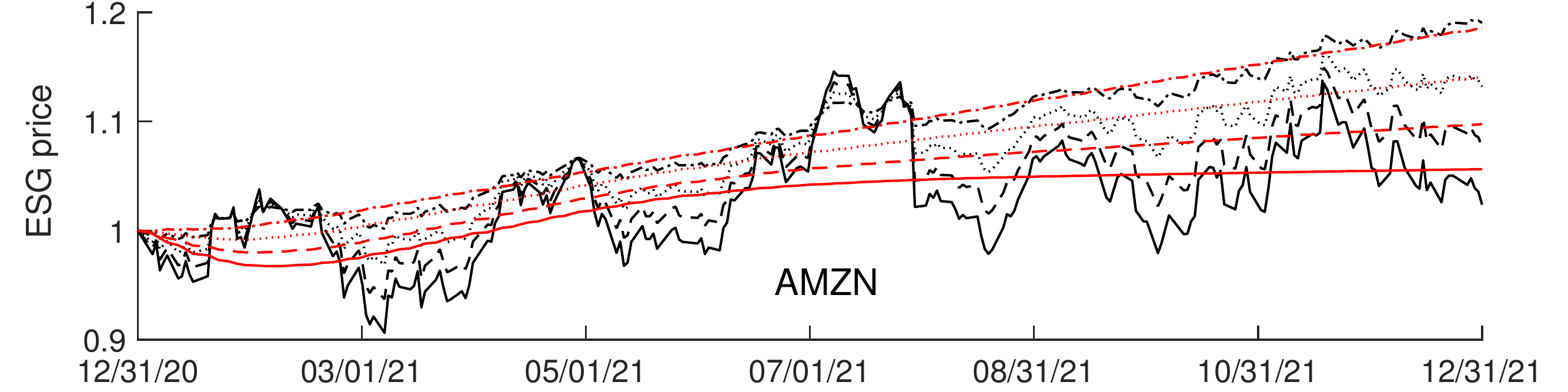}
    \end{subfigure}
    \begin{subfigure}[b]{0.6\textwidth} 
	\includegraphics[width=\textwidth]{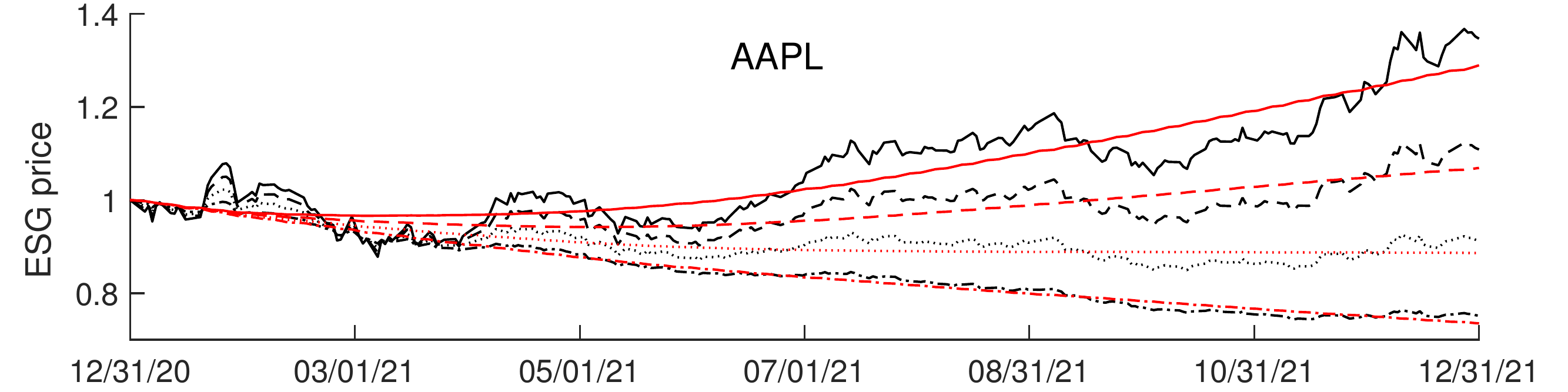}
    \end{subfigure}
	\caption{ESG-valued stock prices (black lines) and the corresponding model fits (red lines) for chosen values of $\lambda$.
		The fits are based upon results for arithmetic returns with $df = 5$.
		For comparison purposes, the initial ESG prices (on 12/30/2020) are scaled to unit value.
		}
	\label{fig:Pfits}
\end{center}
\end{figure}

Fig.~\ref{fig:Pfits} compares the ESG price series for each stock over the historical time period to the
fitted price series \eqref{eq:hist_mdl}.
The fits shown are using results for arithmetic returns with $df = 5$.
The fits based upon log-returns or with $df = 50$ are virtually identical.

\begin{figure}[ht]
\begin{center}
    \begin{subfigure}[b]{0.32\textwidth} 
    	\includegraphics[width=\textwidth]{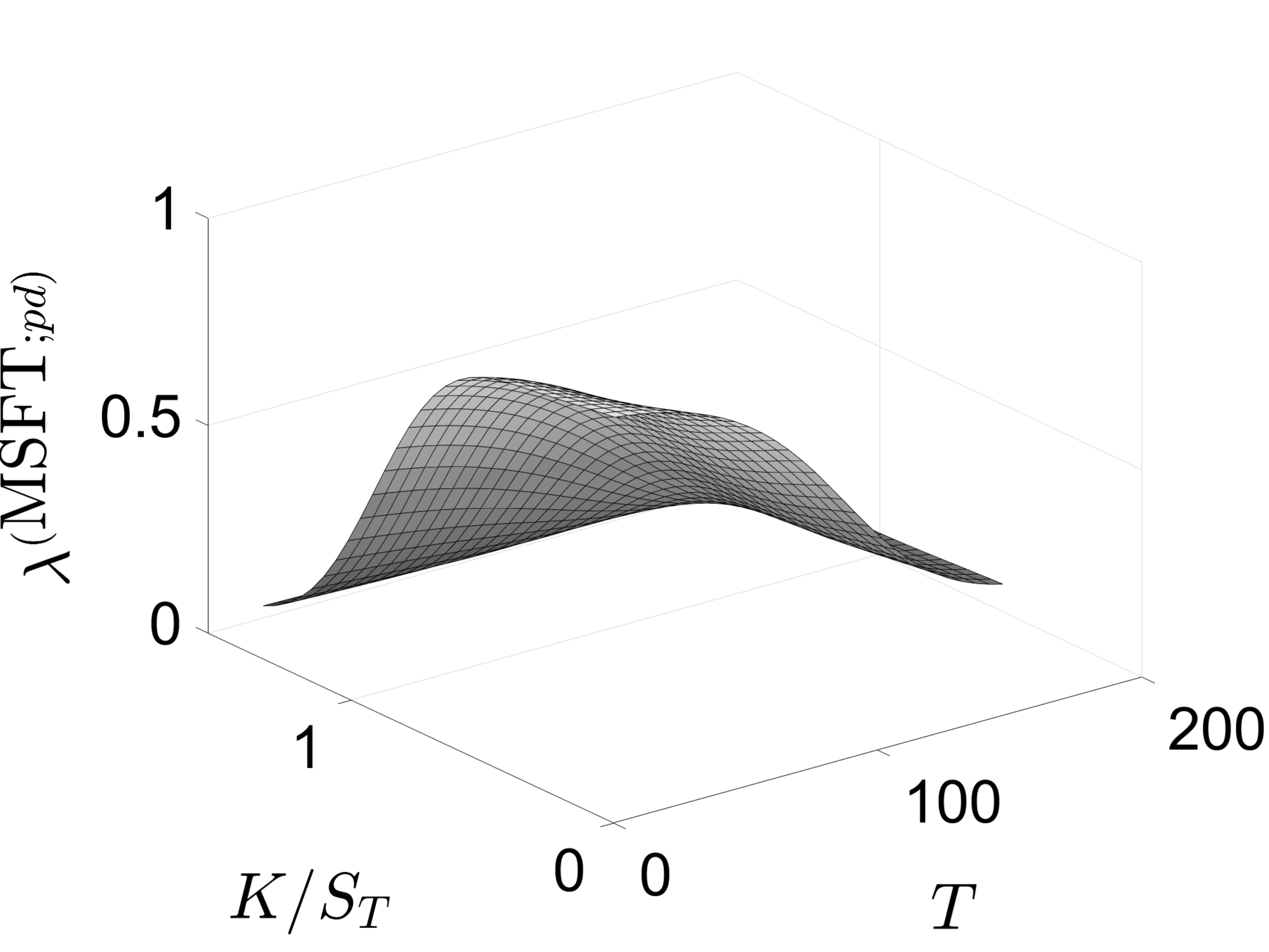}
    \end{subfigure}
    \begin{subfigure}[b]{0.32\textwidth} 
    	\includegraphics[width=\textwidth]{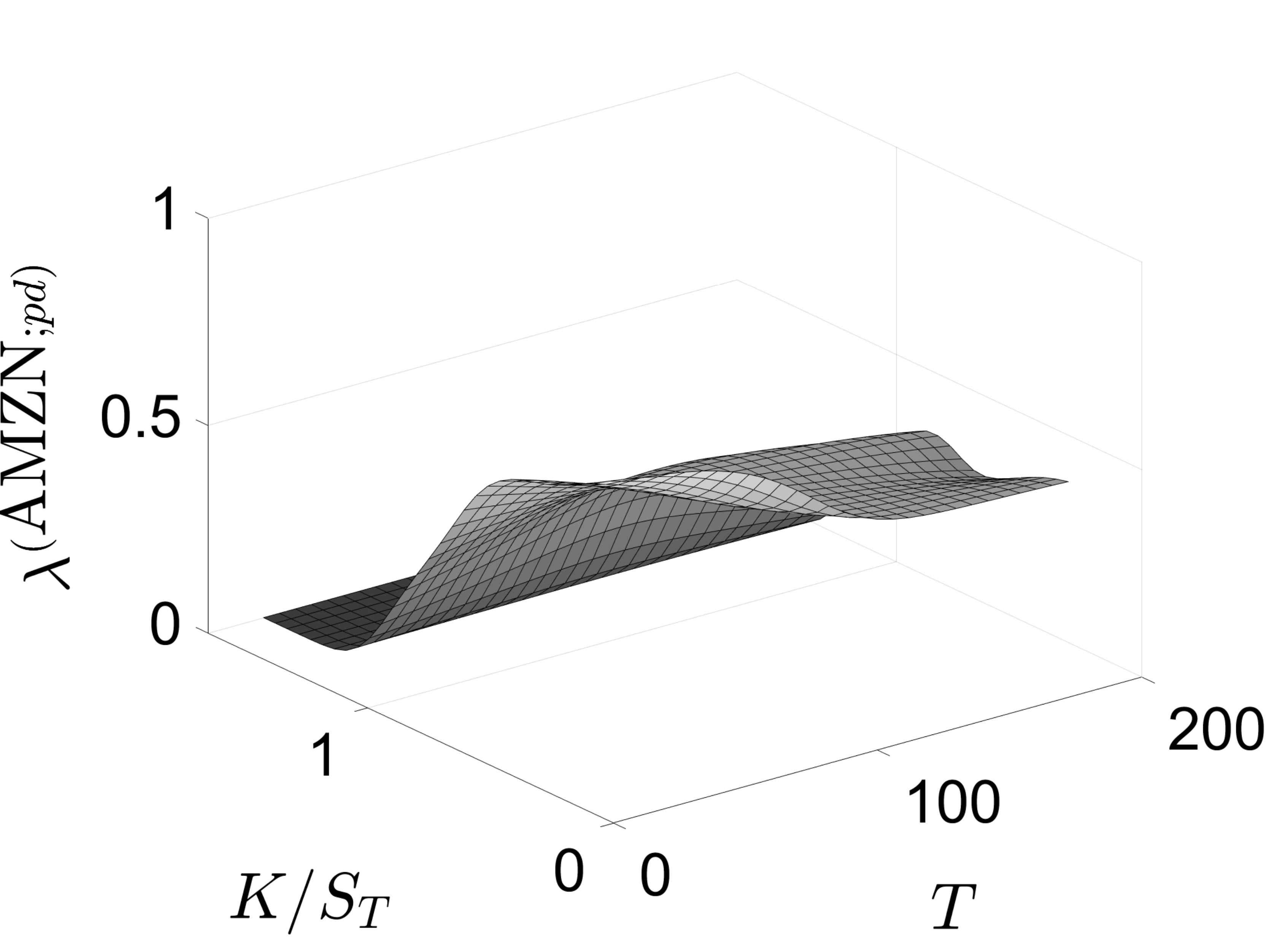}
    \end{subfigure}
    \begin{subfigure}[b]{0.32\textwidth} 
    	\includegraphics[width=\textwidth]{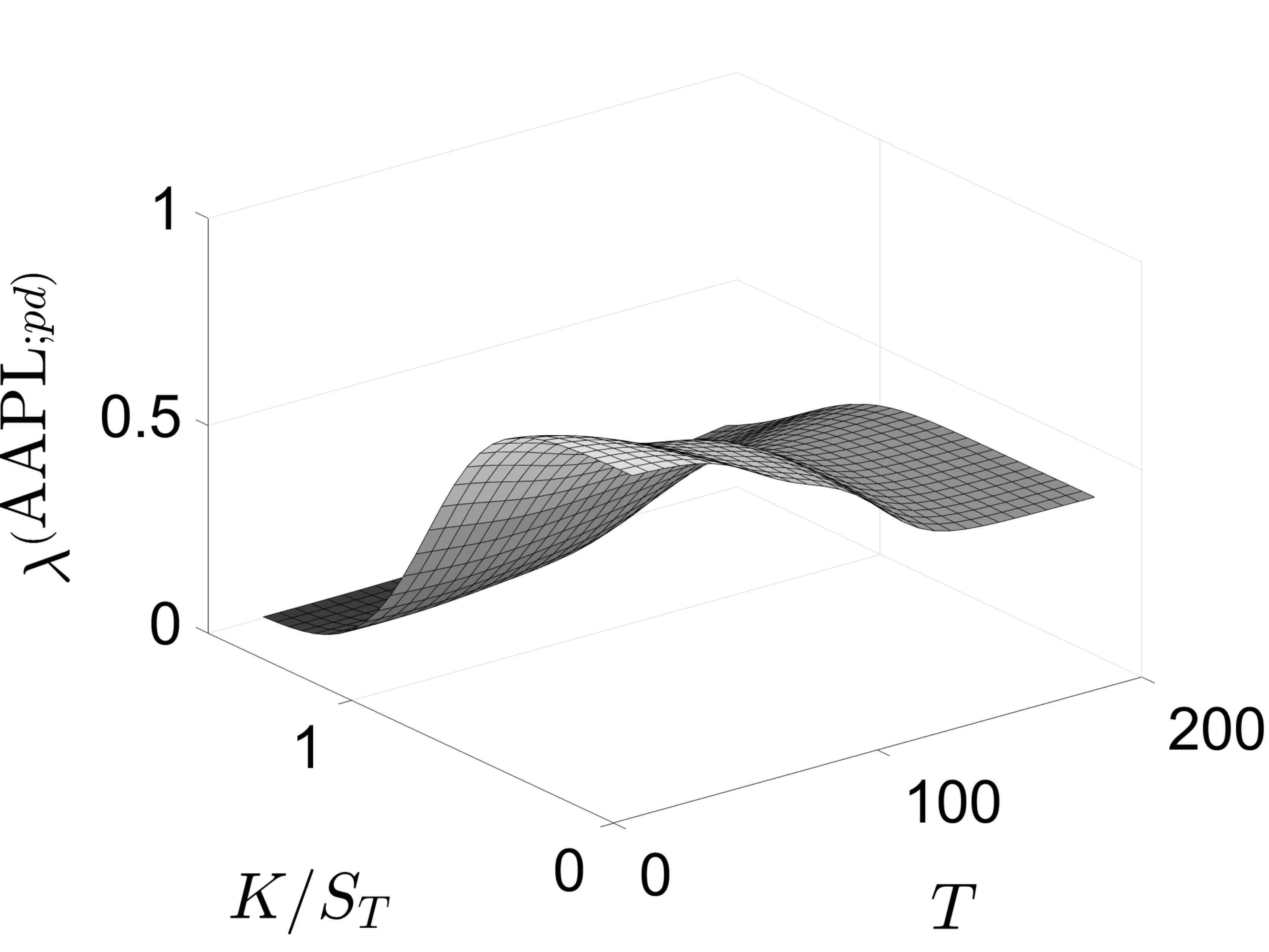}
    \end{subfigure}
        \begin{subfigure}[b]{0.32\textwidth} 
    	\includegraphics[width=\textwidth]{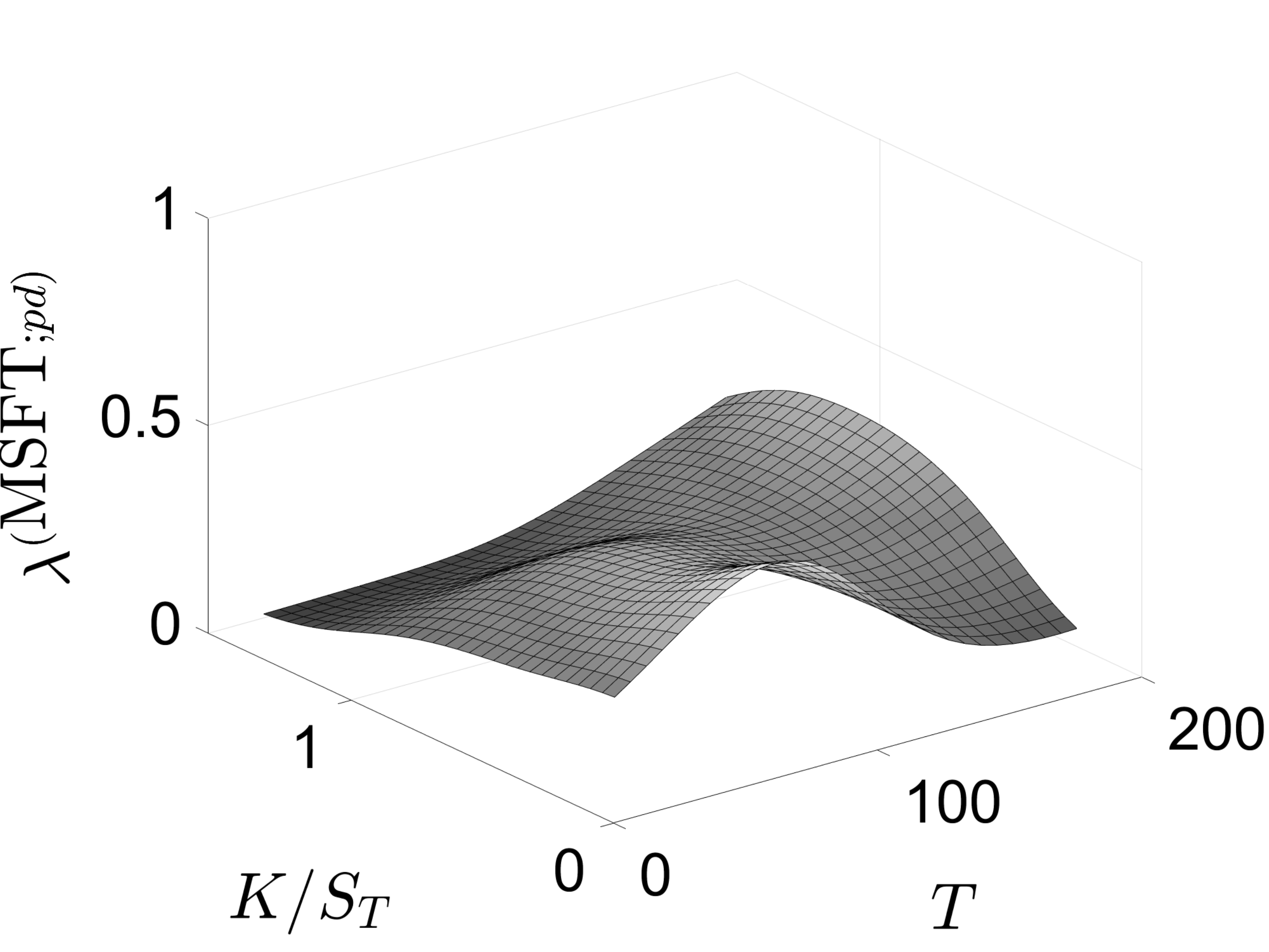}
    \end{subfigure}
    \begin{subfigure}[b]{0.32\textwidth} 
    	\includegraphics[width=\textwidth]{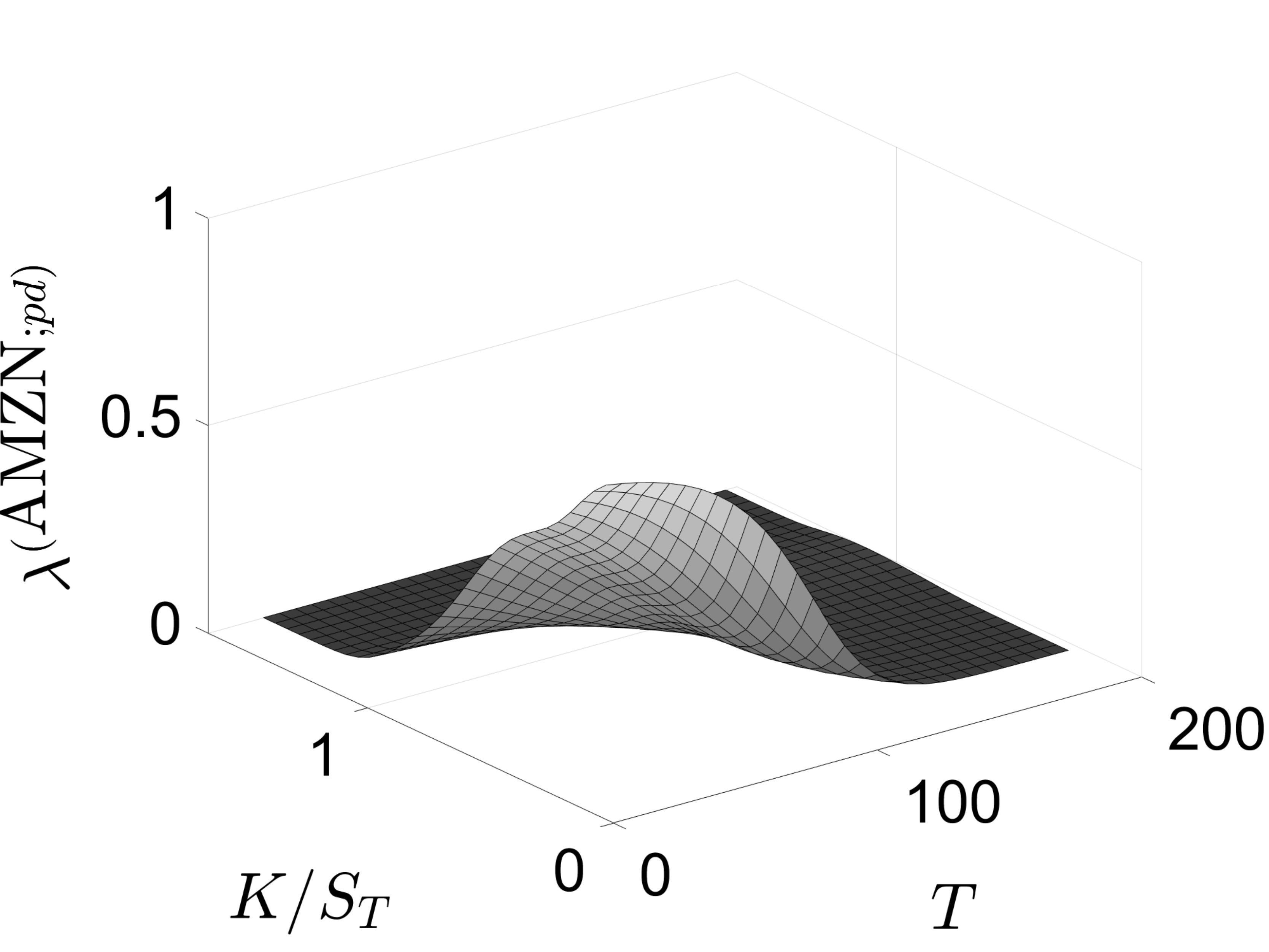}
    \end{subfigure}
    \begin{subfigure}[b]{0.32\textwidth} 
    	\includegraphics[width=\textwidth]{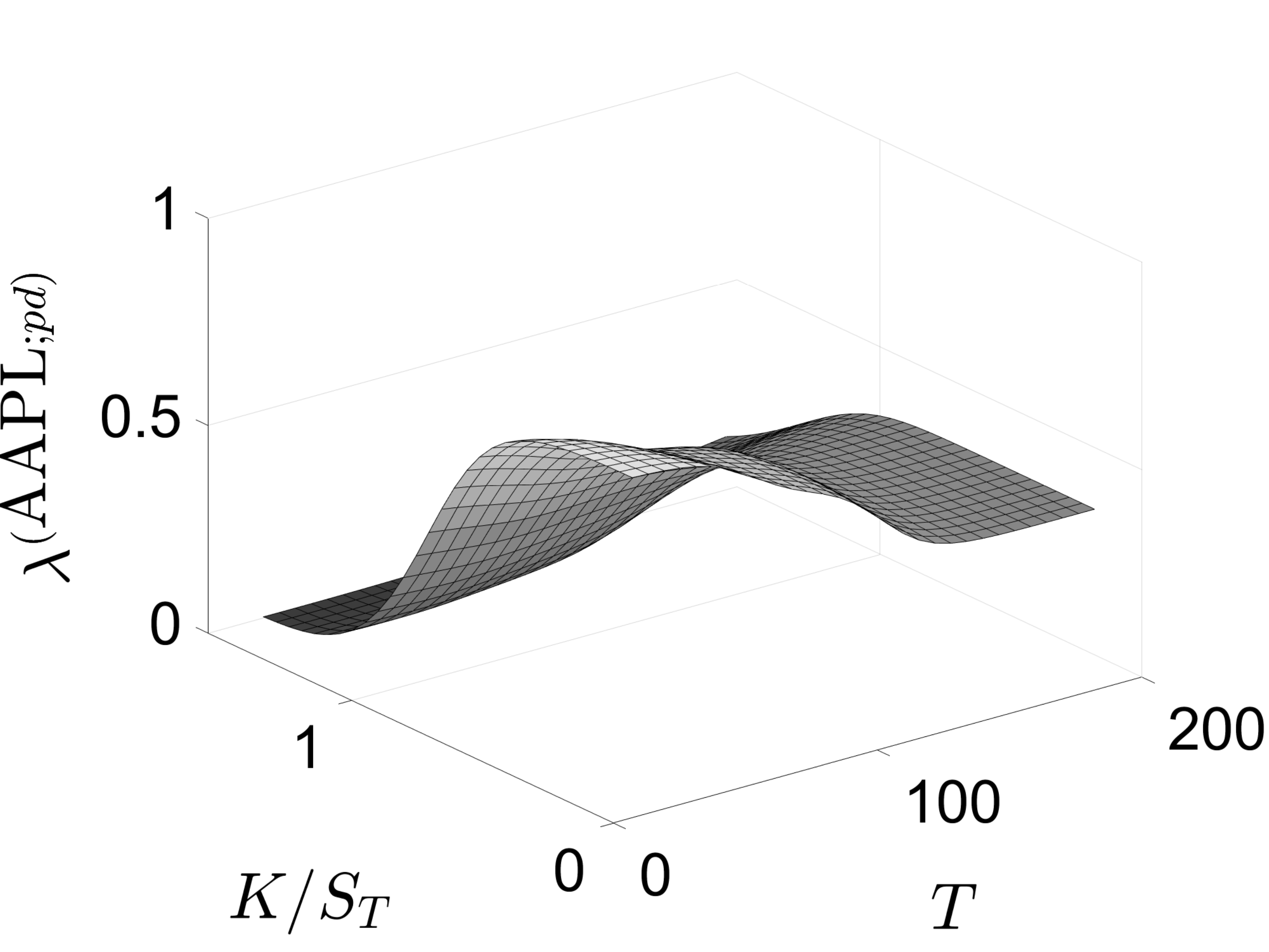}
    \end{subfigure}
    \caption{The implied $\lambda$-surfaces against moneyness $M = K/S_T$ and time to maturity $T$ in days
		for the path dependent, (top row) arithmetic model \eqref{eq:PD_dpm} and (bottom) log-return model
		\eqref{eq:PD_lpm} in which the functions $h(\cdot)$ and $g(\cdot)$
		are modeled as Student's $t$ with $df = 5$.
    }
    \label{fig:PD_ilam}
\end{center}
\end{figure}

\noindent
\textbf{Implied ESG intensity surfaces, path dependent models.}
Using the same market option price data for MSFT, AMZN and APPL as in section \ref{sec:lam_ex}
and the same relative mean-square error minimization \eqref{eq:l_imp_MS},
we computed implied $\lambda$-surfaces for each of these three stocks
(only two of which, MSFT and AAPL, are component assets in our index).
The surfaces computed from the arithmetic and log-return models with $h(\cdot)$ and $g(\cdot)$
modeled as Student's $t$ with $df = 5$ are presented in Fig.~\ref{fig:PD_ilam}.
The path dependent model results in Fig.~\ref{fig:PD_ilam} are markedly different from those
in  Fig.~\ref{fig:impl_lam}.
(Interestingly, as in  Fig.~\ref{fig:impl_lam}, for AAPL there are only slight differences between the $\lambda$
surfaces computed via the two models.)
While values of $\lambda$ generally approach unity (for far into-the-money values of $M$)
in Fig.~\ref{fig:impl_lam},
under the path dependent model, values of $\lambda$ are generally suppressed,
noticeable so over regions of in-the-money values of $M$.
There is also greater dependence on time to maturity $T$ in the path dependent results.
By using Student's $t$ with five degrees of freedom for $h(\cdot)$ and $g(\cdot)$, the path
dependent models are sensitive to differences in the events in the tails of the return distributions
of individual stocks; hence the behavior of the path dependent model results for
(the suppression of the value of) $\lambda(T,M)$ varies with stock.

\section{Discussion}
\label{sec:disc}

\noindent
The formulation of the ESG-valued return \eqref{eq:ESG_rtn}, which places a scaled ESG score on equal footing
(relative to the ESG intensity weight factor) with financial return, has profound financial implication,
both theoretically and practically.
\begin{itemize}
\item
This ESG-valued return is ``true to the market'', in that it is a numeraire that incorporates an ESG component value
that is independent of financial considerations.
This has a secondary benefit.
For a company whose ESG score is currently near or at the maximum rating value, there is little incentive to improve.
However, the ESG-valued return produces an ESG-valued price that can grow with no bound.
\item
Consider the implication of an ESG valuation of a company that is updated on a frequency comparable with return data.
Couple this with the (very real) scenario that multiple agencies are providing valuations for the same company using
different methodologies.
The variance in this ESG valuation, both longitudinally (in time) and in cross section (across rating agencies) produces
an ESG time series whose volatility can be as profound as that of its return.
The ESG-valued return formulation \eqref{eq:ESG_rtn} is capable of analyzing this combined volatility.
\item
Incorporating ESG into a return enables its incorporation into the existing dynamic asset pricing framework,
as we have demonstrated here using discrete, binomial option pricing models.
\item
This ESG-valued return formulation goes directly to the heart of an SRI question: ``Is there a valuation of
SRI investing that is more than just financial return that the marketplace will accept?''
We, of course, don't know the answer to the second part of this question, but our formulation proposes a model
for the first part.
\end{itemize}

By considering discrete option pricing models based (correctly) upon arithmetic returns and (incorrectly) upon log-returns,
we have been able to quantify the type and extent of errors that occur using log-returns.
Implied volatility surfaces can be dramatically different computed using discrete models with log-returns.
In addition, dramatically different implied volatility surfaces result when computed under the continuum assumptions
of the Black-Scholes-Merton model.
When extending the discrete option pricing models to include informed traders,
computation of the implied information intensity was also dramatically different under the arithmetic and log-return models.
While differences in the implied ESG surfaces calculated from the two models in section~\ref{sec:lam_ex} were somewhat
ameliorated,
the implied ESG surface differences increased under the path-dependent formulation in section~\ref{sec:PD_fit}.
Essentially, as additional detail is added to the models, we find that errors derived from the incorrect use
of log-returns increase.

\clearpage
\begin{appendices}

\renewcommand{\theequation}{A.\arabic{equation}}
\setcounter{equation}{0}
\renewcommand{\thefigure}{A\arabic{figure}}
\setcounter{figure}{0}
\renewcommand{\thetable}{A\arabic{table}}
\setcounter{table}{0}

\section{Implied volatility surfaces}
\label{app:A}

\begin{figure}[h!]
\begin{center}

    \begin{subfigure}[b]{0.32\textwidth} 
    	\includegraphics[width=\textwidth]{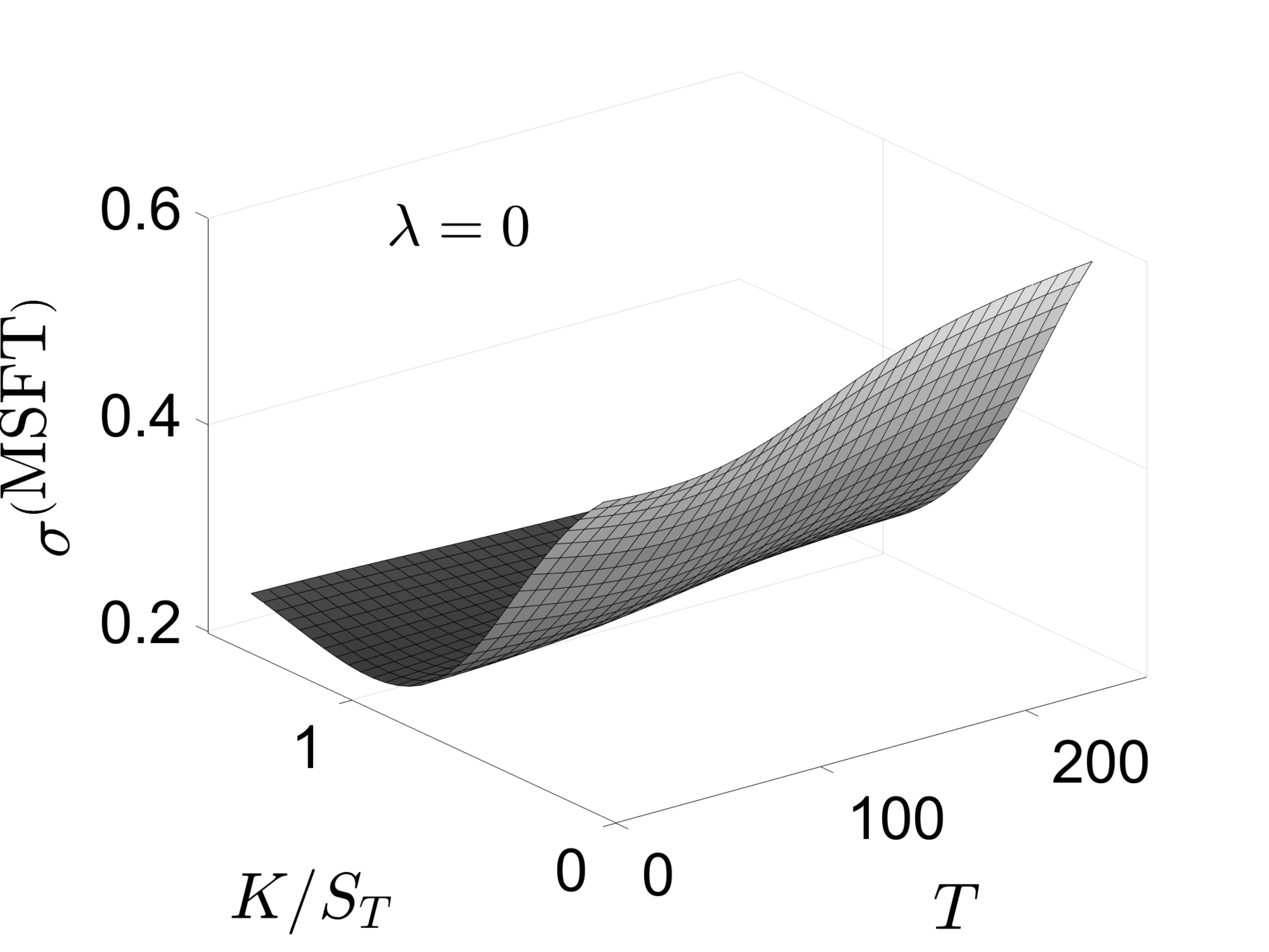}
    \end{subfigure}
    \begin{subfigure}[b]{0.32\textwidth} 
    	\includegraphics[width=\textwidth]{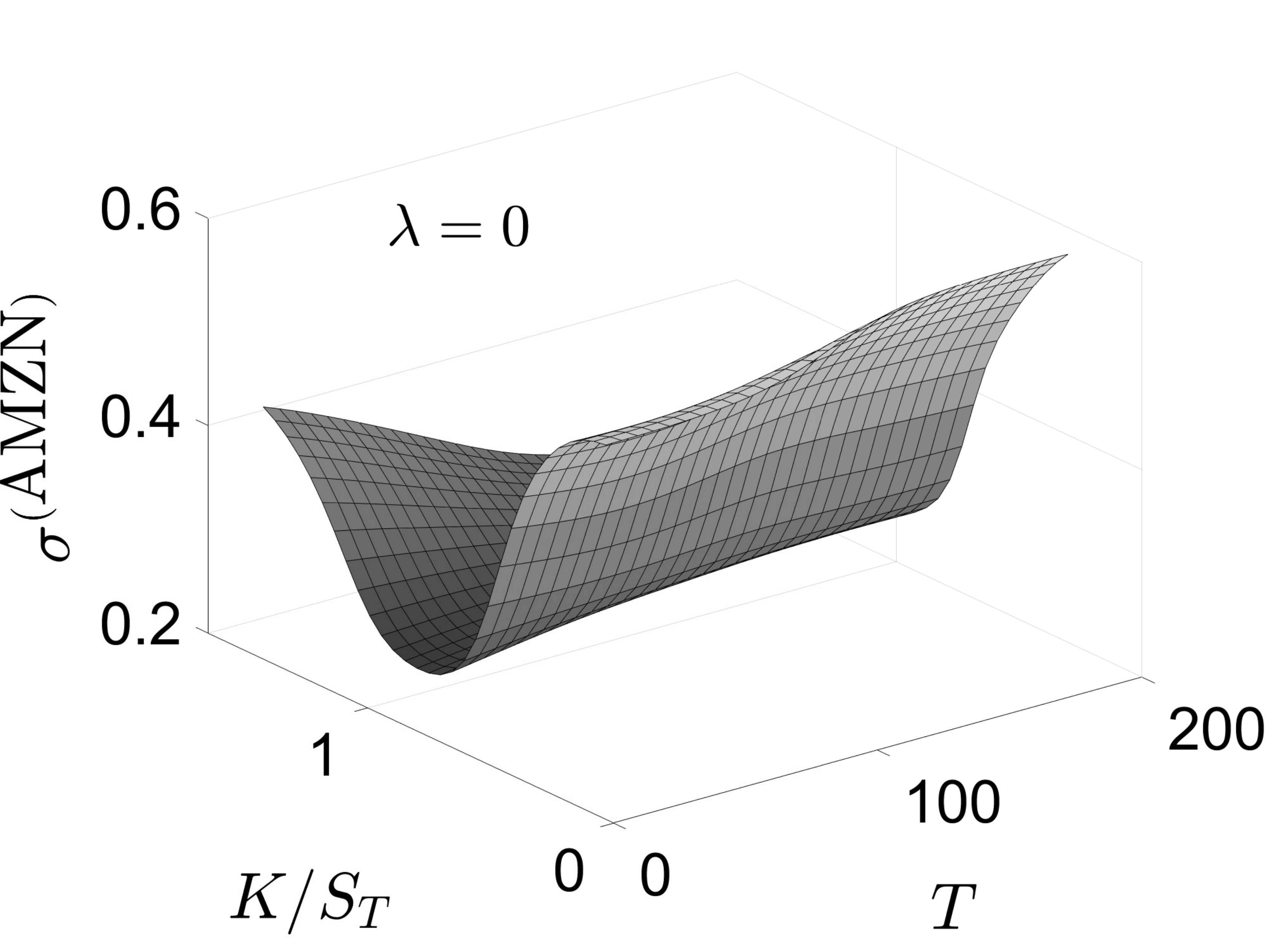}
    \end{subfigure}
        \begin{subfigure}[b]{0.32\textwidth} 
    	\includegraphics[width=\textwidth]{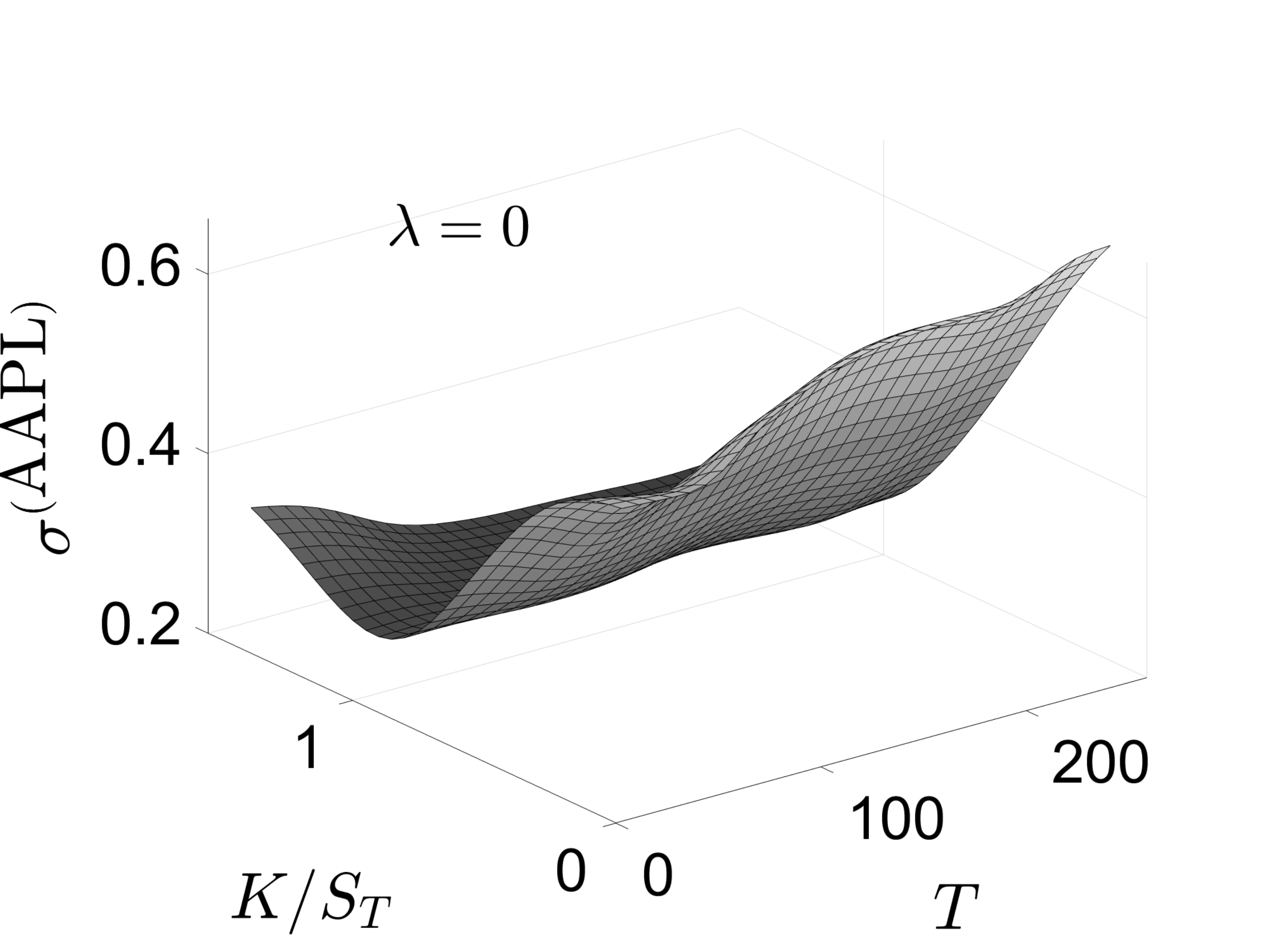}
    \end{subfigure}

    \begin{subfigure}[b]{0.32\textwidth} 
    	\includegraphics[width=\textwidth]{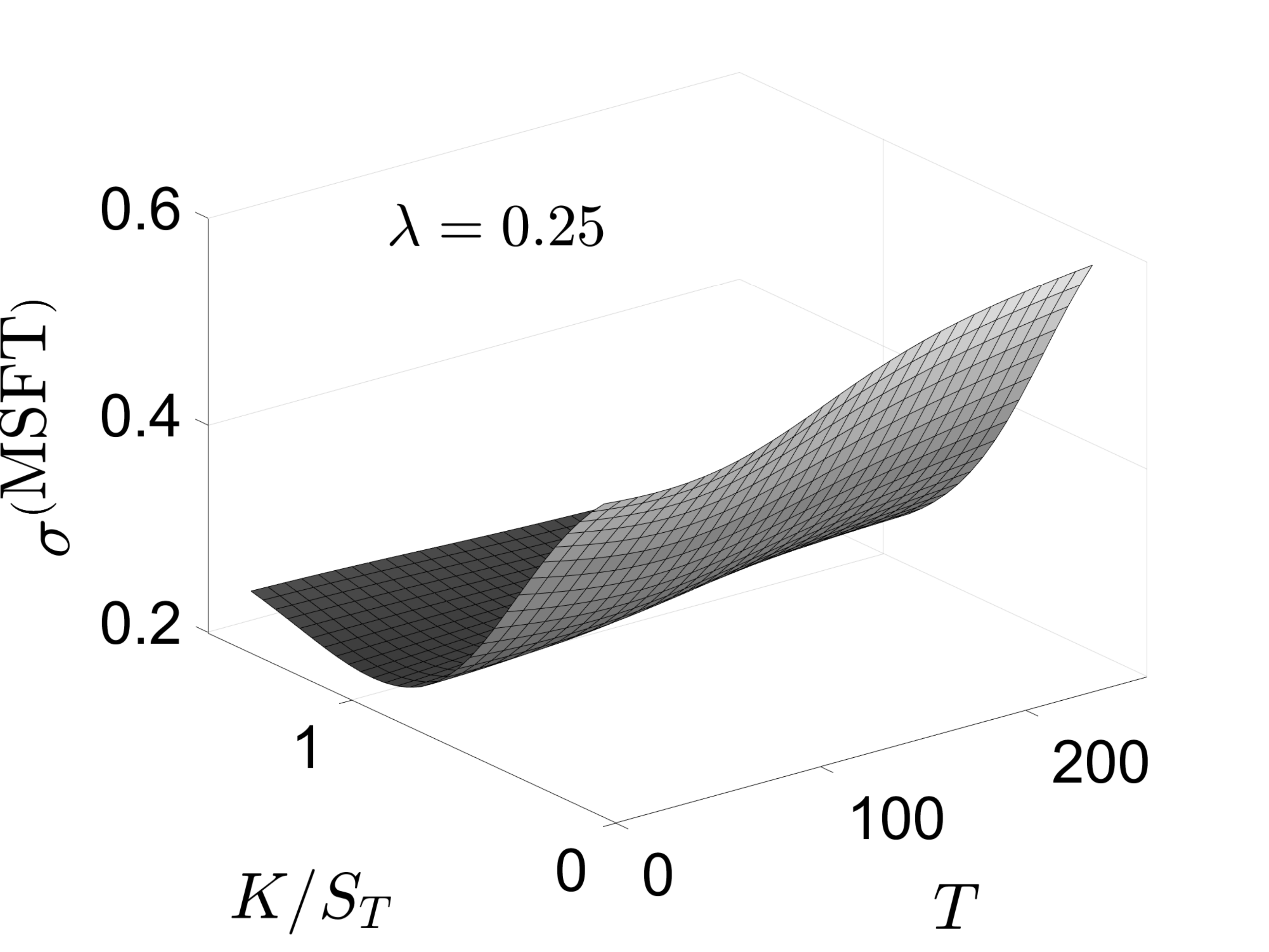}
    \end{subfigure}
    \begin{subfigure}[b]{0.32\textwidth} 
    	\includegraphics[width=\textwidth]{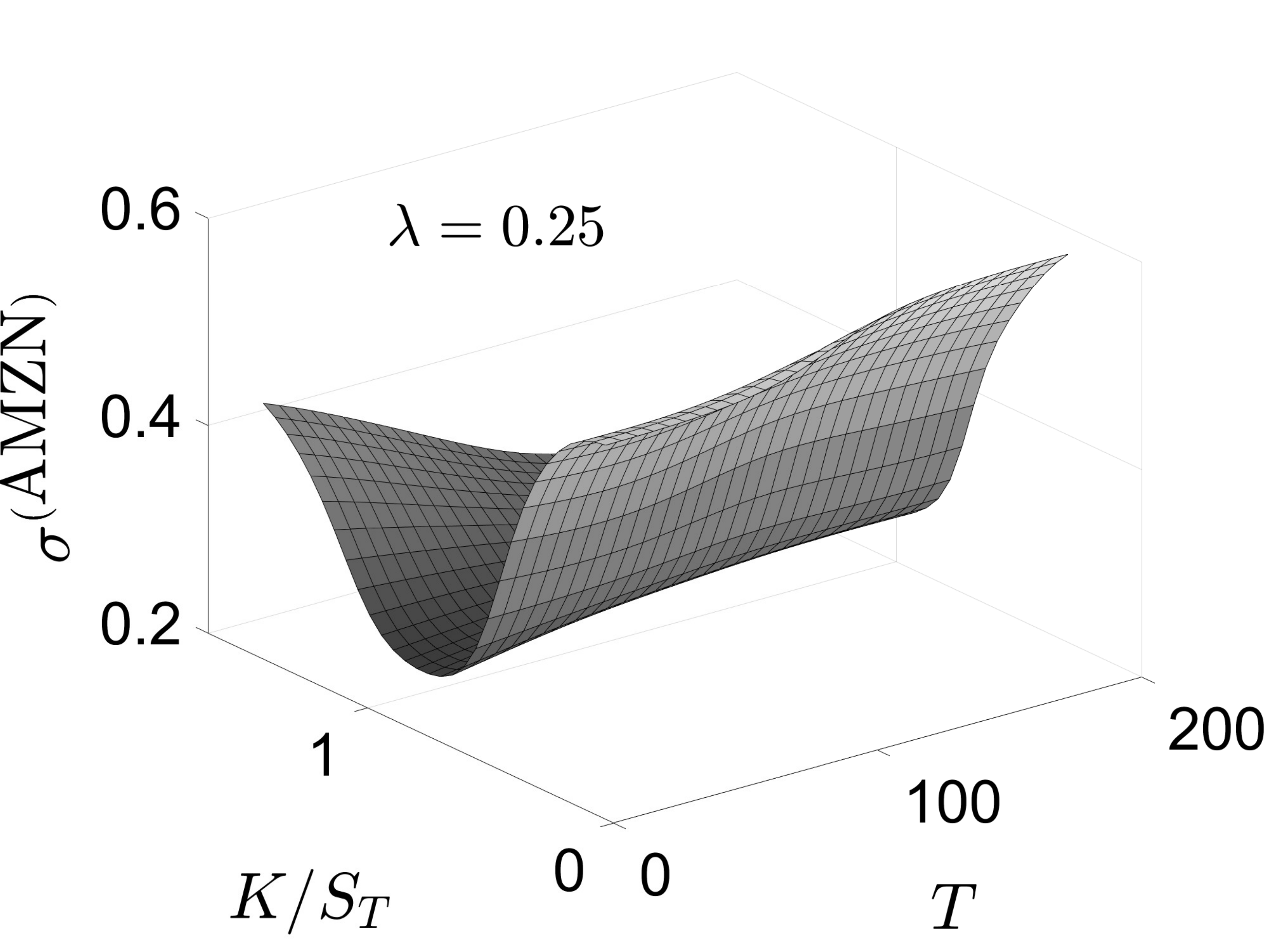}
    \end{subfigure}
        \begin{subfigure}[b]{0.32\textwidth} 
    	\includegraphics[width=\textwidth]{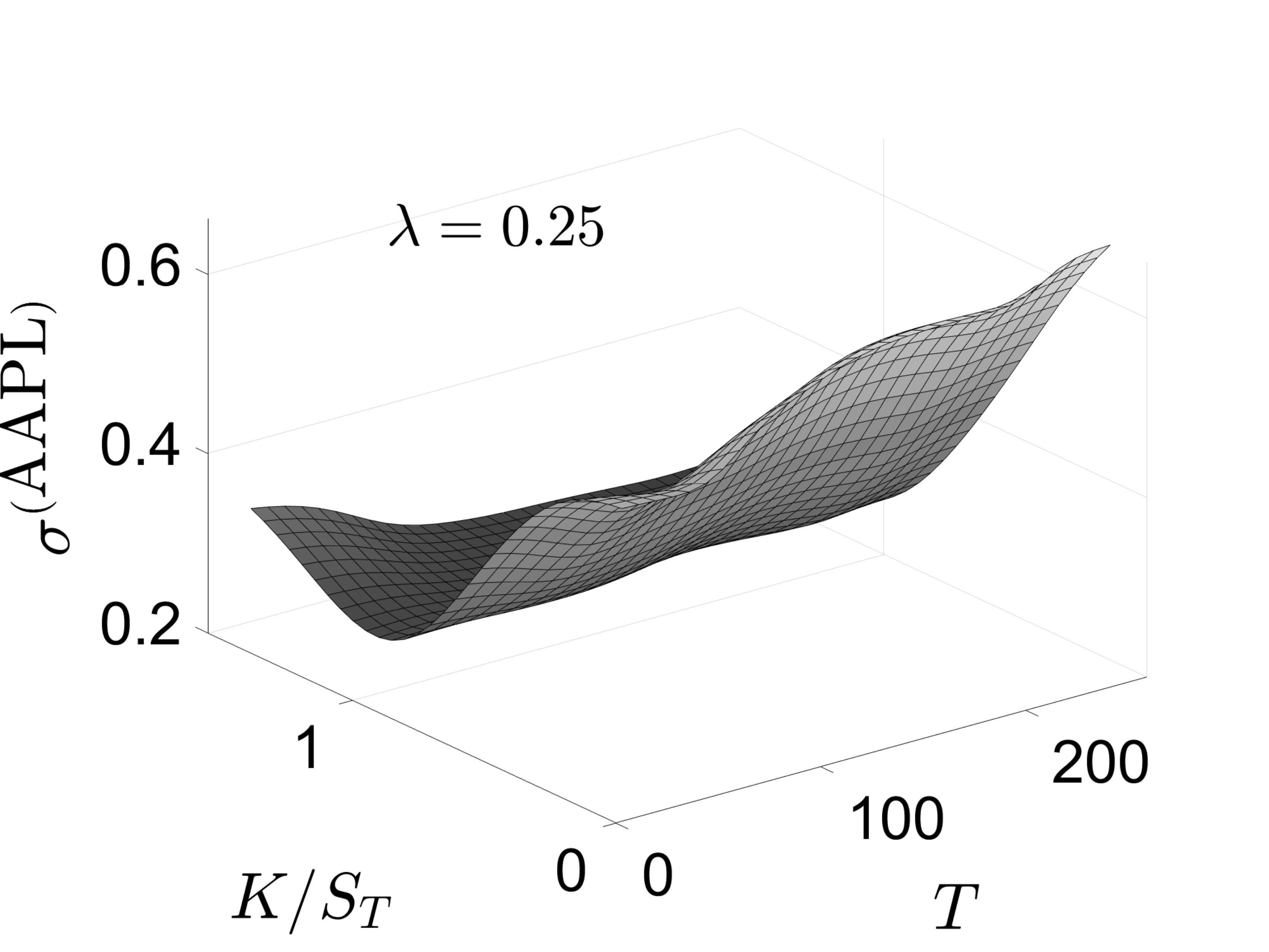}
    \end{subfigure}

    \begin{subfigure}[b]{0.32\textwidth} 
    	\includegraphics[width=\textwidth]{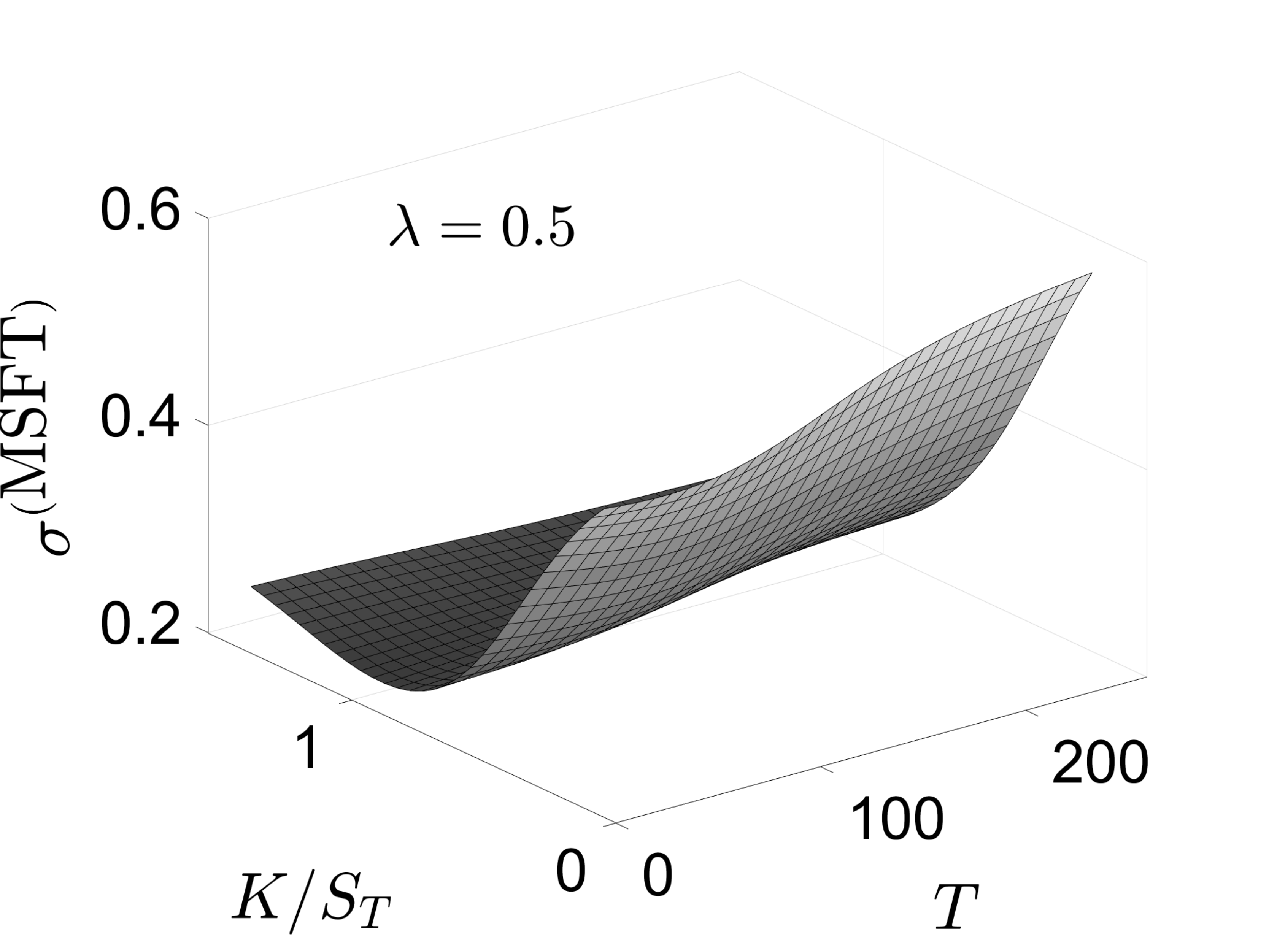}
    \end{subfigure}
    \begin{subfigure}[b]{0.32\textwidth} 
    	\includegraphics[width=\textwidth]{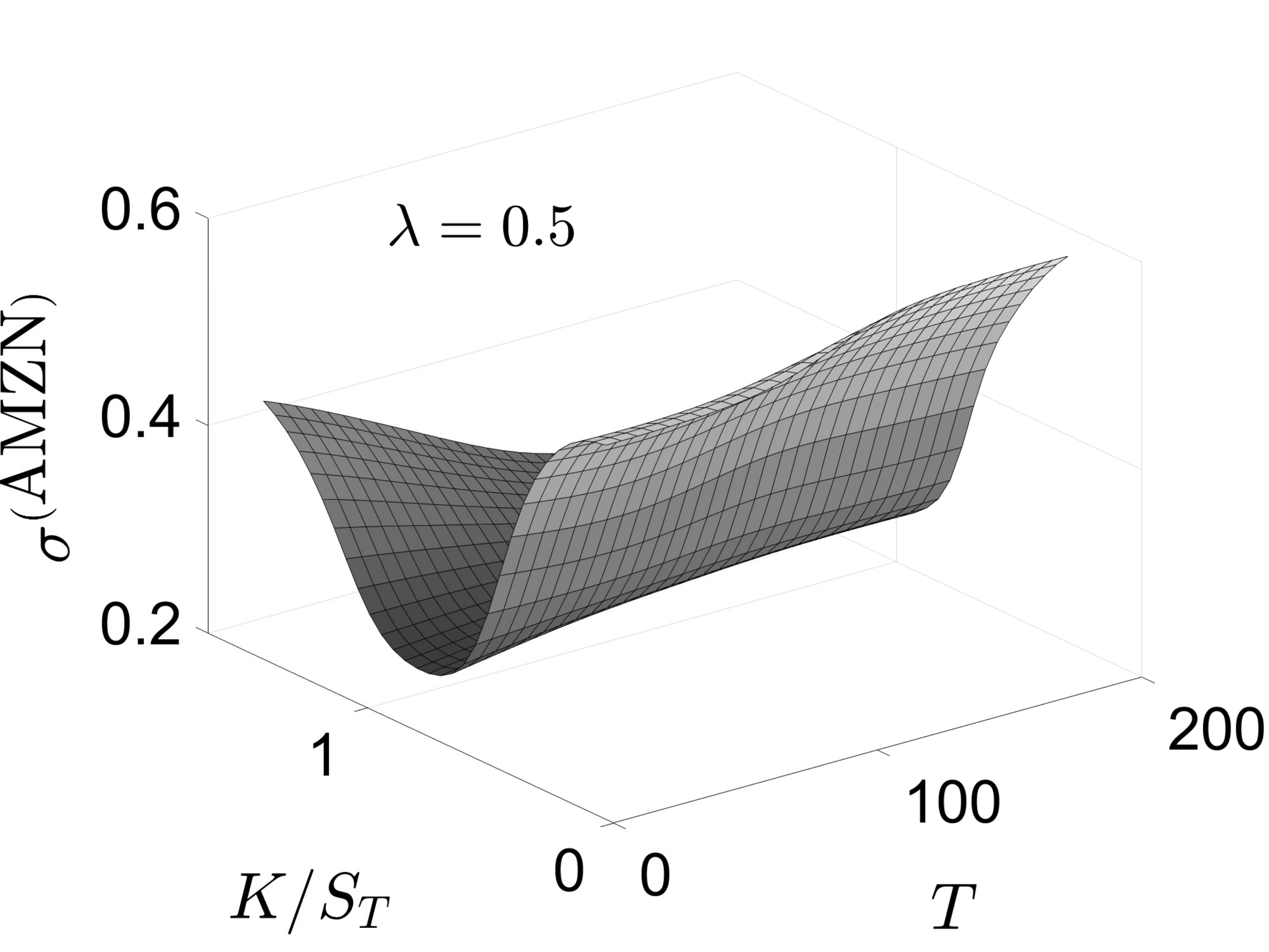}
    \end{subfigure}
    \begin{subfigure}[b]{0.32\textwidth} 
    	\includegraphics[width=\textwidth]{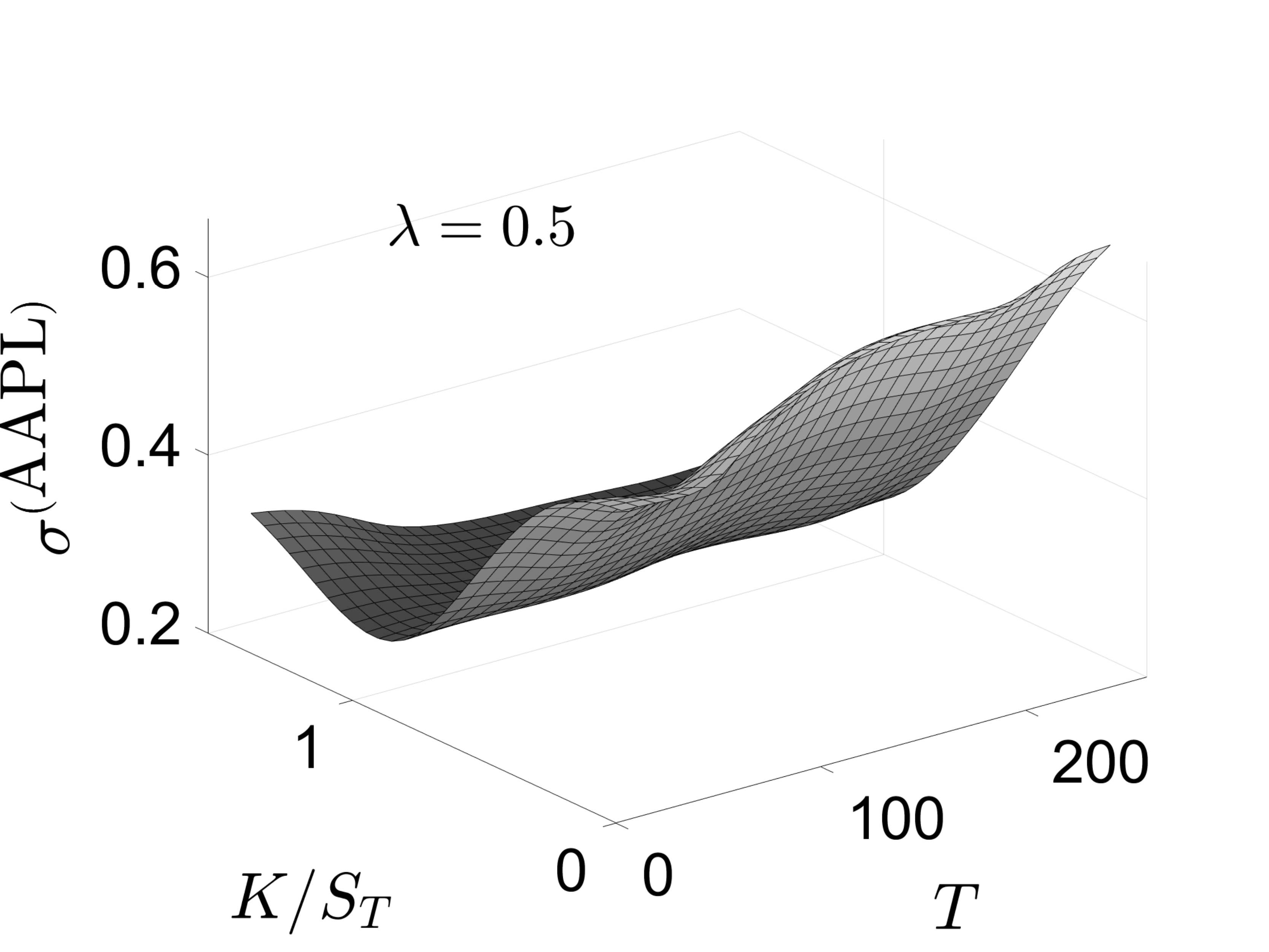}
    \end{subfigure}

    \begin{subfigure}[b]{0.32\textwidth} 
    	\includegraphics[width=\textwidth]{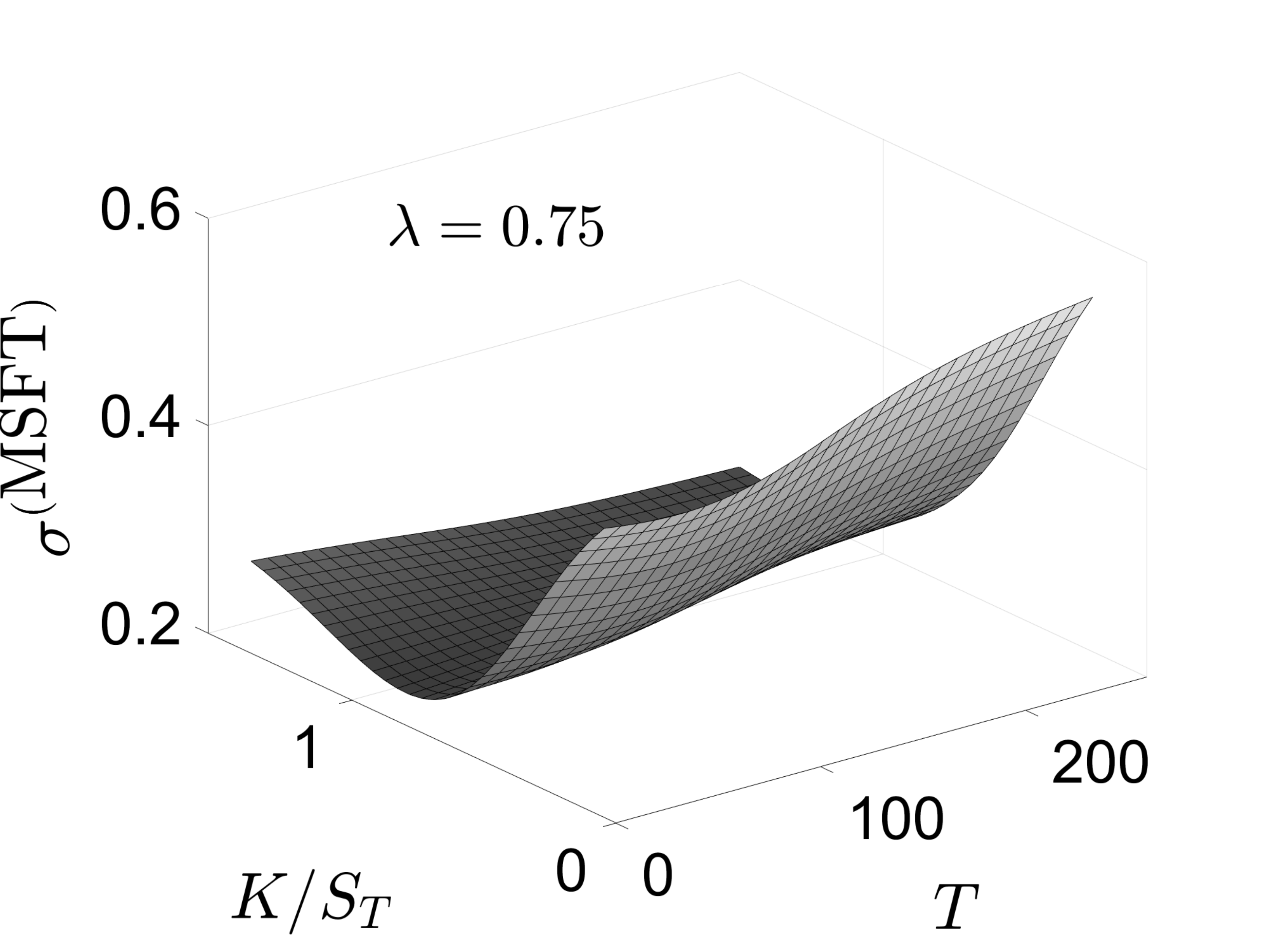}
    \end{subfigure}
    \begin{subfigure}[b]{0.32\textwidth} 
    	\includegraphics[width=\textwidth]{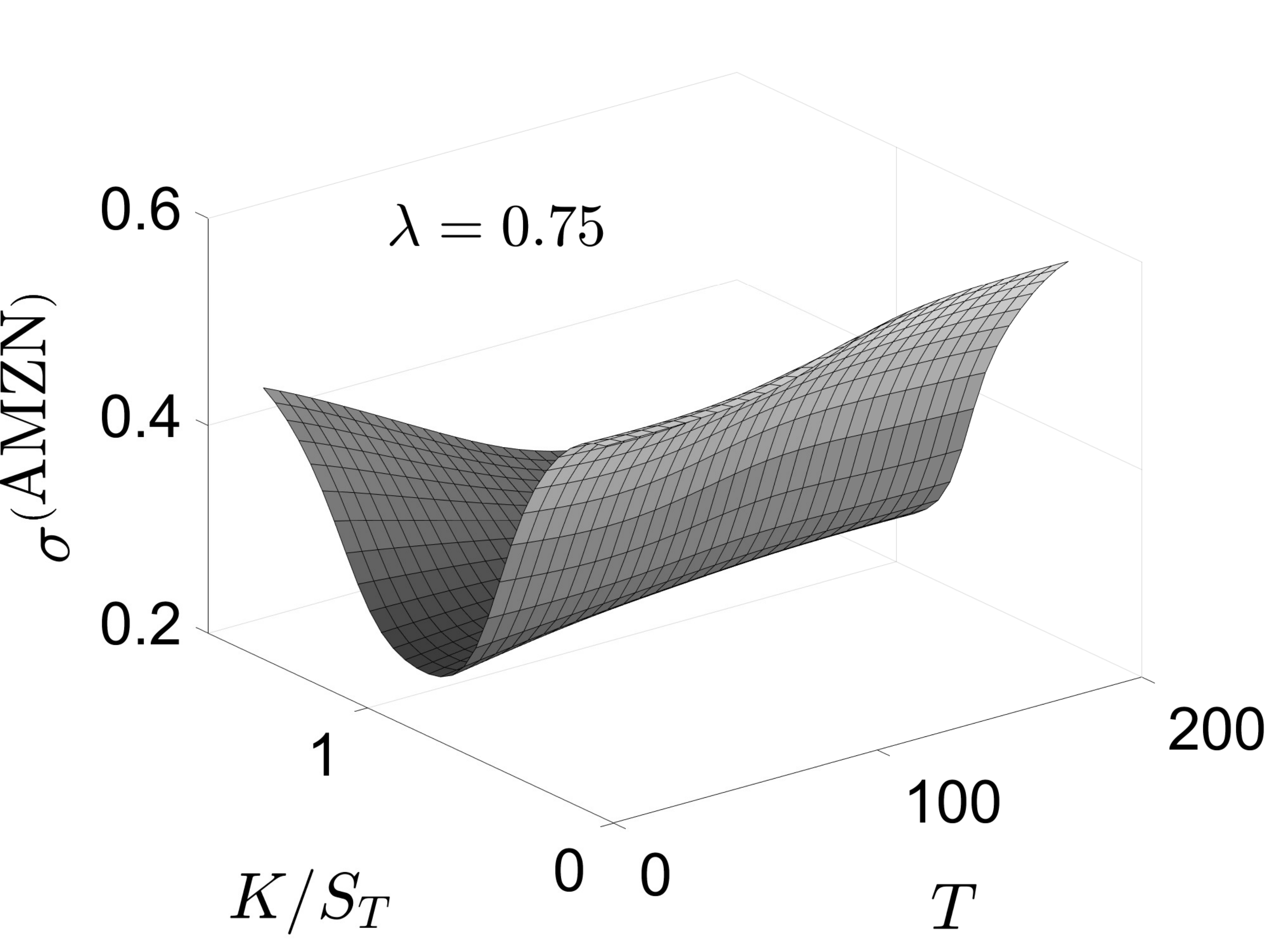}
    \end{subfigure}
        \begin{subfigure}[b]{0.32\textwidth} 
    	\includegraphics[width=\textwidth]{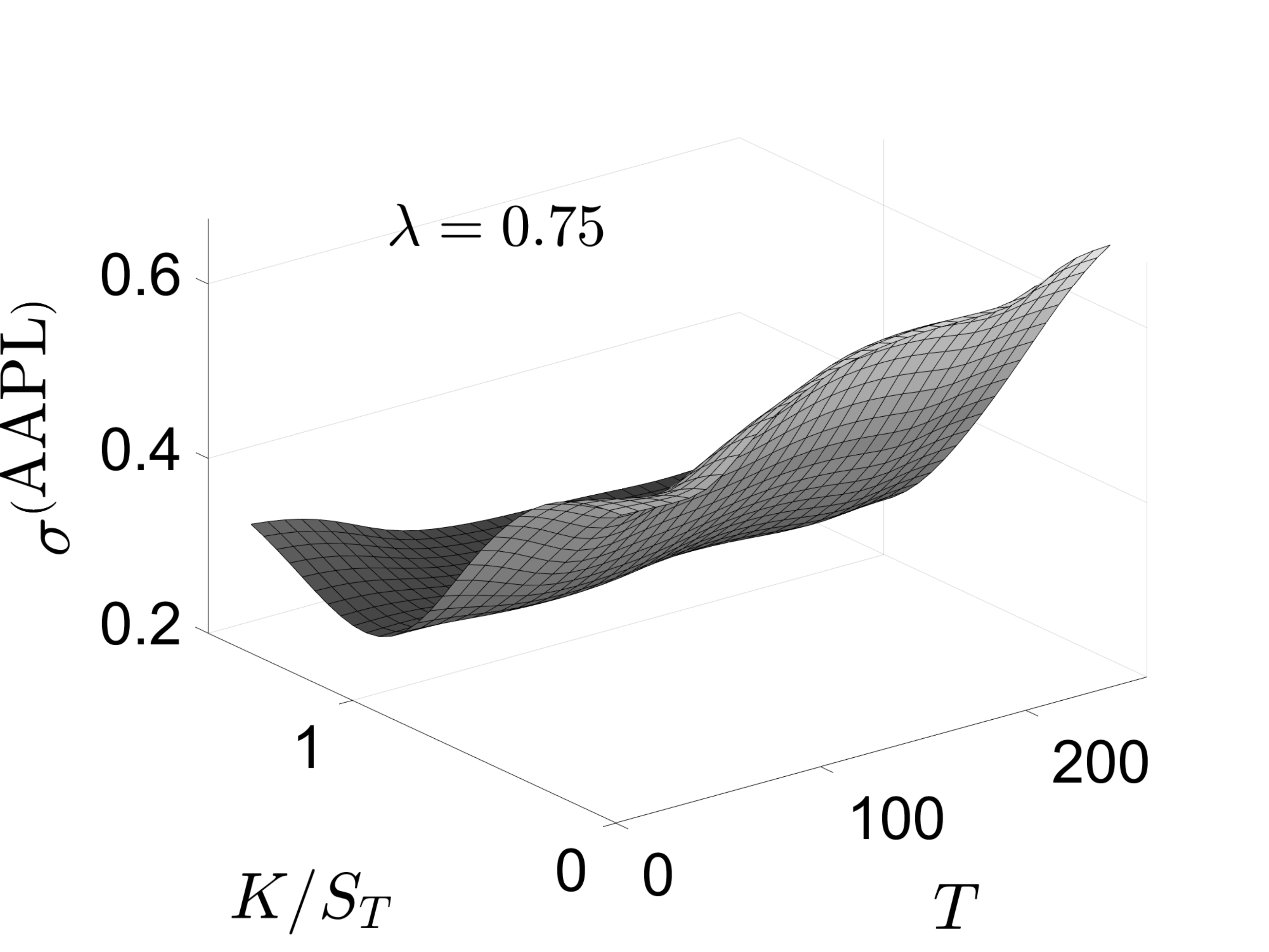}
    \end{subfigure}

    \caption{Implied $\sigma^{(\textrm{stock})}(T,M=K/S_T,\lambda)$ surfaces computed for the
    		arithmetic return model \eqref{eq:ESG_dpm}.}
    \label{fig:impl_vol_arith}
\end{center}
\end{figure}

 \begin{figure}[h!]
\begin{center}
    \begin{subfigure}[b]{0.32\textwidth} 
    	\includegraphics[width=\textwidth]{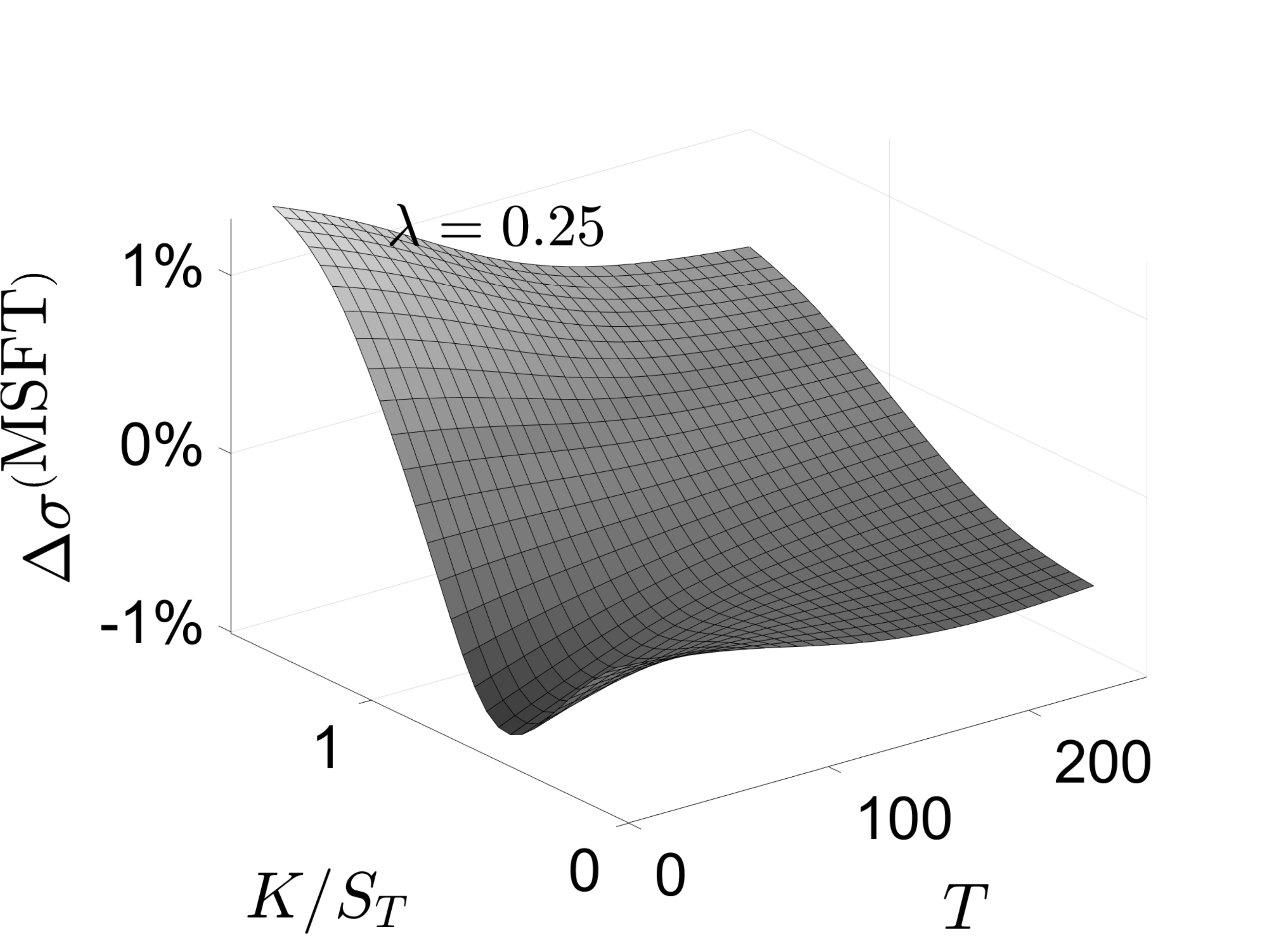}
    \end{subfigure}
    \begin{subfigure}[b]{0.32\textwidth} 
    	\includegraphics[width=\textwidth]{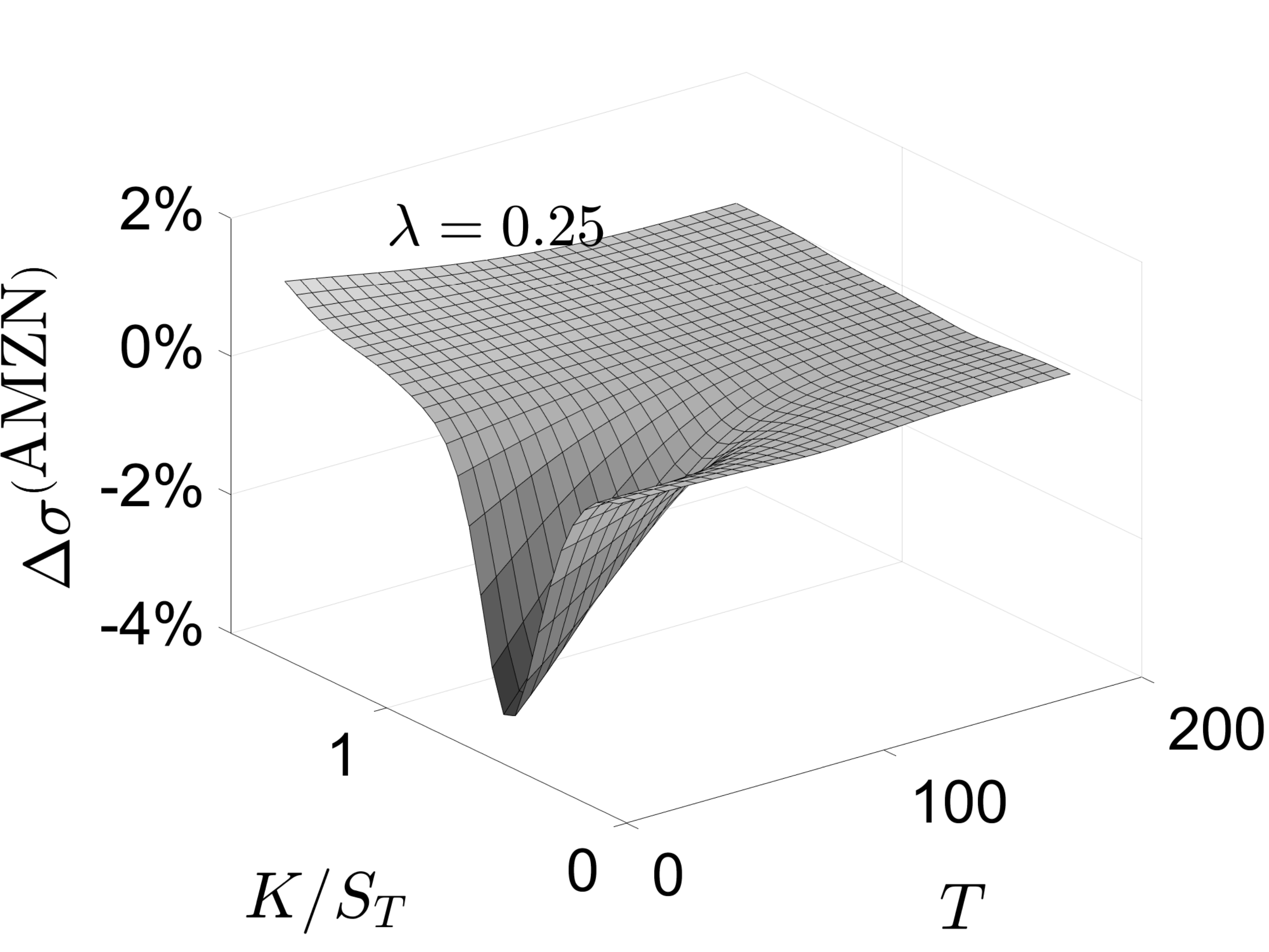}
    \end{subfigure}
    \begin{subfigure}[b]{0.32\textwidth} 
    	\includegraphics[width=\textwidth]{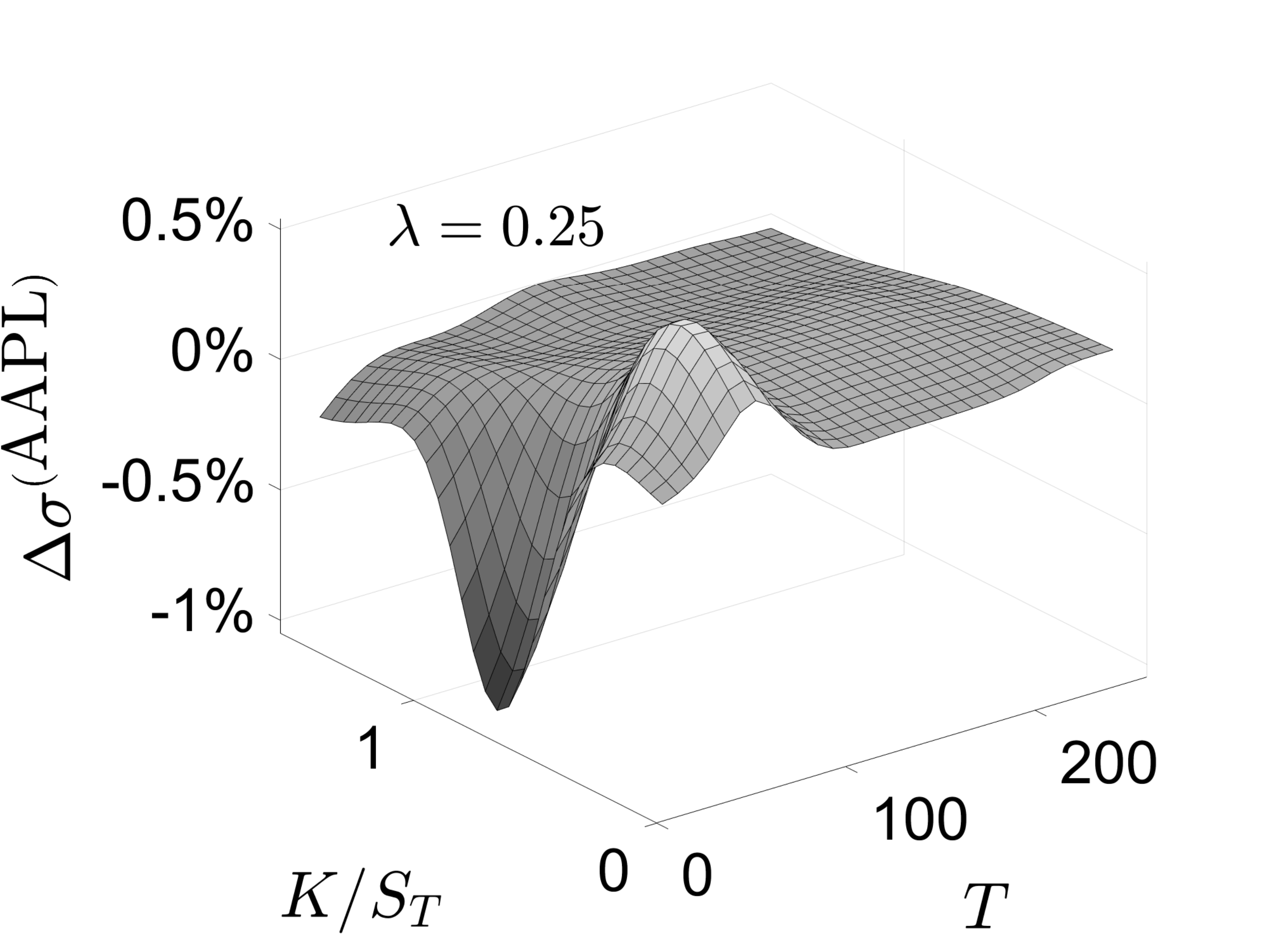}
    \end{subfigure}

    \begin{subfigure}[b]{0.32\textwidth} 
    	\includegraphics[width=\textwidth]{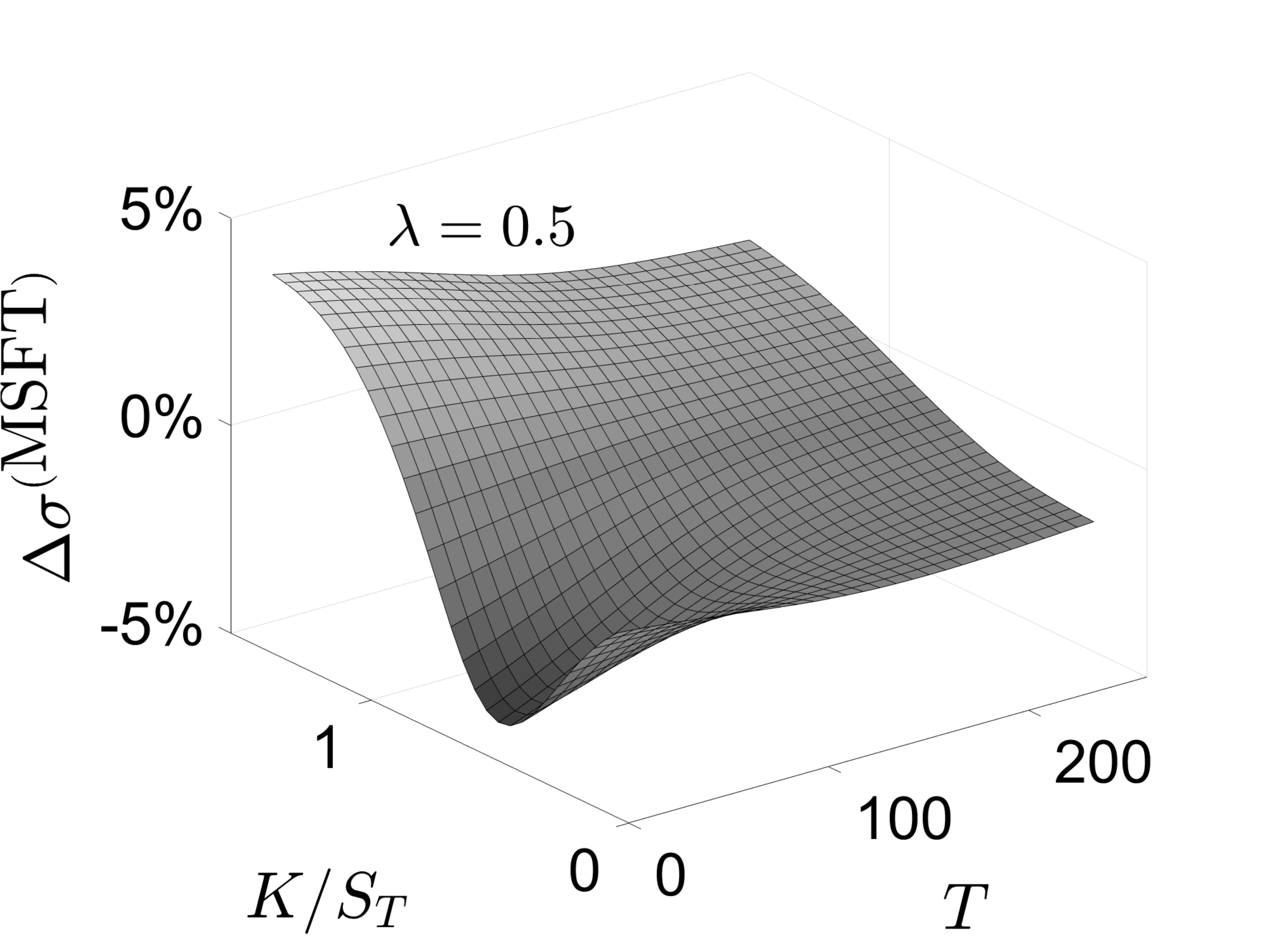}
    \end{subfigure}
    \begin{subfigure}[b]{0.32\textwidth} 
    	\includegraphics[width=\textwidth]{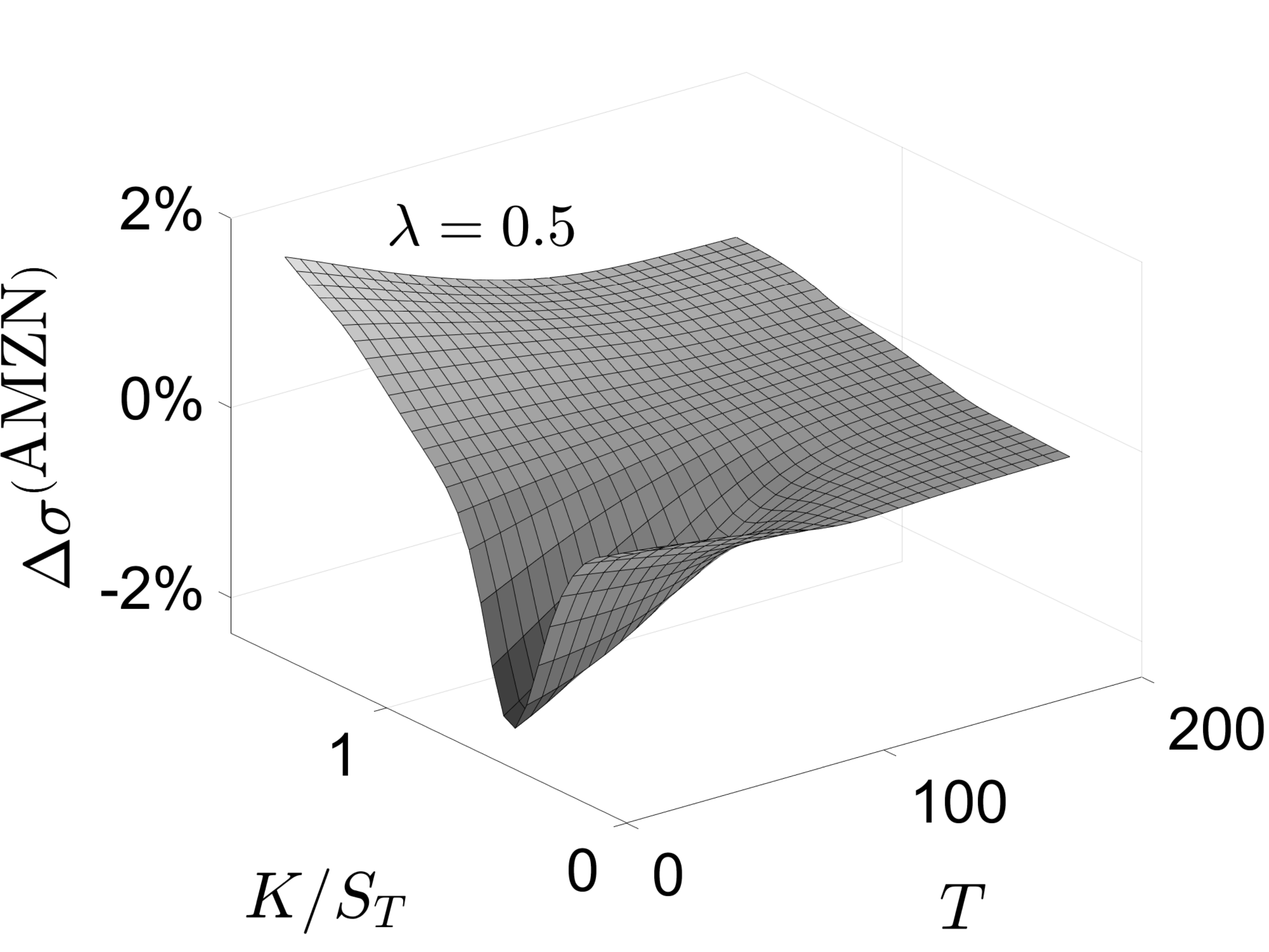}
    \end{subfigure}
        \begin{subfigure}[b]{0.32\textwidth} 
    	\includegraphics[width=\textwidth]{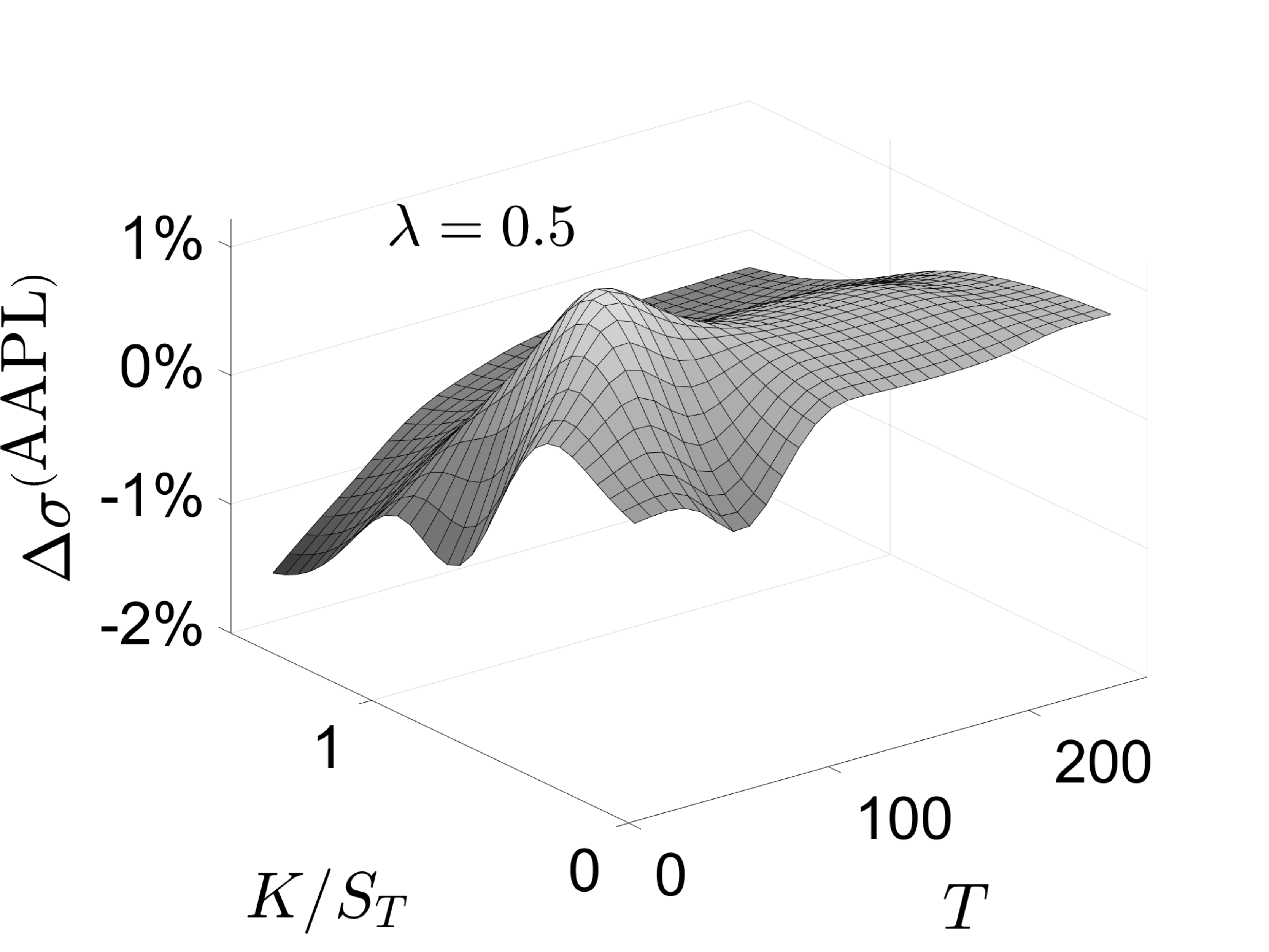}
    \end{subfigure}

    \begin{subfigure}[b]{0.32\textwidth} 
    	\includegraphics[width=\textwidth]{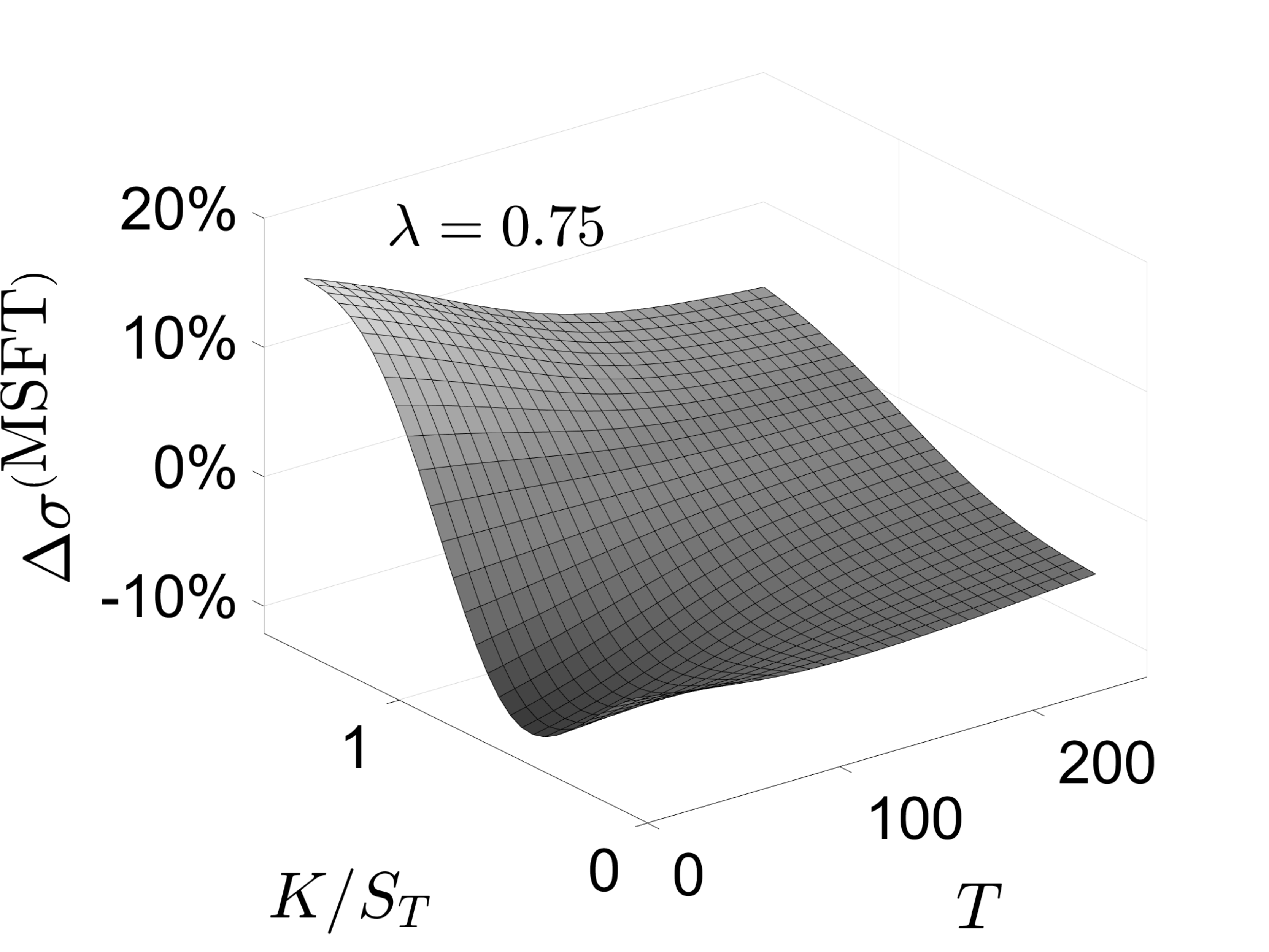}
    \end{subfigure}
    \begin{subfigure}[b]{0.32\textwidth} 
    	\includegraphics[width=\textwidth]{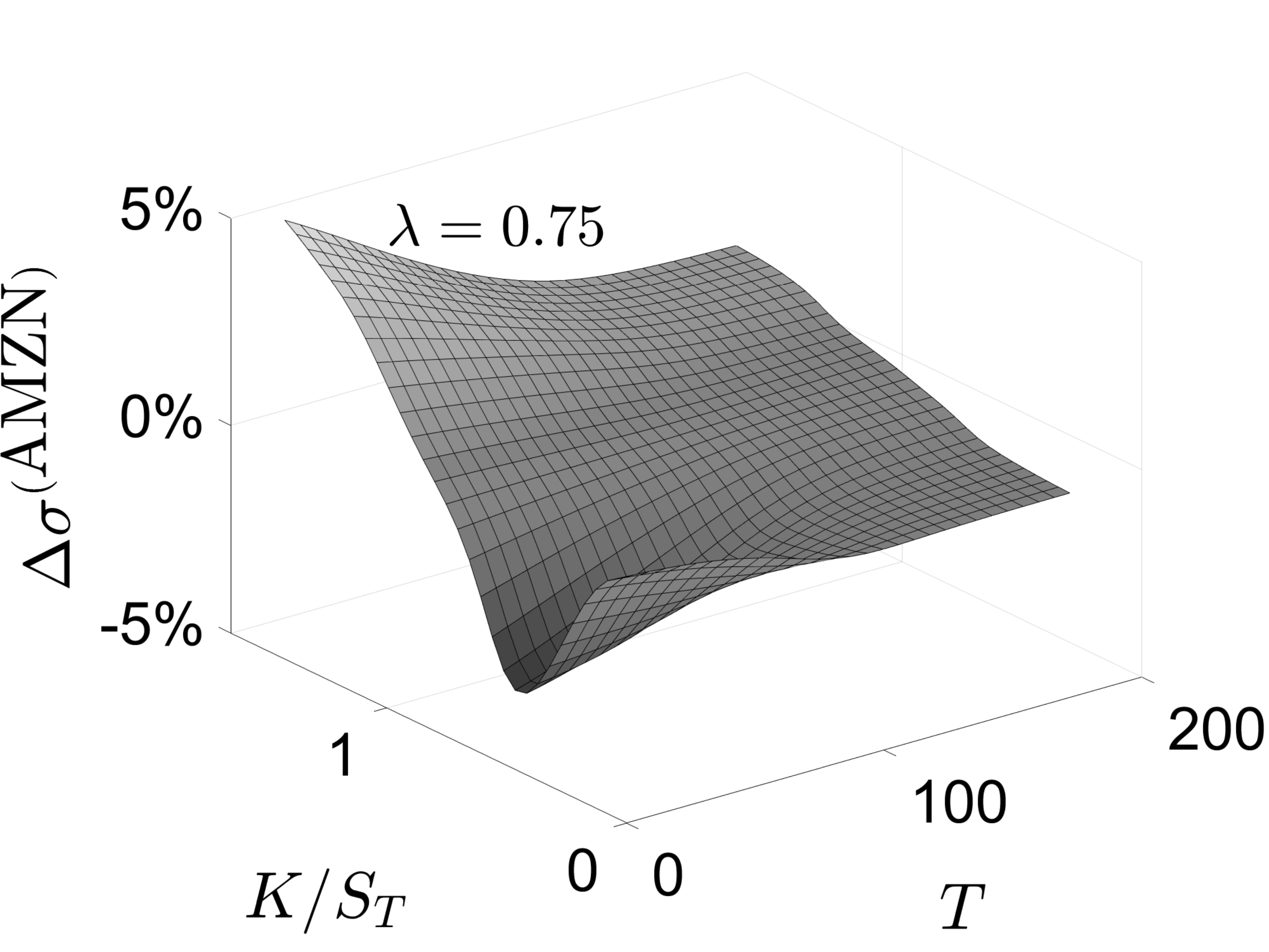}
    \end{subfigure}
        \begin{subfigure}[b]{0.32\textwidth} 
    	\includegraphics[width=\textwidth]{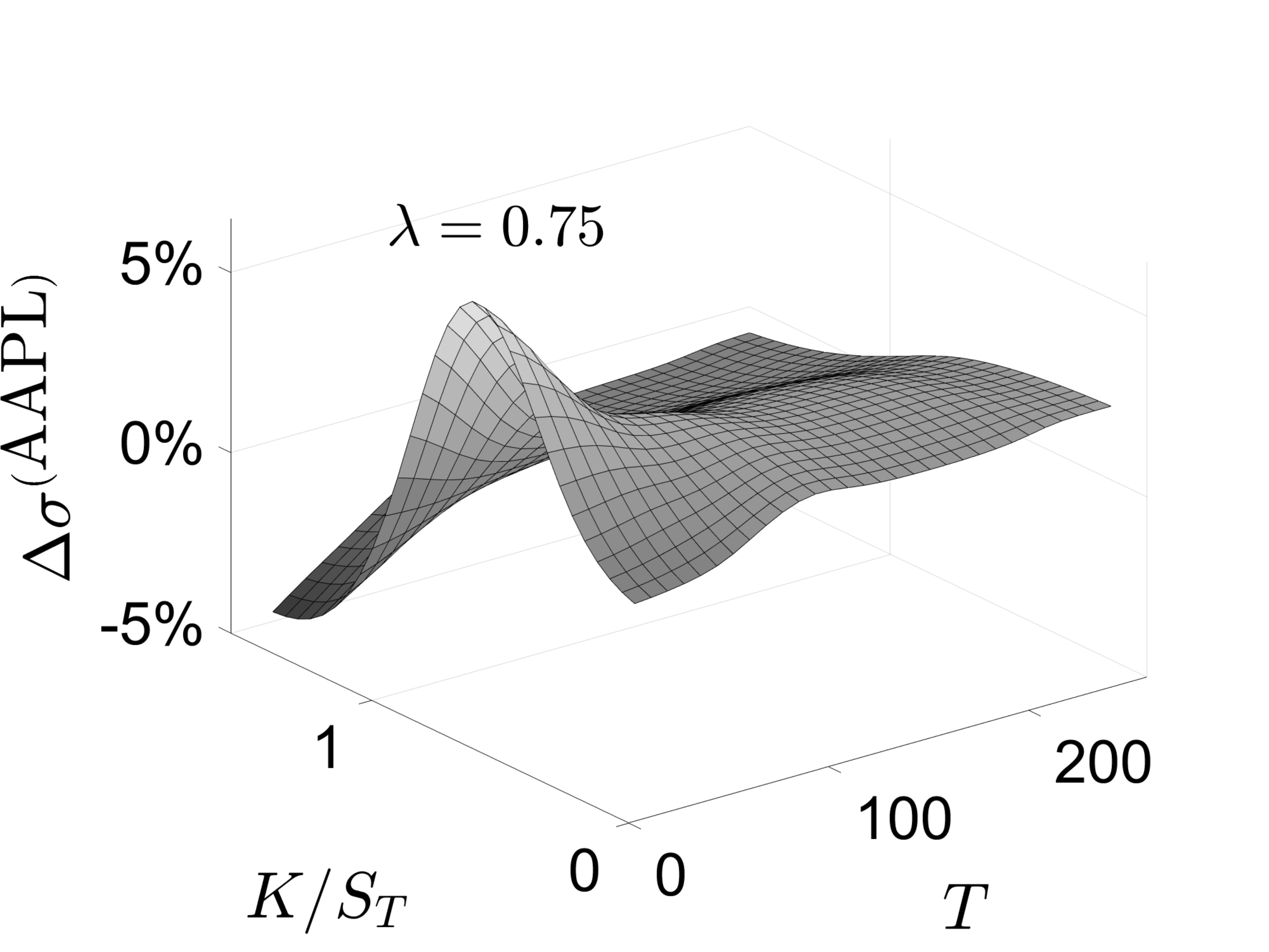}
    \end{subfigure}

    \caption{The percent relative change $\Delta \sigma^{(\text{stock})}(T,M=K/S_T,\lambda)$
    	in the implied volatility surfaces of Fig.~\ref{fig:impl_vol_arith}.}
    \label{fig:impl_vol_arith_dev}
\end{center}
\end{figure}

\begin{figure}[h!]
\begin{center}

    \begin{subfigure}[b]{0.32\textwidth} 
    	\includegraphics[width=\textwidth]{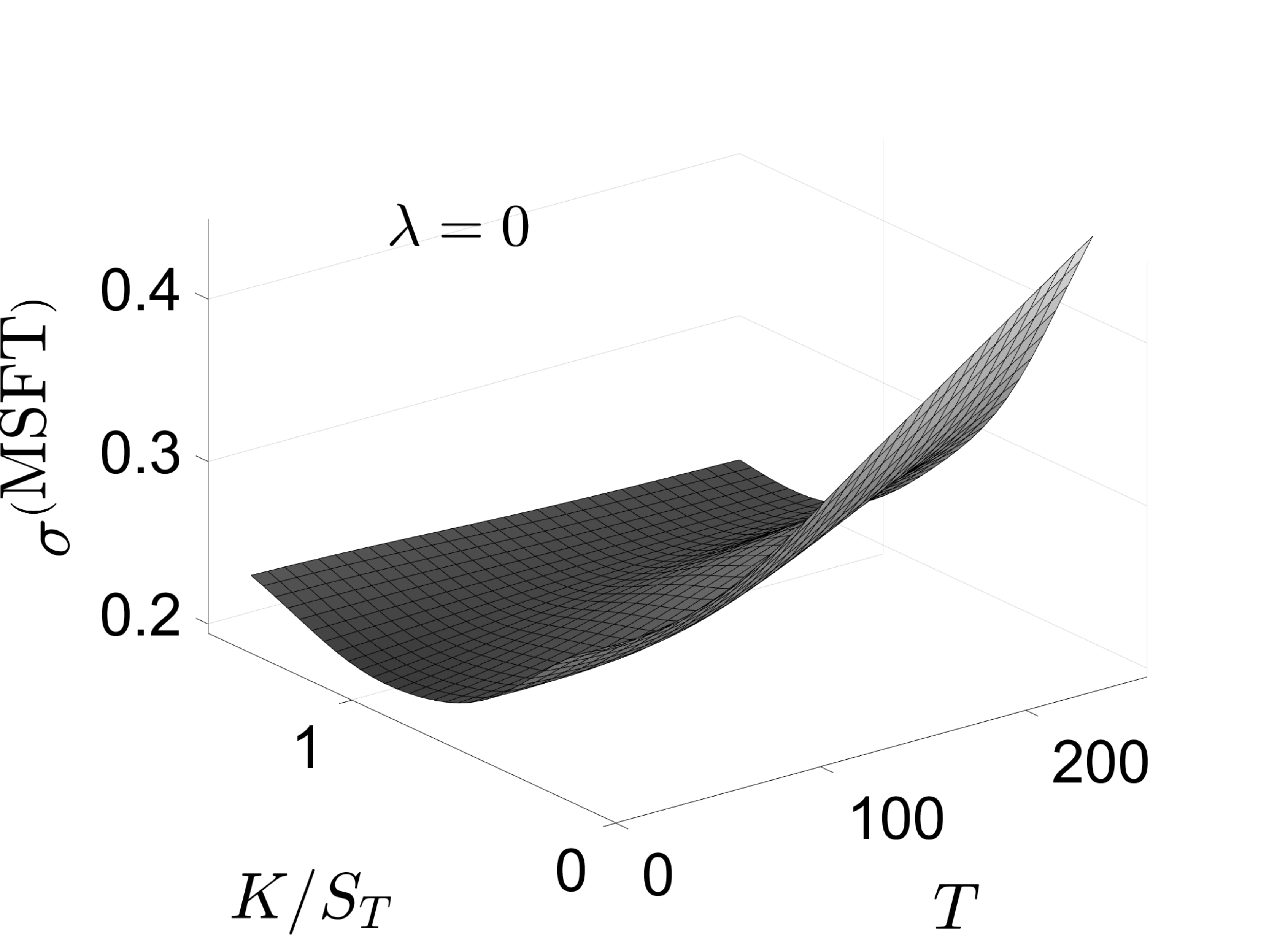}
    \end{subfigure}
    \begin{subfigure}[b]{0.32\textwidth} 
    	\includegraphics[width=\textwidth]{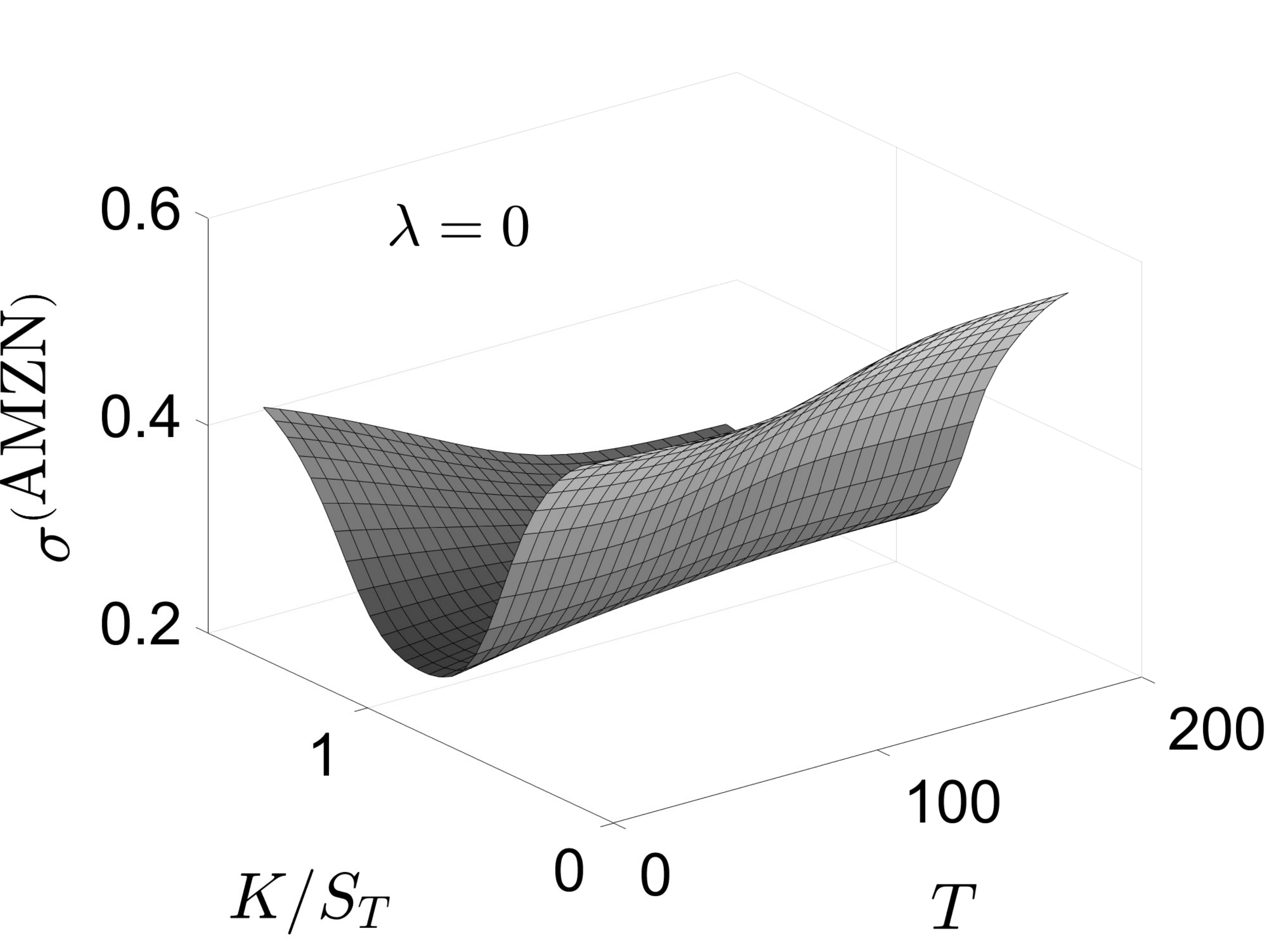}
    \end{subfigure}
        \begin{subfigure}[b]{0.32\textwidth} 
    	\includegraphics[width=\textwidth]{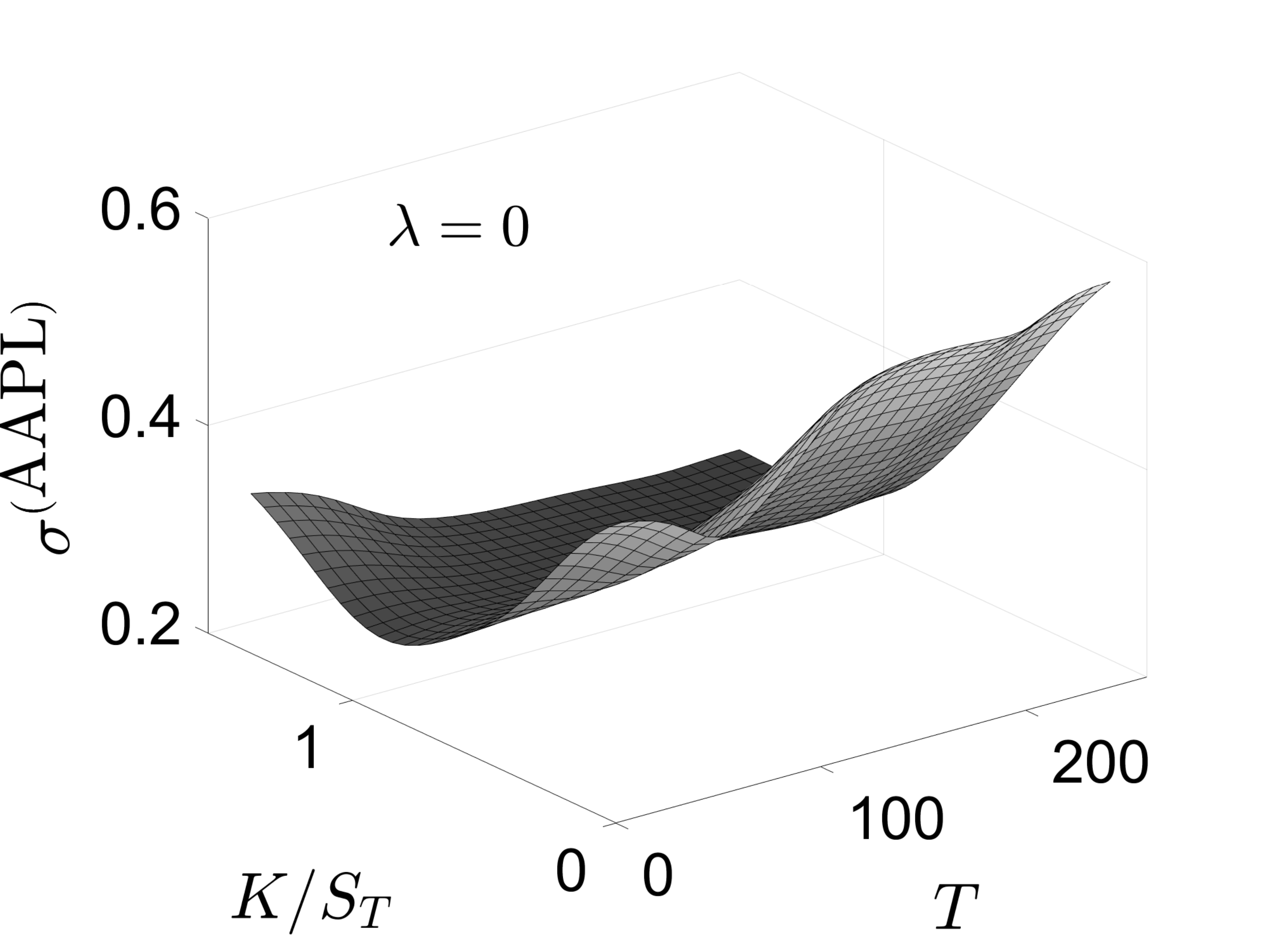}
    \end{subfigure}

    \begin{subfigure}[b]{0.32\textwidth} 
    	\includegraphics[width=\textwidth]{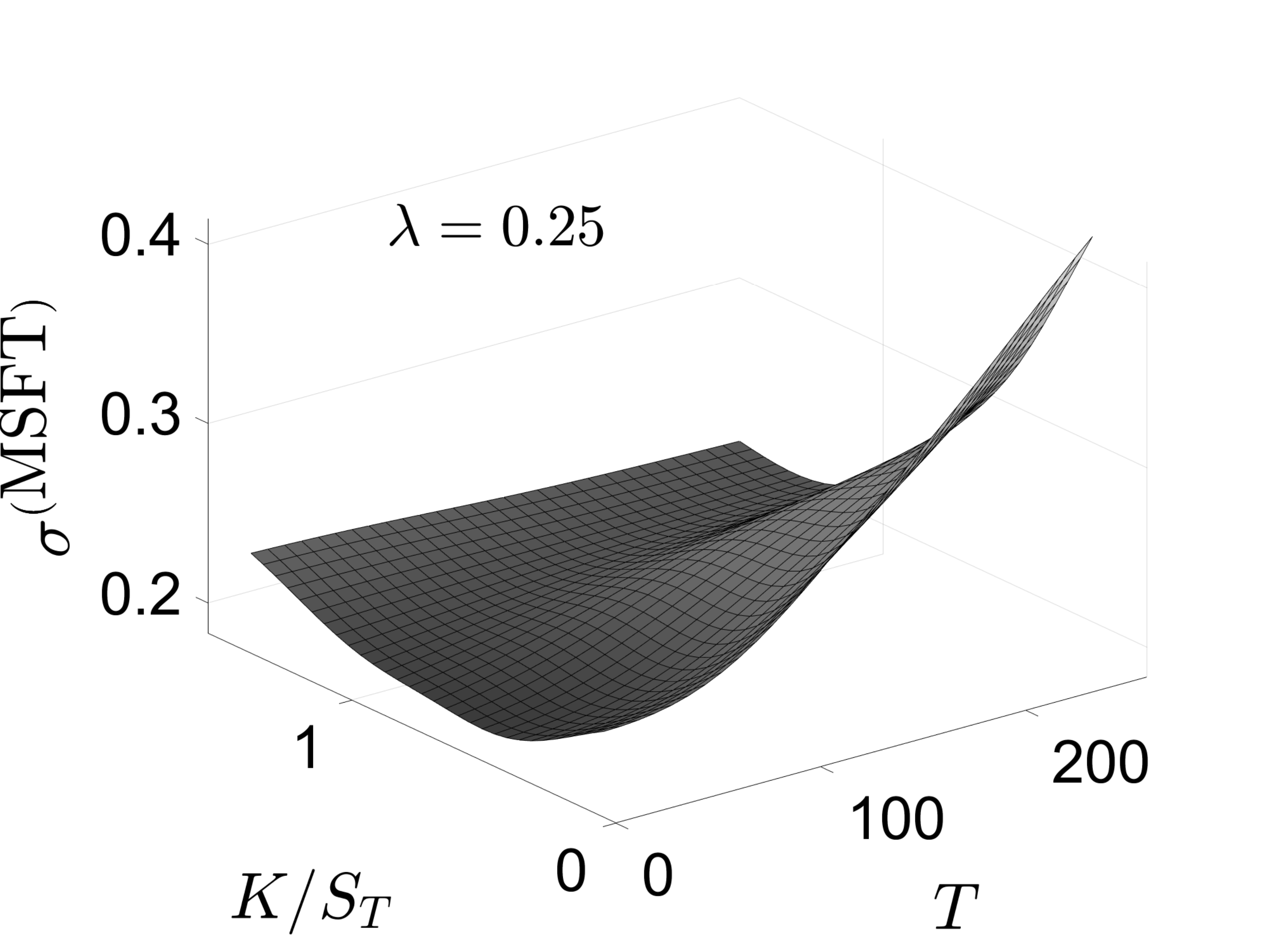}
    \end{subfigure}
    \begin{subfigure}[b]{0.32\textwidth} 
    	\includegraphics[width=\textwidth]{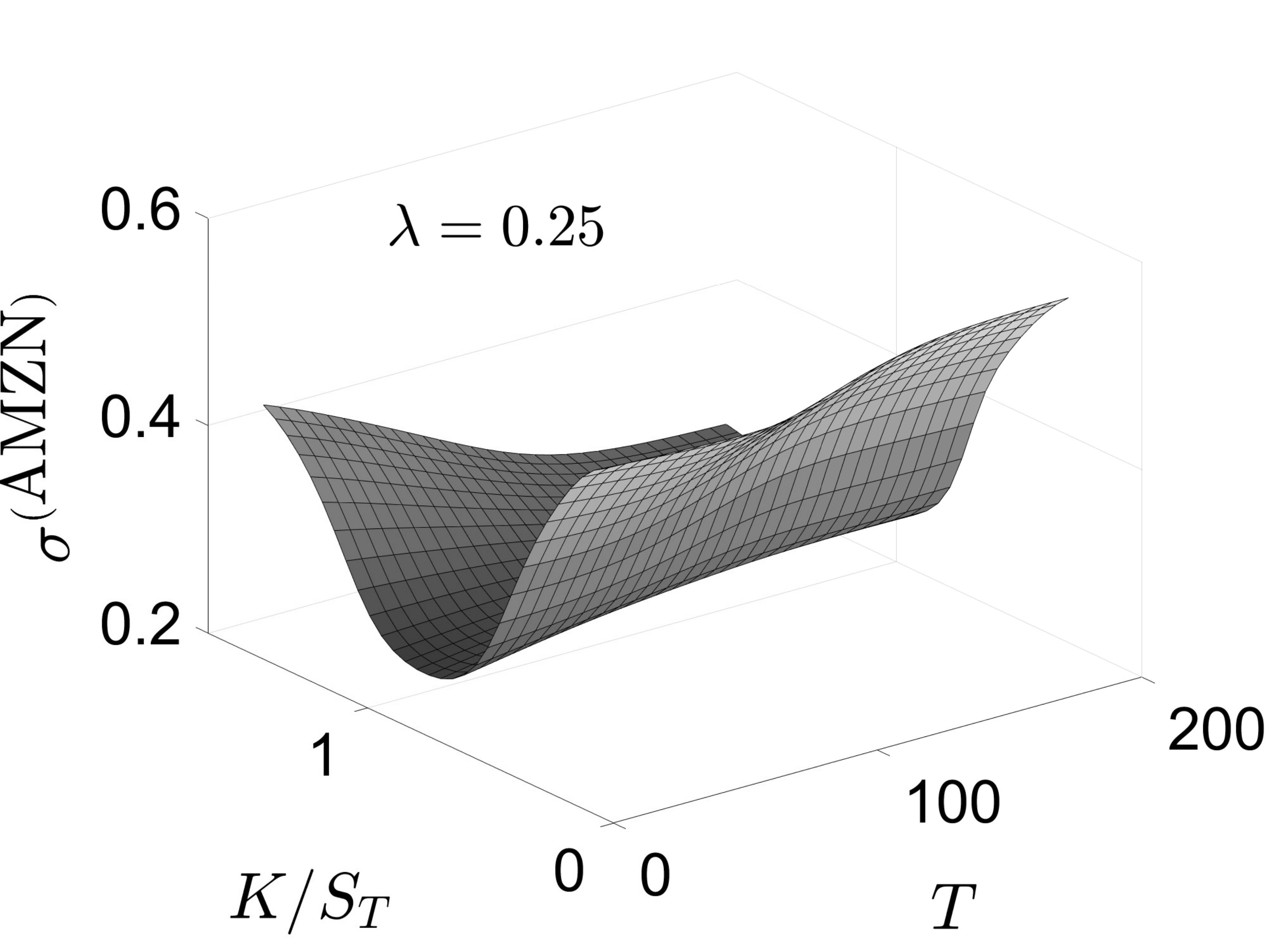}
    \end{subfigure}
        \begin{subfigure}[b]{0.32\textwidth} 
    	\includegraphics[width=\textwidth]{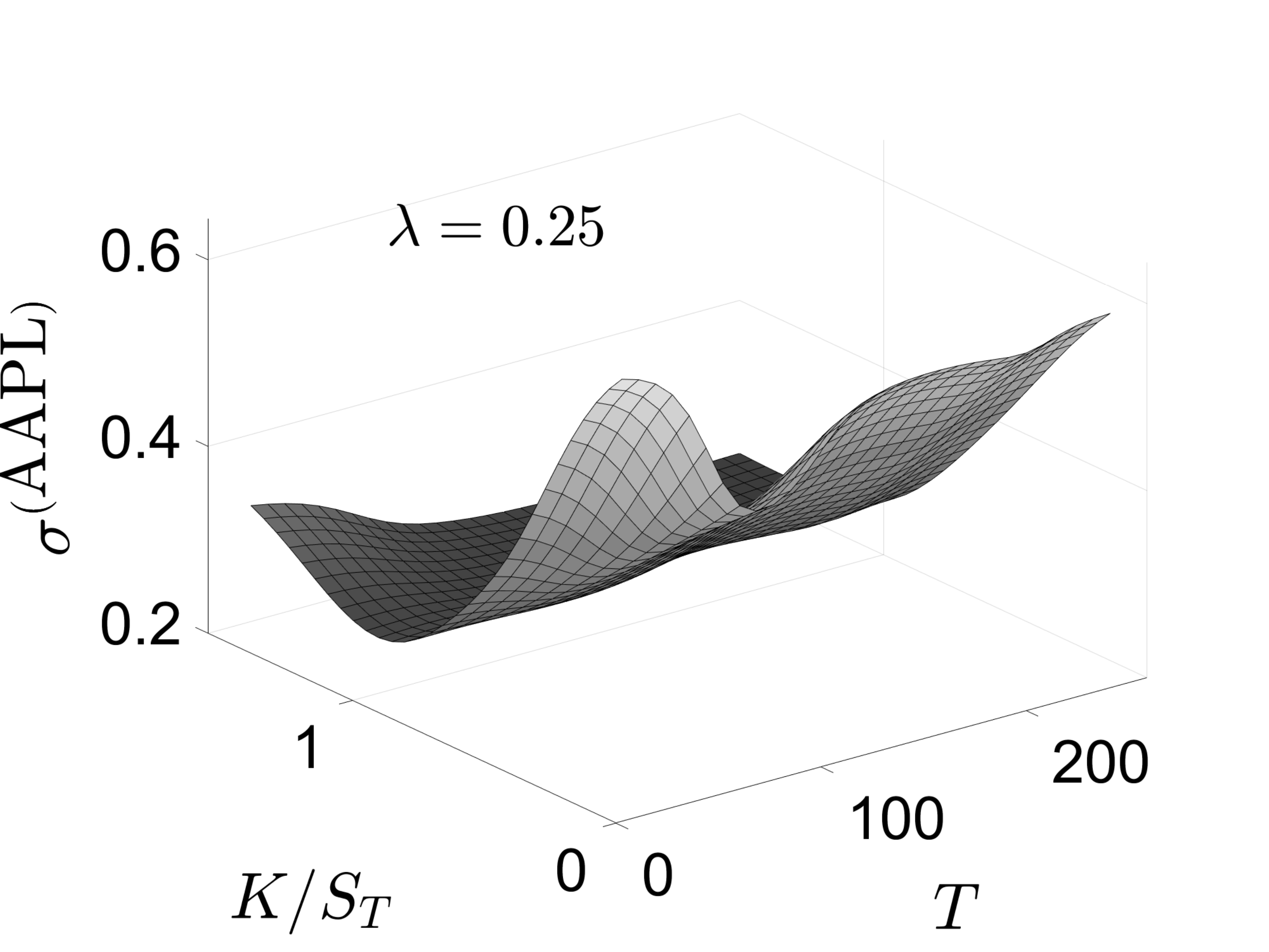}
    \end{subfigure}

    \begin{subfigure}[b]{0.32\textwidth} 
    	\includegraphics[width=\textwidth]{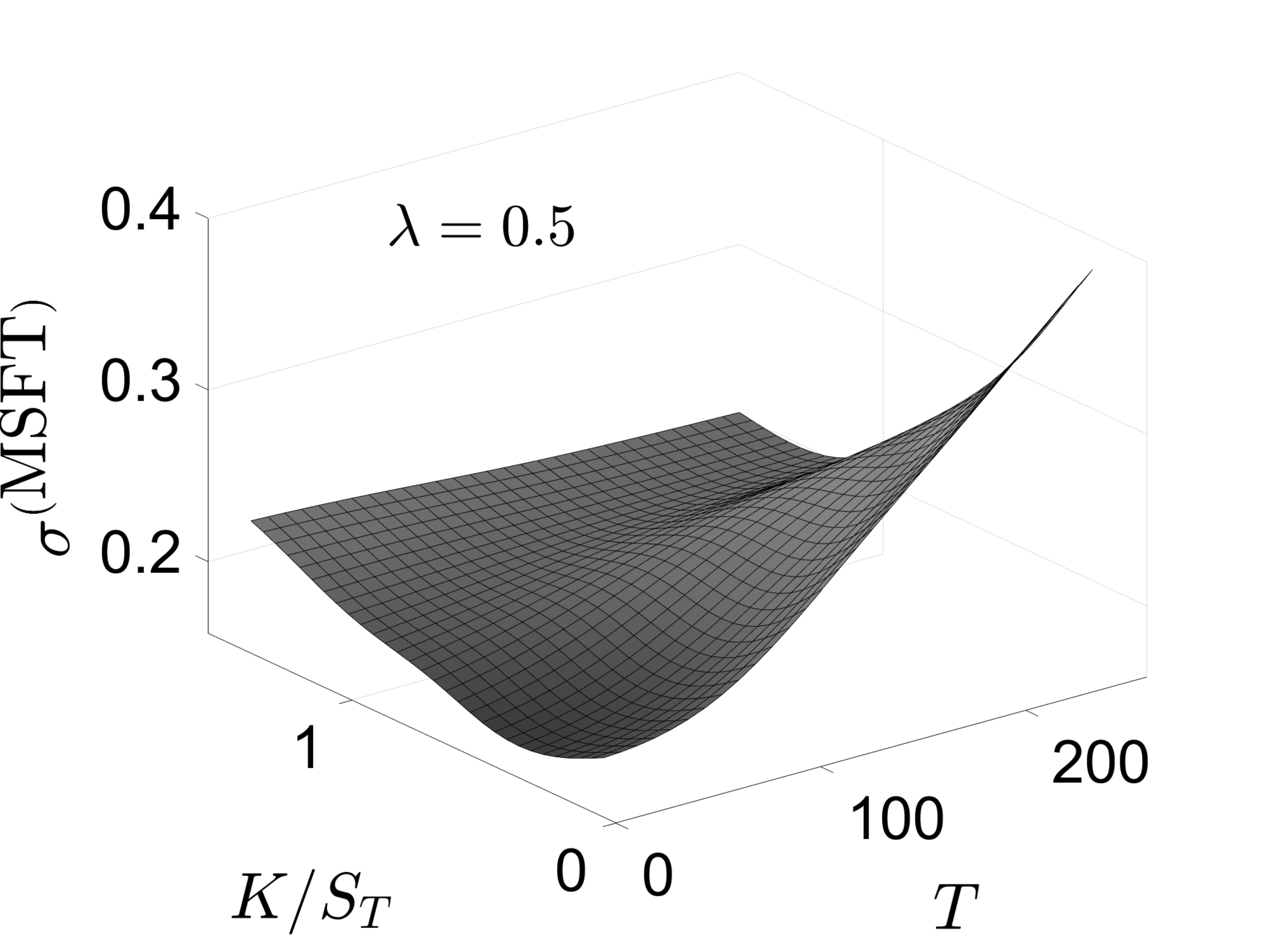}
    \end{subfigure}
    \begin{subfigure}[b]{0.32\textwidth} 
    	\includegraphics[width=\textwidth]{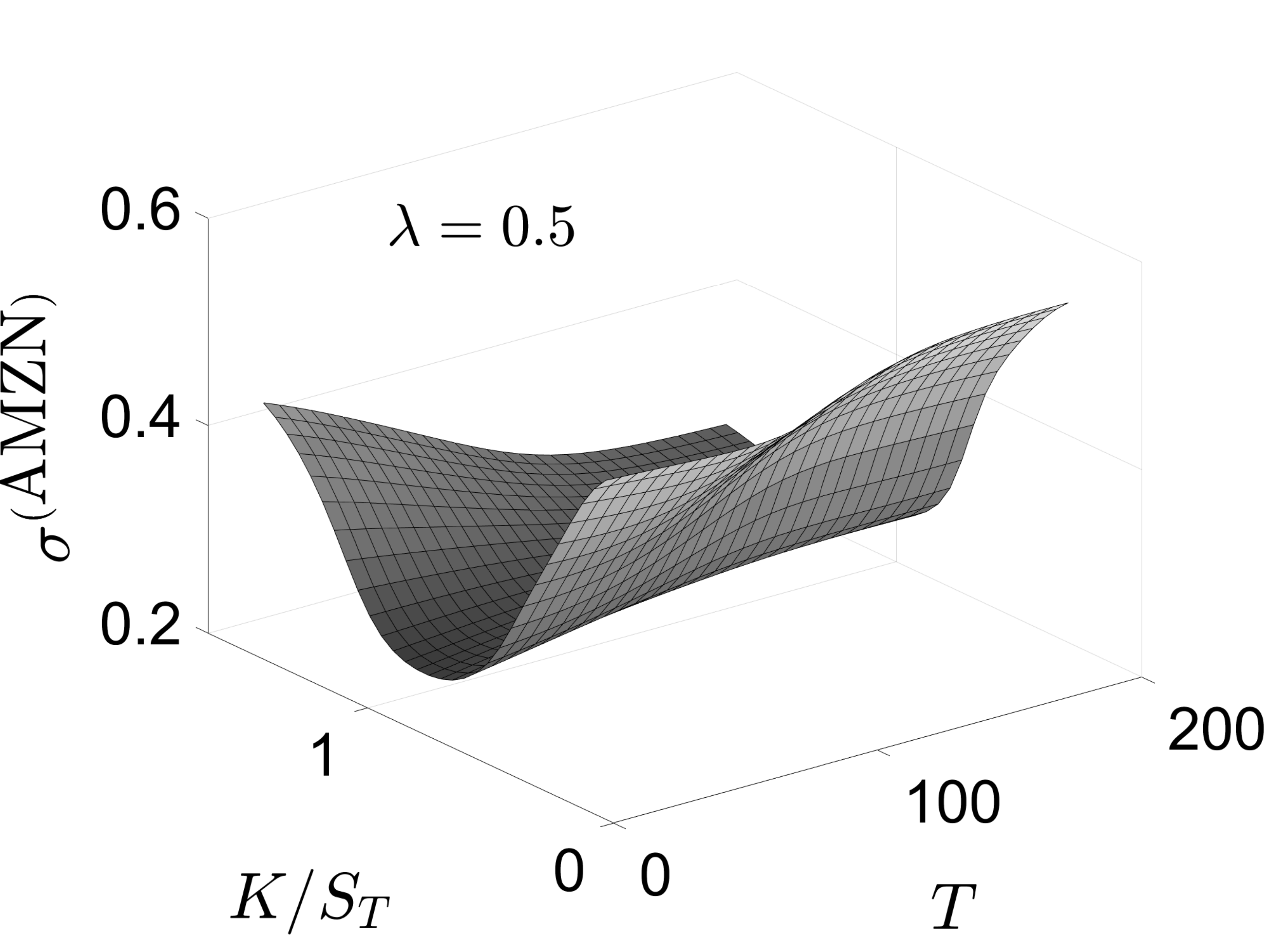}
    \end{subfigure}
    \begin{subfigure}[b]{0.32\textwidth} 
    	\includegraphics[width=\textwidth]{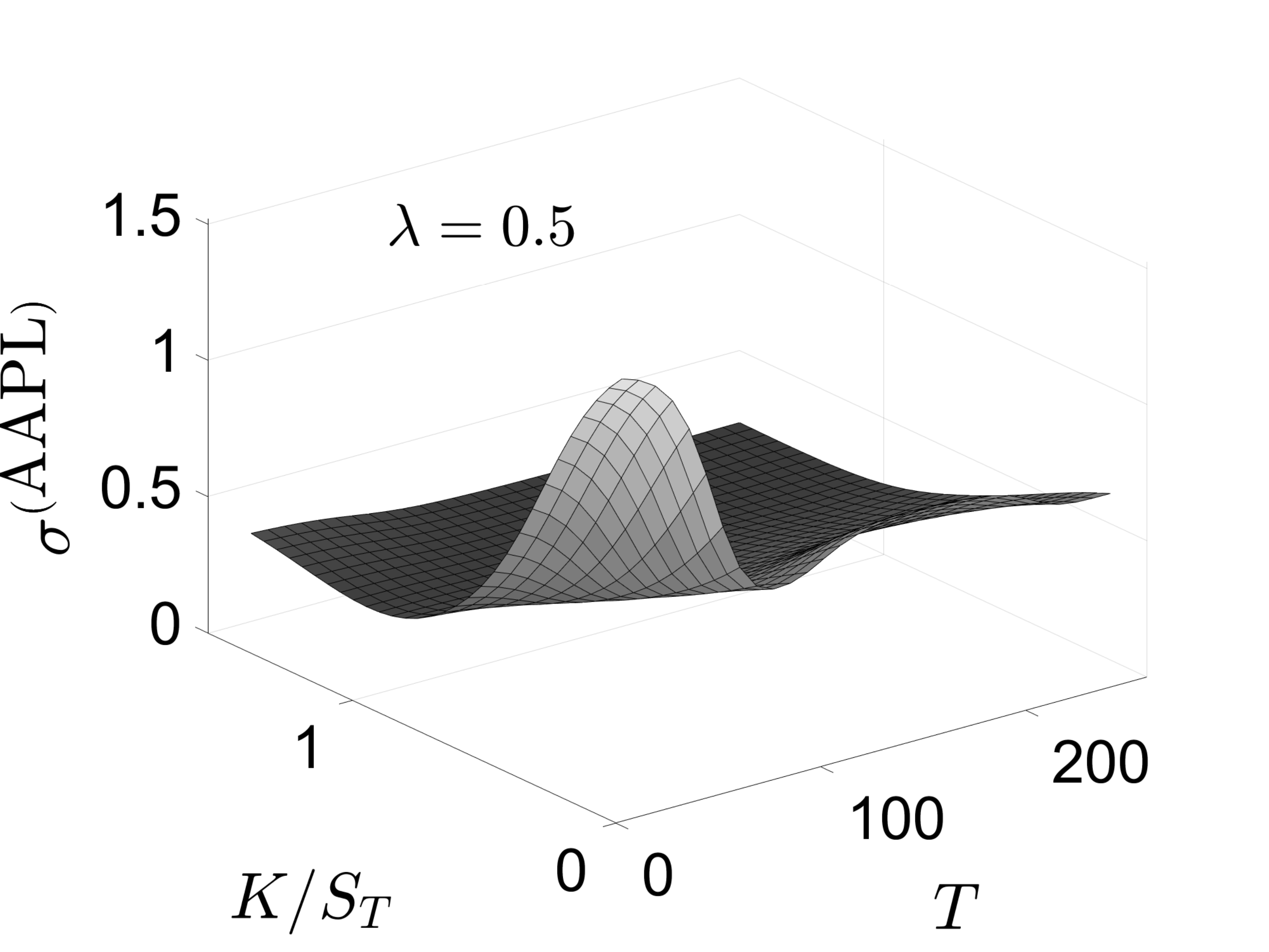}
    \end{subfigure}

    \begin{subfigure}[b]{0.32\textwidth} 
    	\includegraphics[width=\textwidth]{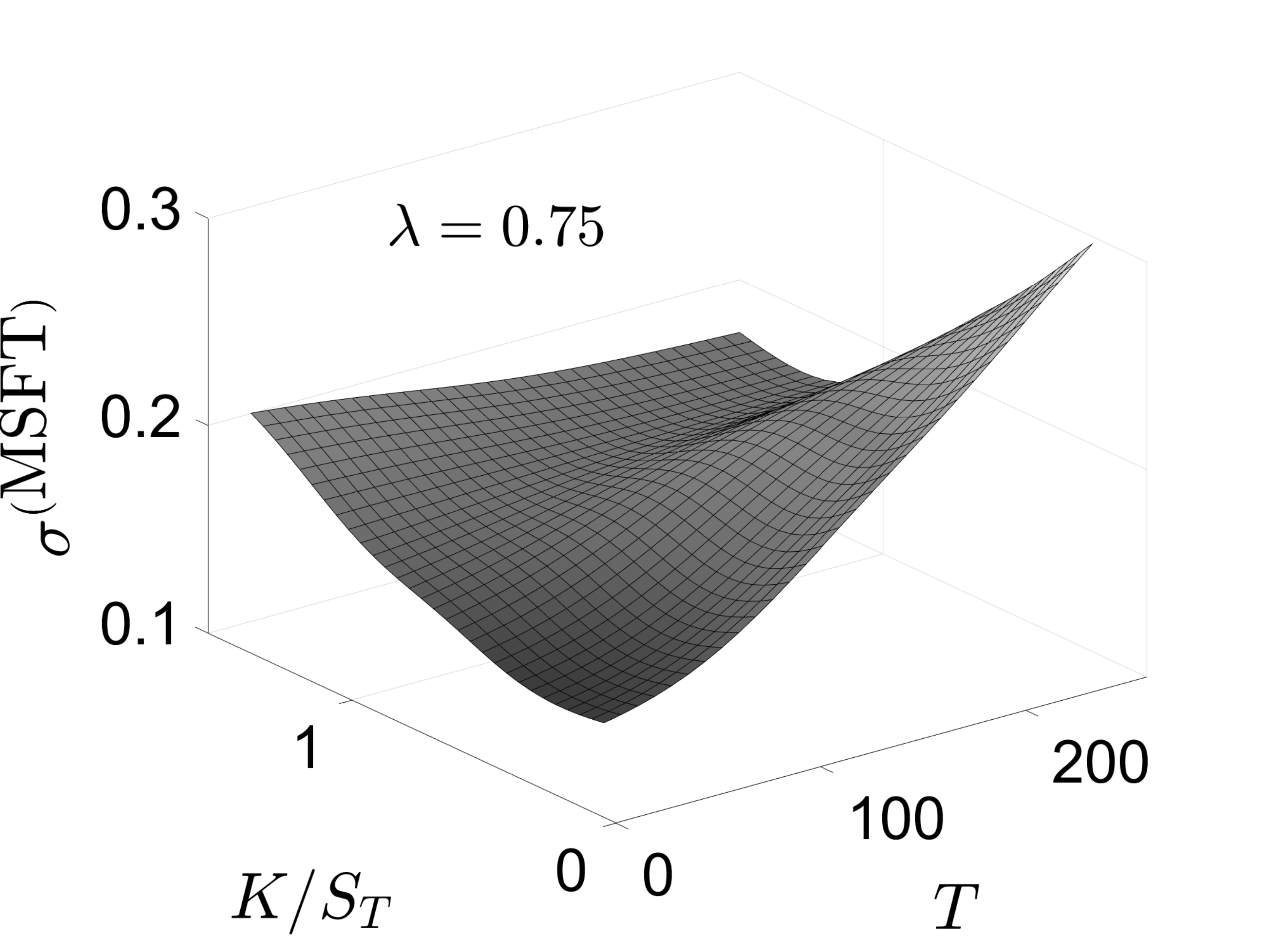}
    \end{subfigure}
    \begin{subfigure}[b]{0.32\textwidth} 
    	\includegraphics[width=\textwidth]{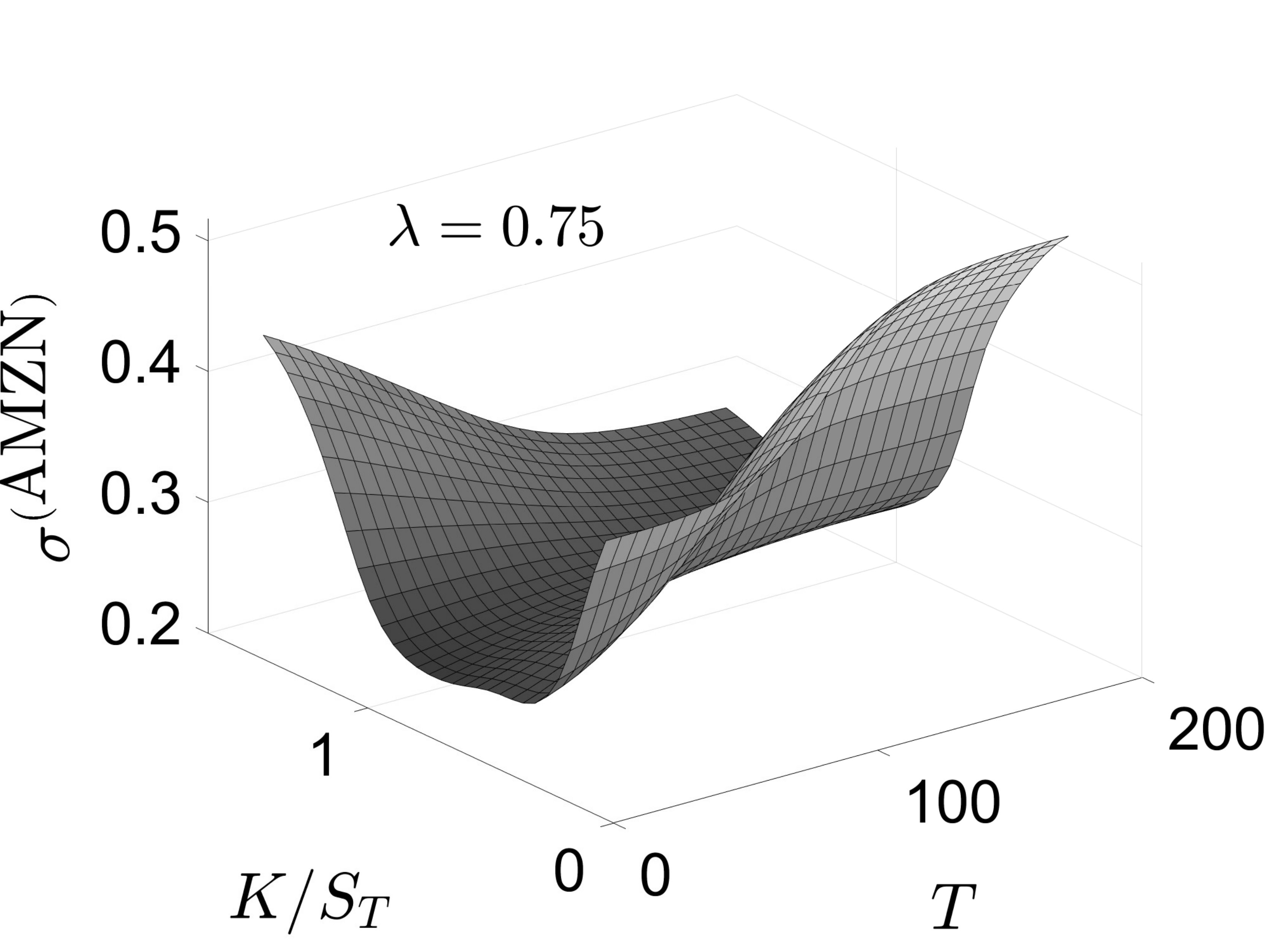}
    \end{subfigure}
        \begin{subfigure}[b]{0.32\textwidth} 
    	\includegraphics[width=\textwidth]{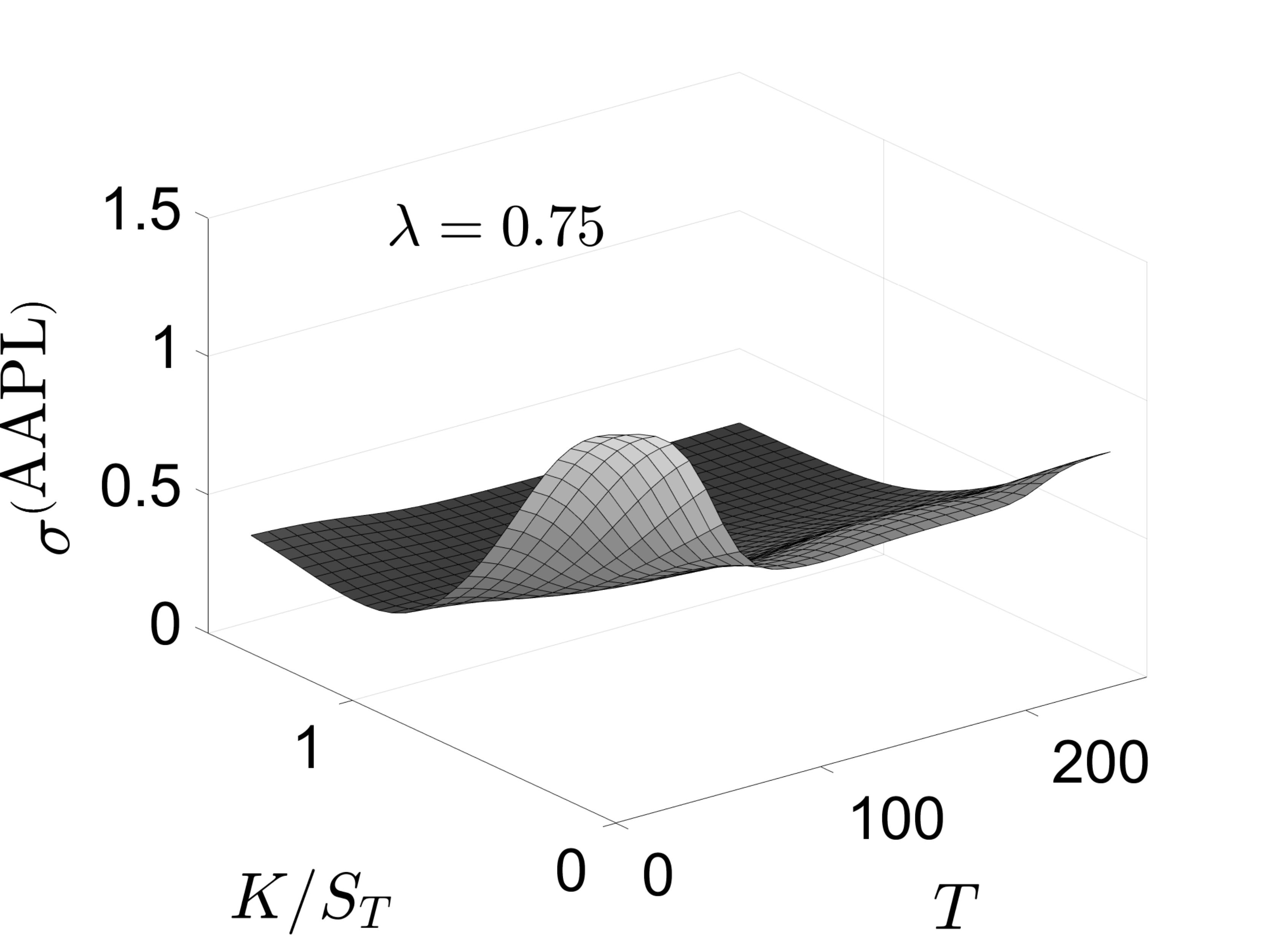}
    \end{subfigure}

    \caption{Implied $\sigma^{(\textrm{stock})}(T,M=K/S_T,\lambda)$ surfaces computed for the
    		log-return model \eqref{eq:ESG_lpm}.}
    \label{fig:impl_vol_ln}
\end{center}
\end{figure}

 \begin{figure}[h!]
\begin{center}   
    \begin{subfigure}[b]{0.32\textwidth} 
    	\includegraphics[width=\textwidth]{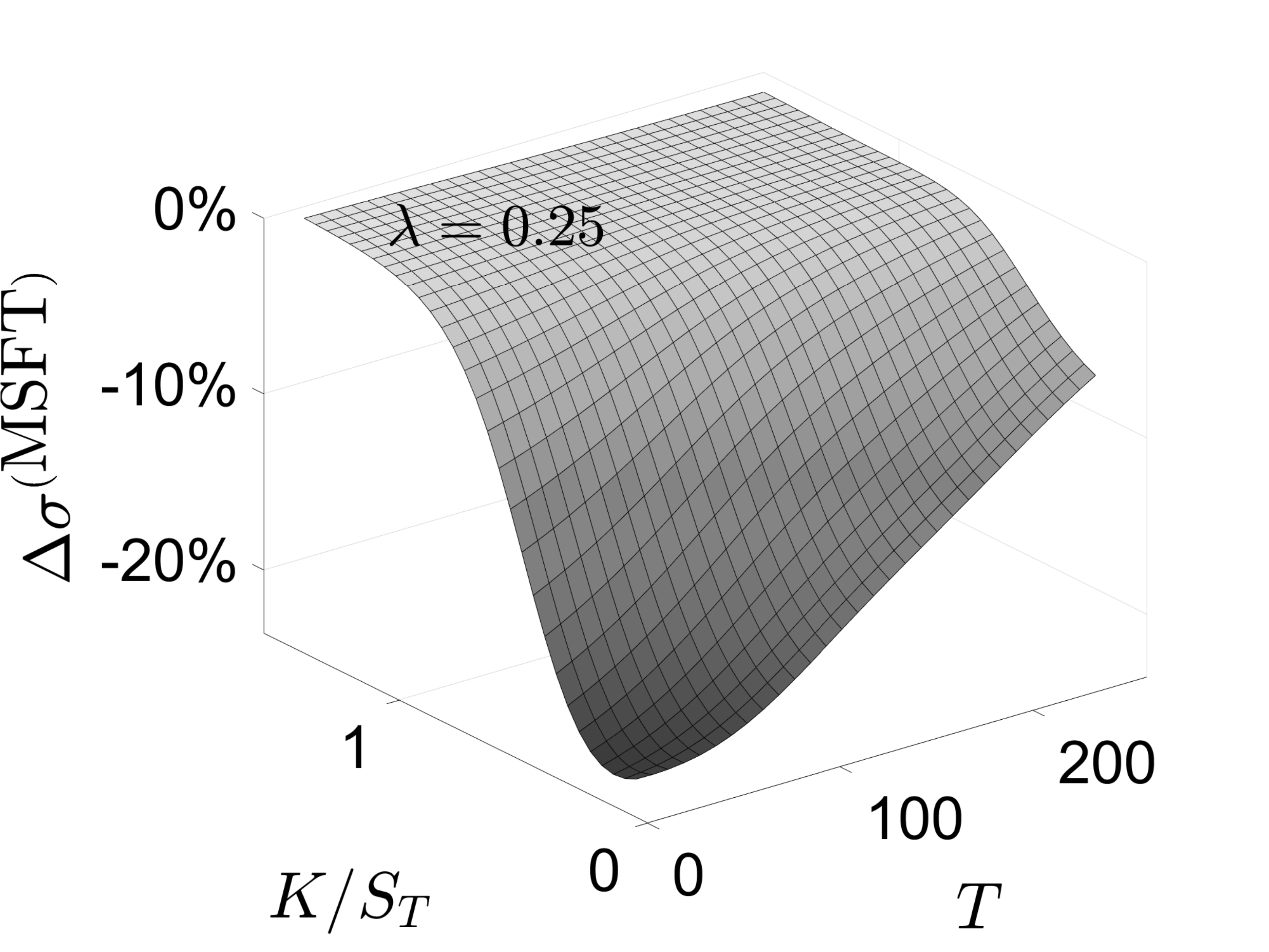}
    \end{subfigure}
    \begin{subfigure}[b]{0.32\textwidth} 
    	\includegraphics[width=\textwidth]{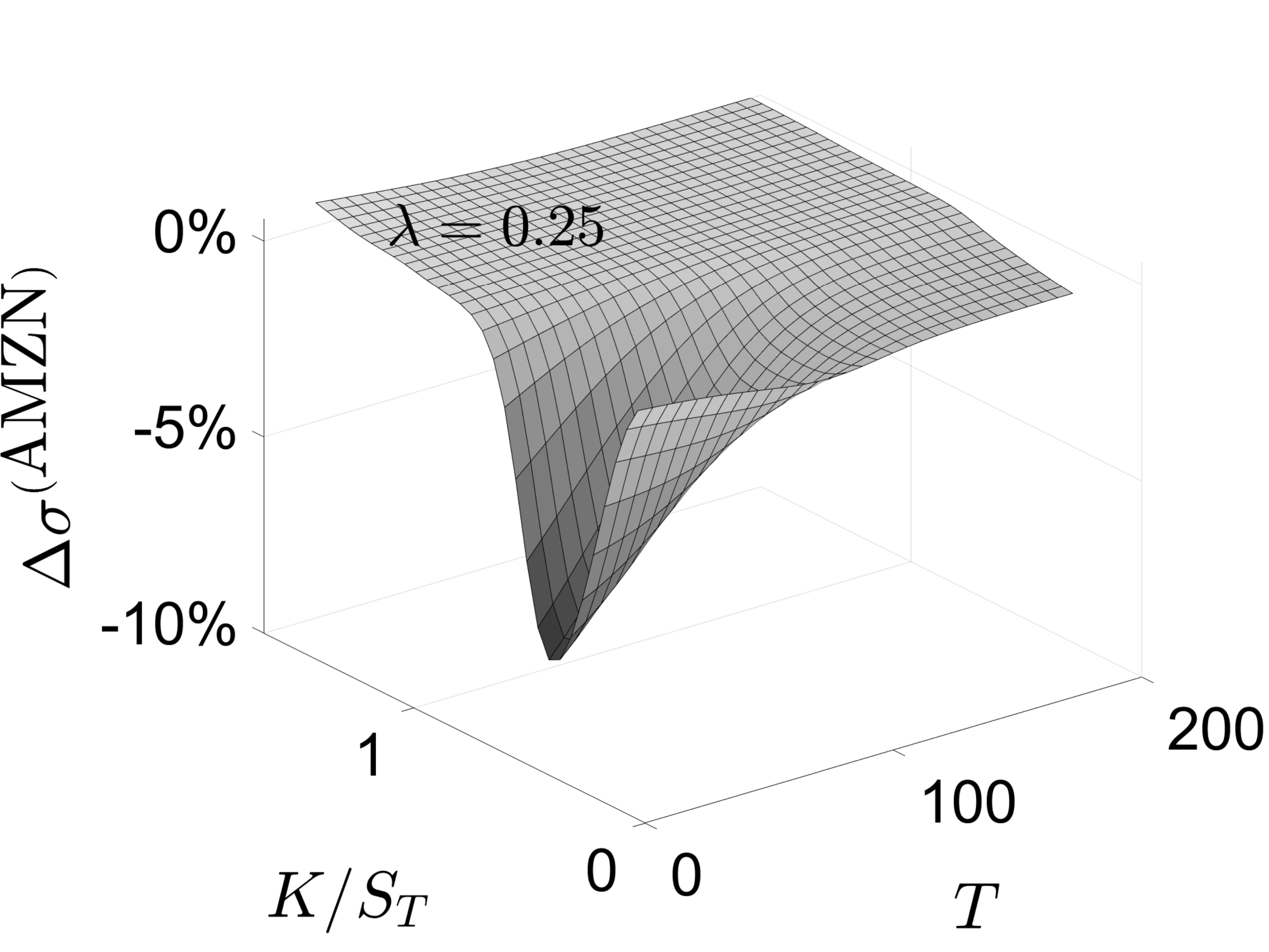}
    \end{subfigure}
    \begin{subfigure}[b]{0.32\textwidth} 
    	\includegraphics[width=\textwidth]{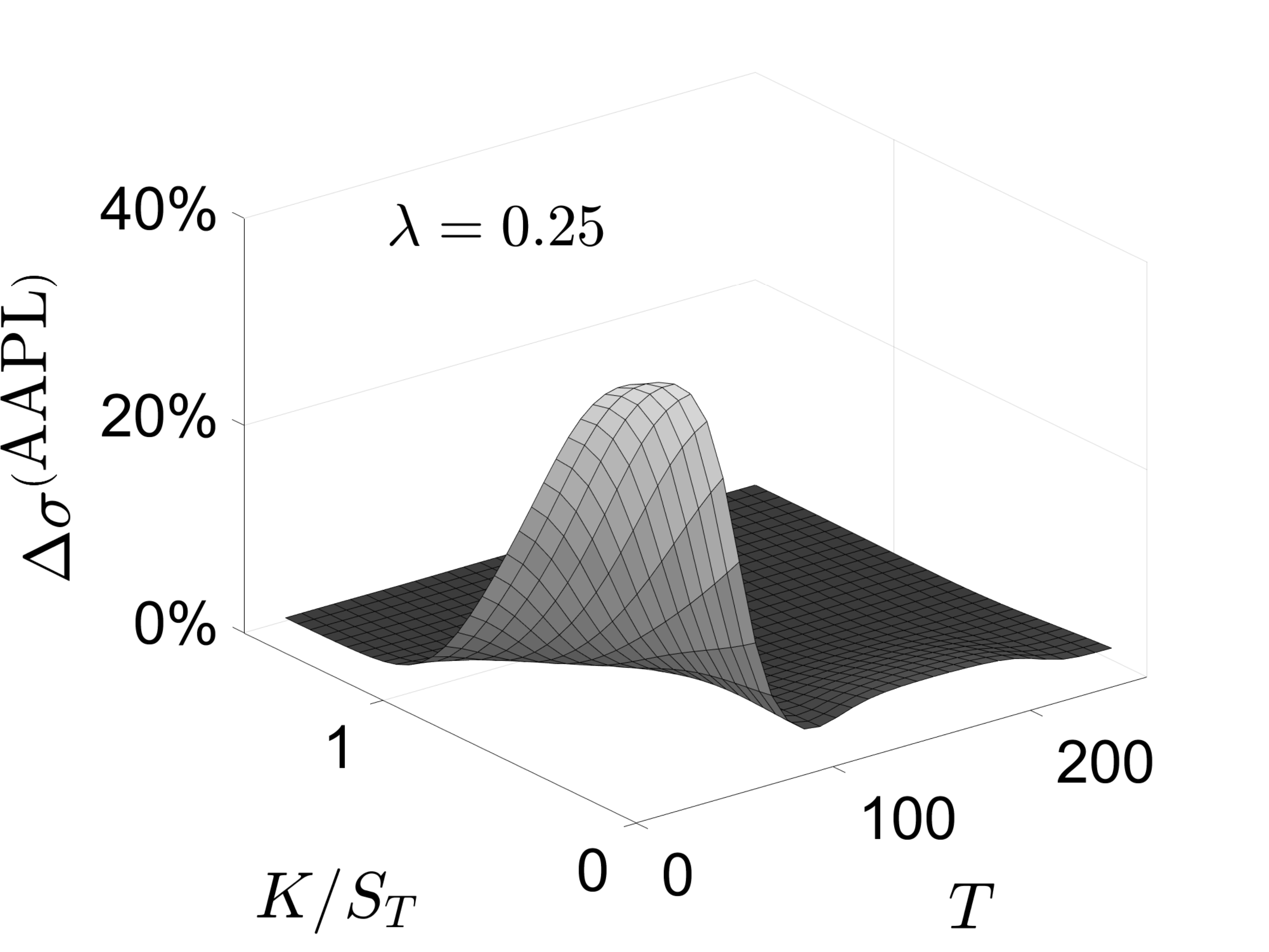}
    \end{subfigure}

    \begin{subfigure}[b]{0.32\textwidth} 
    	\includegraphics[width=\textwidth]{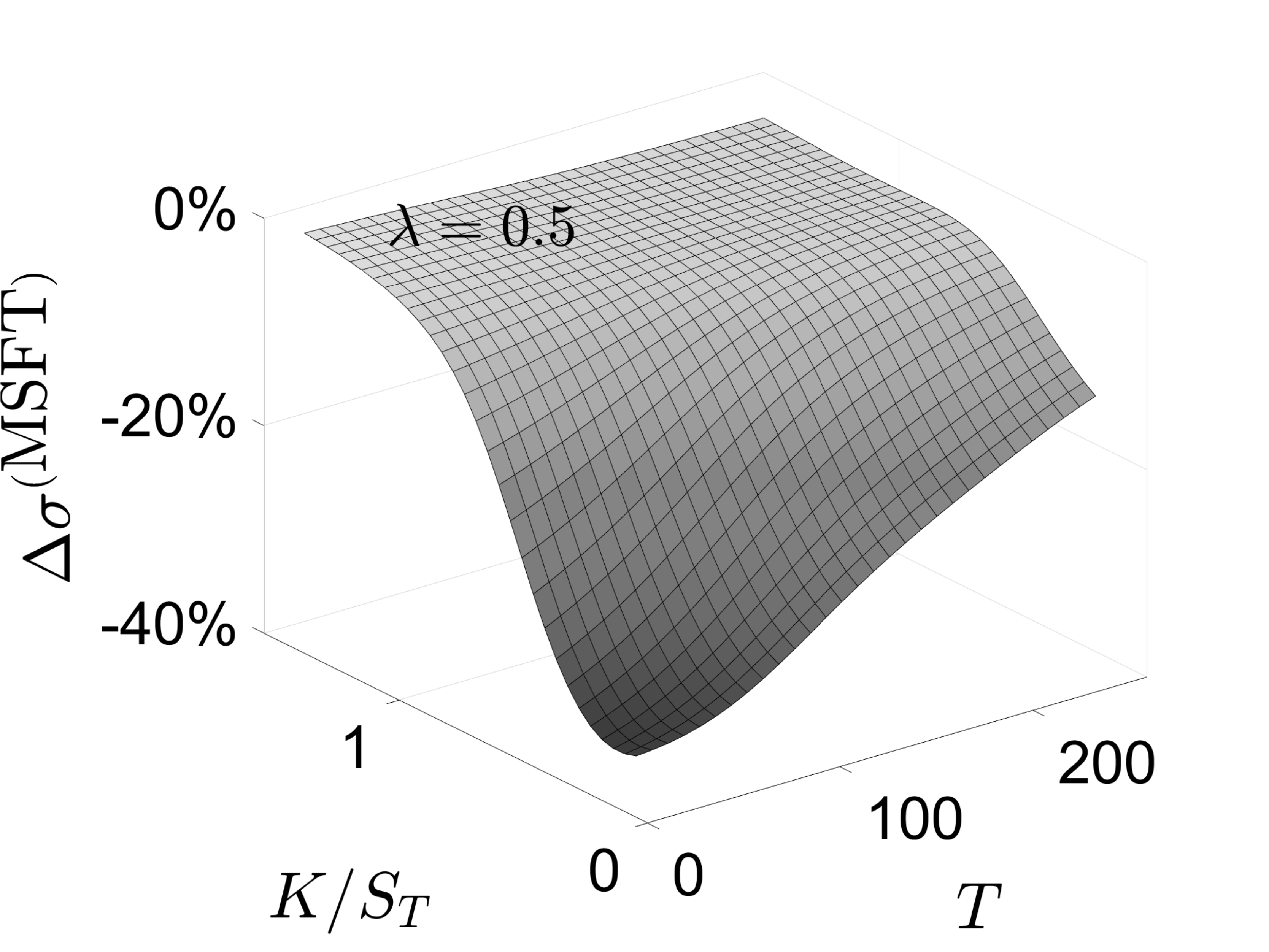}
    \end{subfigure}
    \begin{subfigure}[b]{0.32\textwidth} 
    	\includegraphics[width=\textwidth]{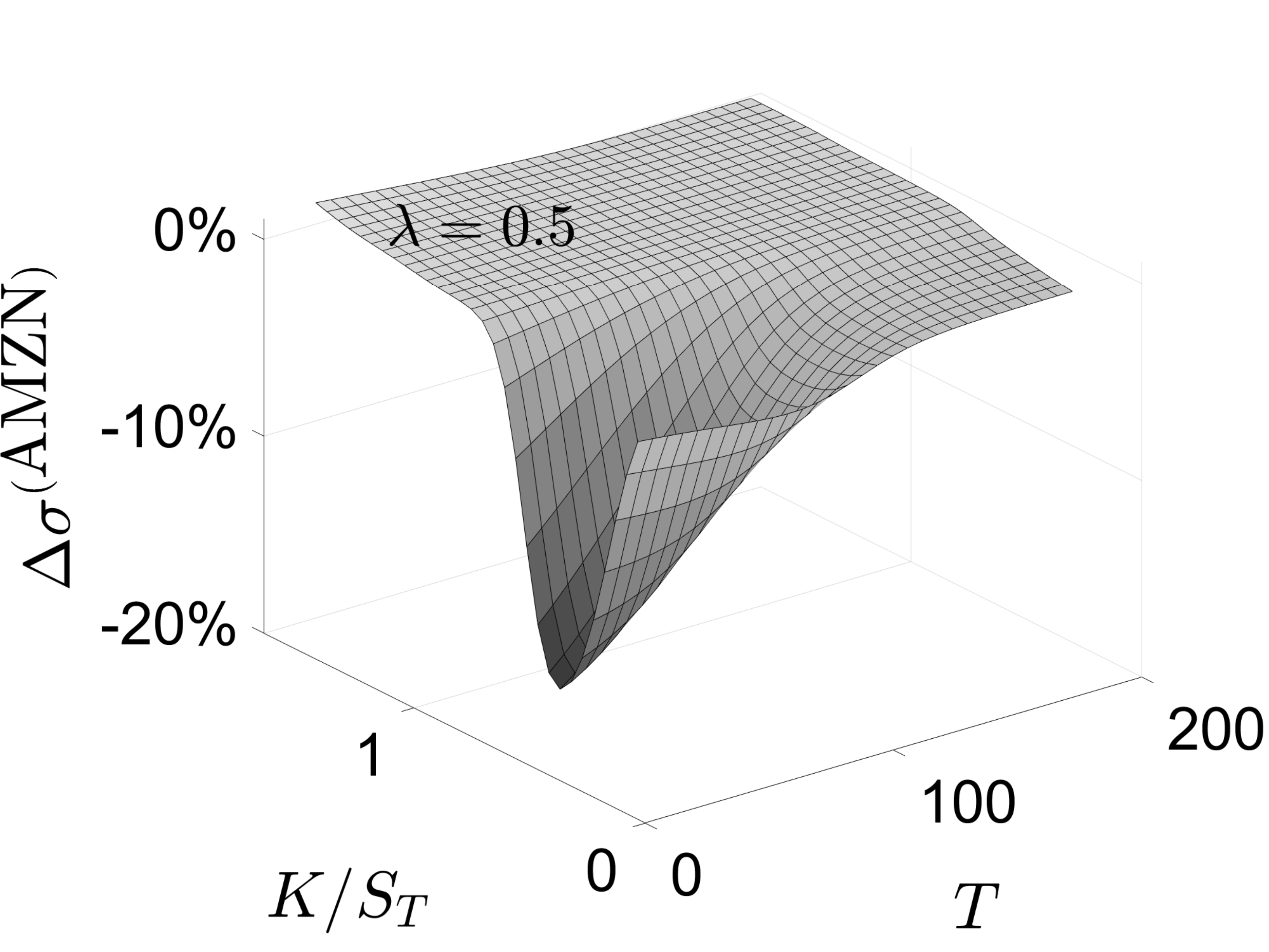}
    \end{subfigure}
        \begin{subfigure}[b]{0.32\textwidth} 
    	\includegraphics[width=\textwidth]{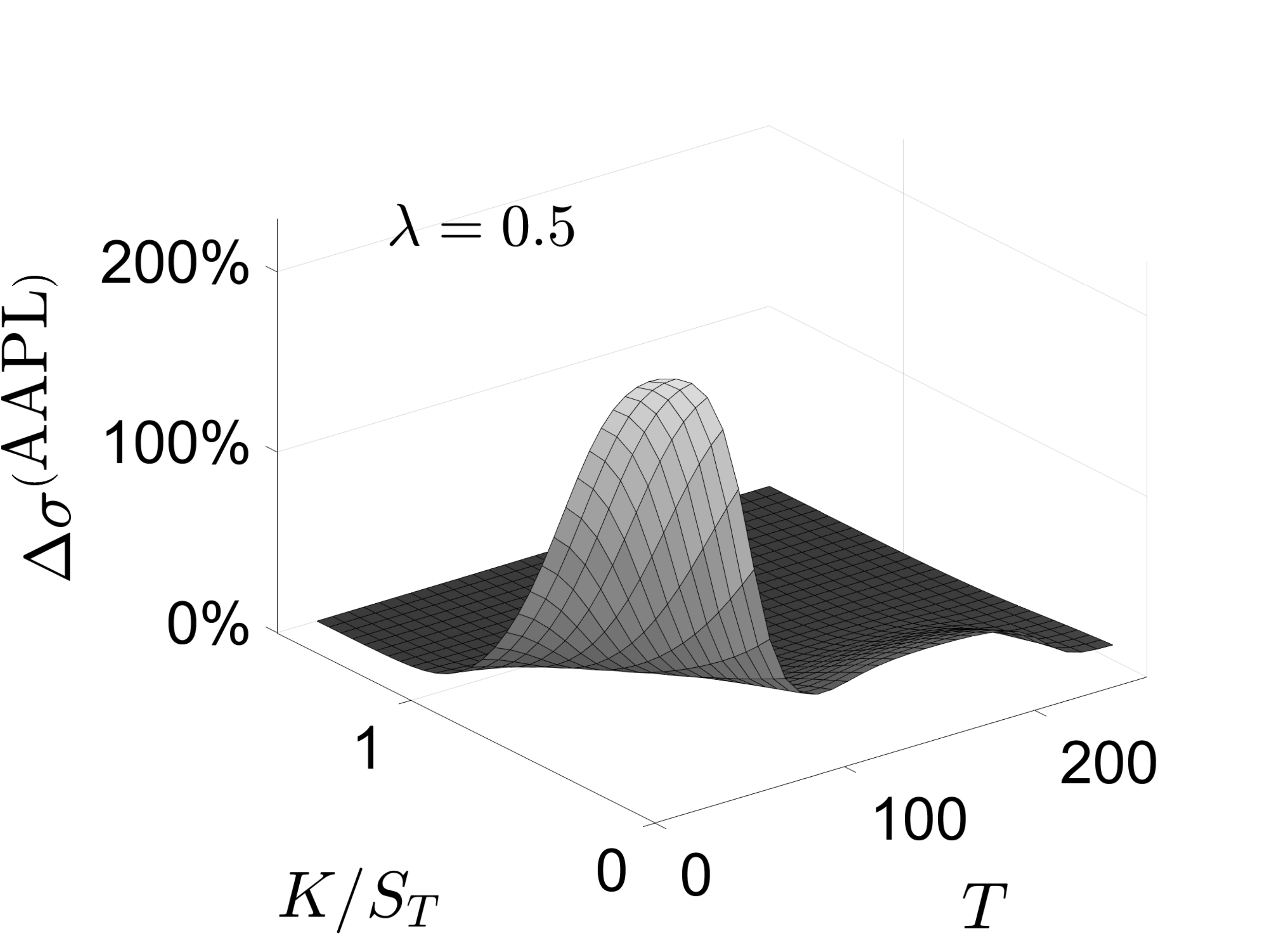}
    \end{subfigure}

    \begin{subfigure}[b]{0.32\textwidth} 
    	\includegraphics[width=\textwidth]{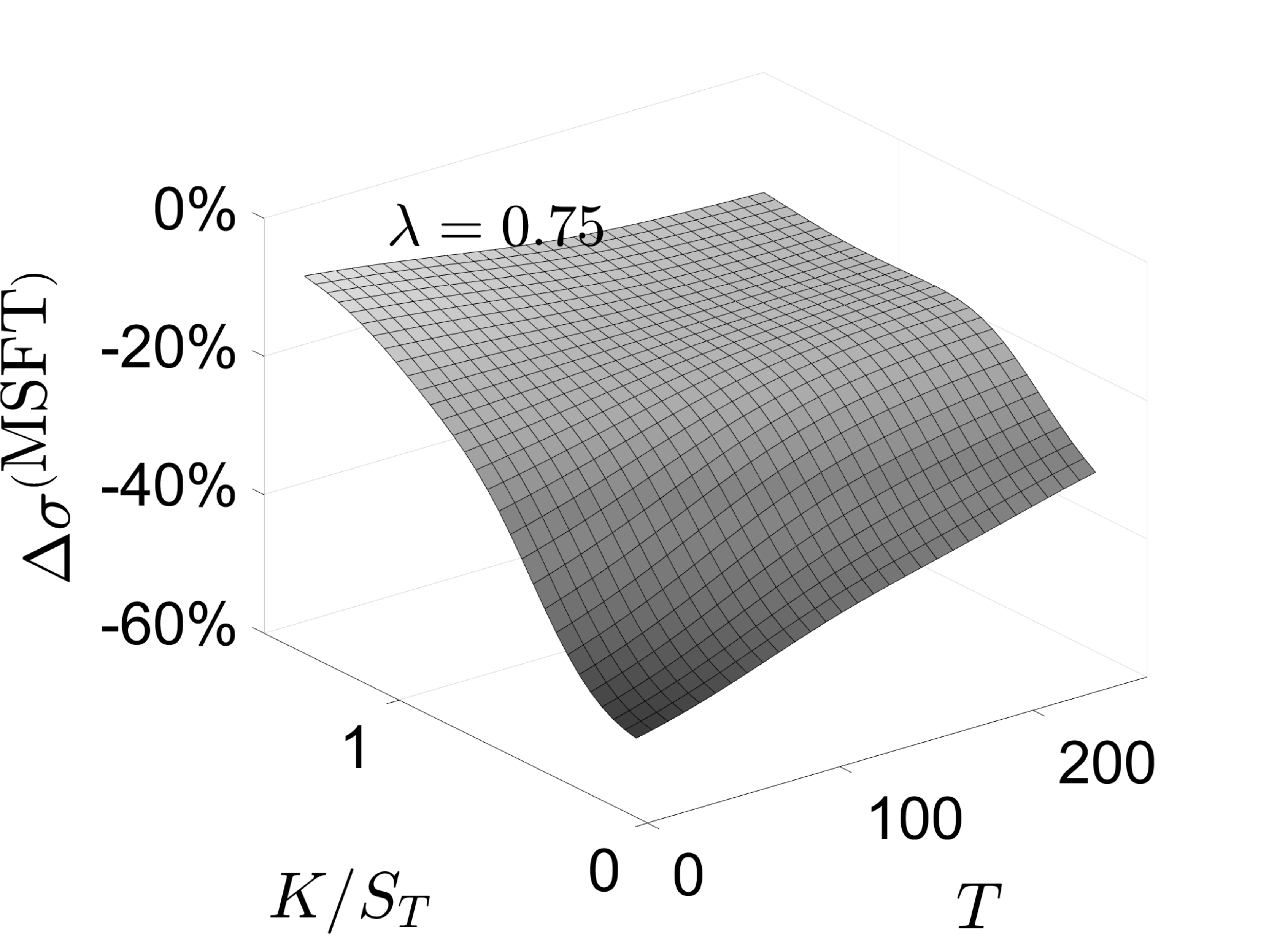}
    \end{subfigure}
    \begin{subfigure}[b]{0.32\textwidth} 
    	\includegraphics[width=\textwidth]{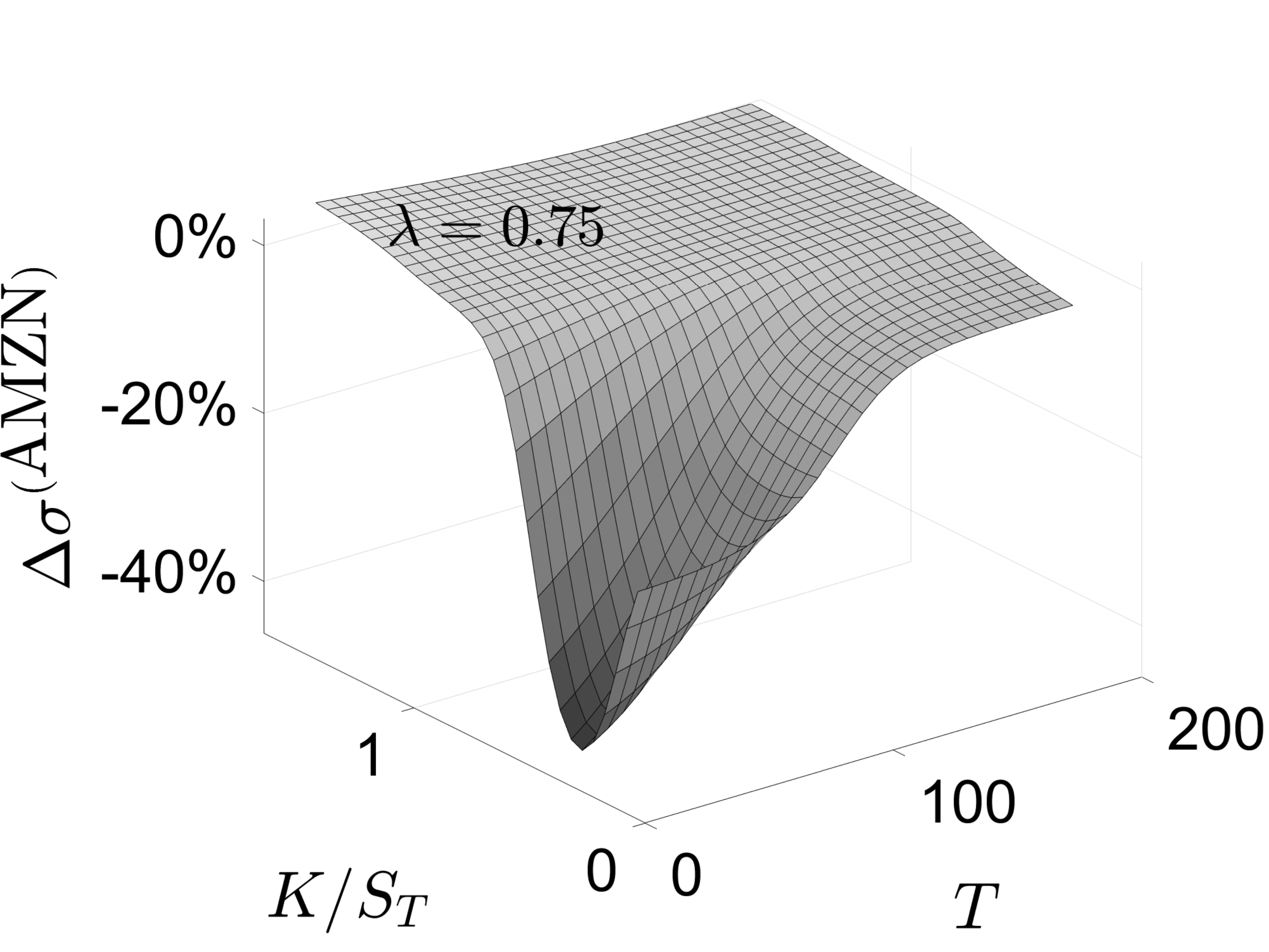}
    \end{subfigure}
        \begin{subfigure}[b]{0.32\textwidth} 
    	\includegraphics[width=\textwidth]{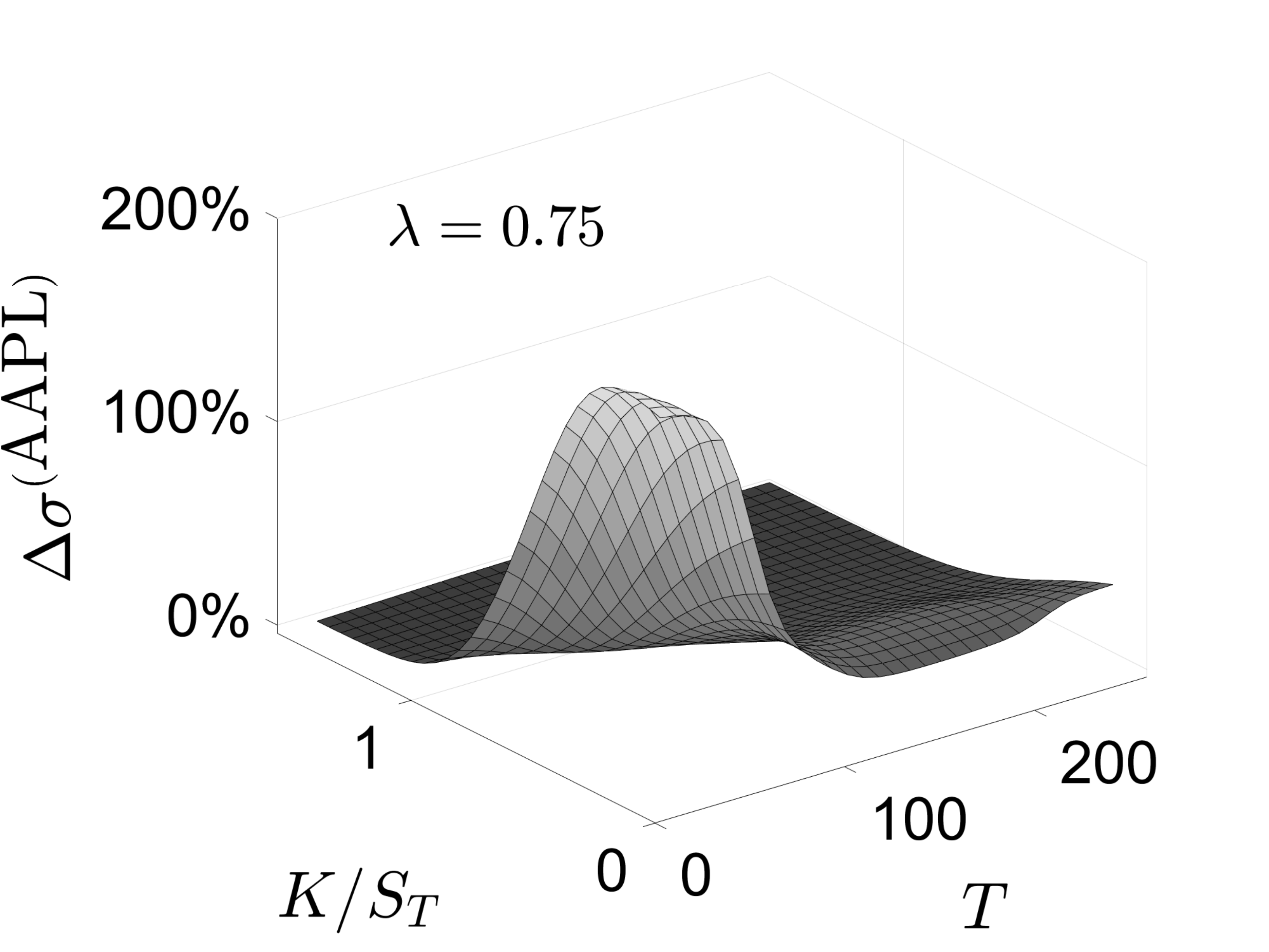}
    \end{subfigure}

    \caption{The percent relative change $\Delta \sigma^{(\text{stock})}(T,M=K/S_T,\lambda)$
    	in the implied volatility surfaces of Fig.~\ref{fig:impl_vol_ln}.}
    \label{fig:impl_vol_ln_dev}
\end{center}
\end{figure}

\begin{figure}[h!]
\begin{center}
    \begin{subfigure}[b]{0.32\textwidth} 
    	\includegraphics[width=\textwidth]{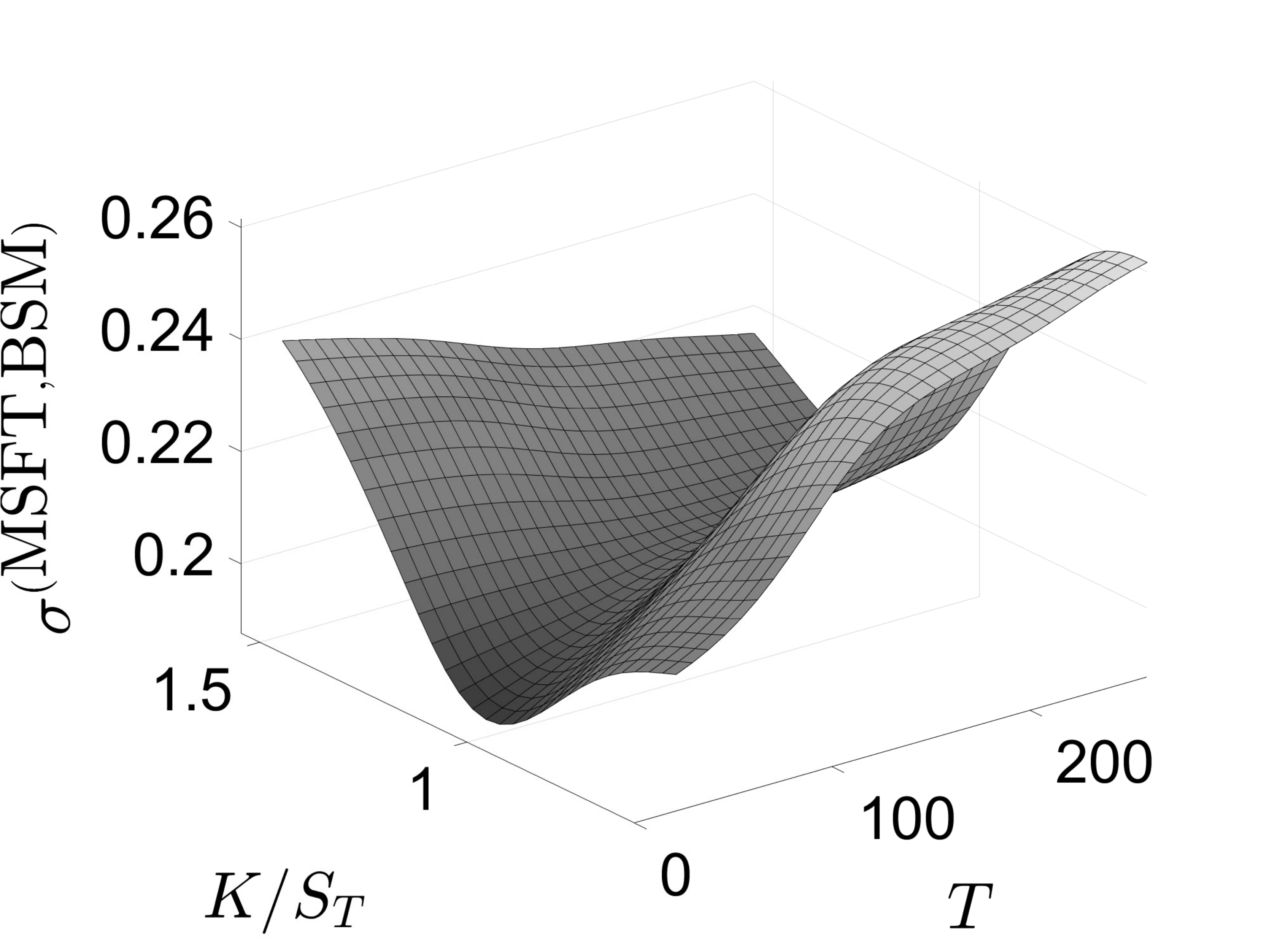}
    \end{subfigure}
    \begin{subfigure}[b]{0.32\textwidth} 
    	\includegraphics[width=\textwidth]{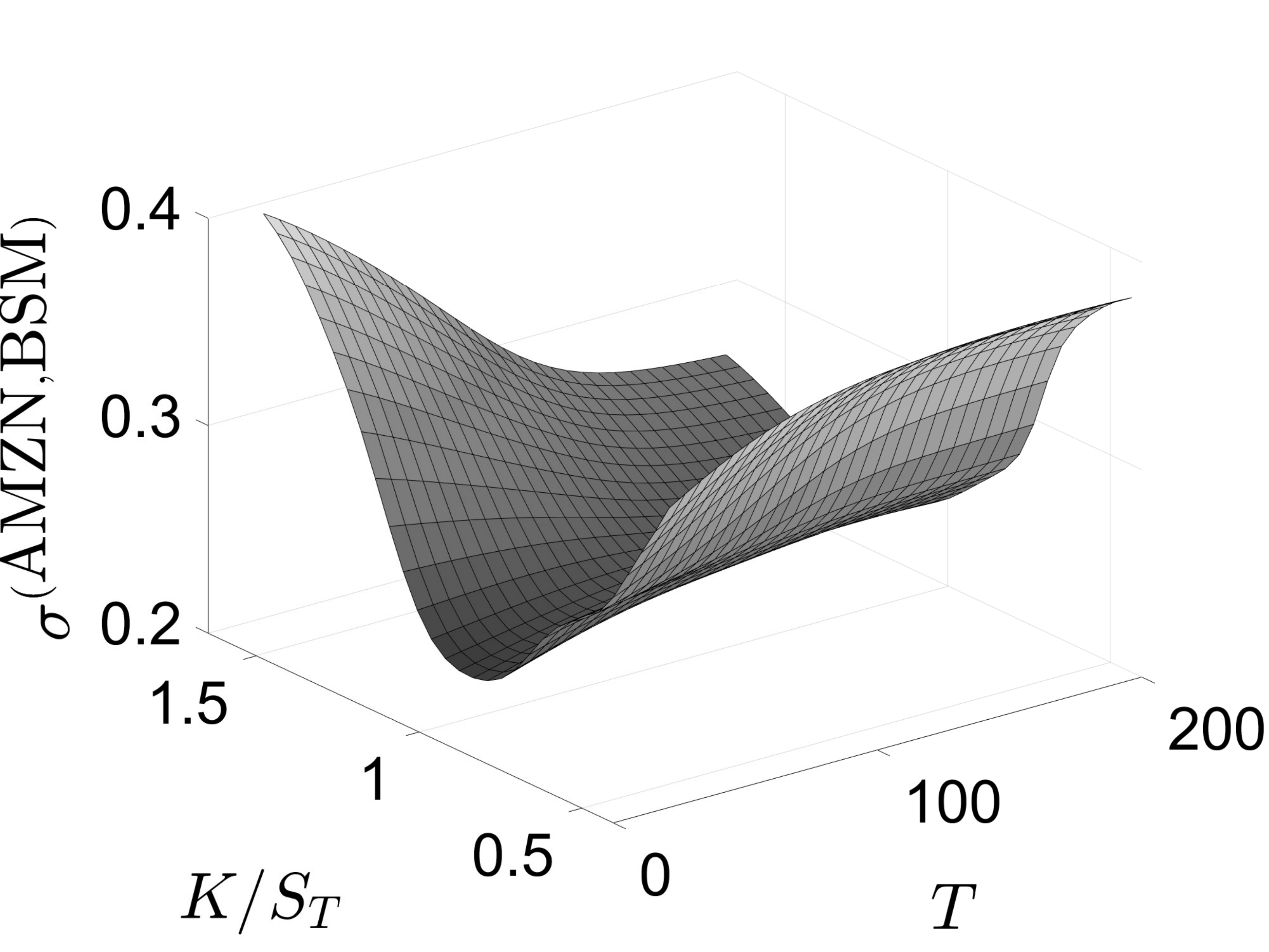}
    \end{subfigure}
        \begin{subfigure}[b]{0.32\textwidth} 
    	\includegraphics[width=\textwidth]{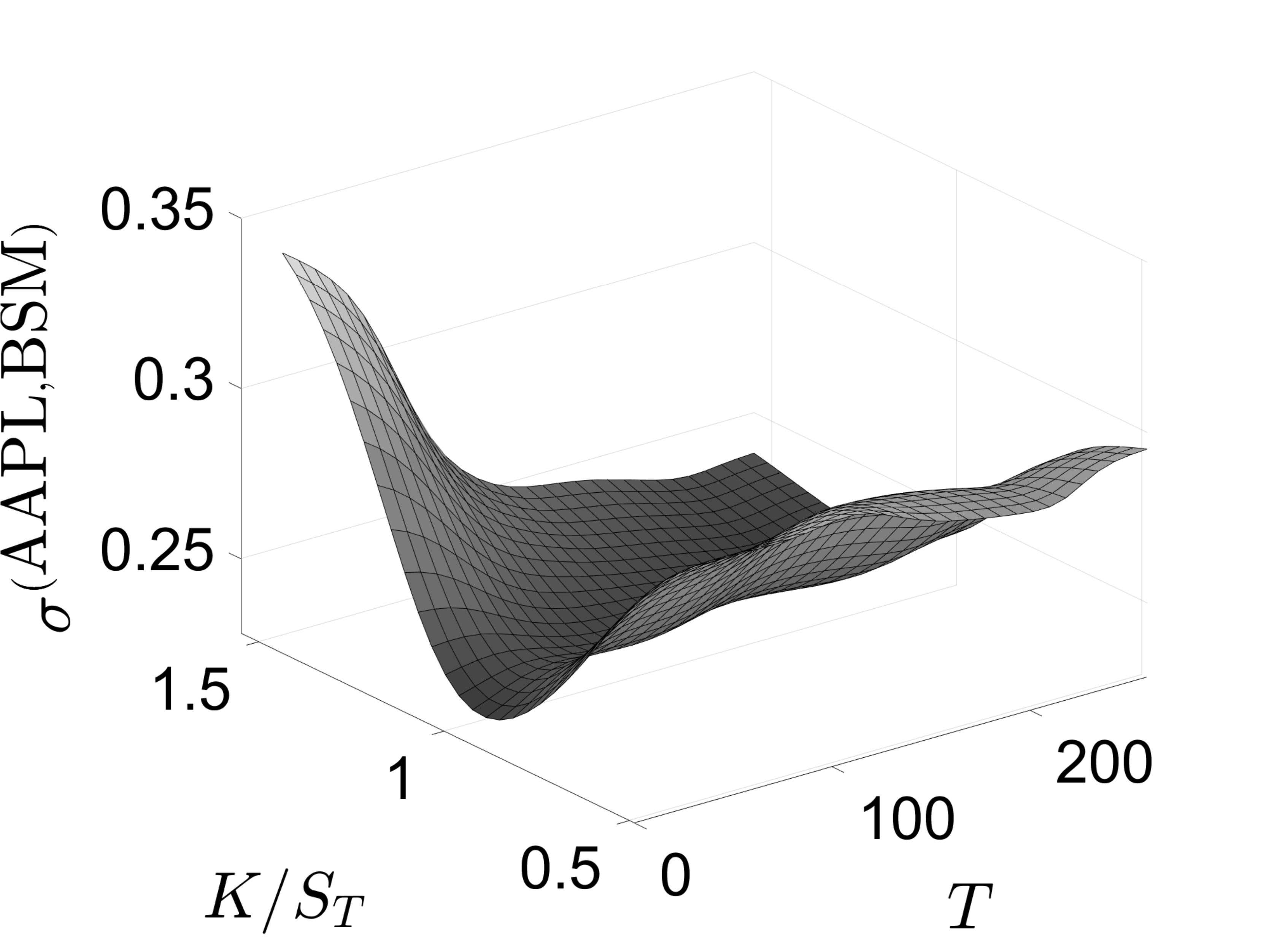}
    \end{subfigure}
    
     \begin{subfigure}[b]{0.32\textwidth} 
    	\includegraphics[width=\textwidth]{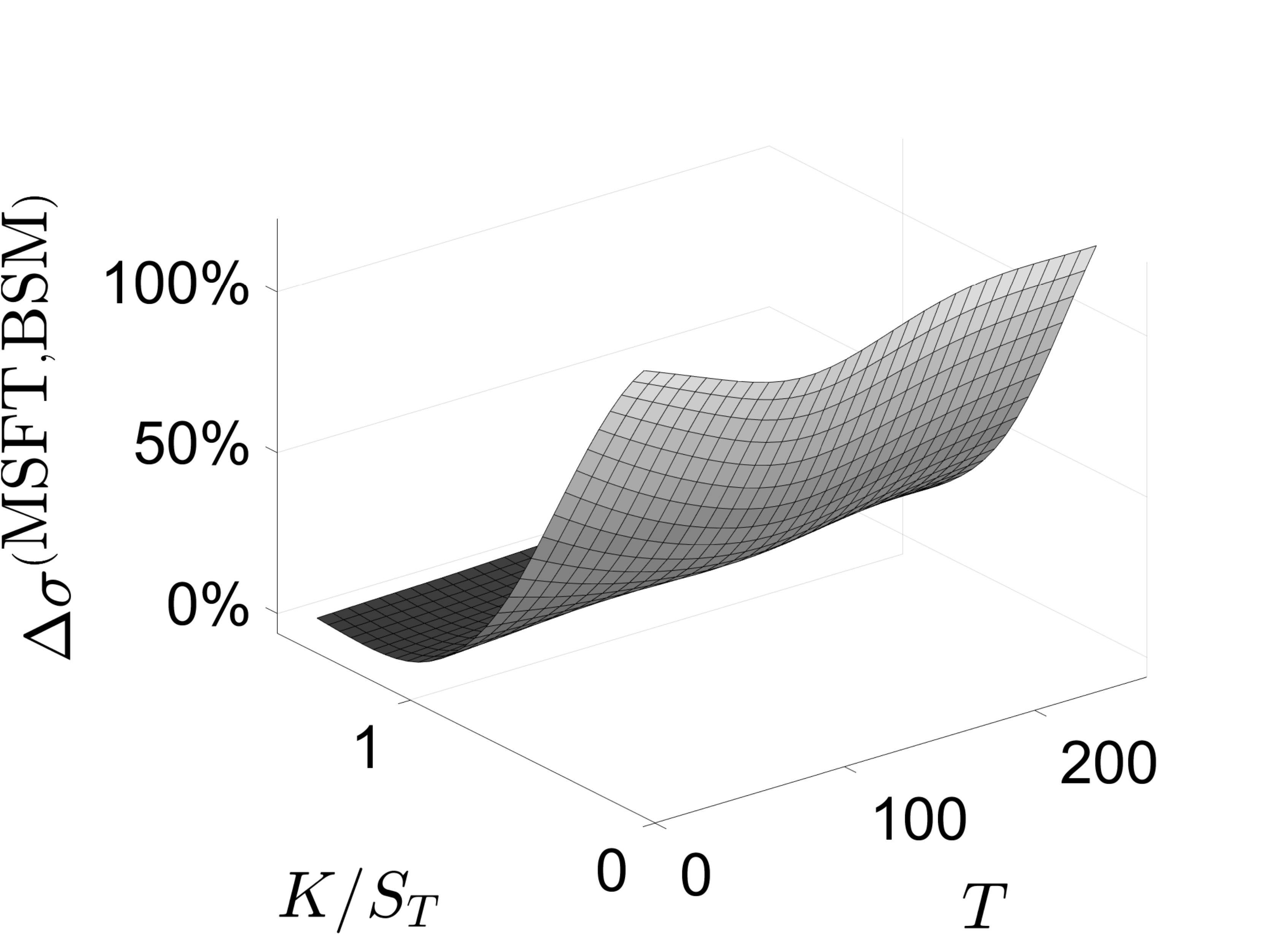}
    \end{subfigure}
    \begin{subfigure}[b]{0.32\textwidth} 
    	\includegraphics[width=\textwidth]{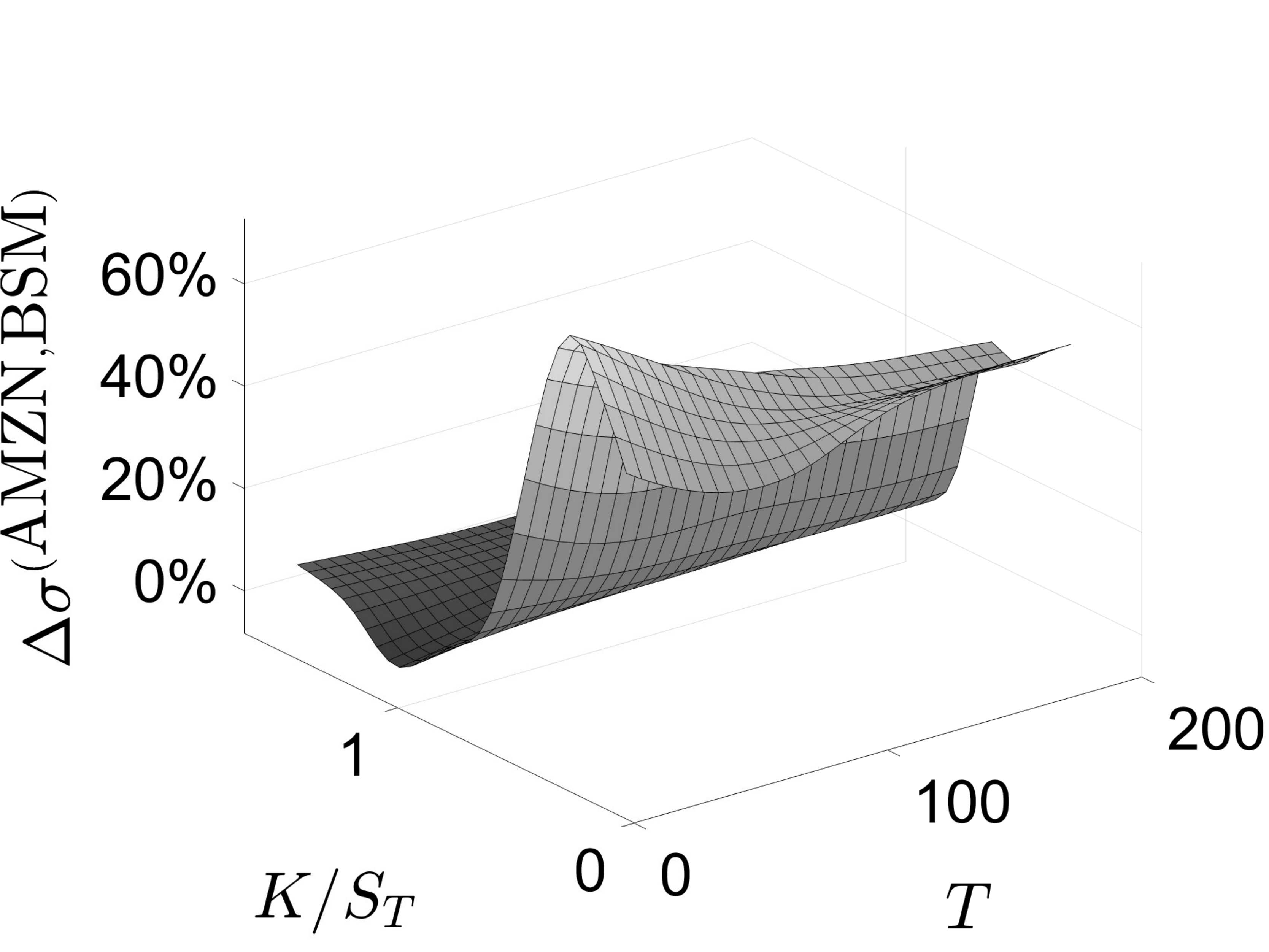}
    \end{subfigure}
    \begin{subfigure}[b]{0.32\textwidth} 
    	\includegraphics[width=\textwidth]{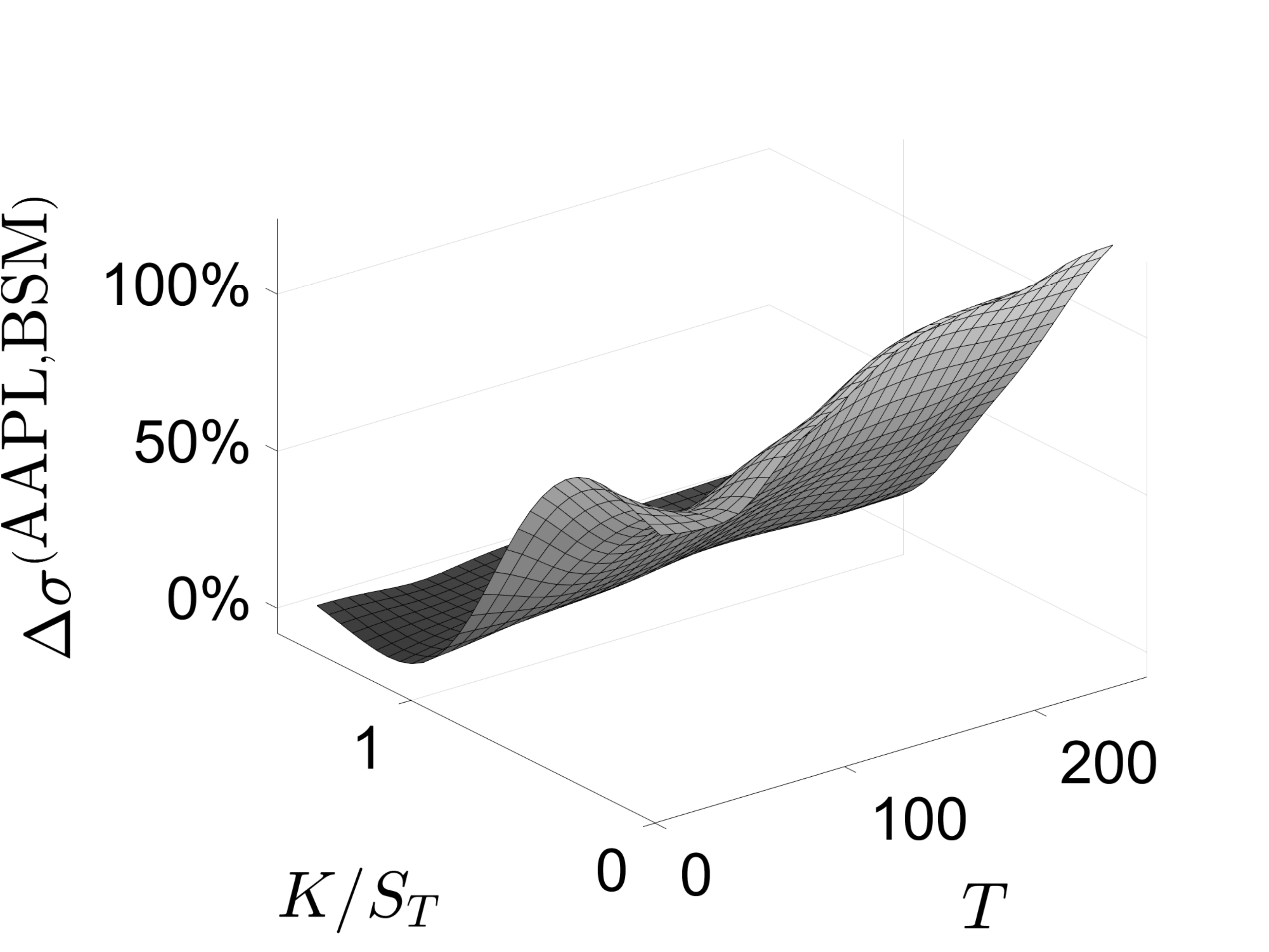}
    \end{subfigure}
    
     \begin{subfigure}[b]{0.32\textwidth} 
    	\includegraphics[width=\textwidth]{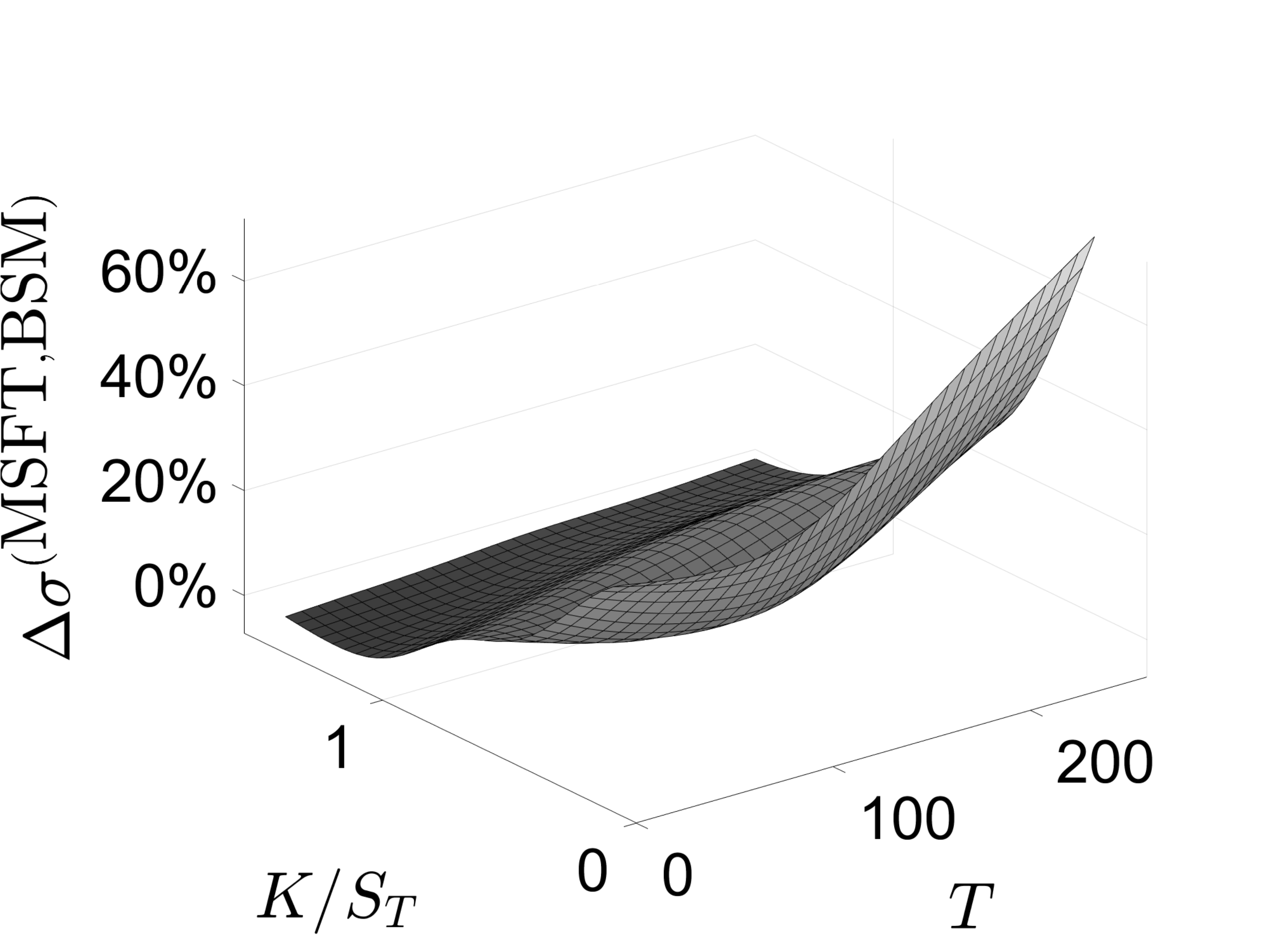}
    \end{subfigure}
    \begin{subfigure}[b]{0.32\textwidth} 
    	\includegraphics[width=\textwidth]{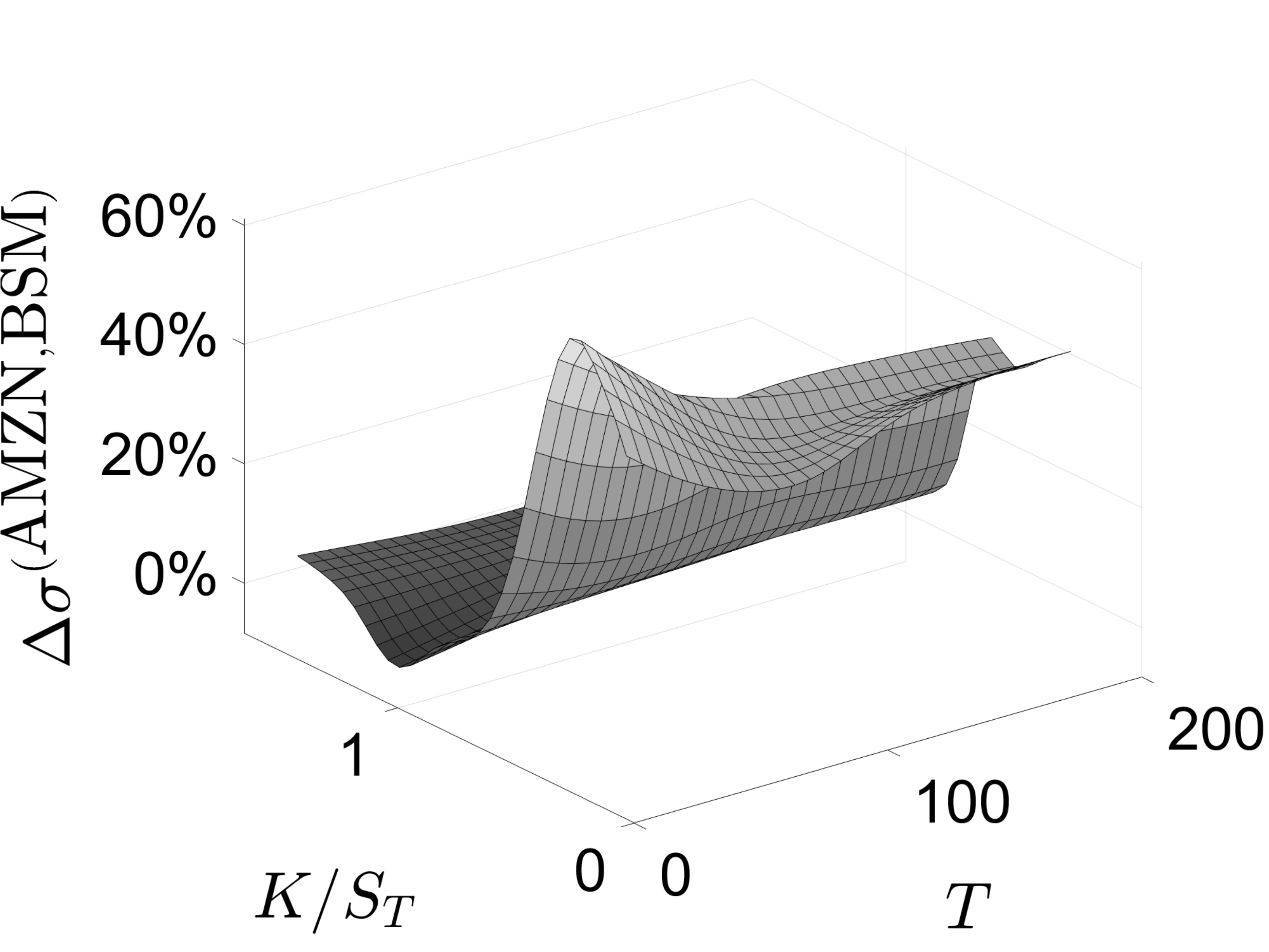}
    \end{subfigure}
    \begin{subfigure}[b]{0.32\textwidth} 
    	\includegraphics[width=\textwidth]{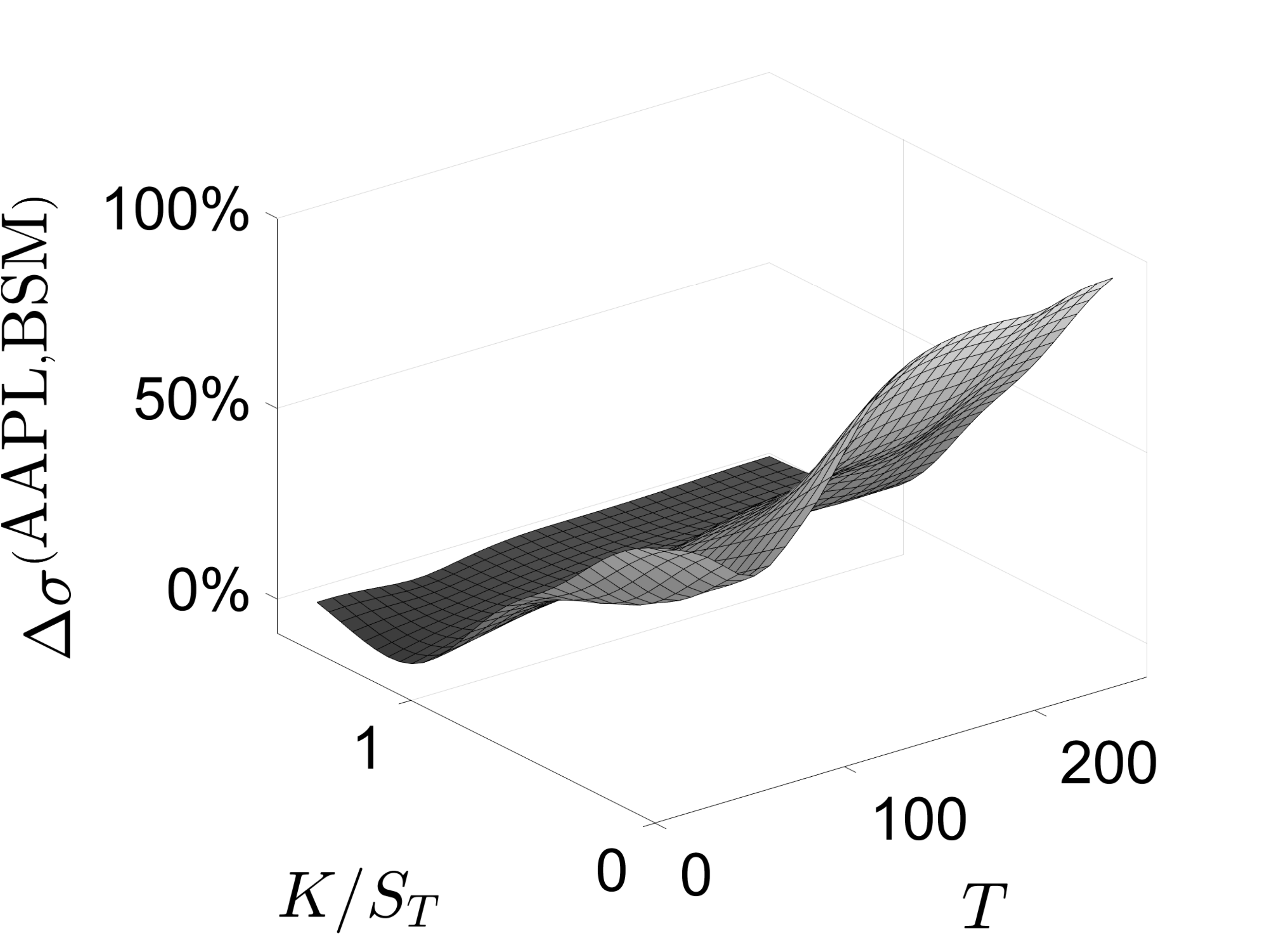}
    \end{subfigure}
    
    \caption{(top) The Black-Scholes-Merton implied volatility surfaces \eqref{eq:impl_vol_BSM}.
		Also shown are the percent relative differences \eqref{eq:impl_vol_BSM_diff} computed
		using the (middle) arithmetic and (bottom) log-return models.}
    \label{fig:impl_vol_BSM}
\end{center}
\end{figure}

\clearpage

\renewcommand{\theequation}{B.\arabic{equation}}
\setcounter{equation}{0}
\renewcommand{\thefigure}{B\arabic{figure}}
\setcounter{figure}{0}
\renewcommand{\thetable}{B\arabic{table}}
\setcounter{table}{0}

\section{Implied information-intensity surfaces}
\label{app:B}

\noindent

 \begin{figure}[h!]
\begin{center}   
    \begin{subfigure}[b]{0.32\textwidth} 
    	\includegraphics[width=\textwidth]{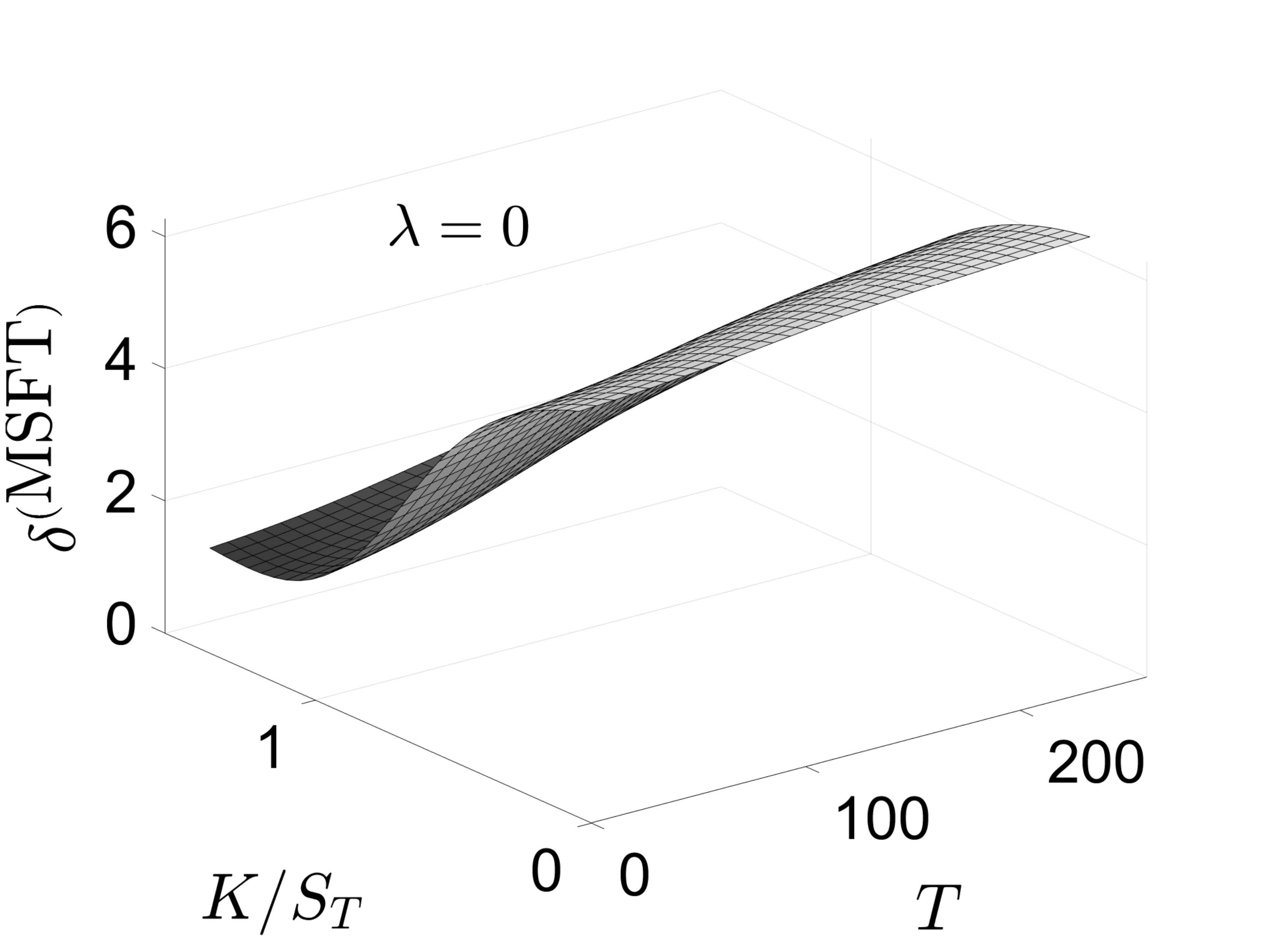}
    \end{subfigure}
    \begin{subfigure}[b]{0.32\textwidth} 
    	\includegraphics[width=\textwidth]{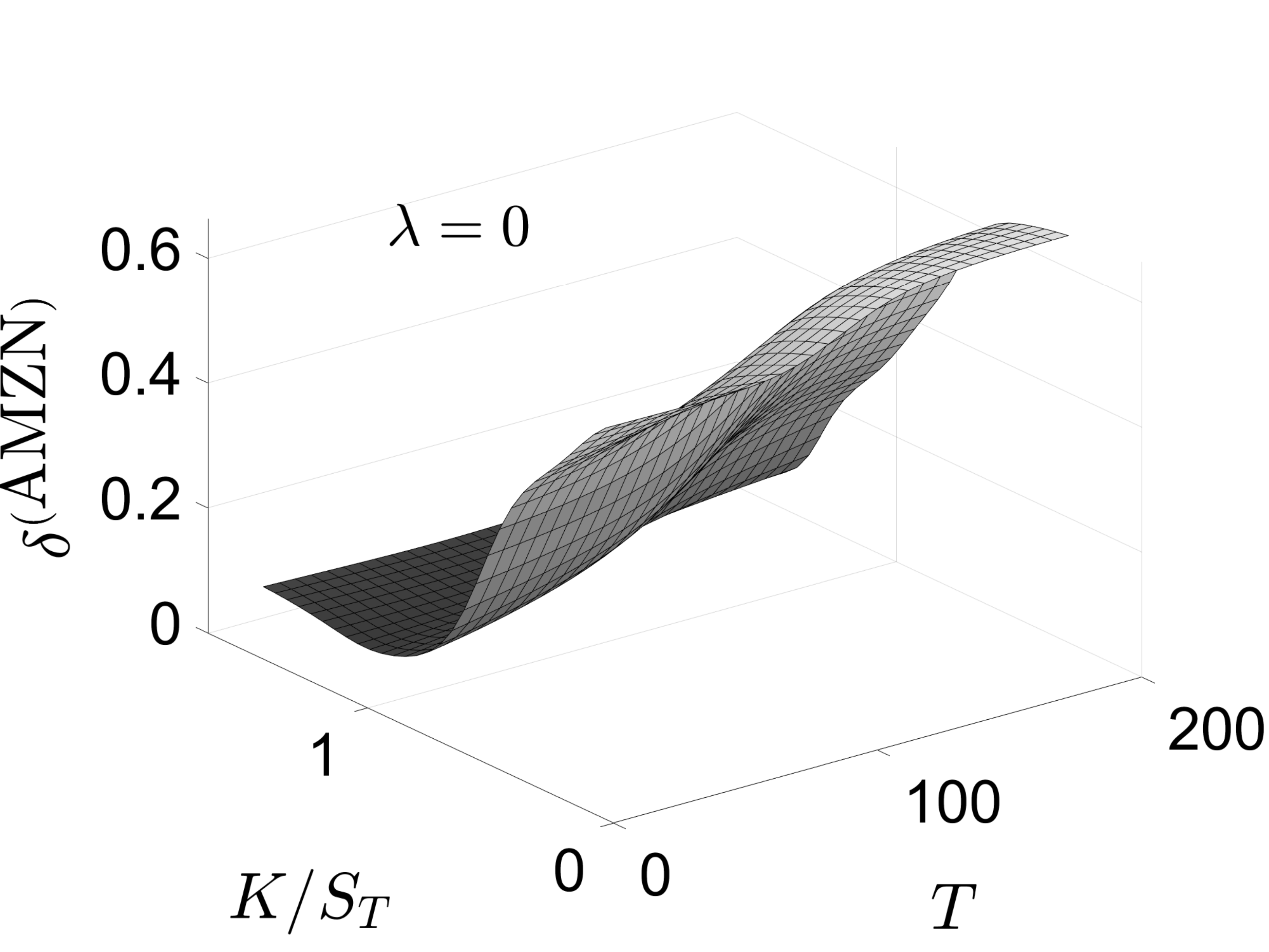}
    \end{subfigure}
    \begin{subfigure}[b]{0.32\textwidth} 
    	\includegraphics[width=\textwidth]{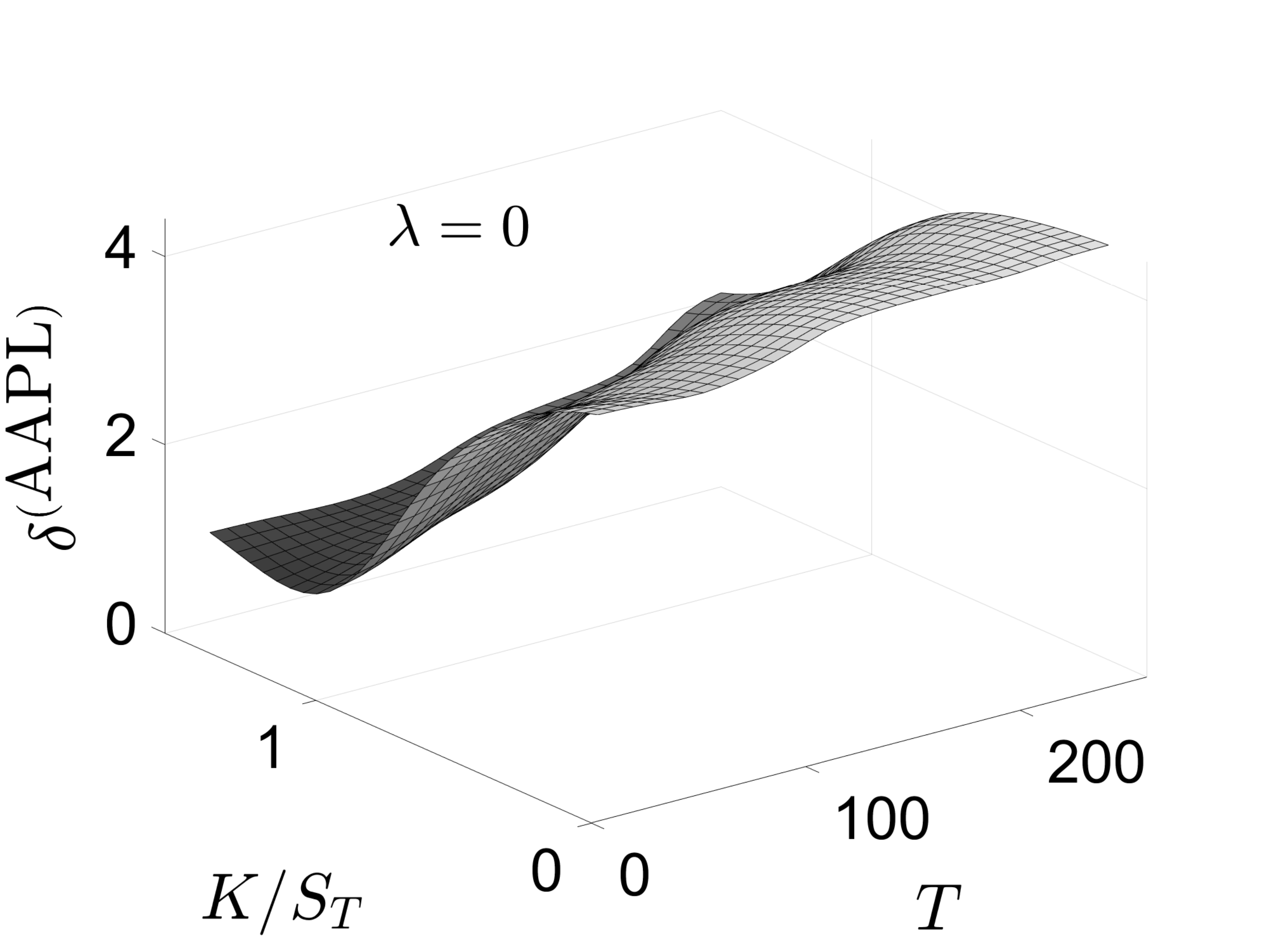}
    \end{subfigure}

    \begin{subfigure}[b]{0.32\textwidth} 
    	\includegraphics[width=\textwidth]{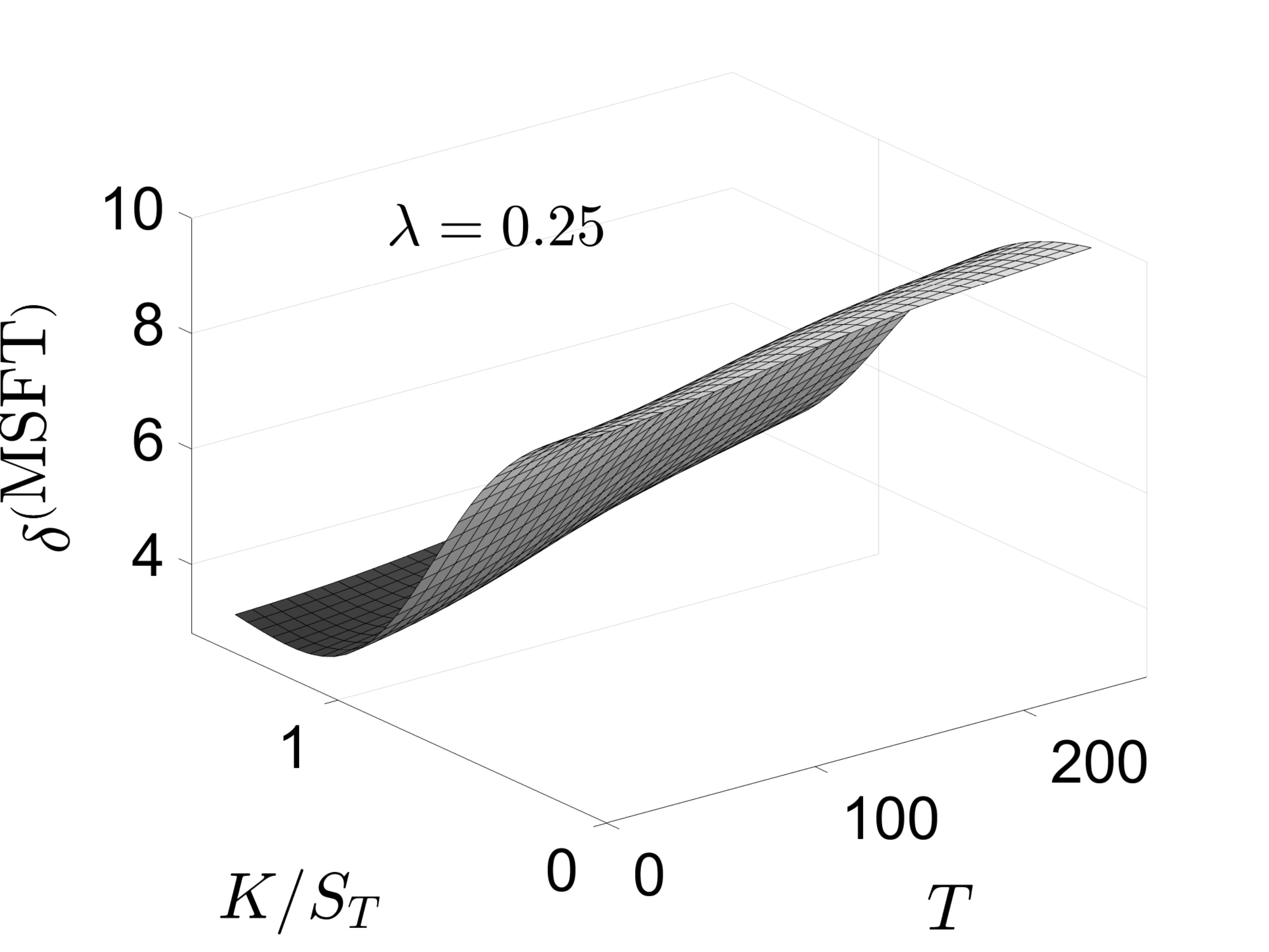}
    \end{subfigure}
    \begin{subfigure}[b]{0.32\textwidth} 
    	\includegraphics[width=\textwidth]{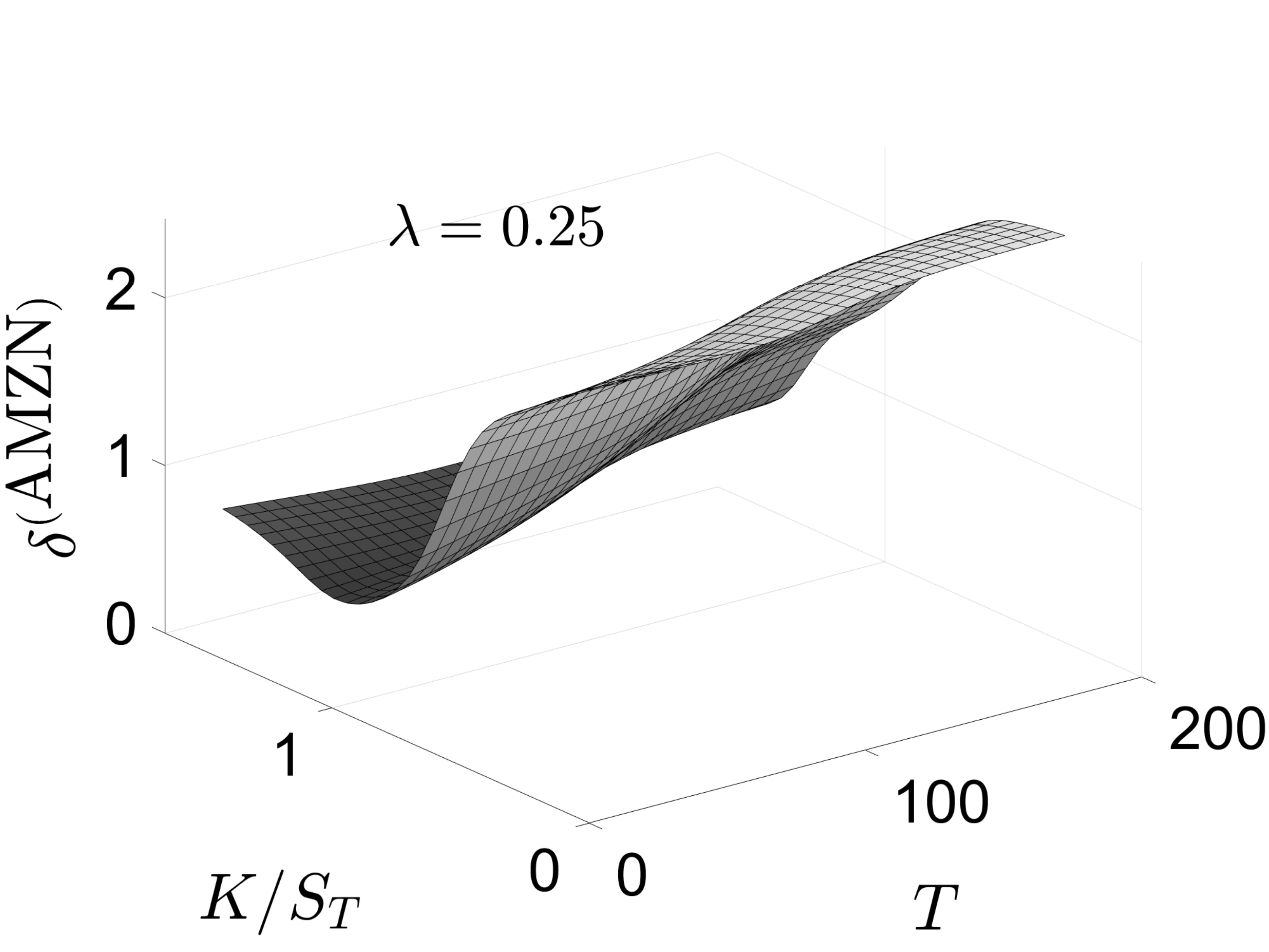}
    \end{subfigure}
    \begin{subfigure}[b]{0.32\textwidth} 
    	\includegraphics[width=\textwidth]{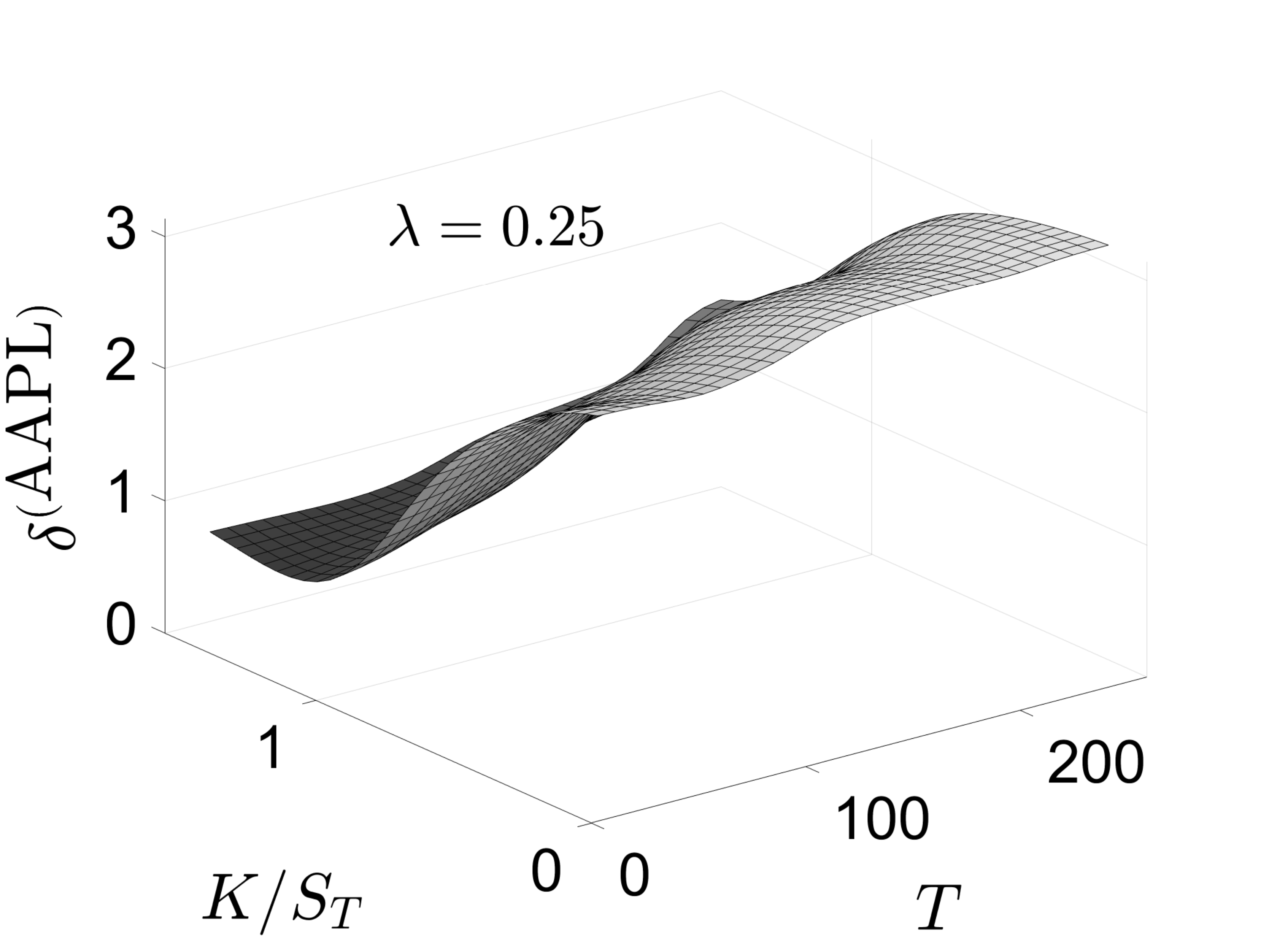}
    \end{subfigure}

    \begin{subfigure}[b]{0.32\textwidth} 
    	\includegraphics[width=\textwidth]{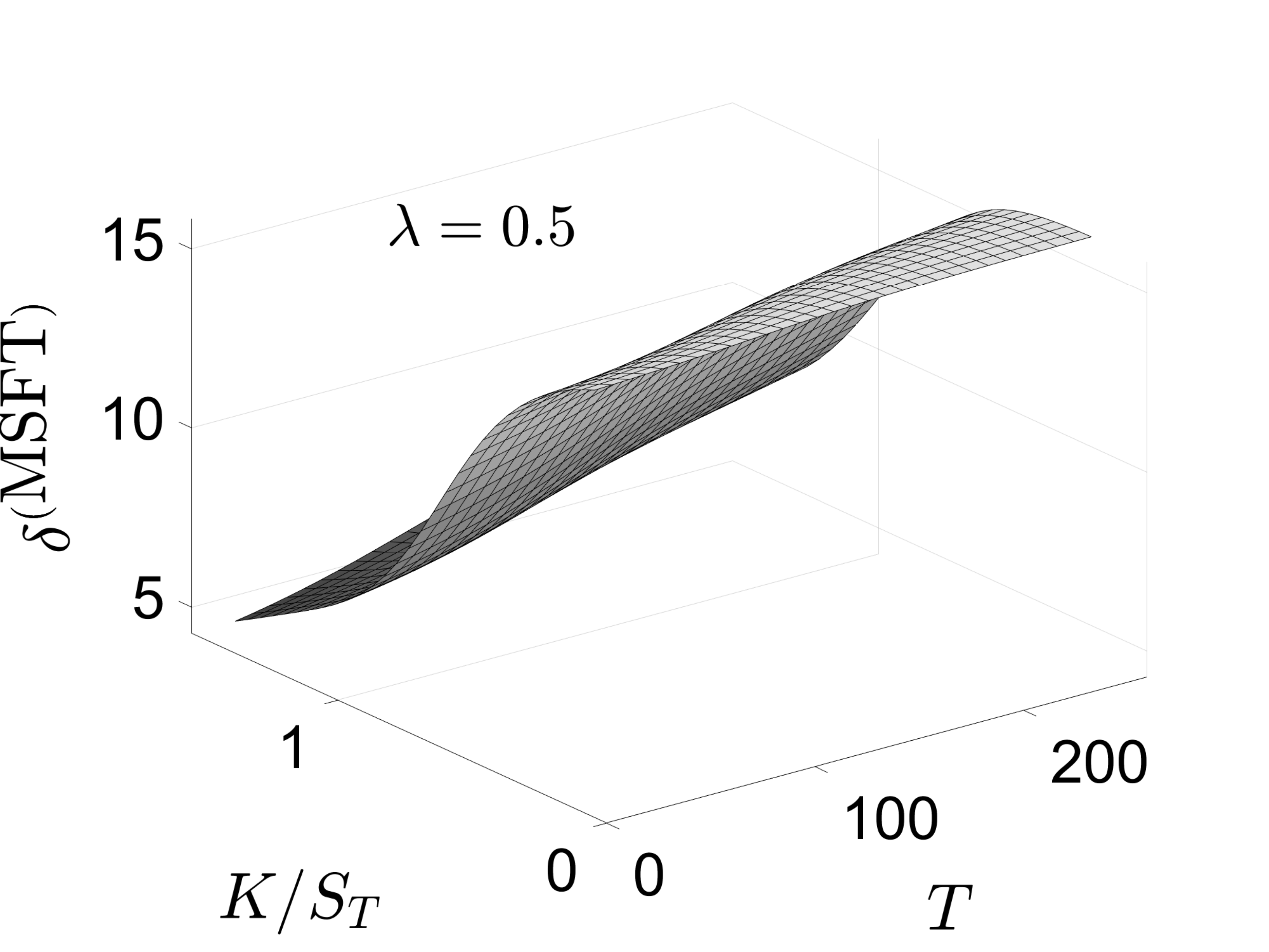}
    \end{subfigure}
    \begin{subfigure}[b]{0.32\textwidth} 
    	\includegraphics[width=\textwidth]{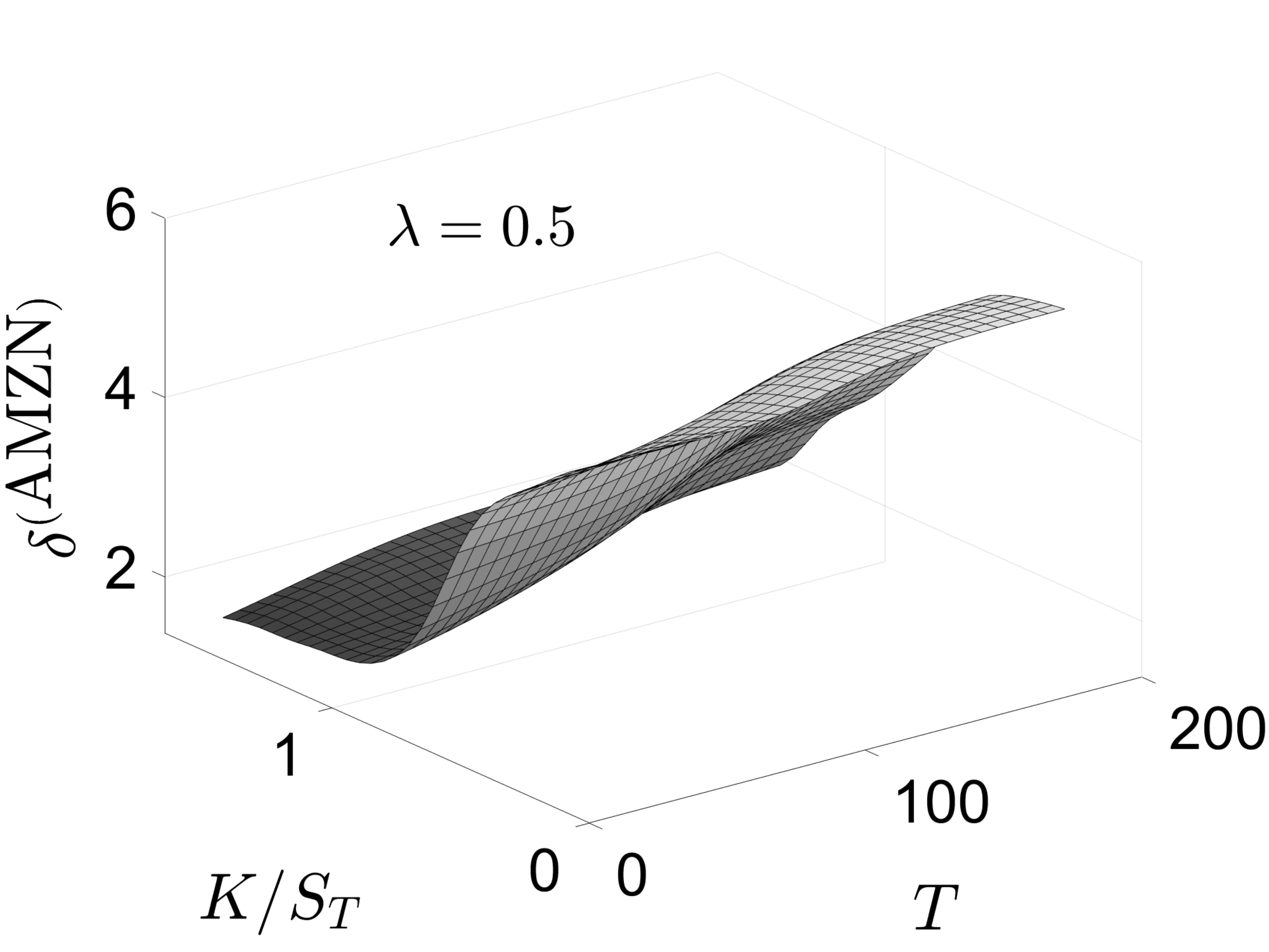}
    \end{subfigure}
    \begin{subfigure}[b]{0.32\textwidth} 
    	\includegraphics[width=\textwidth]{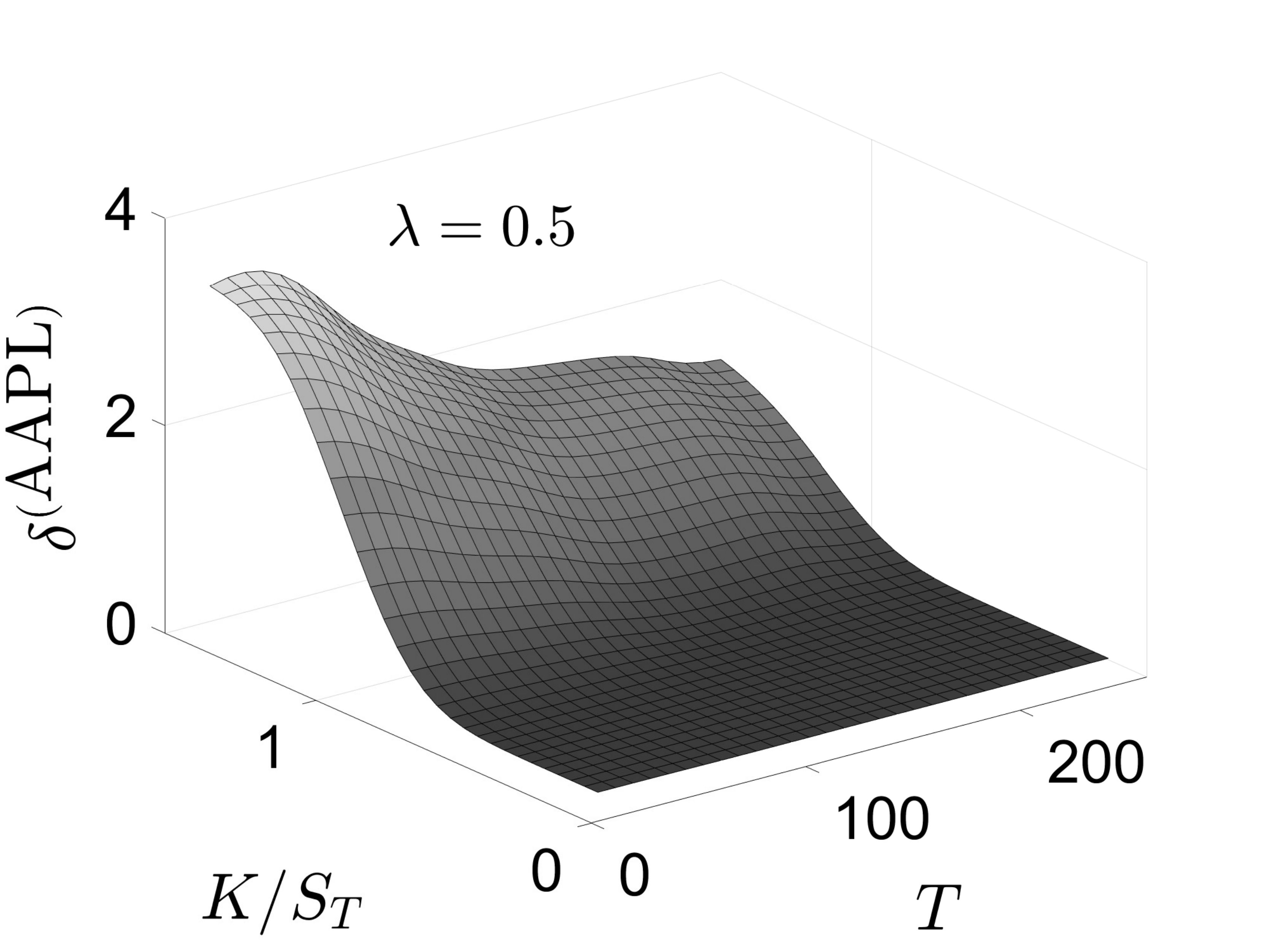}
    \end{subfigure}

    \begin{subfigure}[b]{0.32\textwidth} 
    	\includegraphics[width=\textwidth]{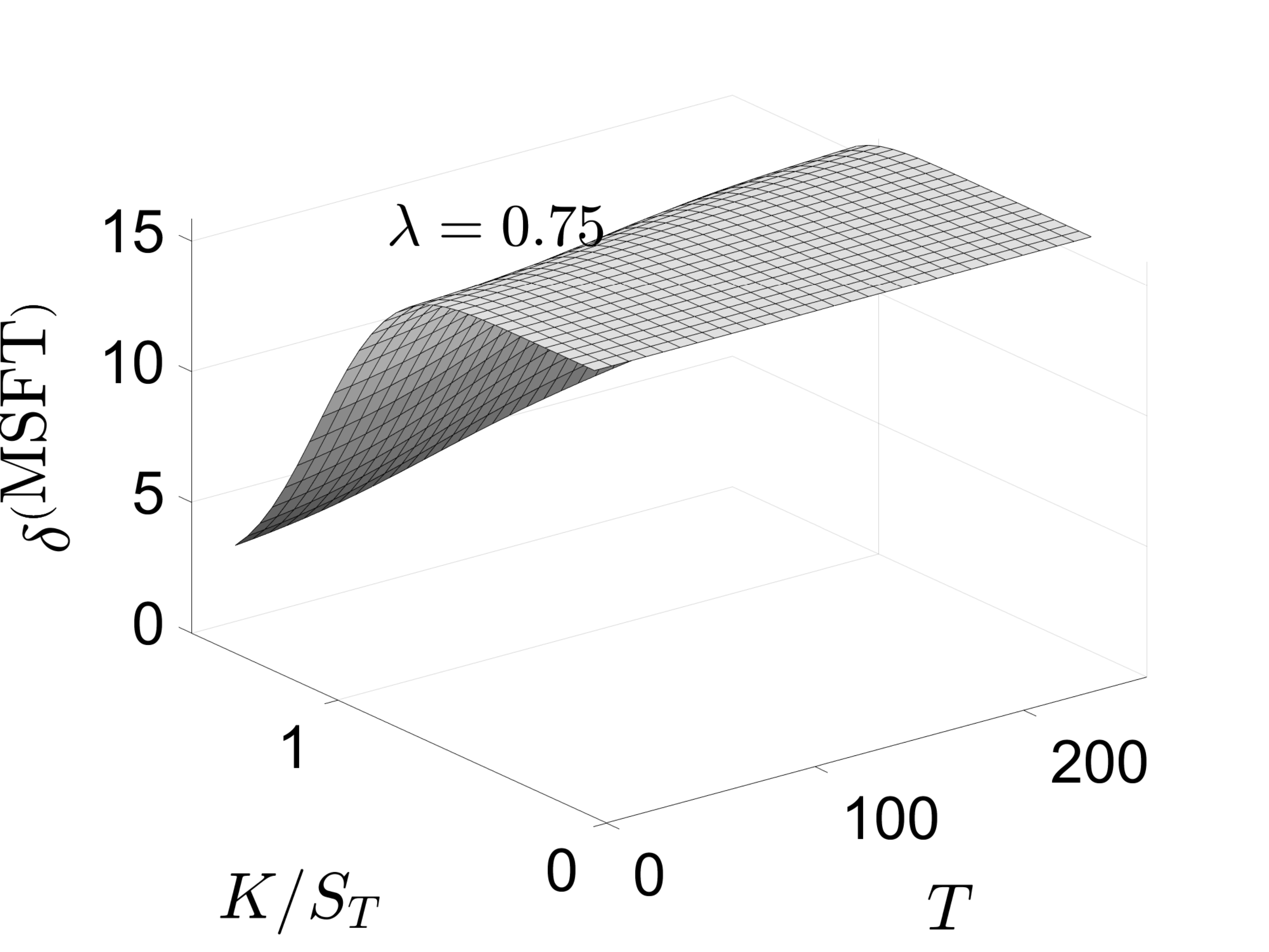}
    \end{subfigure}
    \begin{subfigure}[b]{0.32\textwidth} 
    	\includegraphics[width=\textwidth]{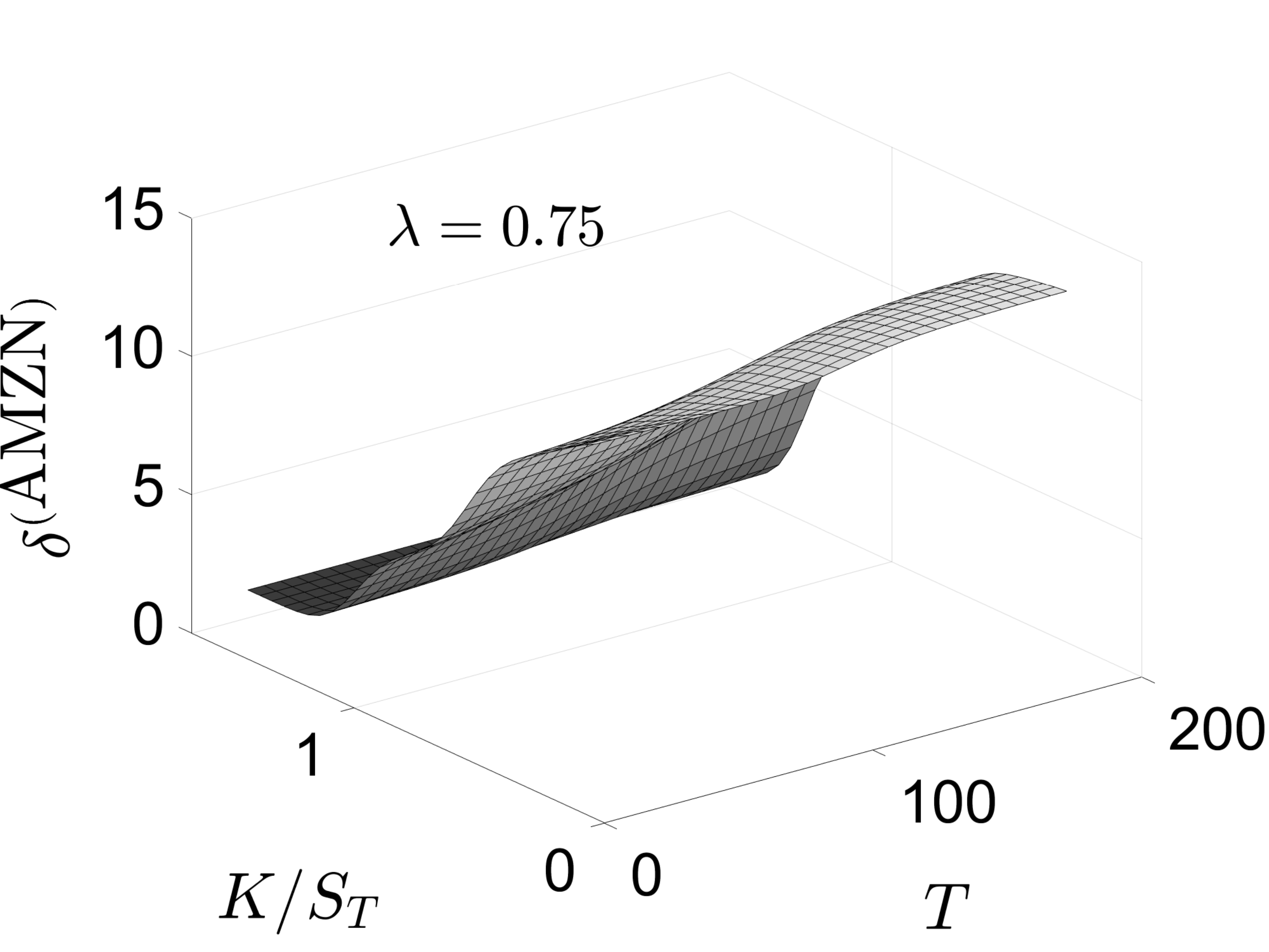}
    \end{subfigure}
    \begin{subfigure}[b]{0.32\textwidth} 
    	\includegraphics[width=\textwidth]{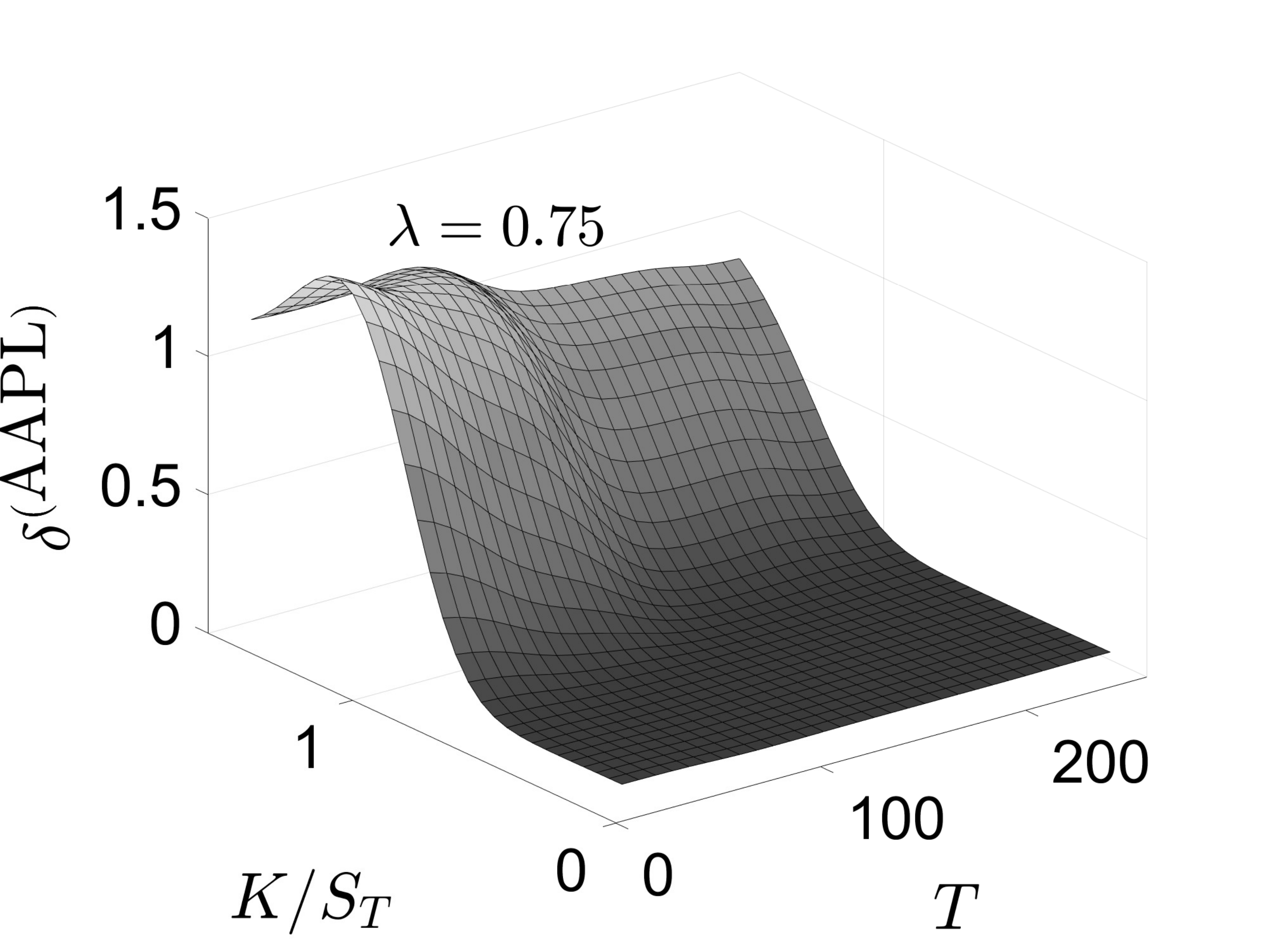}
    \end{subfigure}

    \caption{The implied $\delta^{(\text{stock},\aleph)} (T,K,\lambda)$ surfaces computed for
		the arithmetic return model \eqref{eq:daleph_binomial}, \eqref{eq:daleph_UD}.
    	For clarity, maturity times $T$ are presented in days.}

    \label{fig:dimpl_delta}
\end{center}
\end{figure}

 \begin{figure}[h!]
\begin{center}   
    \begin{subfigure}[b]{0.32\textwidth} 
    	\includegraphics[width=\textwidth]{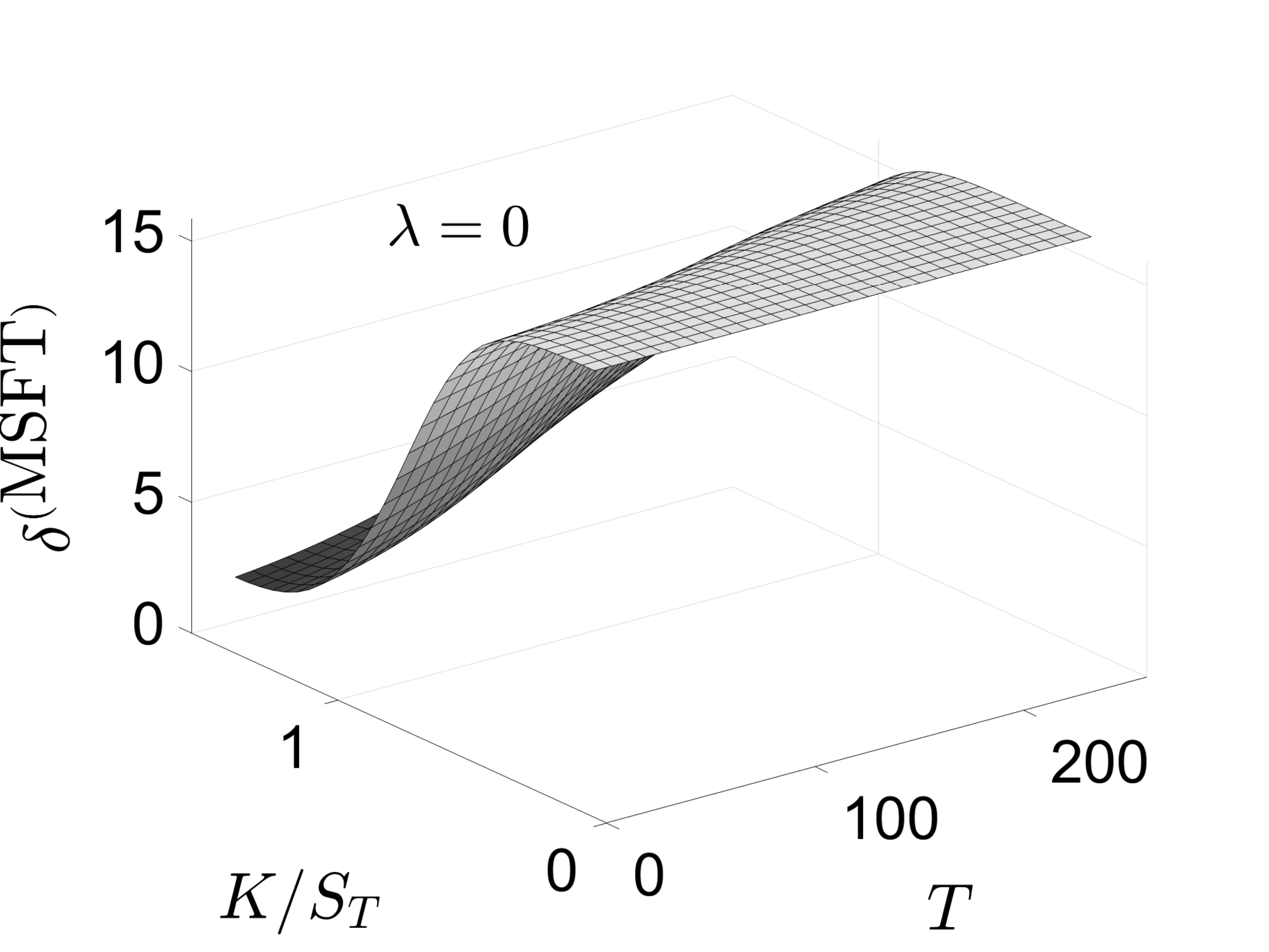}
    \end{subfigure}
    \begin{subfigure}[b]{0.32\textwidth} 
    	\includegraphics[width=\textwidth]{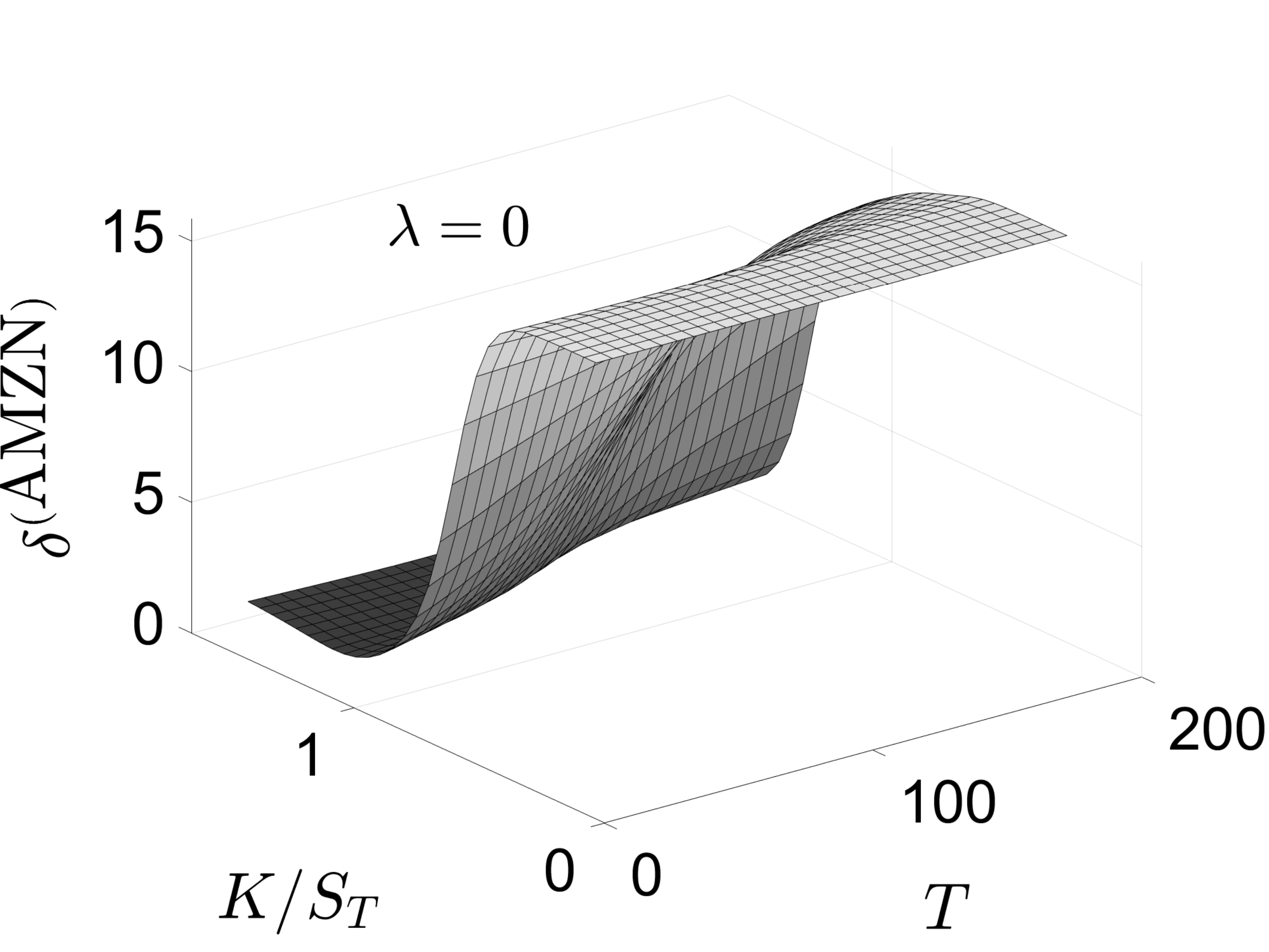}
    \end{subfigure}
    \begin{subfigure}[b]{0.32\textwidth} 
    	\includegraphics[width=\textwidth]{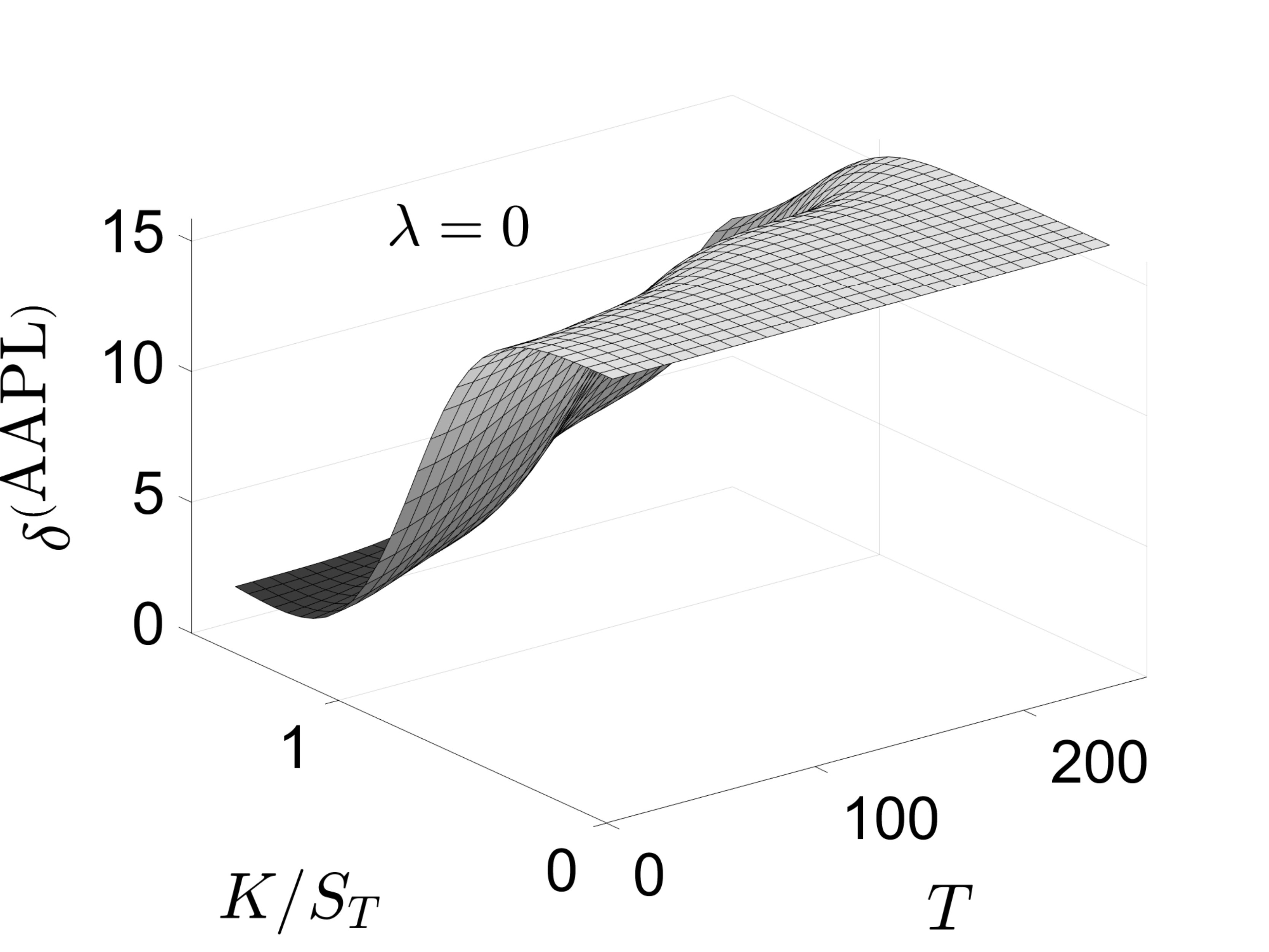}
    \end{subfigure}

    \begin{subfigure}[b]{0.32\textwidth} 
    	\includegraphics[width=\textwidth]{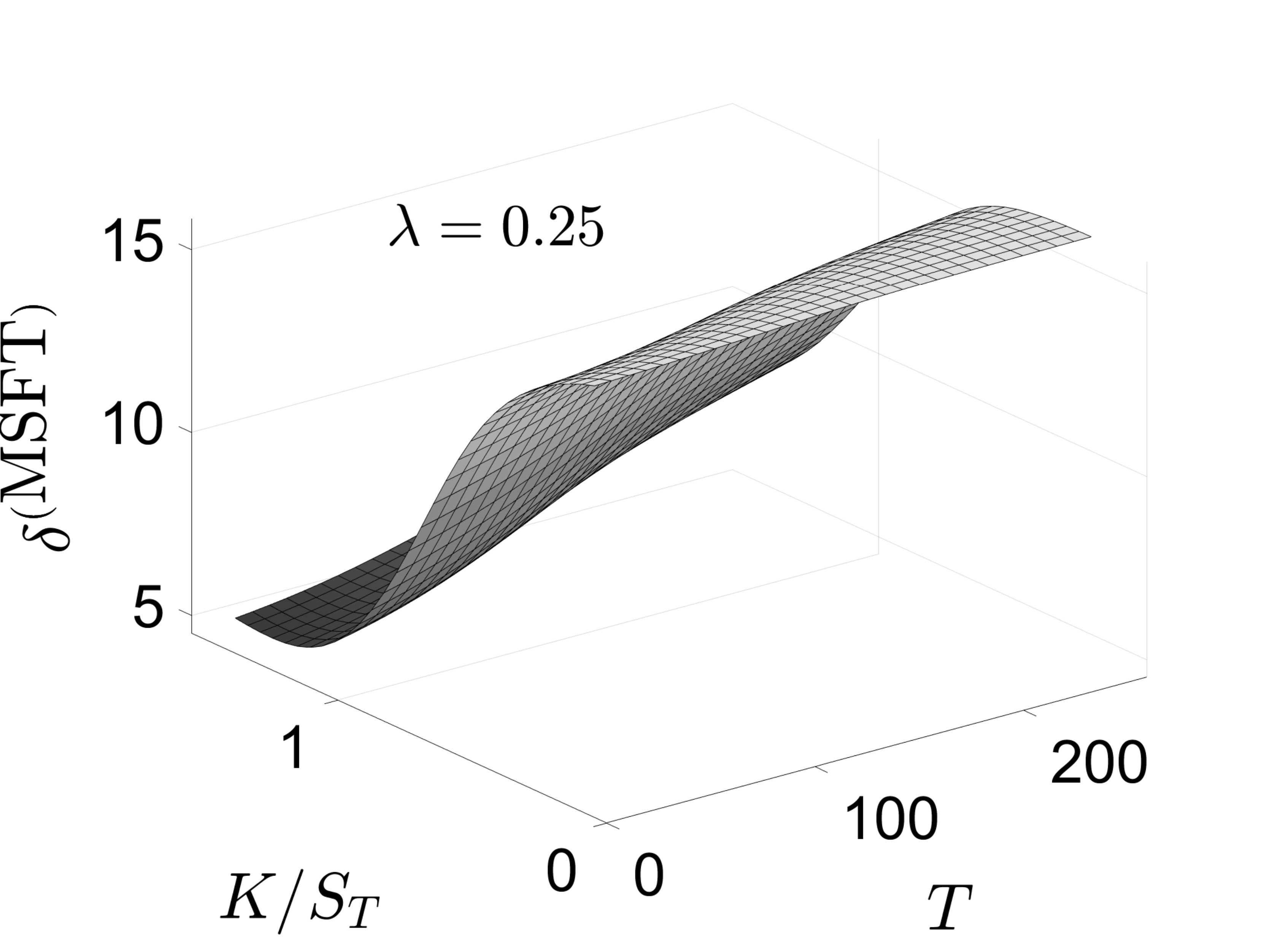}
    \end{subfigure}
    \begin{subfigure}[b]{0.32\textwidth} 
    	\includegraphics[width=\textwidth]{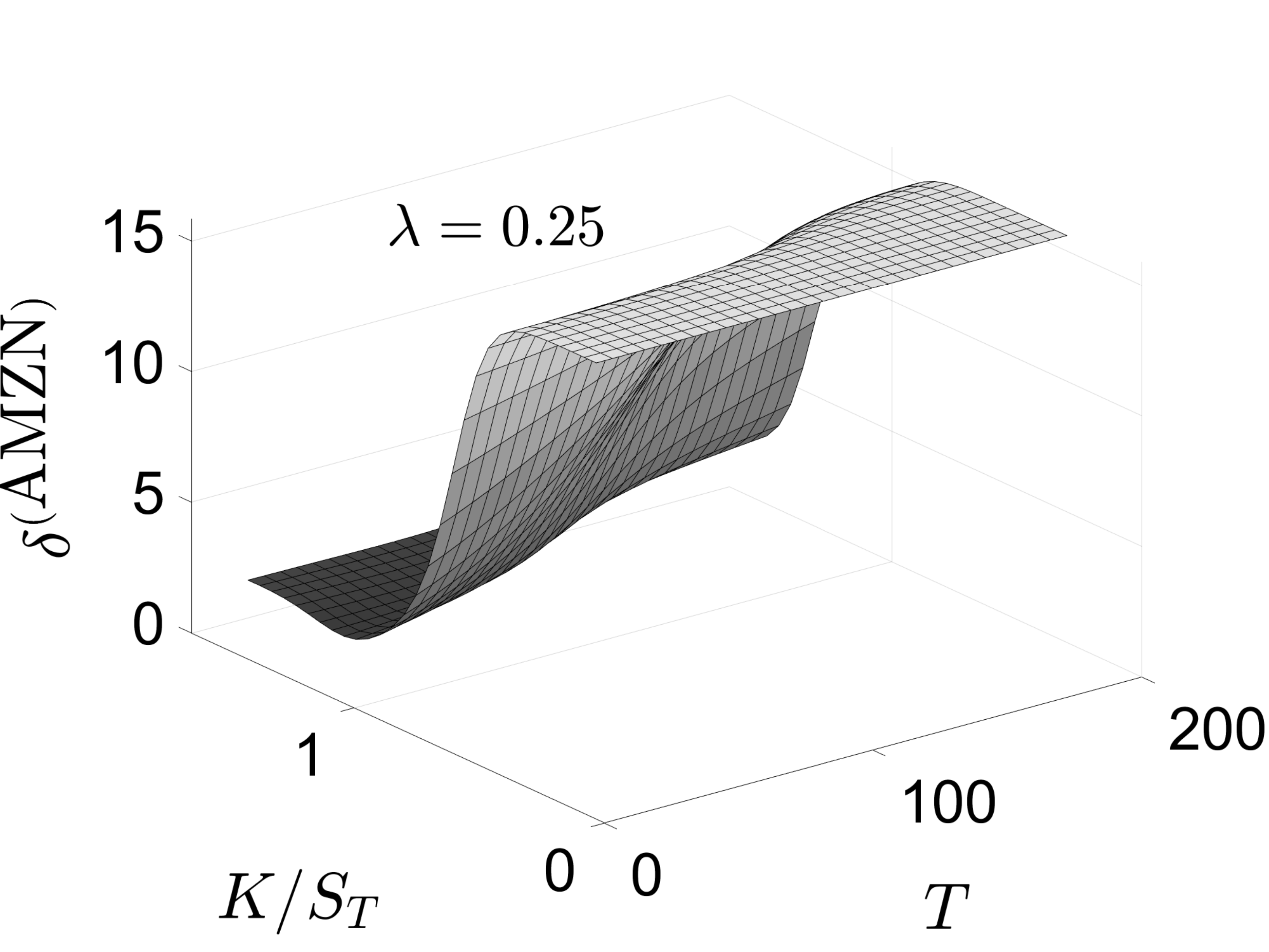}
    \end{subfigure}
    \begin{subfigure}[b]{0.32\textwidth} 
    	\includegraphics[width=\textwidth]{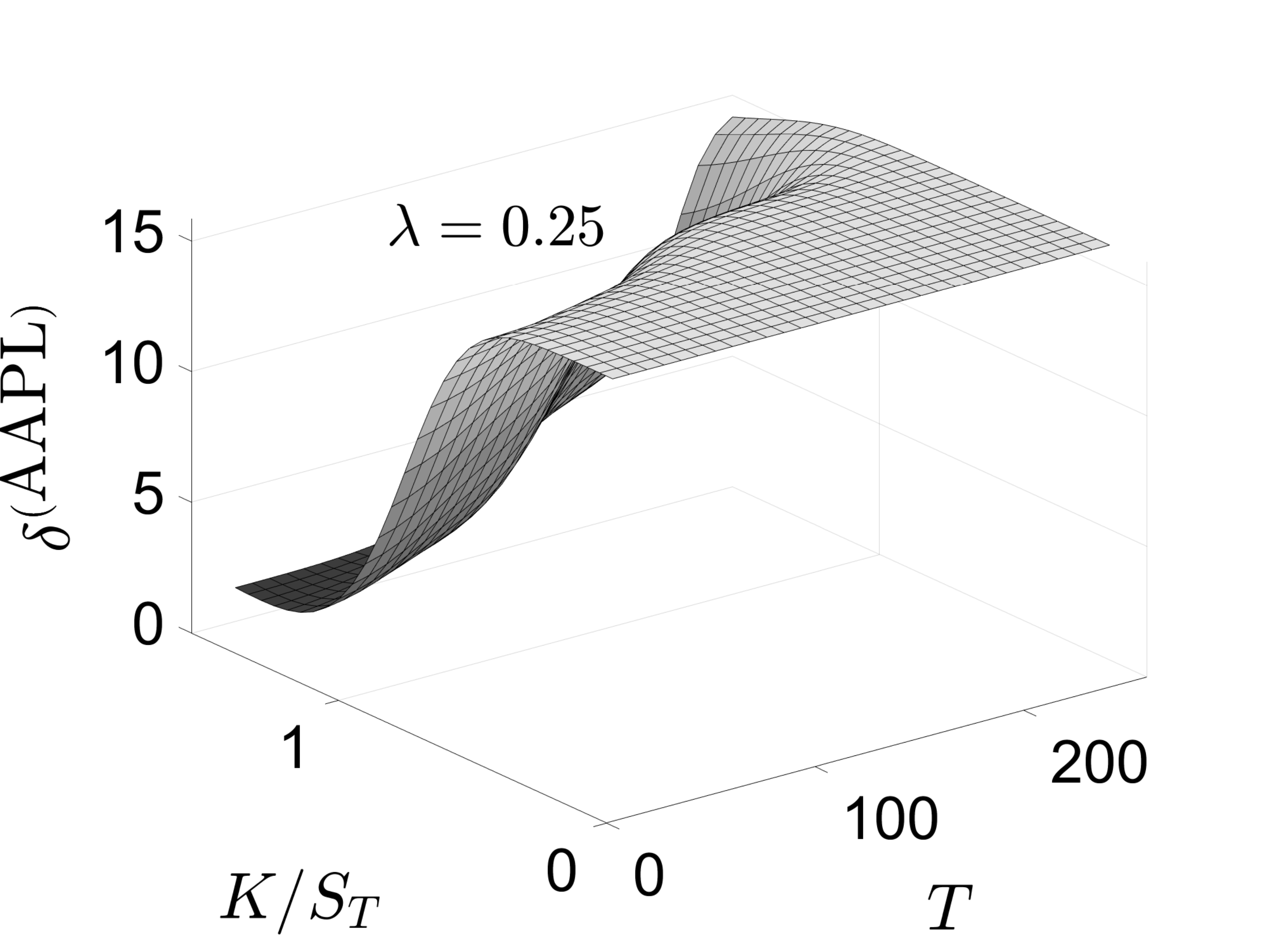}
    \end{subfigure}

    \begin{subfigure}[b]{0.32\textwidth} 
    	\includegraphics[width=\textwidth]{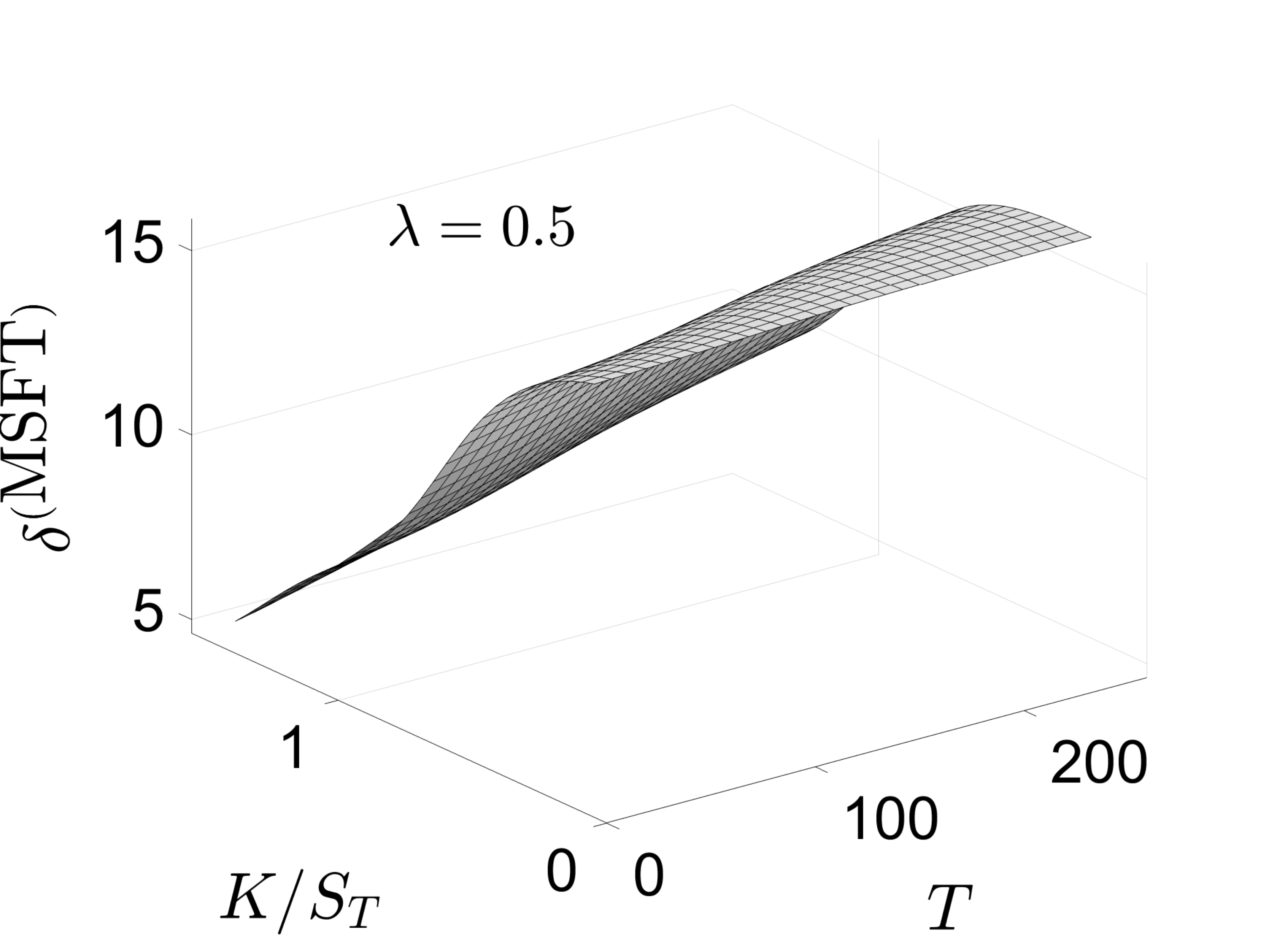}
    \end{subfigure}
    \begin{subfigure}[b]{0.32\textwidth} 
    	\includegraphics[width=\textwidth]{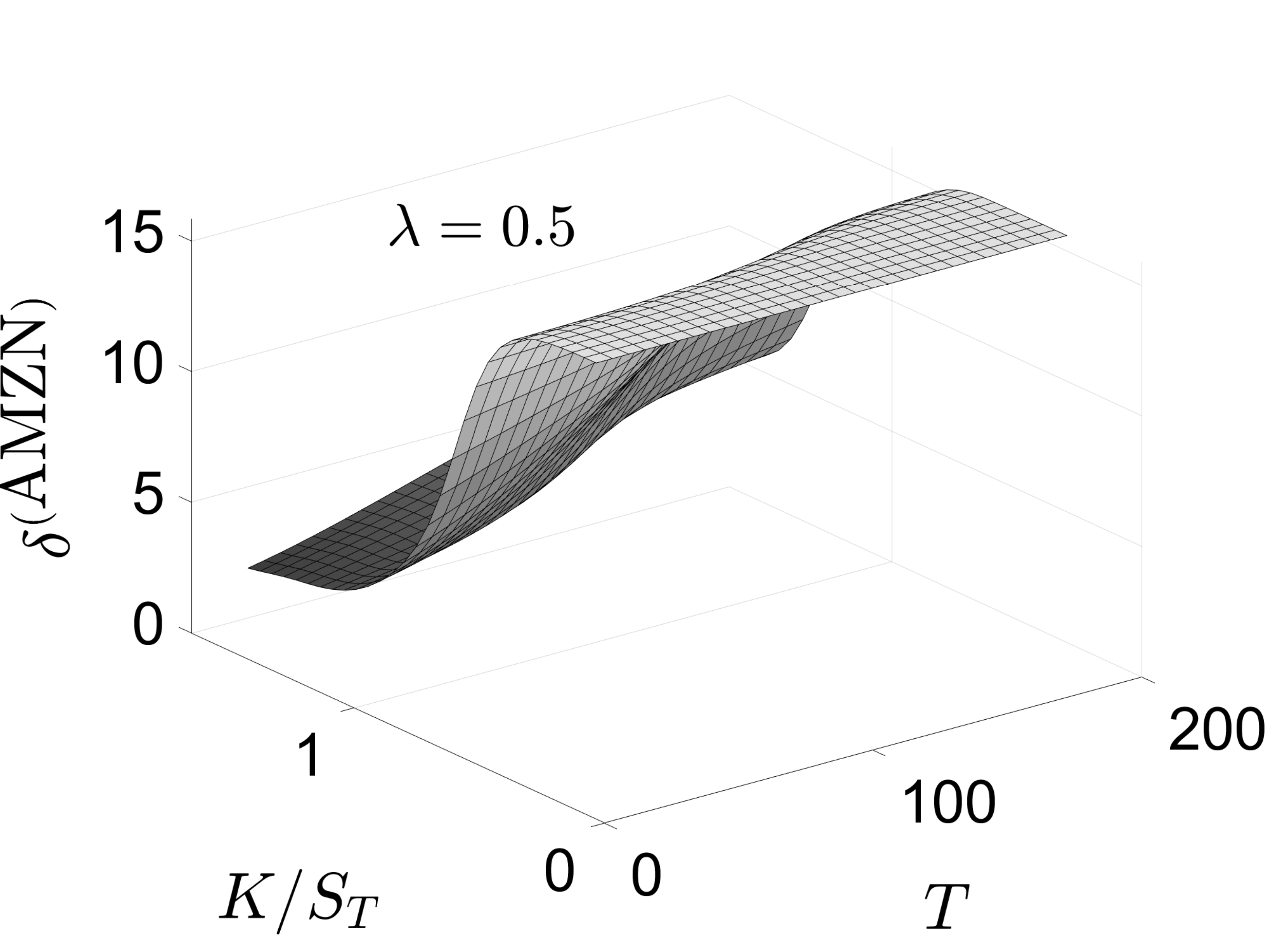}
    \end{subfigure}
    \begin{subfigure}[b]{0.32\textwidth} 
    	\includegraphics[width=\textwidth]{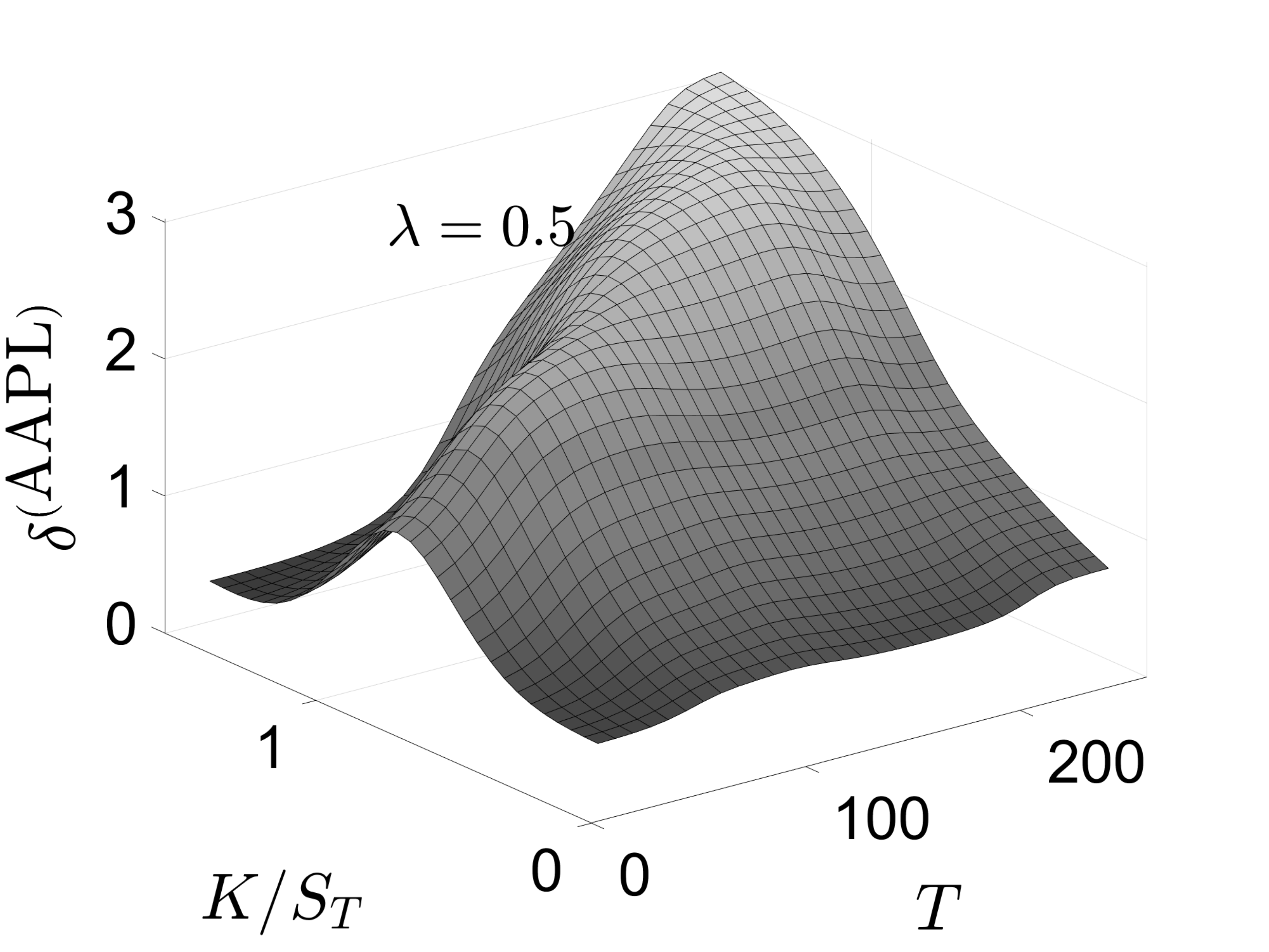}
    \end{subfigure}

    \begin{subfigure}[b]{0.32\textwidth} 
    	\includegraphics[width=\textwidth]{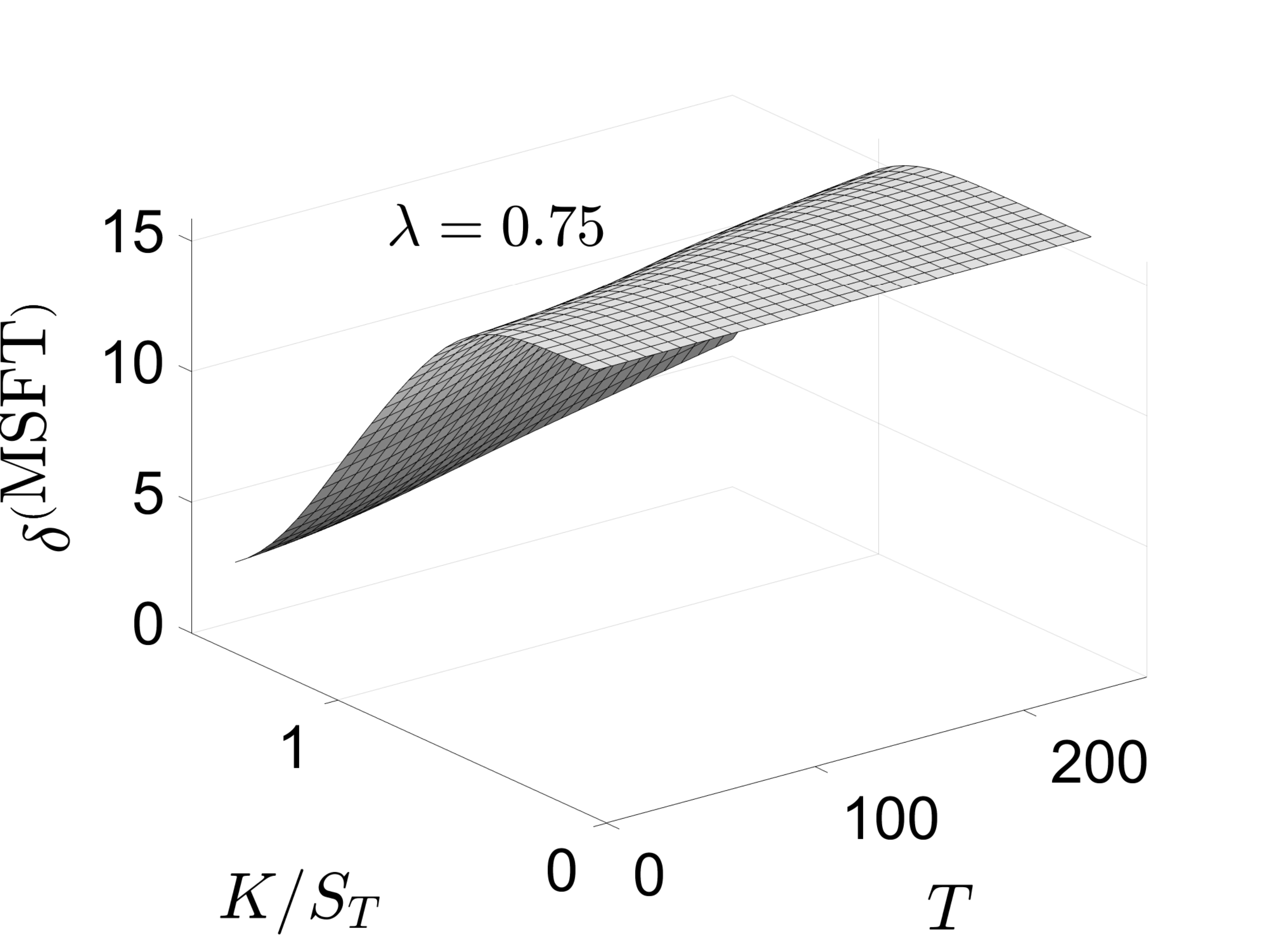}
    \end{subfigure}
    \begin{subfigure}[b]{0.32\textwidth} 
    	\includegraphics[width=\textwidth]{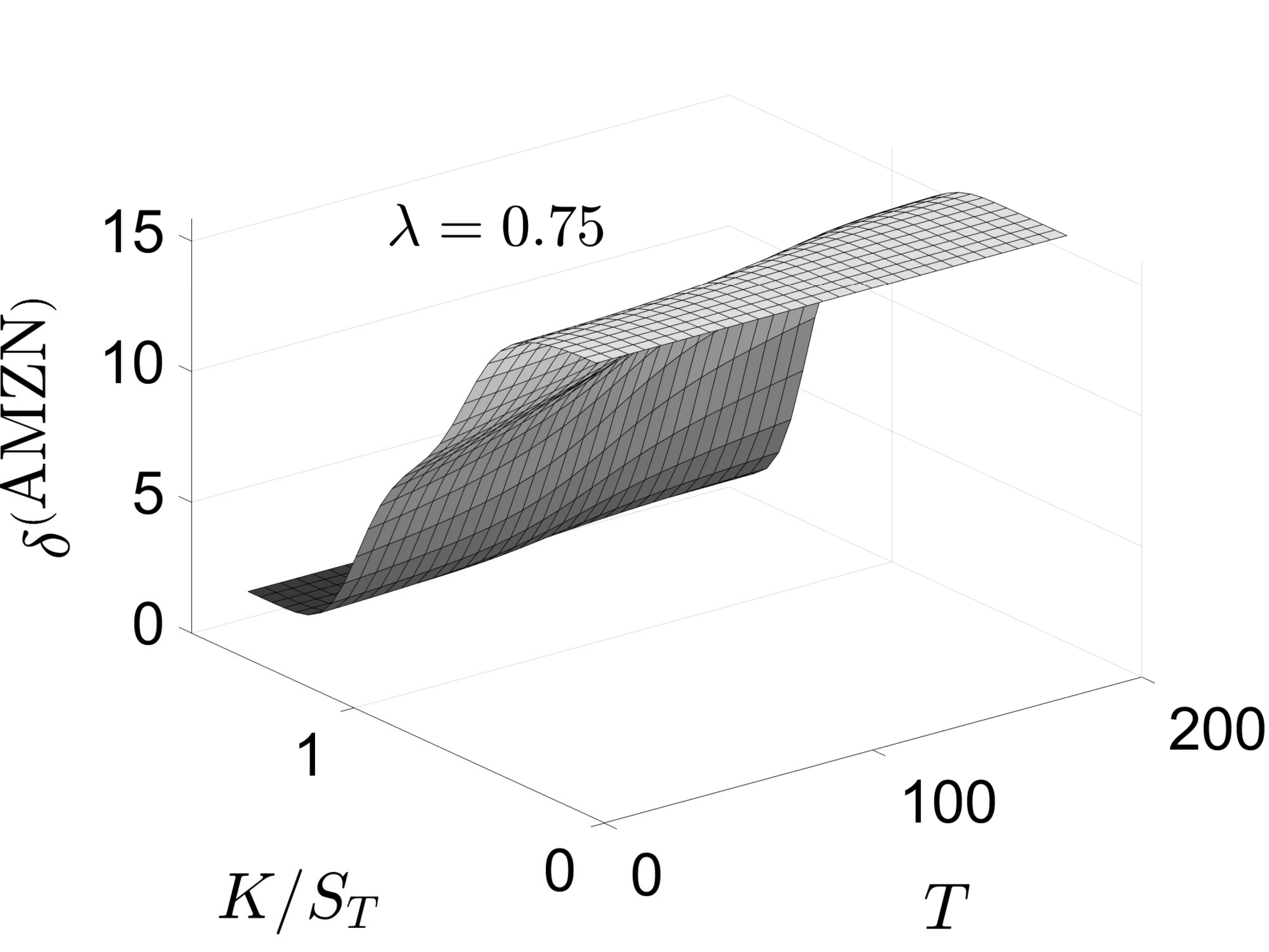}
    \end{subfigure}
    \begin{subfigure}[b]{0.32\textwidth} 
    	\includegraphics[width=\textwidth]{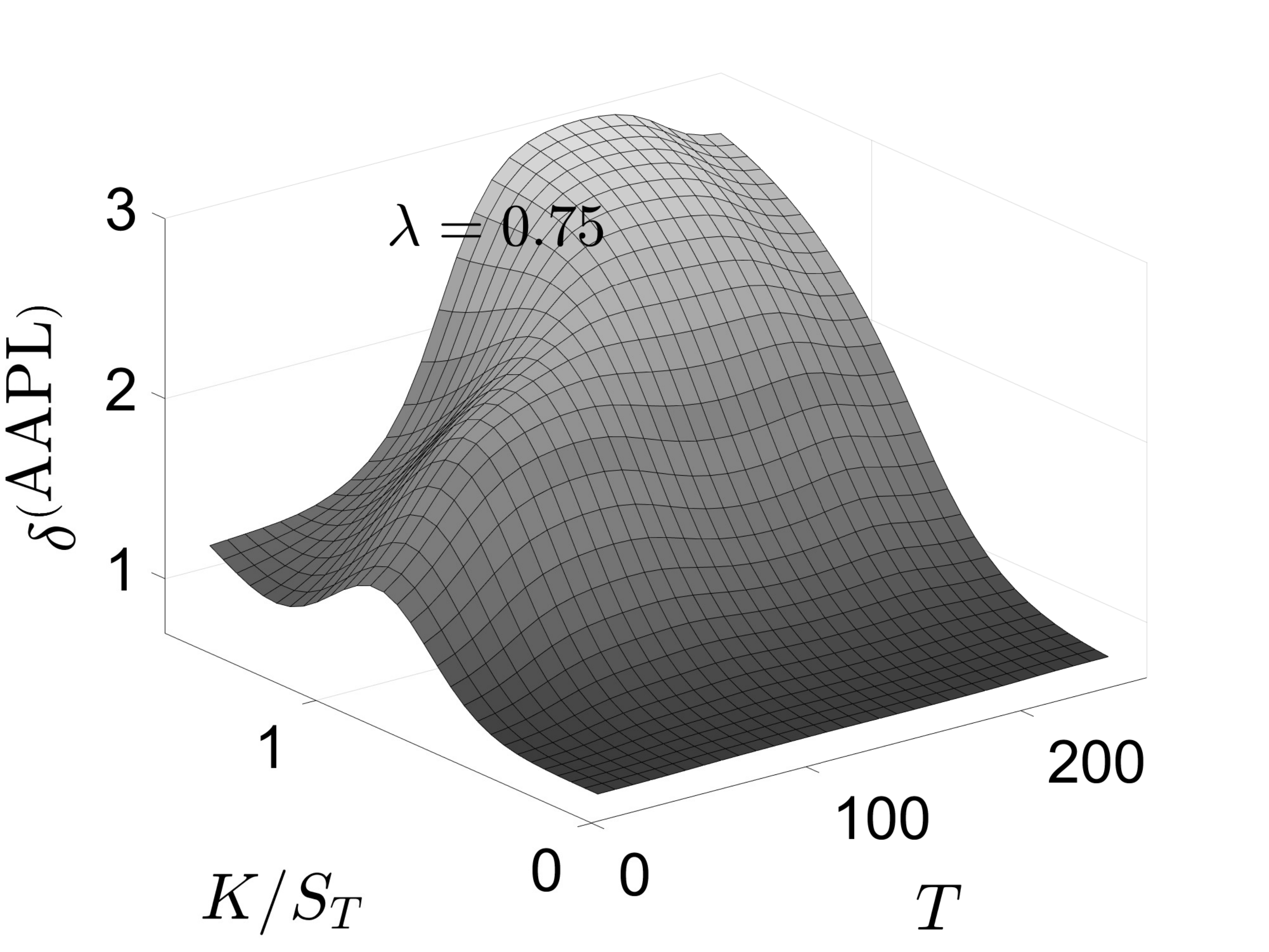}
    \end{subfigure}

    \caption{The implied $\delta^{(\text{stock},\aleph)} (T,K,\lambda)$ surfaces computed for the
		log-return model \eqref{eq:laleph_binomial}, \eqref{eq:laleph_UD}, \eqref{eq:laleph_equiv}.
    	For clarity, maturity times $T$ are presented in days.}

    \label{fig:limpl_delta}
\end{center}
\end{figure}

\end{appendices}

\clearpage

\section*{References}
\noindent
A. Amel-Zadeh \& G. Serafeim (2018)
Why and how investors use {ESG} information: {E}vidence from a global survey,
{\it Financial Analysts Journal}, {\bf 74}(3), 87--103.\\[3pt]
R. B{\'e}nabou \& J. Tirole (2010)
Individual and corporate social responsibility,
{\it Economica}, {\bf 77}(305), 1--19.\\[3pt]
F. Berg, J.F. Koelbel \& R. Rigobon (2019)
{\it Aggregate Confusion: The Divergence of {ESG} Ratings},
Cambridge, MA: MIT Sloan School of Management.\\[3pt]
T.C. Berry \& J.C. Junkus (2013)
Socially responsibleiInvesting: {A}n investor perspective,
{\it Journal of Business Ethics}, {\bf 112}(4), 707--720.\\[3pt]
F. Black \& M. Scholes (1973)
The pricing of options and corporate liabilities, 
{\it Journal of Political Economy} {\bf 81}, 637--654.\\[3pt]
R. Boffo \& R. Patalano (2020)
{ESG} Investing: Practices, Progress and Challenges.
Organization for Economic Cooperation and Development, Paris, France.
\url{www.oecd.org/finance/{ESG}-Investing-Practices-Progress-and-Challenges.pdf} \\[3pt]
A. S. Cherny, A. N. Shiryaev, \& M. Yor (2003)
Limit behavior of the “horizontal-vertical” random walk and some extensions of the
Donsker–Prokhorov invariance principle”, 
{\it Theory of Probability and its Applications}, {\bf 47}(3) (2003), 377--394.\\[3pt]
J. Cox, S. Ross \& M. Rubinstein (1979)
Options pricing:  A simplified approach, 
{\it Journal of Financial Economics} {\bf 7}, 229--263. \\[3pt]
D. Daugaard (2020)
Emerging new themes in environmental, social and governance investing: a systematic literature review,
{\it Accounting \& Finance}{\bf 60}(2), 1501--1530.\\[3pt]
Y. Davydov \& V. Rotar (2008)
On a non-classical invariance principle,
{\it Statistics \& Probability Letters} {\bf 78}, 2031--2038.\\[3pt]
F. Delbaen \& W. Schachmayer (1994)
A general version of the fundamental theorem of asset pricing,
{\it Mathematische Annalen} {\bf 300}, 463--520.\\[3pt]
D. Duffie (2001)
{\it Dynamic Asset Pricing Theory}, Princeton: Princeton University Press.\\[3pt]
S.M. Hartzmark \& A.B. Sussman (2019)
Do investors value sustainability? {A} natural experiment examining ranking and fund flows,
{\it The Journal of Finance}, {\bf 74}(16), 2789--2837.\\[3pt]
Y. Hu, A. Shirvani, Y.S. Kim, F. Fabozzi, \& S. Rachev (2020a)
Option pricing in markets with informed traders, 
{\it International Journal of Theoretical and Applied Finance} {\bf 23} (6), 2050037.\\[3pt]
Y. Hu, A. Shirvani, W.B. Lindquist, F. Fabozzi, \& S. Rachev (2020b)
Option pricing incorporating factor dynamics in complete markets, 
{\it Journal of Risk and Financial Management} {\bf 13} (12), 321.\\[3pt]
J. Hull (2012)
{\it Options, Futures, and Other Derivatives}, Eighth Edition, Boston: Pearson.\\[3pt]
J. Jacod \& Y. Ait-Sahalia (2014)
{\it High Frequency Financial Econometrics},
 Princeton, NJ: Princeton University Press.\\[3pt]
R.A. Jarrow \& A. Rudd (1983)
{\it Option Pricing}, Irwin, Homewood, IL, Dow Jones-Irwin Publishing.\\[3pt]
R.A. Jarrow, P. Protter \& H. Sayit (2009)
No arbitrage without semimartingales,
{\it The Annals of Applied Probability} {\bf 19}(2), 596--616.\\[3pt]
Y.S. Kim, S.V. Stoyanov, S.T. Rachev \& F.J. Fabozzi (2016)
Multi-purpose binomial model: {F}itting all moments to the underlying {B}rownian motion, 
{\it Economics Letters} {\bf 145}, 225--229. \\[3pt]
Y.S. Kim, S.V. Stoyanov, S.T. Rachev \& F.J. Fabozzi (2019)
Enhancing binomial and trinomial option pricing models, 
{\it Finance Research Letters} {\bf 28}, 185--190. \\[3pt]
P. Kr{\"u}ger (2015)
Corporate goodness and shareholder wealth
{\it Journal of Financial Economics}, {\bf 115}(2), 304--329.\\[3pt]
H. Liang \& L. Renneboog (2017)
On the foundations of corporate social responsibility,
{\it The Journal of Finance}, {\bf 72}(2), 853--910.\\[3pt]
K.V. Lins, H. Servaes, Henri \& A. Tamayo (2017)
Social capital, trust, and firm performance: {T}he value of corporate social responsibility during the financial crisis,
{\it The Journal of Finance}, {\bf 72}(4), 1785--1824.\\[3pt]
R.C. Merton (1973) Theory of rational option pricing, 
{\it Bell Journal of Economics and Management Science} {\bf 4}, 141--183. \\[3pt]
S.S. Shreve (2004)
{\it Stochastic Calculus for Finance}, Vol. 1, New York: Springer.\\[3pt]
A.V. Skorokhod (2005)
{\it Basic Principles and Applications of Probability Theory}, Heidelberg: Springer. \\[3pt]
L.T. Starks, P. Venkat \& Q. Zhu (2017)
Corporate {ESG} profiles and investor horizons.
Available at {\it SSRN}: https://ssrn.com/abstract=3049943.

\end{document}